\documentclass[12pt]{article}
\usepackage{enumerate}
\usepackage{url}

\usepackage[font={small}]{caption}
\DeclareMathSizes{10}{9}{6}{5}
\RequirePackage{amsthm,amsmath,amsfonts,amssymb}
\usepackage[authoryear, round, sort&compress]{natbib}
\usepackage{xcolor}

\usepackage[pdftex,bookmarks,colorlinks,breaklinks]{hyperref}
\hypersetup{linkcolor=blue,citecolor=blue,filecolor=blue,urlcolor=blue}
\usepackage{tabularx}
\usepackage{multirow}
\usepackage{mathtools}
\usepackage{bbm}
\usepackage{graphicx}
\usepackage{bm}
\usepackage{longtable}
\usepackage{colortbl}
\usepackage{booktabs}
\usepackage{threeparttable}
\usepackage{float}
\usepackage{adjustbox}
\usepackage{etoolbox}
\AtBeginEnvironment{table}{\def\baselinestretch{1}\normalsize}
\newcommand{\tablefont}{\def\baselinestretch{1}\footnotesize}

\graphicspath{{./}{./art/}}

\def\bx{\mathbf{x}}
\def\bX{\mathbf{X}}

\def\bY{\mathbf{Y}}

\def\bZ{\mathbf{Z}}
\def\bu{\mathbf{u}}
\def\bU{\mathbf{U}}
\def\bA{\mathbf{A}}
\def\bB{\mathbf{B}}
\def\bv{\mathbf{v}}
\def\bc{\mathbf{c}}
\def\bV{\mathbf{V}}

\def\bN{\mathbf{N}}
\def\bL{\mathbf{L}}
\def\bK{\mathbf{K}}
\def\bP{\mathbf{P}}

\def\bI{\mathbf{I}}
\def\bM{\mathbf{M}}

\def\br{\mathbf{r}}

\def\balpha{\boldsymbol{\alpha}}
\def\bvarepsilon{\boldsymbol{\varepsilon}}

\def\bbeta{\boldsymbol{\beta}}
\def\btheta{\boldsymbol{\theta}}

\def\bSigma{\boldsymbol{\Sigma}}

\def\bbetahat{\widehat{\bbeta}}

\def\bzero{\boldsymbol{0}}

\newcommand\ind{\protect\mathpalette{\protect\independenT}{\perp}}
\def\independenT#1#2{\mathrel{\rlap{$#1#2$}\mkern4mu{#1#2}}}

\def\bzero{\mathbf{0}}

\def\bT{\mathbf{T}}

\def\bR{\mathbf{R}}

\def\bS{\mathbf{S}}

\def\bG{\mathbf{G}}

\def\bDelta{\boldsymbol{\Delta}}

\def\bH{\mathbf{H}}
\def\bC{\mathbf{C}}

\def\bW{\mathbf{W}}

\def\bgamma{\boldsymbol{\gamma}}
\def\bdelta{\boldsymbol{\delta}}

\def\bmu{\boldsymbol{\mu}}

\def\bD{\mathbf{D}}

\def\bSigma{\boldsymbol{\Sigma}}

\def \hs2{\hspace{2mm}}

\numberwithin{equation}{section}

\def\boxit#1{\vbox{\hrule\hbox{\vrule\kern6pt  \vbox{\kern6pt#1\kern6pt}\kern6pt\vrule}\hrule}}

\def\bT{\mathbf{T}}
\def\bS{\mathbf{S}}
\def\bG{\mathbf{G}}

\def\bT{\mathbf{T}}

\def\bR{\mathbf{R}}

\def\bOmega{\boldsymbol{\Omega}}

\def\bPhi{\boldsymbol{\Phi}}

\theoremstyle{plain}
\newtheorem{theorem}{Theorem}
\newtheorem{lemma}{Lemma}

\theoremstyle{remark}
\newtheorem{assumption}{Assumption}
\newtheorem{remark}{Remark}

\theoremstyle{definition}
\newtheorem{algo}{Algorithm}

\begin{document}

\def\spacingset#1{\renewcommand{\baselinestretch}%
{#1}\small\normalsize} \spacingset{1}

\date{}
\title{\bf The roles of pre- and post-outcome surrogates in semi-supervised linear regression}
\author{Kai Chen \hspace{.2cm}\\
Institute of Statistics and Big Data, Renmin University of China\\
\href{mailto:kaichen@ruc.edu.cn}{kaichen@ruc.edu.cn}\\
and \\
Zhanhao Yan \hspace{.2cm}\\
Department of Statistics and Operations Research\\
University of North Carolina at Chapel Hill\\
\href{mailto:zhanhaoy@unc.edu}{zhanhaoy@unc.edu}\\
and \\
Yuqian Zhang\thanks{The authors are listed in alphabetical order. The corresponding author is Yuqian Zhang (\texttt{yuqianzhang@ruc.edu.cn}).} \hspace{.2cm}\\
Institute of Statistics and Big Data, Renmin University of China\\
\href{mailto:yuqianzhang@ruc.edu.cn}{yuqianzhang@ruc.edu.cn}
}
\maketitle
\vspace{-2em}

\bigskip
\begin{abstract}
When outcomes are difficult to observe, surrogate variables can provide valuable additional information. Existing surrogate-assisted methods have been developed for both prediction and inference, but have not fully investigated whether and how the temporal roles of surrogates can be exploited to improve the use of additional unlabeled data. We study semi-supervised linear regression with additional surrogate variables, distinguishing between pre-outcome and post-outcome surrogates according to their temporal ordering relative to the outcome. We investigate how these two types of surrogate information can be incorporated and how surrogate misclassification---that is, treating pre-outcome surrogates as post-outcome surrogates or vice versa---as well as surrogate omission affects estimation bias and variance. We consider two target parameters, a direct effect parameter and a prediction parameter, and show that the impact of surrogate classification depends critically on the target of interest. We further propose an optimal weighting strategy to ensure safe use of unlabeled samples under model misspecification. The performance of the proposed methods is demonstrated through simulations and a real-data analysis using the MIMIC-III database.
\end{abstract}

\noindent%
\vfill

\newpage
\spacingset{1.9}

\section{Introduction}\label{sec:introduction}

In many applications in biomedical research and machine learning, the primary outcome of interest, denoted by $Y$, is costly or difficult to obtain and is therefore observed only for a limited subset of samples. For example, labels in image and natural language processing tasks often require labor-intensive manual annotation, while in biomedical studies, long-term clinical outcomes may become available only after extended follow-up. These settings naturally give rise to semi-supervised data, in which a relatively small labeled sample is accompanied by a much larger unlabeled sample. Even when the outcome itself is difficult to observe, related auxiliary measurements, records, or short-term indicators may still be readily available. Such variables are commonly referred to as surrogate variables, denoted by $\bS$, and can serve as observable proxies for the target outcome.

A growing body of literature has investigated the use of surrogate variables in semi-supervised problems. However, the precise scope of what constitutes a surrogate variable remains somewhat ambiguous, as different studies may use the same term to refer to different types of auxiliary variables. To clarify this distinction, we organize existing uses of surrogate variables according to their generation order relative to the outcome.

\emph{Pre-outcome surrogates.} Pre-outcome surrogates are variables observed or generated before the outcome $Y$. This type of surrogate commonly arises in settings where the primary outcome is long-term or costly to observe, while earlier measurements are more readily available. Examples include short-term clinical measurements collected before a long-term endpoint, early biomarkers observed prior to later disease severity, and intermediate responses measured before a final outcome \citep{wang2020model,zhang2022prior,agniel2023doubly,zeng2024continuous,athey2025surrogate,kallus2025role}. Because these variables are available before $Y$ is realized, they can provide useful information about the outcome-generating process and help predict the outcome for unlabeled samples.

\emph{Post-outcome surrogates.} Post-outcome surrogates are variables observed or generated after the outcome $Y$. This setting arises, for example, when $Y$ represents an underlying disease status that requires additional effort to ascertain, while $\bS$ consists of subsequent records related to that status, such as diagnostic codes, physician notes, or other proxy measurements \citep{hou2025efficient,wu2025robust}. In such cases, post-outcome surrogates may also contain useful information about $Y$ and help recover individual outcome realizations. A related strand of the literature treats $\bS$ as a noisy measurement of $Y$ \citep{gronsbell2019automated,hong2019semi,liu2023surrogate}, often under the conditional independence assumption $\bS\perp \bX\mid Y$, where $\bX$ denotes the baseline covariates. However, this assumption can be restrictive in practice, as it essentially requires that the effect of $\bX$ on $\bS$ operate entirely through the outcome $Y$.

Both pre- and post-outcome surrogates contain partial information about the outcome, but they arise from fundamentally different data-generating mechanisms. The existing literature generally does not explicitly emphasize the distinction between these two types of surrogate variables. Although the studies discussed above use the term ``surrogate variables'' to describe auxiliary variables that contain additional information about the outcome, each line of work typically focuses on only one type of surrogate, either pre- or post-outcome. Because their data-generating mechanisms differ, the strategies for extracting and utilizing outcome information from pre- and post-outcome surrogates also differ. Consequently, even when both types of surrogates are simultaneously available, existing methods are generally not designed to fully exploit the auxiliary information contained in both types. Conversely, ignoring their generation order and directly applying a method developed for one type of surrogate to both types may violate the required identification conditions and thereby introduce systematic bias.

Another strand of the literature treats all surrogate variables as additional predictive features, without explicitly modeling their data-generating mechanisms or distinguishing between pre- and post-outcome surrogates \citep{hou2023surrogate,xia2024prediction,tian2024semi,ji2025predictions,cai2025semi}. This raises the question of whether knowledge of the surrogates' temporal roles can improve prediction. Beyond prediction, we also ask whether parameters of interest can still be estimated accurately and efficiently without such knowledge.

Motivated by these considerations, we study semi-supervised linear regression with surrogate variables whose temporal roles relative to the outcome may differ. We distinguish between pre-outcome and post-outcome surrogates and investigate how this distinction affects inference and prediction. Specifically, we ask whether additional surrogates can be used effectively without knowing their temporal roles and how incorrect assignment of these roles affects estimation. We show that the answers depend on the statistical target. For prediction targets, surrogates that provide additional information about the outcome can improve estimation efficiency even when their temporal roles are unknown. When the goal is to understand the data-generating mechanism and estimate direct effects under a causal interpretation, however, misclassifying pre-outcome surrogates as post-outcome surrogates, or vice versa, may introduce systematic estimation bias. Our results suggest that surrogate collection and use should be guided by the statistical target: for prediction, prioritize collecting informative surrogates even when their temporal roles are unknown; for studying the data-generating mechanism and estimating direct effects, account for temporal structure to avoid bias from incorrect surrogate adjustment.

\section{Semi-supervised linear regression with pre- and post-outcome surrogates}\label{sec:setting}

\begin{figure}[htbp] 
 \centering 
 \includegraphics[width=0.5\textwidth]{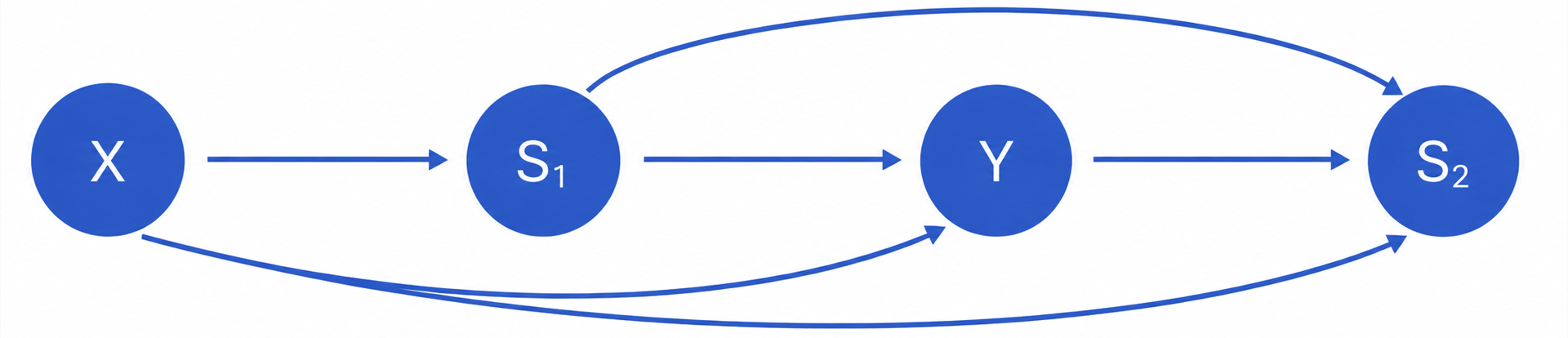}
 \caption{Temporal ordering of the covariates, surrogates, and outcome.}
 \label{fig:DAG} 
\end{figure}

For each subject, let $Y_i\in\mathbb{R}$ denote the outcome variable and $\bX_i\in\mathbb{R}^p$ the covariates. Let $\bS_i$ denote the collection of surrogate variables. According to their temporal ordering relative to $Y_i$, we partition $\bS_i$ into two components, $\bS_{1i}$ and $\bS_{2i}$, where $\bS_{1i}\in\mathbb{R}^{d_1}$ consists of surrogates observed prior to the outcome and $\bS_{2i}\in\mathbb{R}^{d_2}$ consists of surrogates observed after $Y_i$; see Figure~\ref{fig:DAG} for an illustration. Let $d=d_1+d_2$. We assume access to a semi-supervised dataset consisting of a labeled sample $\mathcal{L}=\{(Y_i,\bX_i,\bS_i): i\in \mathcal{I}\}$ and an unlabeled sample $\mathcal{U}=\{(\bX_i,\bS_i): i\in \mathcal{J}\}$, where $|\mathcal{I}|=n$, $|\mathcal{J}|=m$, and $N=n+m$. We are particularly interested in settings with $m\gg n$, but our asymptotic framework also allows $m=o(n)$, including fixed $m$. We assume that the labeled and unlabeled samples are drawn from the same underlying population, so that the outcome variable $Y_i$ is missing completely at random. This is a standard assumption in the semi-supervised inference literature \citep{zhang2019semi,tony2020semisupervised,zhang2022high,deng2023optimal,song2023general,hou2023surrogate}.

Denote $\bar{\bX}_{1i}=(\bX_i^\top,\bS_{1i}^\top)^\top\in \mathbb{R}^{p+d_1}$, $\bar{\bX}_{2i}=(\bar{\bX}_{1i}^\top,Y_i)^\top\in \mathbb{R}^{p+d_1+1}$. We consider the following linear models:
\begin{equation}\label{eq:models}
\begin{split}\bS_{1i}&=\balpha^{*\top}\bX_i+\bvarepsilon_{1i}\quad\text{with}\quad E(\bvarepsilon_{1i})=\mathbf{0}\quad\text{and}\quad E(\bX_i\bvarepsilon_{1i}^\top)=\mathbf{0},\\
Y_i&=\bbeta_1^{*\top}\bX_i+\bbeta_2^{*\top}\bS_{1i}+\varepsilon_{2i}\quad\text{with}\quad E(\varepsilon_{2i})=0\quad\text{and}\quad E(\bar{\bX}_{1i}\varepsilon_{2i})=\mathbf{0},\\
\bS_{2i}&=\bgamma_1^{*\top}\bX_i+\bgamma_2^{*\top}\bS_{1i}+\bgamma_3^* Y_i+\bvarepsilon_{3i}\quad\text{with}\quad E(\bvarepsilon_{3i})=\mathbf{0}\quad\text{and}\quad E(\bar{\bX}_{2i}\bvarepsilon_{3i}^\top)=\mathbf{0}.\end{split}
\end{equation}
We define $\bbeta^*=(\bbeta_1^{*\top},\bbeta_2^{*\top})^\top\in \mathbb{R}^{p+d_1}$, $\bgamma_{12}^*=(\bgamma_1^{*\top},\bgamma_2^{*\top})^\top \in \mathbb{R}^{(p+d_1)\times d_2}$, and $\bgamma^*=(\bgamma_{12}^{*\top},\bgamma_3^*)^\top\in \mathbb{R}^{(p+d_1+1)\times d_2}$. This formulation reflects the intended temporal order: pre-outcome surrogates depend on baseline covariates, the outcome depends on baseline covariates and pre-outcome surrogates, and post-outcome surrogates may additionally depend on the outcome itself.

We focus on the following parameters of interest.

(a) The linear coefficients $\balpha^*\in\mathbb{R}^{p\times d_1}$, $\bbeta_1^*\in\mathbb{R}^{p}$, $\bbeta_2^*\in\mathbb{R}^{d_1}$, $\bgamma_1^*\in\mathbb{R}^{p\times d_2}$, $\bgamma_2^*\in\mathbb{R}^{d_1\times d_2}$, and $\bgamma_3^*\in\mathbb{R}^{d_2}$ characterize the linear relationships in \eqref{eq:models}. Under a causal interpretation of the model, they represent the direct effects of earlier variables on later variables. Here, $\bS_{1i}$ and $\bS_{2i}$ serve as surrogates, so our primary interest lies in parameters describing the relationship between $\bX_i$ and $Y_i$. Other relationships can be studied by assigning the relevant variables to the covariate and response roles. Among these coefficients, we are particularly interested in $\bbeta_1^*$, which represents the direct effect of $\bX_i$ on $Y_i$ through pathways that do not operate through the surrogate variables, viewed here as potential mediators.

(b) The parameter $\btheta^*:=\bbeta_1^*+\balpha^*\bbeta_2^*$ is the coefficient in the following reduced-form $Y_i\sim\bX_i$ model:
\begin{align}\label{eq:Y-X}
Y_i=\btheta^{*\top}\bX_i+\bbeta_2^{*\top}\bvarepsilon_{1i}+\varepsilon_{2i}.
\end{align}
It defines the linear predictor $\btheta^{*\top}\bx$ for a new covariate vector $\bx$, enabling prediction based solely on baseline covariates. Estimating $\btheta^*$ therefore allows predictions to be made without requiring additional surrogate measurements for new individuals, thereby facilitating more timely and cost-effective prediction. Under a causal interpretation of the model, $\btheta^*$ also represents the total effect of $\bX_i$ on $Y_i$, combining the direct effect with the indirect effects mediated through the surrogates.

For both sets of target parameters, we first consider the ideal setting in which the temporal roles of the surrogates are correctly specified. We derive surrogate-assisted semi-supervised estimators and establish their asymptotic efficiency. We then examine the effects of surrogate misclassification and omission to clarify their roles and determine when knowledge of the true temporal structure is beneficial. Because misclassification or omission can change the interpretation of coefficients associated with the surrogates, our analysis of these settings focuses on $\bbeta_1^*$ and $\btheta^*$, both of which characterize the relationship between $\bX_i$ and $Y_i$. We show that surrogate misclassification or omission may introduce asymptotic bias in estimating the direct effect parameter $\bbeta_1^*$. For the prediction parameter $\btheta^*$, surrogate misclassification leaves the estimator unchanged, whereas surrogate omission may reduce its efficiency.

 We next introduce the necessary notation. For a matrix $\bA\in\mathbb{R}^{m\times n}$, ${\rm vec}(\bA)$ denotes the vectorization operator that stacks the columns of $\bA$ into an $(mn)$-dimensional vector, i.e., ${\rm vec}(\bA) = (A_{11}, A_{21}, \cdots, A_{m1}, A_{12},\cdots, A_{mn})^\top$. For matrices $\bA$ and $\bB$, $\bA\otimes \bB$ represents the Kronecker product. For a symmetric matrix $\bSigma\in\mathbb{R}^{d\times d}$, $\lambda_{\rm max}(\bSigma)$ and $\lambda_{\rm min}(\bSigma)$ denote its largest and smallest eigenvalues, ${\rm det}(\bSigma)$ denotes its determinant, and ${\rm tr}(\bSigma)$ denotes its trace. We denote by $\bI_d$ the $d\times d$ identity matrix, and by $\mathbf{0}$ a zero matrix or vector of contextually specified dimension. For two sequences of random variables ${a_n}$ and ${b_n}$, the notation $a_n = O_p(b_n)$ means that $a_n / b_n$ is bounded in probability, whereas $a_n = o_p(b_n)$ means that $a_n / b_n$ converges to zero in probability.
\section{Surrogate-Assisted EM Algorithm}\label{sec:1step_EM}

Under model \eqref{eq:models}, unlabeled observations contain the post-outcome surrogate $\bS_{2i}$ but not the outcome $Y_i$. Their observed-data likelihood contributions therefore involve integrating over the missing $Y_i$, leading to a coupled optimization problem for the parameters of the outcome and post-outcome surrogate models. The Expectation-Maximization (EM) algorithm \citep{balakrishnan2017statistical,wu1983convergence,wang2014high} provides a convenient framework for handling this missing-data structure. Instead of maximizing the observed-data likelihood directly, EM alternates between computing the conditional expectation of the complete-data log-likelihood and maximizing this expected log-likelihood. Specifically, the E-step computes the conditional first and second moments of the missing $Y_i$ given the observed data and the current parameter estimates, while the M-step updates the model parameters through regression-type subproblems.
 
We conduct our analysis under the following Gaussian assumption. Extensions to non-Gaussian settings are discussed in Section~\ref{sec:optimal_lambda}.
\begin{assumption}[Joint Gaussianity]\label{ass:joint_gaussian}
The random vector $(\bX_i^\top,\bS_{1i}^\top,Y_i,\bS_{2i}^\top)^\top$ follows a multivariate Gaussian distribution with a fixed positive definite covariance matrix.
\end{assumption}

Denote $\bSigma_1=\mathrm{var}(\bvarepsilon_{1i})$, $\sigma_2^2=\mathrm{var}(\varepsilon_{2i})$, and $\bSigma_3=\mathrm{var}(\bvarepsilon_{3i})$. In model \eqref{eq:models}, $\balpha^*$ and $\bSigma_1$ can be estimated directly from the combined labeled and unlabeled samples, without EM iterations. The remaining parameters, collected in $\bOmega^*=\{\bbeta^*,\bgamma_{12}^*, \bgamma_3^*, \sigma_2^2, \bSigma_3\}$, are estimated using the EM algorithm described below.

We initialize the procedure with estimates $\bOmega^{(0)}=\{\bbeta^{(0)},\bgamma_{12}^{(0)}, \bgamma_3^{(0)}, (\sigma_2^{(0)})^2, \bSigma_3^{(0)}\}$ obtained using only the labeled observations. The regression coefficients are estimated by ordinary least squares, and the noise parameters are estimated from the corresponding residuals. Under the Gaussian working model, the first E-step computes the conditional moments of each missing outcome $Y_i$ using the conditional distribution $Y_i\mid (\bX_i,\bS_{1i},\bS_{2i},\bOmega^{(0)})\sim \mathcal{N}\{\mu_i^{\bOmega^{(0)}},(\sigma_i^{\bOmega^{(0)}})^2\},$ 
where
\begin{align}
&\mu_i^{\bOmega^{(0)}}=\bbeta^{(0)\top}\bar{\bX}_{1i}+(\sigma_2^{(0)})^2\bgamma_3^{(0)\top}\left\{\bSigma_3^{(0)}+(\sigma_2^{(0)})^2\bgamma_3^{(0)}\bgamma_3^{(0)\top}\right\}^{-1}\left(\bS_{2i}-\bgamma_{12}^{(0)\top}\bar{\bX}_{1i}-\bbeta^{(0)\top}\bar{\bX}_{1i}\bgamma_3^{(0)}\right),\nonumber\\
&(\sigma_i^{\bOmega^{(0)}})^2=(\sigma_2^{(0)})^2-(\sigma_2^{(0)})^4\bgamma_3^{(0)\top}\left\{\bSigma_3^{(0)}+(\sigma_2^{(0)})^2\bgamma_3^{(0)}\bgamma_3^{(0)\top}\right\}^{-1}\bgamma_3^{(0)}.\label{eq:mu_sig}
\end{align}
These conditional moments yield the expected sufficient statistics required for the M-step. Specifically, define the conditional second-moment matrix and mean vector as
 \begin{align}\label{eq:gamma_mu_sig}
 \bSigma_{\bgamma i}^{\bOmega^{(0)}}=\begin{bmatrix}
 \bar{\bX}_{1i}\bar{\bX}_{1i}^\top&\!\! \bar{\bX}_{1i}\mu^{\bOmega^{(0)}}_i\\[1mm]
 \bar{\bX}_{1i}^\top\mu^{\bOmega^{(0)}}_i&\!\! (\mu^{\bOmega^{(0)}}_i)^2+(\sigma^{\bOmega^{(0)}}_i)^2
 \end{bmatrix},\quad {\rm and}\quad 
 \bmu_{\bgamma i}^{\bOmega^{(0)}}=(\bar{\bX}_{1i}^\top,\mu^{\bOmega^{(0)}}_i)^\top.
 \end{align}
In the M-step, maximizing the conditional expectation of the complete-data log-likelihood with respect to the regression coefficients gives the closed-form updates
 \begin{equation}\label{eq:bg_update}
 \begin{split}
 \bbeta^{(1)}&=\Big(\dfrac{1}{N}\sum_{i\in\mathcal{I}\cup\mathcal{J}}\bar{\bX}_{1i}\bar{\bX}_{1i}^\top\Big)^{-1}\Big\{\dfrac{1}{N}\Big(\sum_{i\in\mathcal{I}}\bar{\bX}_{1i}Y_i+\sum_{i\in\mathcal{J}}\bar{\bX}_{1i}\mu^{\bOmega^{(0)}} _i\Big)\Big\},\\
 \!\!\bgamma^{(1)}&=\Big\{\dfrac{1}{N}\Big(\sum_{i\in\mathcal{I}}\bar{\bX}_{2i}\bar{\bX}_{2i}^\top+\sum_{i\in\mathcal{J}}\bSigma_{\bgamma i}^{\bOmega^{(0)}} \Big)\Big\}^{-1}\Big\{\dfrac{1}{N}\Big(\sum_{i\in\mathcal{I}}\bar{\bX}_{2i}\bS_{2i}^\top+\sum_{i\in\mathcal{J}}\bmu_{\bgamma i}^{\bOmega^{(0)}} \bS_{2i}^\top\Big)\Big\}.
 \end{split}
 \end{equation}
We show that the regression estimators $\bbeta^{(1)}$ and $\bgamma^{(1)}$ attain statistical efficiency after this single EM update. Further EM iterations and updates of the noise parameters are therefore not needed to achieve statistical efficiency. We refer to the resulting one-step procedure as the Surrogate-Assisted EM (SAEM) algorithm, summarized in Algorithm~\ref{alg:EM}.
 
 \begin{algo}\label{alg:EM}
 Surrogate-Assisted EM (SAEM) algorithm.
 
 \vspace{4pt}
 \textit{Input:} Labeled observations $(\bX_i,\bS_{1i},Y_i,\bS_{2i})_{i\in\mathcal{I}}$ and unlabeled observations $(\bX_i,\bS_{1i},\bS_{2i})_{i\in\mathcal{J}}$.
 
 \vspace{4pt}
 \textit{Step} 1. Estimate $\bSigma_1$ and $\balpha^*$ using the combined labeled and unlabeled samples:
 \[
 \widehat\bSigma_{\bX}=\dfrac{1}{N}\sum_{i\in\mathcal{I}\cup\mathcal{J}}\bX_i\bX_i^\top,\quad\widehat\balpha=\widehat\bSigma_{\bX}^{-1}\bigg(\dfrac{1}{N}\sum_{i\in\mathcal{I}\cup\mathcal{J}}\bX_i\bS_{1i}^\top\bigg).
 \]
 
 \textit{Step} 2. Compute initial estimates of $\bOmega^*=\{\bbeta^*,\bgamma_{12}^*, \bgamma_3^*, \sigma_2^2, \bSigma_3\}$ using the labeled observations:
 \begin{align*}
 \bbeta^{(0)}&=\Big(\dfrac{1}{n}\sum_{i\in\mathcal{I}}\bar{\bX}_{1i}\bar{\bX}_{1i}^\top\Big)^{-1}\Big(\dfrac{1}{n}\sum_{i\in\mathcal{I}}\bar{\bX}_{1i}Y_i\Big),\quad (\sigma_2^{(0)})^2=\dfrac{1}{n}\sum_{i\in\mathcal{I}}(Y_i-\bbeta^{(0)\top} \bar{\bX}_{1i})^2,\\
 \bgamma^{(0)}&=(\bgamma_{12}^{(0)\top},
 \bgamma_3^{(0)})^\top=\Big(\dfrac{1}{n}\sum_{i\in\mathcal{I}}\bar{\bX}_{2i}\bar{\bX}_{2i}^\top\Big)^{-1}\Big(\dfrac{1}{n}\sum_{i\in\mathcal{I}}\bar{\bX}_{2i}\bS_{2i}^\top\Big),\\
 \bSigma_3^{(0)}&=\dfrac{1}{n}\sum_{i\in\mathcal{I}}(\bS_{2i}-\bgamma^{(0)\top}\bar{\bX}_{2i})(\bS_{2i}-\bgamma^{(0)\top}\bar{\bX}_{2i})^\top.
 \end{align*}
 
 \textit{Step} 3. For each $i\in\mathcal{J}$, compute $\mu_i^{\bOmega^{(0)}}$, $\sigma_i^{\bOmega^{(0)}}$, $\bSigma_{\bgamma i}^{\bOmega^{(0)}}$, and $\bmu_{\bgamma i}^{\bOmega^{(0)}}$ using \eqref{eq:mu_sig} and \eqref{eq:gamma_mu_sig}.
 
 \vspace{4pt}
 \textit{Step} 4. Compute the regression updates $\bbeta^{(1)}$ and $\bgamma^{(1)}$ using \eqref{eq:bg_update}.
 
 \vspace{4pt}
 \textit{Output:} $(\widehat{\balpha},\;\widehat{\bbeta}=\bbeta^{(1)},\;\widehat{\bgamma}=\bgamma^{(1)})$.
 \end{algo}

Algorithm~\ref{alg:EM} yields the SAEM estimators $\widehat{\balpha}$, $\widehat{\bbeta}=(\widehat{\bbeta}_1^\top,\widehat{\bbeta}_2^\top)^\top$, and $\widehat{\bgamma}= (\widehat{\bgamma}_{12}^\top,\widehat{\bgamma}_3)^\top$. Here, $\widehat{\bbeta}_1$ and $\widehat{\bbeta}_2$ estimate the coefficients associated with $\bX_i$ and $\bS_{1i}$, respectively, in the second equation of model~\eqref{eq:models}. In the third equation, $\widehat{\bgamma}_{12}=(\widehat{\bgamma}_1^\top,\widehat{\bgamma}_2^\top)^\top$ estimates the coefficient matrix associated with $\bar{\bX}_{1i}$, while $\widehat{\bgamma}_3$ estimates the coefficient vector associated with $Y_i$. We further define $\widehat{\btheta}=\widehat{\bbeta}_1+\widehat{\balpha}\widehat{\bbeta}_2$, which serves as the SAEM estimator of the prediction parameter $\btheta^*$.

\section{Theory}\label{sec:theory}
\subsection{Upper and lower bounds}\label{sec:upper_bound}

Let $\bSigma_{\bX}=E(\bX_i\bX_i^\top)$, $\bSigma_{\bX,1}=E(\bar{\bX}_{1i}\bar{\bX}_{1i}^\top)$, and $\bSigma_{\bX,2}=E(\bar{\bX}_{2i}\bar{\bX}_{2i}^\top)$. Throughout the analysis, we assume that the dimensions $p,d_1,d_2$ and the joint distribution of $(\bX_i,\bS_{1i},Y_i,\bS_{2i})$ are fixed, $n\to\infty, m\geq 0$, $N=n+m$, and $m/N\to \tau\in [0,1]$. %
We vectorize the matrix-valued estimators and consider ${\rm vec}(\widehat{\balpha})$, $\widehat{\bbeta}$, $\widehat\btheta$, and the stacked vector $({\rm vec}(\widehat{\bgamma}_{12})^\top,\widehat{\bgamma}_3^\top)^\top$. 

The following theorem establishes the asymptotic normality of these estimators.
 
 \begin{theorem}\label{thm:upper_bound}
 Let Assumption \ref{ass:joint_gaussian} hold. Then as $n\to \infty$, the following results hold:
 
 (i) $\sqrt{N}\{{\rm vec}(\widehat\balpha)-{\rm vec}(\balpha^*)\}\xrightarrow{\rm d}\mathcal{N}(\mathbf{0},\bV_{\balpha})$, where $\bV_{\balpha}=\bSigma_1\otimes \bSigma_{\bX}^{-1}$.
 
 (ii) $\sqrt{n} (\widehat\bbeta - \bbeta^*) 
 \xrightarrow{\rm d}\mathcal{N}(\mathbf{0},\bV_{\bbeta})$, where
 $\bV_{\bbeta}=\sigma_2^2(1-\kappa \tau)\bSigma_{\bX,1}^{-1}$ and $\kappa=\sigma_2^2\bgamma_3^{*\top}\bSigma_3^{-1}\bgamma_3^*/(1+\sigma_2^2\bgamma_3^{*\top}\bSigma_3^{-1}\bgamma_3^*)\in[0,1).$ As a result, $\sqrt{n} (\widehat\bbeta_1 - \bbeta_1^*) 
 \xrightarrow{\rm d}\mathcal{N}(\mathbf{0},\bV_{\bbeta_1})$, where
 $\bV_{\bbeta_1}=\sigma_2^2(1-\kappa \tau)(\bSigma_{\bX}^{-1}+\balpha^*\bSigma_1^{-1}\balpha^{*\top})$.
 
 (iii) $\sqrt{n}(\widehat\btheta - \btheta^*)
 \xrightarrow{\rm d}\mathcal{N}(\mathbf{0},\bV_{\btheta})$, where $\bV_{\btheta}=\{\sigma_2^2(1-\kappa\tau)+(1-\tau)\bbeta_2^{*\top}\bSigma_1\bbeta_2^*\}\bSigma_{\bX}^{-1}.$
 
 (iv) $\sqrt{n}\{ ({\rm vec}(\widehat{\bgamma}_{12})^\top,\widehat{\bgamma}_3^\top)^\top -({\rm vec}(\bgamma_{12}^*)^\top,\bgamma_3^{*\top})^\top\}\xrightarrow{\rm d} \mathcal{N}(\mathbf{0},\bV_{\bgamma})$, where
 \begin{equation*}
 \begin{split}
 \bV_{\bgamma}&= \begin{bmatrix}
 \bV_{\bgamma,2}+\bV_{\bgamma,1}\otimes (\bbeta^*\bbeta^{*\top}) & -\bV_{\bgamma,1}\otimes \bbeta^*\\
 -\bV_{\bgamma,1}\otimes \bbeta^{*\top} & \bV_{\bgamma,1}
 \end{bmatrix},\\
 \bV_{\bgamma,1}&=\dfrac{1}{\sigma_2^2}(1-\kappa\tau)\bSigma_3-\tau\dfrac{(1-\sigma_2^2\bgamma_3^{*\top}\bSigma_{3}^{-1}\bgamma_3^*)}{(1+\sigma_2^2\bgamma_3^{*\top}\bSigma_{3}^{-1}\bgamma_3^*)^2}\bgamma_3^*\bgamma_3^{*\top},\\
 \bV_{\bgamma,2}&=\bigg\{(1-\tau)\bSigma_3+\dfrac{\tau \sigma_2^2}{1+\sigma_2^2 \bgamma_3^{*\top}\bSigma_{3}^{-1}\bgamma_3^*}\bgamma_3^*\bgamma_3^{*\top}\bigg\}\otimes \bSigma_{\bX,1}^{-1}.
 \end{split}
 \end{equation*}
 \end{theorem}

We next derive the Cramér--Rao lower bounds for the surrogate-assisted semi-supervised setting under the Gaussian model.
 
\begin{theorem}\label{thm:lower_bound}
Let Assumption~\ref{ass:joint_gaussian} hold. The covariance matrices of any unbiased estimators of ${\rm vec}(\balpha^*)$, $\bbeta^*$, $\btheta^*$, and $({\rm vec}(\bgamma_{12}^*)^\top,\bgamma_3^{*\top})^\top$ are bounded below, in the positive semidefinite order, by
\[
N^{-1}\bV_{\balpha},\qquad
n^{-1}\bV_{\bbeta},\qquad
n^{-1}\bV_{\btheta},\qquad
n^{-1}\bV_{\bgamma},
\]
respectively, with $\tau$ replaced by the actual sample fraction $m/N$ in the covariance formulas of Theorem~\ref{thm:upper_bound}.
\end{theorem}

The bounds in Theorem~\ref{thm:lower_bound} depend on the actual sample sizes $n$ and $m$, whereas the limiting covariance matrices in Theorem~\ref{thm:upper_bound} are expressed in terms of $\tau$. Since their appropriately normalized limits coincide, the SAEM estimators asymptotically attain the corresponding Cram\'{e}r--Rao lower bounds. Thus, a single EM update suffices to achieve asymptotic efficiency, and further EM iterations are unnecessary for this purpose.

\begin{remark}[Comparison with the supervised estimators]\label{remark:compare-sup}
We next compare the efficiency of the SAEM estimators with that of the supervised maximum likelihood estimators (MLEs), which use only the labeled observations. We restrict the discussion below to the case $\tau>0$. When $\tau=0$, the number of unlabeled observations is asymptotically negligible relative to the number of labeled observations, and the asymptotic variances of all SAEM estimators coincide with those of their supervised MLE counterparts.

\emph{Efficiency gain for estimating $\balpha^*$.} Since $(\bX_i,\bS_{1i})$ is observed for all individuals, the SAEM estimator of $\balpha^*$ utilizes all $N$ observations, whereas the supervised estimator uses only the $n$ labeled observations. To compare their asymptotic efficiencies on the same scale, we consider the $\sqrt{n}$-normalized estimators. By Theorem~\ref{thm:upper_bound}(i), the limiting covariance matrix of the SAEM estimator on this scale is $(1-\tau)\bV_{\balpha}$, compared with $\bV_{\balpha}$ for the supervised estimator. Thus, the reduction in the limiting covariance is $\tau \bV_{\balpha}$, which is strictly positive definite when $\tau>0$.

\emph{Efficiency gain for estimating $\bbeta^*$ and $\bbeta_1^*$}. For estimating $\bbeta^*$ or $\bbeta_1^*$, incorporating unlabeled observations can reduce the asymptotic variance, although the convergence rate remains governed by the labeled sample size $n$. Under $\sqrt{n}$ normalization, the supervised MLEs for $\bbeta^*$ and its sub-vector $\bbeta_1^*$ have limiting covariance matrices $\bV_{\bbeta}^{\rm sup}=\sigma_2^2\bSigma_{\bX,1}^{-1}$ and $\bV_{\bbeta_1}^{\rm sup}=\sigma_2^2(\bSigma_{\bX}^{-1}+\balpha^*\bSigma_1^{-1}\balpha^{*\top})$, respectively. By Theorem~\ref{thm:upper_bound}(ii), the reductions in the limiting covariance matrices are
\begin{align*}
\bV_{\bbeta}^{\rm sup}-\bV_{\bbeta}&=\tau\kappa\sigma_2^2\bSigma_{\bX,1}^{-1}=\tau\kappa\bV_{\bbeta}^{\rm sup}\succeq \mathbf{0},\\
\bV_{\bbeta_1}^{\rm sup}-\bV_{\bbeta_1}&=\tau\kappa\sigma_2^2(\bSigma_{\bX}^{-1}+\balpha^*\bSigma_1^{-1}\balpha^{*\top})=\tau\kappa\bV_{\bbeta_1}^{\rm sup}\succeq \mathbf{0}.
\end{align*}
Thus, the relative efficiency gains are determined by the unlabeled fraction $\tau$ and the factor $\kappa$. To interpret $\kappa$, we combine the second and third equations in \eqref{eq:models} to obtain
\[
\bS_{2i}=\bgamma_{12}^{*\top}\bar{\bX}_{1i}+\bgamma_3^* \bbeta^{*\top}\bar{\bX}_{1i}+\bv_i,\;\; \bv_i=
\varepsilon_{2i}\bgamma_3^*+\bvarepsilon_{3i}.
\]
Conditional on $\bar{\bX}_{1i}$, the residual components of $\bS_{2i}$ and $Y_i$ are $\bv_i$ and $\varepsilon_{2i}$, respectively. Hence, $\kappa$ is the population coefficient of determination for the linear regression $\varepsilon_{2i}\sim \bv_i$. Equivalently, it measures the proportion of the residual variation in $Y_i$ explained by $\bS_{2i}$ after conditioning on $\bar{\bX}_{1i}$.

In particular, when $\bgamma_3^*=\mathbf{0}$, the post-outcome surrogate $\bS_{2i}$ provides no additional information about $Y_i$ beyond $\bar{\bX}_{1i}$, implying $\kappa=0$. Consequently, the additional unlabeled observations of $(\bX_i,\bS_{1i})$ provide no efficiency gain for estimating $\bbeta^*$. The same conclusion holds for the direct effect parameter $\bbeta_1^*$, even when $\bS_{1i}$ is informative about $Y_i$ and $\bbeta_2^*\neq\mathbf{0}$. This result is also consistent with the existing understanding of the role of unlabeled samples in semi-supervised inference. When the uninformative post-outcome surrogates $\bS_{2i}$ are omitted, the problem reduces to semi-supervised linear regression of $Y_i$ on $(\bX_i,\bS_{1i})$. Under the correctly specified Gaussian linear model, the unlabeled observations $(\bX_i,\bS_{1i})$ provide no additional likelihood information about the regression coefficients $\bbeta^*$. Therefore, the supervised MLE remains the MLE in the semi-supervised setting, as noted by \cite{zhang2000value,seeger2000learning,chen2025semi}, among others. Consequently, $\bV_{\bbeta}=\bV_{\bbeta}^{\rm sup}$ and $\bV_{\bbeta_1}=\bV_{\bbeta_1}^{\rm sup}$.

In contrast, when $\bgamma_3^*\neq\mathbf{0}$, $\bS_{2i}$ provides additional information about the noise component $\varepsilon_{2i}$ of $Y_i$, leading to $\kappa>0$. Consequently, incorporating the unlabeled observations yields a strictly positive definite reduction in asymptotic variance, with $\bV_{\bbeta}\prec\bV_{\bbeta}^{\rm sup}$ and $\bV_{\bbeta_1}\prec\bV_{\bbeta_1}^{\rm sup}$. These observations highlight a key distinction between the roles of pre- and post-outcome surrogates. The post-outcome surrogate $\bS_{2i}$ carries downstream information about the realized outcome $Y_i$, including its noise component $\varepsilon_{2i}$, whereas the pre-outcome surrogate $\bS_{1i}$ is observed before $Y_i$ and is independent of $\varepsilon_{2i}$. Additional information about $\varepsilon_{2i}$ enables more accurate estimation of the coefficients in the same regression model, namely $\bbeta^*$. Therefore, unlike the pre-outcome surrogate $\bS_{1i}$, the post-outcome surrogate $\bS_{2i}$ can improve estimation of how $Y_i$ is generated from $\bX_i$ and $\bS_{1i}$.

\emph{Efficiency gain for estimating $\btheta^*$.} The supervised MLE for $\btheta^*$ has asymptotic variance $\bV_{\btheta}^{\rm sup}=(\sigma_2^2+\bbeta_2^{*\top}\bSigma_1\bbeta_2^*)\bSigma_{\bX}^{-1}$. Incorporating unlabeled observations reduces the asymptotic variance by
\begin{align*}
\bV_{\btheta}^{\rm sup}-\bV_{\btheta}=\tau(\kappa\sigma_2^2+ \bbeta_2^{*\top}\bSigma_1\bbeta_2^*)\bSigma_{\bX}^{-1}\succeq \mathbf{0}.
\end{align*}
The two terms in this reduction reflect distinct sources of information. As discussed above, the condition $\bgamma_3^*\neq\mathbf{0}$, which implies $\kappa\neq0$, indicates that the post-outcome surrogate $\bS_{2i}$ provides useful downstream information about $Y_i$ through $\varepsilon_{2i}$. In contrast, the condition $\bbeta_2^*\neq\mathbf{0}$ indicates that the pre-outcome surrogate $\bS_{1i}$ provides useful upstream information about $Y_i$ through $\bvarepsilon_{1i}$. To see the latter role, $\btheta^*$ is the regression coefficient in the model \eqref{eq:Y-X}, where the residual variation in $Y_i$ arises from both $\bvarepsilon_{1i}$ and $\varepsilon_{2i}$. Additional observations of $\bS_{1i}$ and $\bS_{2i}$ can therefore improve efficiency by providing information about these noise components. Whenever either $\bbeta_2^*$ or $\bgamma_3^*$ is nonzero, at least one type of surrogate provides information about the residual variation in $Y_i$ beyond the original covariates, yielding $\bV_{\btheta}\prec\bV_{\btheta}^{\rm sup}$.

On the other hand, when $\bbeta_2^*=\mathbf{0}$ and $\bgamma_3^*=0$, neither $\bS_1$ nor $\bS_2$ provides additional information about $Y$, and hence no efficiency gain is available, with $\bV_{\btheta}=\bV_{\btheta}^{\rm sup}$. This also includes settings in which the unlabeled sample contains only the baseline covariates $\bX$ without any surrogate variables. Under such circumstances, $\btheta^*$ and $\bbeta_1^*$ coincide, and their SAEM estimator reduces to the supervised MLE. Since SAEM attains the Cram\'{e}r--Rao lower bound, this implies that additional unlabeled observations of $\bX$ alone provide no efficiency gain for estimating $\btheta^*$ under the correctly specified Gaussian model, again consistent with the observations in \cite{zhang2000value,seeger2000learning,chen2025semi}. Efficiency gains therefore require additional informative surrogate variables, either pre- or post-outcome.

\emph{Efficiency gain for estimating $\bgamma^*$.} Finally, the supervised MLE for the stacked parameter vector
$({\rm vec}(\bgamma_{12}^*)^\top,\bgamma_3^{*\top})^\top$ has asymptotic variance
\[
\bV_{\bgamma}^{\rm sup}
=
\left[\begin{matrix}
\bV_{\bgamma,2}^{\rm sup}+\bV_{\bgamma,1}^{\rm sup}\otimes (\bbeta^*\bbeta^{*\top})
& -\bV_{\bgamma,1}^{\rm sup}\otimes \bbeta^* \\[2mm]
-\bV_{\bgamma,1}^{\rm sup}\otimes \bbeta^{*\top}
& \bV_{\bgamma,1}^{\rm sup}
\end{matrix}\right],
\]
where $\bV_{\bgamma,1}^{\rm sup}=\bSigma_3/\sigma_2^2$ and $\bV_{\bgamma,2}^{\rm sup}=\bSigma_3\otimes \bSigma_{\bX,1}^{-1}$. The SAEM estimator satisfies $\bV_{\bgamma}\preceq\bV_{\bgamma}^{\rm sup}$ and is therefore at least as efficient as the supervised MLE. When $\tau\neq 0$ and $\bgamma_3^*\neq\mathbf{0}$, the covariance difference is positive definite, yielding strict efficiency gains for estimating both $\bgamma_{12}^*$ and $\bgamma_3^*$.

When $\bgamma_3^*=\mathbf{0}$, additional unlabeled observations do not change the asymptotic variance for estimating $\bgamma_3^*$. Nevertheless, the covariance reduction for estimating $\bgamma_{12}^*$ is $\tau\bSigma_3\otimes\bSigma_{\bX,1}^{-1}\succ\mathbf{0}$. Thus, unlabeled observations still improve estimation of the coefficients associated with the fully observed covariates, even when the post-outcome surrogate provides no additional information about the outcome noise.
\end{remark}

\subsection{The impact of surrogate misclassification and omission}\label{sec:misclassification_S}

We next examine the effects of misclassifying pre-outcome surrogates as post-outcome surrogates, or vice versa, and of omitting surrogates from the analysis. For this purpose, we introduce a general framework that accommodates both misclassification and omission.

We assume that the true data-generating mechanism is given by \eqref{eq:models}, but allow the surrogate groups used in the analysis to differ from the true groups. Let $\bS_i=(\bS_{1i}^\top,\bS_{2i}^\top)^\top$ collect all surrogate variables. Define three row-selection matrices $\bP_a\in\{0,1\}^{d_a^\prime\times d}$ for $a\in\{0,1,2\}$ and set
\[
\bS_{1i}^\prime=\bP_1\bS_i\in\mathbb R^{d_1^\prime},\qquad\bS_{2i}^\prime=\bP_2\bS_i\in\mathbb R^{d_2^\prime},\qquad\bS_{0i}^\prime=\bP_0\bS_i\in\mathbb R^{d_0^\prime}.
\]
Here, $\bS_{1i}^\prime$ and $\bS_{2i}^\prime$ are treated as pre- and post-outcome surrogates, respectively. The variables included in $\bS_{1i}^\prime$ and $\bS_{2i}^\prime$ may have different temporal roles under the true data-generating mechanism, while $\bS_{0i}^\prime$ contains the surrogates omitted from the analysis. We assume that each surrogate variable belongs to exactly one of these three groups, so that $d_0^\prime+d_1^\prime+d_2^\prime=d=d_1+d_2$ and $\bP_{\rm all}=(\bP_1^\top,\bP_2^\top,\bP_0^\top)^\top$ is a permutation matrix.

We consider the following working models based on the selected surrogate groups:
\begin{equation}\label{eq:models_mis}
\begin{split}
\bS_{1i}^\prime&=\bar{\balpha}^{{\rm p}\top}\bX_i+\bvarepsilon_{1i}^\prime,\\
Y_i&=\bar{\bbeta}_1^{{\rm p}\top}\bX_i+\bar{\bbeta}_2^{{\rm p}\top}\bS_{1i}^\prime+\varepsilon_{2i}^\prime,\\
\bS_{2i}^\prime&=\bar{\bgamma}_1^{{\rm p}\top}\bX_i+\bar{\bgamma}_2^{{\rm p}\top}\bS_{1i}^\prime+\bar{\bgamma}_3^{\rm p}Y_i+\bvarepsilon_{3i}^\prime.
\end{split}
\end{equation}
All coefficients in \eqref{eq:models_mis} are defined as the corresponding population least-squares regression coefficients. For example, in the first equation, we define $\bar{\balpha}^{\rm p}=\mathop{\arg\min}_{\balpha\in\mathbb R^{p\times d_1^{\prime}}} E\{\|\bS_{1i}^\prime-\balpha^\top\bX_i\|_2^2\}$ and $\bvarepsilon_{1i}^\prime=\bS_{1i}^\prime-\bar{\balpha}^{{\rm p}\top}\bX_i$. The coefficients and residuals in the second and third equations are defined analogously.

Applying Algorithm~\ref{alg:EM} to the labeled observations $(\bX_i, \bS_{1i}^\prime, Y_i, \bS_{2i}^\prime)_{i \in \mathcal{I}}$ and the unlabeled observations $(\bX_i, \bS_{1i}^\prime, \bS_{2i}^\prime)_{i \in \mathcal{J}}$ yields the one-step EM estimators $\{\widehat{\balpha}^{\rm p},\widehat{\bbeta}^{\rm p},\widehat{\bgamma}^{\rm p}\}$. Since the selection scheme $\rm p=\{\bP_1,\bP_2,\bP_0\}$ is unknown, we cannot generally map the estimated surrogate coefficients back to their counterparts in the true model. We therefore focus on estimating $\bbeta_1^*$ and $\btheta^*$, both of which characterize the relationship between $\bX_i$ and $Y_i$. We write $\widehat{\bbeta}^{\rm p}=(\widehat{\bbeta}_1^{{\rm p}\top},\widehat{\bbeta}_2^{{\rm p}\top})^\top$ and $\widehat{\btheta}^{\rm p}=\widehat{\bbeta}_1^{\rm p}+\widehat{\balpha}^{\rm p}\widehat{\bbeta}_2^{\rm p}$.
Let $\bSigma_1^{\prime}=\mathrm{var}(\bvarepsilon_{1i}^\prime)$, $\sigma_2^{\prime 2}=\mathrm{var}(\varepsilon_{2i}^\prime)$, and $\bSigma_3^{\prime}=\mathrm{var}(\bvarepsilon_{3i}^\prime)$, and define $\bar{\bX}_{1i}^\prime=(\bX_i^\top,\bS_{1i}^{\prime\top})^\top$ and $\bSigma_{\bX,1}^\prime=E(\bar{\bX}_{1i}^\prime\bar{\bX}_{1i}^{\prime\top})$. 

We first examine the asymptotic properties of $\widehat{\bbeta}_1^{\rm p}$.

\begin{theorem}\label{thm:thm_beta1}
Let Assumption~\ref{ass:joint_gaussian} hold. Define $\bdelta=E(\bS_{1i}^\prime\bS_{1i}^{\prime\top})^{-1}E(\bS_{1i}^\prime\bX_i^\top)$ and $\bG=\{E(\bX_i\bX_i^\top)-E(\bX_i\bS_{1i}^{\prime\top})E(\bS_{1i}^\prime\bS_{1i}^{\prime\top})^{-1}E(\bS_{1i}^\prime\bX_i^\top)\}^{-1}$. Then, as $n\to \infty$,
$\sqrt{n}(\widehat{\bbeta}_1^{\rm p}-\bbeta_1^*-\bB_{\bbeta_1}^{\rm p})\xrightarrow{\rm d}\mathcal{N}(\mathbf{0},\bV_{\bbeta_1}^{\rm p}),$
where the asymptotic covariance matrix and bias are given by
\begin{align*}
\bV_{\bbeta_1}^{\rm p}&=[\bI_p,\mathbf{0}_{p\times d_1^\prime}]\sigma_2^{\prime 2}\left(1-\tau\frac{\sigma_2^{\prime 2}\bar{\bgamma}_3^{{\rm p}\top}\bSigma_3^{\prime-1}\bar{\bgamma}_3^{\rm p}}{1+\sigma_2^{\prime 2}\bar{\bgamma}_3^{{\rm p}\top}\bSigma_3^{\prime-1}\bar{\bgamma}_3^{\rm p}}\right)\bSigma_{\bX,1}^{\prime-1}[\bI_p,\mathbf{0}_{p\times d_1^\prime}]^\top,\\
\bB_{\bbeta_1}^{\rm p}&=\bG\left[E\left\{(\bX_i-\bdelta^\top\bS_{1i}^\prime)\bS_{1i}^\top\right\}\bbeta_2^*- \bdelta^\top E(\bS_{1i}^\prime\varepsilon_{2i})\right].
\end{align*}
\end{theorem}

\begin{remark}[Bias and variance for estimating $\bbeta_1^*$]\label{remark:beta1}
We examine how surrogate misclassification and omission affect the bias and variance when estimating $\bbeta_1^*$.

\emph{Estimation bias.} The first component of the bias $\bB_{\bbeta_1}^{\rm p}$ arises when true pre-outcome surrogates are excluded from $\bS_{1i}^\prime$. The matrix $\bdelta$ contains the population regression coefficients of $\bX_i$ on $\bS_{1i}^\prime$ and satisfies the orthogonality condition $E\{(\bX_i-\bdelta^\top \bS_{1i}^\prime)\bS_{1i}^{\prime\top}\} = \mathbf{0}$. Consequently, if some components of $\bS_{1i}$ are assigned to $\bS_{2i}^\prime$ or $\bS_{0i}^\prime$, this bias term generally does not vanish. Under a causal interpretation of the true model, $\bbeta_1^*$ represents the direct effect of $\bX_i$ on $Y_i$. Excluding true pre-outcome surrogates from the outcome regression may cause their indirect effects to be absorbed into the estimated coefficient of $\bX_i$, biasing estimation of the direct effect. One exception occurs when $\bbeta_2^*=\mathbf{0}$: $\bS_{1i}$ does not contribute to $Y_i$ in the outcome model, so omitting $\bS_{1i}$ alone does not introduce asymptotic bias.

The second component of $\bB_{\bbeta_1}^{\rm p}$ captures the bias caused by including true post-outcome surrogates $\bS_{2i}$ in $\bS_{1i}^\prime$. Under the true model, the orthogonality condition $E(\bS_{1i}\varepsilon_{2i})=\mathbf 0$ holds. However, if $\bS_{1i}^\prime$ includes components of $\bS_{2i}$, then $E(\bS_{1i}^\prime\varepsilon_{2i})\neq\mathbf 0$ in general, because these components can be correlated with $\varepsilon_{2i}$ through $Y_i$. Treating these variables as pre-outcome surrogates can therefore introduce regressors correlated with the true outcome error and bias estimation of the direct effect of $\bX_i$ on $Y_i$. Another exception occurs when $\bgamma_3^*=\mathbf{0}$. In this case, $\bS_{2i}$ carries no additional information about $Y_i$ conditional on $(\bX_i,\bS_{1i})$, so treating these variables as pre-outcome surrogates does not introduce asymptotic bias.

Together, these two components show that misclassifying surrogate roles or omitting pre-outcome surrogates $\bS_{1i}$ may introduce asymptotic bias in estimating the direct effect $\bbeta_1^*$. Indeed, to identify the direct effect of $\bX_i$ on $Y_i$, one should adjust for all and only the pre-outcome surrogates $\bS_{1i}$ that lie on the pathways from $\bX_i$ to $Y_i$, while avoiding adjustment for post-outcome surrogates $\bS_{2i}$. In contrast, if $\bS_{1i}^\prime=\bS_{1i}$ and $\bS_{2i}^\prime$ is any sub-vector of $\bS_{2i}$, then $\bB_{\bbeta_1}^{\rm p}=\mathbf{0}$ and $\widehat{\bbeta}_1^{\rm p}$ is asymptotically unbiased. Thus, omitting post-outcome surrogates alone does not introduce asymptotic bias. We next examine its effect on the variance.

\emph{Asymptotic variance.} Suppose $\bS_{1i}^\prime=\bS_{1i}$. Then $\bB_{\bbeta_1}^{\rm p}=\mathbf{0}$ and 
\begin{align*}\bV_{\bbeta_1}^{\rm p}=\sigma_2^2(1-\kappa^{\rm p} \tau)\bigg(\bSigma_{\bX}^{-1}+\balpha^*\bSigma_1^{-1}\balpha^{*\top}\bigg),\quad\kappa^{\rm p}=\frac{\sigma_2^2\bar{\bgamma}_3^{\rm p\top}\bSigma_3^{\prime-1}\bar{\bgamma}_3^{\rm p}}{1+\sigma_2^2\bar{\bgamma}_3^{\rm p\top}\bSigma_3^{\prime-1}\bar{\bgamma}_3^{\rm p}}\in[0,1).\end{align*}
Applying the variance comparison in Remark~\ref{remark:compare-sup} to the sub-vector $\bbeta_1^*$ gives
\begin{align*}\bV_{\bbeta_1}=\sigma_2^2(1-\kappa \tau)\bigg(\bSigma_{\bX}^{-1}+\balpha^*\bSigma_1^{-1}\balpha^{*\top}\bigg),\quad\bV_{\bbeta_1}^{\rm sup}=\sigma_2^2\bigg(\bSigma_{\bX}^{-1}+\balpha^*\bSigma_1^{-1}\balpha^{*\top}\bigg),\end{align*} where $\bV_{\bbeta_1}$ and $\bV_{\bbeta_1}^{\rm sup}$ are the asymptotic covariance matrices of the SAEM estimator $\bbetahat_1$ under correct surrogate classification with all surrogates retained and of the supervised MLE, respectively.

Since $\kappa,\kappa^{\rm p}\in[0,1)$ and $\kappa-\kappa^{\rm p}=\{{\rm var}(Y_i\mid\bX_i,\bS_{1i},\bS_{2i}^\prime)-{\rm var}(Y_i\mid\bX_i,\bS_{1i},\bS_{2i})\}/\sigma_2^2\geq0,$ it follows that $\bV_{\bbeta_1}=(1-\kappa \tau)\bV_{\bbeta_1}^{\rm sup}\preceq(1-\kappa^{\rm p}\tau)\bV_{\bbeta_1}^{\rm sup}=\bV_{\bbeta_1}^{\rm p}\preceq\bV_{\bbeta_1}^{\rm sup}.$
Thus, omitting post-outcome surrogates may reduce efficiency relative to the SAEM estimator using all surrogates, while the resulting estimator remains asymptotically at least as efficient as the supervised MLE.

In summary, estimation of the direct-effect parameter $\bbeta_1^*$ is sensitive to the choice of adjustment variables. Misclassifying surrogates or omitting true pre-outcome surrogates may introduce a nonvanishing asymptotic bias, regardless of the limiting unlabeled fraction. In contrast, omitting post-outcome surrogates alone preserves consistency, although it may reduce efficiency when the omitted surrogates are informative.
\end{remark}

We next examine how surrogate misclassification and omission affect the estimation of $\btheta^*$.

\begin{theorem}\label{thm:thm_theta}
Let Assumption~\ref{ass:joint_gaussian} hold. Define $\nu^2={\rm var}(Y_i\mid\bX_i)$ and $\nu_{\rm p}^2={\rm var}(Y_i\mid\bX_i, \bS_{1i}^\prime, \bS_{2i}^\prime)$. Then, for any fixed selection scheme $\rm p$, $\sqrt{n}(\widehat{\btheta}^{\rm p}-\btheta^*)\xrightarrow{\rm d}\mathcal{N}(\mathbf{0},\bV_{\btheta}^{\rm p}),$ where $\bV_{\btheta}^{\rm p}=\{(1-\tau)\nu^2+\tau\nu_{\rm p}^2\}\bSigma_{\bX}^{-1}.$ The following properties also hold.

\noindent{\rm (i)} Consider two selection schemes, $\rm p$ and $\rm q$, that retain the same set of surrogate variables but may assign them differently as pre- or post-outcome surrogates. Whenever the relevant sample Gram matrices are nonsingular, the identity $\widehat{\btheta}^{\rm p}=\widehat{\btheta}^{\rm q}$ holds exactly.

\noindent{\rm (ii)} Suppose the surrogate variables retained under scheme $\rm p$ form a subset of those retained under scheme $\rm q$, and define $\nu_{\rm q}^2$ analogously to $\nu_{\rm p}^2$. Then $\bV_{\btheta}^{\rm p}-\bV_{\btheta}^{\rm q}=\tau(\nu_{\rm p}^2-\nu_{\rm q}^2)\bSigma_{\bX}^{-1}\succeq\mathbf{0}.$
\end{theorem}

\begin{remark}[Bias and variance for estimating $\btheta^*$]\label{remark:theta}
Theorem~\ref{thm:thm_theta} shows that neither surrogate misclassification nor surrogate omission introduces asymptotic bias in estimating $\btheta^*$. Moreover, when the retained surrogate set is fixed, misclassification leaves the SAEM estimator unchanged even in finite samples. Thus, unlike estimation of the direct effect parameter $\bbeta_1^*$, estimation of the prediction parameter $\btheta^*$ does not benefit from knowledge of the temporal ordering of the surrogates in this setting. This observation also provides theoretical justification for the strand of literature that treats all surrogate variables as additional predictive features \citep{hou2023surrogate,xia2024prediction,tian2024semi,ji2025predictions,cai2025semi}. In particular, we show that even when the temporal ordering of the surrogates relative to the outcome is known, incorporating such information provides no additional benefit for estimating $\btheta^*$ under the considered framework. On the other hand, omitting either pre- or post-outcome surrogates may reduce efficiency when the omitted variables contain additional information about $Y_i$ conditional on the baseline covariates and retained surrogates. Therefore, when prediction is the primary objective, one should collect as many informative surrogates as feasible to improve efficiency, without requiring knowledge of their temporal ordering relative to the outcome.
\end{remark}

\subsection{Extension to non-Gaussian settings}\label{sec:optimal_lambda}

Our analysis so far assumes Gaussian designs. We now extend the analysis to non-Gaussian settings to examine the validity of the proposed methods beyond this assumption. In this setting, the coefficients in \eqref{eq:models} are defined as population least-squares projection coefficients. For example, in the first equation, we define $\balpha^*=\mathop{\arg\min}_{\balpha\in\mathbb{R}^{p\times d_1}} E\{\|\bS_{1i}-\balpha^\top\bX_i\|_2^2\}$ and $\bvarepsilon_{1i}=\bS_{1i}-\balpha^{*\top}\bX_i$. The coefficients and residuals in the other equations are defined analogously. These definitions allow the linear models for the conditional means to be misspecified; for example, $E(\bvarepsilon_{1i}\mid \bX_i)\neq \mathbf{0}$ is permitted.

Under the Gaussian model, Section~\ref{sec:upper_bound} shows that SAEM has a limiting covariance matrix no larger than that of supervised estimation. Taking estimation of $\bbeta^*$ as an example, this efficiency gain arises because, conditional on $\bar{\mathbf X}_{1i}$, the residual component of the post-outcome surrogate, $\mathbf v_i=\varepsilon_{2i}\boldsymbol{\gamma}_3^*+\boldsymbol{\varepsilon}_{3i}$, contains useful linear information about the outcome residual $\varepsilon_{2i}$. Under the Gaussian setting, the E-step exploits this information in an optimal linear way, so the resulting imputation does not increase the asymptotic variance. However, under non-Gaussian designs or misspecified linear models, this favorable variance reduction is no longer guaranteed. In such settings, the imputed correction may introduce additional noise relative to the information it carries about $\varepsilon_{2i}$. Consequently, the SAEM estimator is no longer guaranteed to have asymptotic variance no larger than that of the supervised estimator and may even suffer an efficiency loss.

To address this issue, we introduce a weighting parameter $\lambda\in\mathbb{R}$ to adaptively balance the supervised and semi-supervised information. Let $\bbeta^{(0)}$ and $\widehat{\bbeta}$ denote the supervised and SAEM estimators of $\bbeta^*$ in Algorithm~\ref{alg:EM}, let $\widehat{\btheta}=\widehat{\bbeta}_1+\widehat{\balpha}\widehat{\bbeta}_2$ denote the SAEM estimator of $\btheta^*$, and let $\widehat{\btheta}_{\rm sup}=(\sum_{i\in\mathcal I}\bX_i\bX_i^\top)^{-1}\sum_{i\in\mathcal I}\bX_iY_i$ denote the supervised estimator of $\btheta^*$, that is, the labeled-only marginal ordinary least squares (OLS) estimator obtained by regressing $Y$ on $\bX$. We define
\begin{align}\label{eq:weighted_est}
\widehat{\bbeta}_{\lambda}=\bbeta^{(0)}+\lambda(\widehat{\bbeta}-\bbeta^{(0)}),\qquad
\widehat{\btheta}_{\lambda}=\widehat{\btheta}_{\rm sup}+\lambda(\widehat{\btheta}-\widehat{\btheta}_{\rm sup}).
\end{align}
Writing $\widehat{\bbeta}_{\lambda}=(\widehat{\bbeta}_{\lambda,1}^\top,\widehat{\bbeta}_{\lambda,2}^\top)^\top$, we estimate the two target parameters $\bbeta_1^*$ and $\btheta^*$ by $\widehat{\bbeta}_{\lambda,1}$ and $\widehat{\btheta}_{\lambda}$, respectively. Each estimator adds to the supervised estimator a multiple $\lambda$ of the semi-supervised correction, that is, of the difference between the SAEM and supervised estimators. In particular, $\lambda=0$ gives the supervised estimators, whereas $\lambda=1$ recovers the SAEM estimators.

We choose the weighting parameter to minimize the trace of the corresponding limiting covariance matrix. Define $\bv_i=\varepsilon_{2i}\bgamma_3^{*}+\bvarepsilon_{3i}$, $u_i=\bvarepsilon_{1i}^\top\bbeta_2^*$, $e_i=\varepsilon_{2i}+u_i$, $\bc^*=\sigma_2^2(\bSigma_3+\sigma_2^2\bgamma_3^{*}\bgamma_3^{*\top})^{-1}\bgamma_3^*$, and $q_i=u_i+\bc^{*\top}\bv_i$. Set $\bV_{\varepsilon_2}=E(\bar{\bX}_{1i}\bar{\bX}_{1i}^\top\varepsilon_{2i}^2)$, $\bV_{\bv}=E\{\bar{\bX}_{1i}\bar{\bX}_{1i}^\top(\bc^{*\top}\bv_i)^2\}$, $\bV_{\bv\varepsilon_2}=E\{\bar{\bX}_{1i}\bar{\bX}_{1i}^\top\bc^{*\top}\bv_i\varepsilon_{2i}\}$, $\bV_{e}=E(\bX_i\bX_i^\top e_i^2)$, $\bV_q=E(\bX_i\bX_i^\top q_i^2)$, and $\bV_{eq}=E(\bX_i\bX_i^\top e_iq_i)$. Let $\bV_{\bbeta_1,\lambda}$ and $\bV_{\btheta,\lambda}$ denote the limiting covariance matrices of $\sqrt n(\widehat{\bbeta}_{\lambda,1}-\bbeta_1^*)$ and $\sqrt n(\widehat{\btheta}_{\lambda}-\btheta^*)$, respectively. Then, $\bV_{\bbeta_1,\lambda}=[\bI_p,\mathbf{0}_{p\times d_1}]\bSigma_{\bX,1}^{-1}(\bV_{\varepsilon_2}+\tau\lambda^2\bV_{\bv}-2\tau\lambda\bV_{\bv\varepsilon_2})\bSigma_{\bX,1}^{-1}[\bI_p,\mathbf{0}_{p\times d_1}]^\top$ and $\bV_{\btheta,\lambda}=\bSigma_{\bX}^{-1}(\bV_{e}+\tau\lambda^2\bV_q-2\tau\lambda\bV_{eq})\bSigma_{\bX}^{-1}.$
For $\tau>0$, minimizing the traces of these matrices yields the oracle weighting parameters
\begin{align*}
\lambda_{\bbeta_1}^*
=\frac{\operatorname{tr}\left\{[\bI_p,\mathbf{0}_{p\times d_1}]\bSigma_{\bX,1}^{-1}\bV_{\bv\varepsilon_2}\bSigma_{\bX,1}^{-1}[\bI_p,\mathbf{0}_{p\times d_1}]^\top\right\}}
{\operatorname{tr}\left\{[\bI_p,\mathbf{0}_{p\times d_1}]\bSigma_{\bX,1}^{-1}\bV_{\bv}\bSigma_{\bX,1}^{-1}[\bI_p,\mathbf{0}_{p\times d_1}]^\top\right\}},\quad
\lambda_{\btheta}^*
=\frac{\operatorname{tr}(\bSigma_{\bX}^{-1}\bV_{eq}\bSigma_{\bX}^{-1})}
{\operatorname{tr}(\bSigma_{\bX}^{-1}\bV_q\bSigma_{\bX}^{-1})}.
\end{align*}
If the denominator defining either oracle weight is zero, we set that weight to zero. This includes $\bgamma_3^*=\mathbf{0}$ for $\lambda_{\bbeta_1}^*$ and $\bbeta_2^*=\bgamma_3^*=\mathbf{0}$ for $\lambda_{\btheta}^*$. Finally, when $\tau=0$, the limiting covariance matrices do not depend on $\lambda$, and the displayed ratios are retained as conventions whenever they are well-defined.

In practice, the optimal weighting parameters $\lambda^*_{\bbeta_1}$ and $\lambda_{\btheta}^*$ can be estimated from the observed data. Using the initial estimates $\{\bbeta^{(0)},(\sigma_2^{(0)})^2, \bgamma^{(0)},\bSigma_3^{(0)}\}$ defined in Algorithm~\ref{alg:EM}, we first set $\bc^{(0)}=(\sigma_2^{(0)})^2\{\bSigma_3^{(0)}+(\sigma_2^{(0)})^2\bgamma_3^{(0)}\bgamma_3^{(0)\top}\}^{-1}\bgamma_3^{(0)}$. For each labeled observation $i\in\mathcal{I}$, define the residuals $\hat{\varepsilon}_{2i}=Y_i-\bar{\bX}_{1i}^\top\bbeta^{(0)}$, $\hat{\bvarepsilon}_{3i}=\bS_{2i}-\bgamma^{(0)\top}\bar{\bX}_{2i}$, and $\hat{\bv}_i=\hat{\varepsilon}_{2i}\bgamma_3^{(0)}+\hat{\bvarepsilon}_{3i}$. Let $\widehat{\bSigma}_{\bX,1}=n^{-1}\sum_{i\in\mathcal{I}}\bar{\bX}_{1i}\bar{\bX}_{1i}^\top$, $\widehat{\bV}_{\bv\varepsilon_2}=n^{-1}\sum_{i\in\mathcal{I}}\bar{\bX}_{1i}\bar{\bX}_{1i}^\top\bc^{(0)\top}\hat{\bv}_i\hat{\varepsilon}_{2i}$, and $\widehat{\bV}_{\bv}=n^{-1}\sum_{i\in\mathcal{I}}\bar{\bX}_{1i}\bar{\bX}_{1i}^\top(\bc^{(0)\top}\hat{\bv}_i)^2$. We also define $\widehat{\bSigma}_{\bX,L}=n^{-1}\sum_{i\in\mathcal{I}}\bX_i\bX_i^\top$. For estimating $\btheta^*$, write $\bbeta^{(0)}=(\bbeta_1^{(0)\top},\bbeta_2^{(0)\top})^\top$ and let $\widehat\balpha_L=\widehat\bSigma_{\bX,L}^{-1}n^{-1}\sum_{i\in\mathcal I}\bX_i\bS_{1i}^\top$, $\hat u_i=(\bS_{1i}-\widehat\balpha_L^\top\bX_i)^\top\bbeta_2^{(0)}$, $\hat e_i=Y_i-\bX_i^\top(\bbeta_1^{(0)}+\widehat\balpha_L\bbeta_2^{(0)})$, and $\hat q_i=\hat u_i+\bc^{(0)\top}\hat\bv_i$. Define $\widehat\bV_q=n^{-1}\sum_{i\in\mathcal I}\bX_i\bX_i^\top\hat q_i^2$, and $\widehat\bV_{eq}=n^{-1}\sum_{i\in\mathcal I}\bX_i\bX_i^\top\hat e_i\hat q_i$. Write $\widehat D_{\bbeta_1}=\operatorname{tr}\{[\bI_p,\mathbf{0}_{p\times d_1}]\widehat\bSigma_{\bX,1}^{-1}\widehat\bV_{\bv}\widehat\bSigma_{\bX,1}^{-1}[\bI_p,\mathbf{0}_{p\times d_1}]^\top\}$ and $\widehat D_{\btheta}=\operatorname{tr}(\widehat\bSigma_{\bX,L}^{-1}\widehat\bV_q\widehat\bSigma_{\bX,L}^{-1})$ for the two estimated denominators. The optimal weighting parameters are then estimated by $\hat\lambda_{\bbeta_1}=1_{\widehat D_{\bbeta_1}>a_n}\operatorname{tr}\{[\bI_p,\mathbf{0}_{p\times d_1}]\widehat{\bSigma}_{\bX,1}^{-1}\widehat{\bV}_{\bv\varepsilon_2}\widehat{\bSigma}_{\bX,1}^{-1}[\bI_p,\mathbf{0}_{p\times d_1}]^\top\}/\widehat D_{\bbeta_1}$ and $\hat\lambda_{\btheta}=1_{\widehat D_{\btheta}>a_n}\operatorname{tr}(\widehat\bSigma_{\bX,L}^{-1}\widehat\bV_{eq}\widehat\bSigma_{\bX,L}^{-1})/\widehat D_{\btheta}$,
where $a_n$ satisfies $a_n\to 0$ and $na_n\to \infty$. The threshold $a_n$ is introduced to handle degenerate cases and guard against small denominators, thereby improving numerical stability.

The Optimally Weighted Surrogate-Assisted EM (OW-SAEM) estimators of $\bbeta_1^*$ and $\btheta^*$ are obtained by substituting the estimated weights into \eqref{eq:weighted_est}: they are $\widehat{\bbeta}_{\hat{\lambda}_{\bbeta_1},1}$, the first $p$ components of $\bbeta^{(0)}+\hat{\lambda}_{\bbeta_1}(\widehat{\bbeta}-\bbeta^{(0)})$, and $\widehat{\btheta}_{\hat{\lambda}_{\btheta}}=\widehat{\btheta}_{\rm sup}+\hat{\lambda}_{\btheta}(\widehat{\btheta}-\widehat{\btheta}_{\rm sup})$, respectively.

For the theoretical analysis in this subsection, we replace the Gaussian regularity condition by the following weaker moment condition.

\begin{assumption}[Non-Gaussian regularity]\label{ass:non_gaussian}
Suppose $\lambda_{\rm min}\{E(\bar{\bX}_{2i}\bar{\bX}_{2i}^\top)\}>0$, $E(\varepsilon_{2i}^2\mid \bar{\bX}_{1i})>0$, $\lambda_{\rm min}\{E(\bvarepsilon_{3i}\bvarepsilon_{3i}^\top\mid \bar{\bX}_{2i})\}>0$, and $\lambda_{\rm min}\{E(\bv_{i}\bv_{i}^\top\mid \bar{\bX}_{1i})\}>0$. Assume further that $E(\|\bar{\bX}_{2i}\|_2^4)<\infty$, $E(\varepsilon_{2i}^4\mid \bar{\bX}_{1i})\le C$, and $E(\|\bvarepsilon_{3i}\|_2^4\mid \bar{\bX}_{2i})\le C$ for some constant $C$ almost surely.
\end{assumption}
 
The following theorem establishes the theoretical properties of the OW-SAEM estimators.
 
\begin{theorem}\label{thm:weighted_targets}
Suppose Assumption~\ref{ass:non_gaussian} holds. For $T\in\{\bbeta_1,\btheta\}$, let $\widehat T_{\lambda}$ denote $\widehat{\bbeta}_{\lambda,1}$ when $T=\bbeta_1$ and $\widehat{\btheta}_{\lambda}$ when $T=\btheta$. Then,
$\sqrt n(\widehat T_{\hat\lambda_T}-T^*)\xrightarrow{\rm d}\mathcal N(\mathbf 0,\bV_{T,\lambda_T^*}).$
In particular, $\operatorname{tr}(\bV_{T,\lambda_T^*})\le\operatorname{tr}(\bV_{T,0})$, where $\bV_{T,0}$ denotes the asymptotic covariance matrix of the corresponding supervised estimator. The improvement is strict when $\tau>0$ and the numerator of the corresponding oracle ratio is nonzero. Moreover, if the denominator of the oracle ratio for $T$ is zero, then $P(\hat\lambda_T=0)\to1$. %
\end{theorem}

Theorem~\ref{thm:weighted_targets} shows that the OW-SAEM estimators attain the same limiting covariance matrices as their oracle counterparts that use the optimal weights $\lambda_T^*$. In terms of the trace of the limiting covariance matrix, OW-SAEM is no less efficient than the supervised estimator for both $\bbeta_1^*$ and $\btheta^*$. Thus, the proposed weighting strategy enables safe use of the additional unlabeled samples by guaranteeing asymptotic efficiency no worse than that of the corresponding supervised estimators, even under non-Gaussian distributions or misspecification of the linear working models.

For inference, we estimate $\bV_{T,\lambda}$ by $\widehat{\bV}_{T,\lambda}$, obtained by replacing $\tau$ with $m/N$ and the population moments in $\bV_{T,\lambda}$ with the labeled-sample moments defined above, together with $\widehat{\bV}_{\varepsilon_2}=n^{-1}\sum_{i\in\mathcal{I}}\bar{\bX}_{1i}\bar{\bX}_{1i}^\top\hat{\varepsilon}_{2i}^2$ and $\widehat{\bV}_{e}=n^{-1}\sum_{i\in\mathcal{I}}\bX_i\bX_i^\top\hat e_i^2$. The choices $\lambda=0$, $\lambda=1$, and $\lambda=\hat\lambda_T$ give the covariance estimators for the supervised, SAEM, and OW-SAEM estimators, respectively.

\section{Numerical Studies}
\subsection{Simulation Studies}\label{sec:simulation}
We evaluate the proposed methods under two data-generating scenarios. Setting~1 considers a correctly specified linear model under a Gaussian design and is used to assess the efficiency gains of the SAEM estimator introduced in Section~\ref{sec:1step_EM}, as well as the impact of surrogate misclassification and omission. Setting~2 considers model misspecification and is designed to represent a setting unfavorable to the original SAEM method, thereby further assessing whether the OW-SAEM estimators developed in Section~\ref{sec:optimal_lambda} can safely leverage additional unlabeled data. For comparison, we also implement a supervised learning (SL) estimator obtained by fitting OLS using only the labeled observations. Additional simulation results are provided in Section~\ref{sec:numerical} of the Supplementary Material. Each configuration is evaluated over 1,000 Monte Carlo replications.

\textbf{Setting 1 (Gaussian design):} Set $p=4$, $d_1=2$, and $d_2=3$. Generate $\bX_i\sim\mathcal{N}(\mathbf{0},\mathbf{I}_4)$, $\bvarepsilon_{1i}\sim\mathcal{N}(\mathbf{0},\mathbf{I}_2)$, $\varepsilon_{2i}\sim\mathcal{N}(0,1)$, and $\bvarepsilon_{3i}\sim\mathcal{N}(\mathbf{0},\mathbf{I}_3)$. The data-generating process follows \eqref{eq:models} with $\balpha^*=\mathbf{1}_{4\times 2}$, $\bbeta_1^*=\mathbf{1}_{4}$, $\bbeta_2^*=\mathbf{1}_{2}$, $\bgamma_1^*=\mathbf{1}_{4\times 3}$, $\bgamma_2^*=\mathbf{1}_{2\times 3}$, $\bgamma_3^*=\mathbf{1}_{3}$. %
 
To examine the effect of incorrect surrogate partitioning, we additionally consider three misclassified variants: SAEM-mis1, which swaps the roles of $\bS_1$ and $\bS_2$; SAEM-mis2, which assigns all surrogate variables to the post-outcome group, i.e., $\bS_1^\prime=\emptyset$ and $\bS_2^\prime=(\bS_1,\bS_2)$; and SAEM-mis3, which assigns all surrogate variables to the pre-outcome group, i.e., $\bS_1^\prime=(\bS_1,\bS_2)$ and $\bS_2^\prime=\emptyset$. We also consider two surrogate-omission variants: SAEM-omi1, which omits $\bS_1$ and sets $\bS_1^\prime=\emptyset$ and $\bS_2^\prime=\bS_2$, and SAEM-omi2, which omits $\bS_2$ and sets $\bS_1^\prime=\bS_1$ and $\bS_2^\prime=\emptyset$.
We report the mean squared error (MSE), the Cram\'{e}r--Rao lower bound (CRLB), the volume of the $95\%$ confidence ellipsoid (Volume), and the empirical coverage probability of the $95\%$ confidence ellipsoid (Coverage). %

\begin{table}[!htbp]
\centering
\caption{Simulation results under Setting 1 with fixed $m/n=10$ and varying $n$.}
\label{tab:table1}
\setlength{\tabcolsep}{3pt}
\renewcommand{\arraystretch}{1.10}
\begin{threeparttable}
\resizebox{\textwidth}{!}{%
\begin{tabular}{cl cccc cccc cccc}
\toprule
Target & Method & \multicolumn{4}{c}{$n=200$} & \multicolumn{4}{c}{$n=500$} & \multicolumn{4}{c}{$n=5000$} \\
\cmidrule(lr){3-6}\cmidrule(lr){7-10}\cmidrule(lr){11-14}
 & & MSE & CRLB & Volume & Coverage (\%) & MSE & CRLB & Volume & Coverage (\%) & MSE & CRLB & Volume & Coverage (\%) \\
\midrule
\multirow{7}{*}{$\bbeta_1^*$} & SL & 0.0608 & 0.0600 & 0.0332 & 94.2 & 0.0247 & 0.0240 & 0.0053 & 94.9 & 0.0024 & 0.0024 & $5.32e-5$ & 94.8 \\
 & SAEM & 0.0189 & 0.0191 & 0.0032 & 92.4 & 0.0075 & 0.0076 & $5.29e-4$ & 94.2 & $7.49e-4$ & $7.64e-4$ & $5.38e-6$ & 94.3 \\
 & SAEM-mis1 & 5.917 & 0.0191 & 0.0028 & 0.0 & 5.906 & 0.0076 & $4.63e-4$ & 0.0 & 5.893 & $7.64e-4$ & $4.72e-6$ & 0.0 \\
 & SAEM-mis2 & 16.013 & 0.0191 & 0.0027 & 0.0 & 16.006 & 0.0076 & $4.38e-4$ & 0.0 & 15.997 & $7.64e-4$ & $4.43e-6$ & 0.0 \\
 & SAEM-mis3 & 9.020 & 0.0191 & 0.0030 & 0.0 & 9.019 & 0.0076 & $4.97e-4$ & 0.0 & 8.991 & $7.64e-4$ & $5.07e-6$ & 0.0 \\
 & SAEM-omi1 & 16.013 & 0.0191 & 0.0034 & 0.0 & 16.007 & 0.0076 & $5.60e-4$ & 0.0 & 15.997 & $7.64e-4$ & $5.66e-6$ & 0.0 \\
 & SAEM-omi2 & 0.0608 & 0.0191 & 0.0318 & 92.7 & 0.0247 & 0.0076 & 0.0052 & 94.8 & 0.0024 & $7.64e-4$ & $5.31e-5$ & 94.5 \\
\addlinespace[5pt]
\multirow{7}{*}{$\btheta^*$} & SL & 0.0617 & 0.0600 & 0.100 & 95.1 & 0.0244 & 0.0240 & 0.0159 & 94.4 & 0.0024 & 0.0024 & $1.60e-4$ & 94.6 \\
 & SAEM & 0.0109 & 0.0100 & 0.0027 & 92.4 & 0.0040 & 0.0040 & $4.38e-4$ & 94.9 & $3.95e-4$ & $4.00e-4$ & $4.43e-6$ & 95.0 \\
 & SAEM-mis1 & 0.0109 & 0.0100 & 0.0027 & 92.4 & 0.0040 & 0.0040 & $4.38e-4$ & 94.9 & $3.95e-4$ & $4.00e-4$ & $4.43e-6$ & 95.0 \\
 & SAEM-mis2 & 0.0109 & 0.0100 & 0.0027 & 92.4 & 0.0040 & 0.0040 & $4.38e-4$ & 94.9 & $3.95e-4$ & $4.00e-4$ & $4.43e-6$ & 95.0 \\
 & SAEM-mis3 & 0.0109 & 0.0100 & 0.0027 & 92.4 & 0.0040 & 0.0040 & $4.38e-4$ & 94.9 & $3.95e-4$ & $4.00e-4$ & $4.43e-6$ & 95.0 \\
 & SAEM-omi1 & 0.0124 & 0.0100 & 0.0034 & 92.0 & 0.0045 & 0.0040 & $5.60e-4$ & 94.4 & $4.50e-4$ & $4.00e-4$ & $5.66e-6$ & 94.9 \\
 & SAEM-omi2 & 0.0252 & 0.0100 & 0.0149 & 93.3 & 0.0093 & 0.0040 & 0.0024 & 95.0 & $9.50e-4$ & $4.00e-4$ & $2.48e-5$ & 93.8 \\
\addlinespace[5pt]
\bottomrule
\end{tabular}%
}
\end{threeparttable}
\end{table}

\begin{table}[!htbp]
\centering
\caption{Simulation results under Setting 1 with fixed $n=500$ and varying $m$.}
\label{tab:table2}
\setlength{\tabcolsep}{3pt}
\renewcommand{\arraystretch}{1.10}
\begin{threeparttable}
\resizebox{\textwidth}{!}{%
\begin{tabular}{cl cccc cccc cccc}
\toprule
Target & Method & \multicolumn{4}{c}{$m=500$} & \multicolumn{4}{c}{$m=5000$} & \multicolumn{4}{c}{$m=50000$} \\
\cmidrule(lr){3-6}\cmidrule(lr){7-10}\cmidrule(lr){11-14}
 & & MSE & CRLB & Volume & Coverage (\%) & MSE & CRLB & Volume & Coverage (\%) & MSE & CRLB & Volume & Coverage (\%) \\
\midrule
\multirow{7}{*}{$\bbeta_1^*$} & SL & 0.0247 & 0.0240 & 0.0053 & 94.9 & 0.0247 & 0.0240 & 0.0053 & 94.9 & 0.0247 & 0.0240 & 0.0053 & 94.9 \\
 & SAEM & 0.0153 & 0.0150 & 0.0021 & 94.6 & 0.0075 & 0.0076 & $5.29e-4$ & 94.2 & 0.0061 & 0.0062 & $3.44e-4$ & 94.1 \\
 & SAEM-mis1 & 5.909 & 0.0150 & $5.82e-4$ & 0.0 & 5.906 & 0.0076 & $4.63e-4$ & 0.0 & 5.909 & 0.0062 & $4.41e-4$ & 0.0 \\
 & SAEM-mis2 & 16.017 & 0.0150 & 0.0047 & 0.0 & 16.006 & 0.0076 & $4.38e-4$ & 0.0 & 15.997 & 0.0062 & $1.33e-4$ & 0.0 \\
 & SAEM-mis3 & 9.019 & 0.0150 & $5.03e-4$ & 0.0 & 9.019 & 0.0076 & $4.97e-4$ & 0.0 & 9.019 & 0.0062 & $4.95e-4$ & 0.0 \\
 & SAEM-omi1 & 16.019 & 0.0150 & 0.0049 & 0.0 & 16.007 & 0.0076 & $5.60e-4$ & 0.0 & 15.997 & 0.0062 & $2.11e-4$ & 0.0 \\
 & SAEM-omi2 & 0.0247 & 0.0150 & 0.0053 & 94.9 & 0.0247 & 0.0076 & 0.0052 & 94.8 & 0.0247 & 0.0062 & 0.0052 & 94.7 \\
\addlinespace[5pt]
\multirow{7}{*}{$\btheta^*$} & SL & 0.0244 & 0.0240 & 0.0159 & 94.4 & 0.0244 & 0.0240 & 0.0159 & 94.4 & 0.0244 & 0.0240 & 0.0159 & 94.4 \\
 & SAEM & 0.0130 & 0.0130 & 0.0047 & 95.1 & 0.0040 & 0.0040 & $4.38e-4$ & 94.9 & 0.0022 & 0.0022 & $1.33e-4$ & 94.4 \\
 & SAEM-mis1 & 0.0130 & 0.0130 & 0.0047 & 95.1 & 0.0040 & 0.0040 & $4.38e-4$ & 94.9 & 0.0022 & 0.0022 & $1.33e-4$ & 94.4 \\
 & SAEM-mis2 & 0.0130 & 0.0130 & 0.0047 & 95.1 & 0.0040 & 0.0040 & $4.38e-4$ & 94.9 & 0.0022 & 0.0022 & $1.33e-4$ & 94.4 \\
 & SAEM-mis3 & 0.0130 & 0.0130 & 0.0047 & 95.1 & 0.0040 & 0.0040 & $4.38e-4$ & 94.9 & 0.0022 & 0.0022 & $1.33e-4$ & 94.4 \\
 & SAEM-omi1 & 0.0132 & 0.0130 & 0.0049 & 95.1 & 0.0045 & 0.0040 & $5.60e-4$ & 94.4 & 0.0028 & 0.0022 & $2.11e-4$ & 95.0 \\
 & SAEM-omi2 & 0.0160 & 0.0130 & 0.0070 & 94.8 & 0.0093 & 0.0040 & 0.0024 & 95.0 & 0.0080 & 0.0022 & 0.0018 & 94.6 \\
\addlinespace[5pt]
\bottomrule
\end{tabular}%
}
\end{threeparttable}
\end{table}

Table~\ref{tab:table1} compares SL, SAEM, and the SAEM variants with misclassified or omitted surrogates as the labeled sample size $n$ increases, while holding the ratio $m/n=10$ fixed. For estimation of both $\bbeta_1^*$ and $\btheta^*$, when the surrogate roles are correctly classified, SAEM consistently reduces the empirical MSE and the volume of the $95\%$ confidence ellipsoid relative to SL, with the empirical MSE approaching the corresponding Cram\'{e}r--Rao lower bound as $n$ increases. The misclassified variants exhibit markedly different patterns for $\bbeta_1^*$ and $\btheta^*$. For $\bbeta_1^*$, incorrect surrogate partitioning leads to substantially larger MSE and essentially zero coverage. In contrast, for $\btheta^*$, all three misclassified variants yield exactly the same results as the correctly specified SAEM estimator. This occurs because these variants and SAEM use the same set of surrogate variables and therefore produce the same estimator, as established in Theorem~\ref{thm:thm_theta}. The surrogate-omission variants also exhibit distinct behavior. For estimating $\bbeta_1^*$, SAEM-omi1 omits $\bS_1$ and therefore converges to $\btheta^*$ rather than $\bbeta_1^*$, resulting in large MSE and zero coverage. SAEM-omi2 omits $\bS_2$ and has the same asymptotic behavior as SL for estimating $\bbeta_1^*$, and is therefore less efficient than SAEM with the full set of surrogates, as discussed in Remark~\ref{remark:beta1}; indeed, its finite-sample performance is similar to that of SL. For $\btheta^*$, both omission variants remain consistent but have larger MSE than SAEM, consistent with the results in Theorem~\ref{thm:thm_theta}(ii).

Table~\ref{tab:table2} fixes $n=500$ and varies the number of unlabeled observations. For estimation of both $\bbeta_1^*$ and $\btheta^*$, SAEM consistently achieves smaller MSE than SL, with larger unlabeled sample sizes $m$ leading to further reductions in MSE. Taking estimation of $\bbeta_1^*$ as an illustration, increasing $m$ from 500 to 5000 reduces the MSE by approximately $50.98\%$, whereas increasing $m$ further from 5000 to 50000 yields only a reduction of approximately $18.67\%$. A similar pattern is observed for estimating $\btheta^*$: increasing $m$ produces more substantial MSE reductions when the unlabeled sample size is relatively small. This observation is consistent with Theorem~\ref{thm:upper_bound} and the discussion in Remark~\ref{remark:compare-sup}. For estimation of both $\bbeta_1^*$ and $\btheta^*$, incorporating unlabeled observations can reduce the asymptotic variance, while the convergence rate remains governed by the labeled sample size $n$. Regarding the misclassified SAEM variants, they again yield identical estimators for $\btheta^*$, whereas estimation of $\bbeta_1^*$ exhibits substantially larger errors. The omission variants show the same pattern for $\bbeta_1^*$ at every $m$. For $\btheta^*$, their MSE gap to SAEM widens as $m$ increases, since the efficiency loss in Theorem~\ref{thm:thm_theta}(ii) is proportional to $\tau$.

\begin{table}[!htbp]
\centering
\caption{Simulation results under Setting 2 with fixed $m/n=10$ and varying $(n,c_{\bS_2})$.}
\label{tab:table3}
\setlength{\tabcolsep}{2.5pt}
\renewcommand{\arraystretch}{1.05}
\resizebox{\textwidth}{!}{%
\begin{tabular}{ll ccc ccc ccc}
\toprule
 & & \multicolumn{3}{c}{$c_{\bS_2}=0$} & \multicolumn{3}{c}{$c_{\bS_2}=2.5$} & \multicolumn{3}{c}{$c_{\bS_2}=4$} \\
\cmidrule(lr){3-5}\cmidrule(lr){6-8}\cmidrule(lr){9-11}
Size & Method & Coverage (\%) & MSE & Volume & Coverage (\%) & MSE & Volume & Coverage (\%) & MSE & Volume \\
\midrule
\multicolumn{11}{l}{\textit{Target: $\bbeta_1^*$}} \\
$n=500$ & SL & 93.6 & 0.0244 & 0.0095 & 93.6 & 0.0244 & 0.0095 & 93.6 & 0.0244 & 0.0095 \\
 & SAEM & 95.5 & 0.0027 & $1.27e-4$ & 89.8 & 0.0302 & 0.0051 & 90.7 & 0.0394 & 0.0102 \\
 & OW-SAEM & 93.9 & 0.0027 & $1.19e-4$ & 92.3 & 0.0247 & 0.0070 & 93.5 & 0.0235 & 0.0103 \\
 & \footnotesize Avg. $\hat\lambda_{\bbeta_1}$ & \multicolumn{3}{c}{\footnotesize 1.011} & \multicolumn{3}{c}{\footnotesize 0.316} & \multicolumn{3}{c}{\footnotesize $-0.536$} \\
\addlinespace[3pt]
$n=2000$ & SL & 95.2 & 0.0064 & $6.36e-4$ & 95.2 & 0.0064 & $6.36e-4$ & 95.2 & 0.0064 & $6.36e-4$ \\
 & SAEM & 95.1 & $6.68e-4$ & $8.03e-6$ & 92.7 & 0.0077 & $3.85e-4$ & 93.4 & 0.0095 & $7.50e-4$ \\
 & OW-SAEM & 94.8 & $6.67e-4$ & $7.89e-6$ & 94.4 & 0.0063 & $4.96e-4$ & 94.3 & 0.0059 & $6.79e-4$ \\
 & \footnotesize Avg. $\hat\lambda_{\bbeta_1}$ & \multicolumn{3}{c}{\footnotesize 1.012} & \multicolumn{3}{c}{\footnotesize 0.266} & \multicolumn{3}{c}{\footnotesize $-0.575$} \\
\addlinespace[3pt]
$n=5000$ & SL & 93.0 & 0.0026 & $1.04e-4$ & 93.0 & 0.0026 & $1.04e-4$ & 93.0 & 0.0026 & $1.04e-4$ \\
 & SAEM & 94.6 & $2.68e-4$ & $1.29e-6$ & 94.6 & 0.0029 & $6.40e-5$ & 93.6 & 0.0037 & $1.24e-4$ \\
 & OW-SAEM & 94.5 & $2.68e-4$ & $1.28e-6$ & 92.8 & 0.0025 & $8.12e-5$ & 93.6 & 0.0025 & $1.10e-4$ \\
 & \footnotesize Avg. $\hat\lambda_{\bbeta_1}$ & \multicolumn{3}{c}{\footnotesize 1.012} & \multicolumn{3}{c}{\footnotesize 0.262} & \multicolumn{3}{c}{\footnotesize $-0.582$} \\
\addlinespace[3pt]
\addlinespace[4pt]
\multicolumn{11}{l}{\textit{Target: $\btheta^*$}} \\
$n=500$ & SL & 94.1 & 0.0238 & 0.0103 & 94.1 & 0.0238 & 0.0103 & 94.1 & 0.0238 & 0.0103 \\
 & SAEM & 95.1 & 0.0024 & $1.22e-4$ & 89.6 & 0.0276 & 0.0036 & 90.9 & 0.0355 & 0.0070 \\
 & OW-SAEM & 93.0 & 0.0024 & $1.15e-4$ & 91.7 & 0.0231 & 0.0056 & 93.9 & 0.0240 & 0.0098 \\
 & \footnotesize Avg. $\hat\lambda_{\btheta}$ & \multicolumn{3}{c}{\footnotesize 1.011} & \multicolumn{3}{c}{\footnotesize 0.416} & \multicolumn{3}{c}{\footnotesize 0.080} \\
\addlinespace[3pt]
$n=2000$ & SL & 95.1 & 0.0062 & $6.78e-4$ & 95.1 & 0.0062 & $6.78e-4$ & 95.1 & 0.0062 & $6.78e-4$ \\
 & SAEM & 94.2 & $6.31e-4$ & $7.71e-6$ & 92.8 & 0.0071 & $2.74e-4$ & 93.4 & 0.0085 & $5.14e-4$ \\
 & OW-SAEM & 94.2 & $6.30e-4$ & $7.59e-6$ & 93.7 & 0.0059 & $4.05e-4$ & 94.9 & 0.0062 & $6.72e-4$ \\
 & \footnotesize Avg. $\hat\lambda_{\btheta}$ & \multicolumn{3}{c}{\footnotesize 1.012} & \multicolumn{3}{c}{\footnotesize 0.371} & \multicolumn{3}{c}{\footnotesize 0.018} \\
\addlinespace[3pt]
$n=5000$ & SL & 93.5 & 0.0024 & $1.10e-4$ & 93.5 & 0.0024 & $1.10e-4$ & 93.5 & 0.0024 & $1.10e-4$ \\
 & SAEM & 94.4 & $2.48e-4$ & $1.24e-6$ & 94.3 & 0.0027 & $4.56e-5$ & 94.3 & 0.0032 & $8.50e-5$ \\
 & OW-SAEM & 93.8 & $2.48e-4$ & $1.23e-6$ & 93.3 & 0.0023 & $6.61e-5$ & 93.7 & 0.0024 & $1.10e-4$ \\
 & \footnotesize Avg. $\hat\lambda_{\btheta}$ & \multicolumn{3}{c}{\footnotesize 1.012} & \multicolumn{3}{c}{\footnotesize 0.368} & \multicolumn{3}{c}{\footnotesize 0.010} \\
\bottomrule
\end{tabular}%
}
\end{table}

\textbf{Setting 2 (Non-Gaussian design and model misspecification):} Set $p=4$, $d_1=2$, $d_2=3$. Let $X_{i1}=D_iR_i/\sqrt{0.05}$, where $D_i\sim\mathrm{Bernoulli}(0.05)$ and $R_i$ takes the values $\pm1$ with probability $1/2$ each, and generate $X_{i2},X_{i3},X_{i4}\sim\mathrm{Unif}(-\sqrt{3},\sqrt{3})$, $\bvarepsilon_{1i}\sim\mathcal{N}(\mathbf{0},\mathbf{I}_2)$, $\xi_{2i}\sim\mathrm{Unif}(-\sqrt{3},\sqrt{3})$, and $\boldsymbol{\xi}_{3i}\sim\mathcal{N}(\mathbf{0},\mathbf{I}_3)$, all mutually independent. The data are generated from
\begin{equation}\label{eq:model2_dgp}
 \begin{split}
 \bS_{1i}&=\balpha^{*\top}\bX_i+\bvarepsilon_{1i},\\
 Y_i&=\bbeta_1^{*\top}\bX_i+\bbeta_2^{*\top}\bS_{1i}+c_Y h(\bX_i)+\xi_{2i},\\
 \bS_{2i}&=\mathbf{1}_{4\times 3}^{\top}\bX_i+\mathbf{1}_{2\times 3}^{\top}\bS_{1i}+3Y_i\mathbf{1}_{3}-c_{\bS_2} h(\bX_i)\mathbf{1}_{3}+\boldsymbol{\xi}_{3i},
 \end{split}
\end{equation}
where $h(\bX_i)=X_{i1}X_{i2}$, $\balpha^*=0.5\cdot\mathbf{1}_{4\times 2}$, $\bbeta_1^*=\mathbf{1}_{4}$, and $\bbeta_2^*=0.5\cdot\mathbf{1}_{2}$. We fix $c_Y=0.5$ and vary $c_{\bS_2}\in\{0,2.5,4\}$ to control the strength of the nonlinear signal.

Table~\ref{tab:table3} examines the misspecified setting with $m/n=10$ and $n\in\{500,2000,5000\}$. When $c_{\bS_2}=0$, the SAEM estimators of both $\bbeta_1^*$ and $\btheta^*$ achieve smaller MSE than their SL counterparts. In this case, the estimated weights $\hat\lambda_{\bbeta_1}$ and $\hat\lambda_{\btheta}$ are both close to one, so the OW-SAEM estimators nearly coincide with the corresponding SAEM estimators. When $c_{\bS_2}\in\{2.5,4\}$, however, the SAEM estimators of both $\bbeta_1^*$ and $\btheta^*$ exhibit larger MSE than the corresponding SL estimators. In these cases, the estimated weights move away from one, and OW-SAEM achieves MSE comparable to or smaller than that of SL for both targets. Notably, for estimating $\bbeta_1^*$ when $c_{\bS_2}=4$, the average $\hat\lambda_{\bbeta_1}$ is negative, indicating that the semi-supervised correction should enter with the opposite sign and that directly incorporating the unlabeled samples may be detrimental. Consistent with this observation, SAEM exhibits substantially larger MSE than SL, whereas OW-SAEM achieves smaller MSE than SL, particularly when $n$ is small. In contrast, for estimating $\btheta^*$ when $c_{\bS_2}=4$, the average $\hat\lambda_{\btheta}$ is close to zero, suggesting that little or no information from the unlabeled data should be incorporated. Accordingly, SAEM again exhibits larger MSE than SL, whereas the near-zero weight makes OW-SAEM perform similarly to SL, thereby maintaining safe use of the unlabeled samples.

\subsection{Real-data Applications}\label{sec:real_data}
In this section, we apply the proposed estimators to the Medical Information Mart for Intensive Care III (MIMIC-III) database~\citep{johnson2016mimic} to illustrate their empirical performance. MIMIC-III is a large, publicly available critical care database containing de-identified health records for more than 40,000 ICU patients admitted to Beth Israel Deaconess Medical Center between 2001 and 2012. Following the data-processing pipeline of \cite{komorowski2018artificial}, we identify a sepsis cohort of 18,911 ICU stays, with clinical records represented as discrete time series measured at 4-hour intervals. Hereafter, each 4-hour interval is referred to as a block.

The baseline covariates $\bX$ are extracted from block 1 and include demographic characteristics and initial clinical measurements. The outcome $Y$ is the Sequential Organ Failure Assessment (SOFA) score \citep{vincent1996sofa} at block 6, corresponding to approximately one day after the first record and serving as a mid-stay measure of disease severity. We focus on two questions: (a) how the baseline variables directly affect the SOFA score at block 6, and (b) how well the SOFA score at block 6 can be predicted using only the baseline variables. These questions correspond to estimation of $\bbeta_1^*$ and $\btheta^*$, respectively. Additional longitudinal information collected at other stages is incorporated as surrogate information to improve estimation of these quantities. The pre-outcome surrogates $\bS_1$ are constructed by averaging severity scores and treatment values over blocks 2--5, corresponding to the first 20 hours, while the post-outcome surrogates $\bS_2$ are derived from blocks 7--12 and capture the subsequent clinical trajectory. A summary of the variables is provided in Table~\ref{tab:mimic_variables}. We restrict the analysis to ICU stays of patients who survived to hospital discharge and whose records include block 1, block 6, and at least one block in each of the windows 2--5 and 7--12. This yields a final analysis sample of 15,543 ICU stays. All variables are standardized before analysis, and each surrogate is averaged over the observed blocks within its corresponding window.

\begin{table}[!htbp]
\centering
\caption{Variable construction in the MIMIC sepsis data analysis.}
\label{tab:mimic_variables}
\small
\setlength{\tabcolsep}{4pt}
\renewcommand{\arraystretch}{1.12}
\begin{tabularx}{0.96\textwidth}{@{}c l >{\raggedright\arraybackslash}X c@{}}
\toprule
\textbf{Variable} & \textbf{Time period} & \textbf{Contents} & \textbf{Dimension} \\
\midrule
$\bX$ 
& Block 1 (baseline) 
& Demographics, vital signs, and laboratory values 
& $p=47$ \\
$\bS_1$ 
& Blocks 2--5 
& Mean SOFA, SIRS, total input, and mechanical ventilation 
& $d_1=4$ \\
$Y$ 
& Block 6 
& SOFA score 
& $1$ \\
$\bS_2$ 
& Blocks 7--12 
& Mean SOFA, SIRS, total input, and mechanical ventilation 
& $d_2=4$ \\
\bottomrule
\end{tabularx}
\end{table}

\begin{table}[!htbp]
\centering
\caption{MIMIC sepsis data: estimation performance across missing rates.}
\label{tab:MIMIC_results_merged}
\small
\setlength{\tabcolsep}{6pt}
\renewcommand{\arraystretch}{1.08}
\begin{tabular}{llrrrr}
\toprule
Quantity & Method & Missing 10\% & Missing 30\% & Missing 50\% & Missing 70\% \\
\midrule
$\mathrm{MSE}(\widehat{\bbeta}_1)\times 10^3$ & SL & 0.025 & 0.098 & 0.234 & 0.572 \\
 & SAEM & 0.020 & 0.077 & 0.186 & 0.448 \\
 & OW-SAEM & 0.020 & 0.077 & 0.185 & 0.449 \\
\addlinespace[3pt]
$\mathrm{MSE}(\widehat{\btheta})\times 10^3$ & SL & 0.034 & 0.136 & 0.376 & 0.956 \\
 & SAEM & 0.013 & 0.050 & 0.122 & 0.296 \\
 & OW-SAEM & 0.013 & 0.050 & 0.122 & 0.297 \\
\midrule
Avg. $\hat\lambda_{\bbeta_1}$ & OW-SAEM & 0.974 & 0.972 & 0.971 & 0.973 \\
Avg. $\hat\lambda_{\btheta}$ & OW-SAEM & 1.004 & 1.002 & 0.997 & 0.995 \\
\bottomrule
\end{tabular}
\end{table}

We simulate a semi-supervised setting by artificially masking the outcome $Y$ for a proportion of ICU stays. This design provides a full-data SL benchmark and allows us to systematically evaluate the performance of the semi-supervised methods across different labeled-to-unlabeled ratios. We consider missing rates of $10\%$, $30\%$, $50\%$, and $70\%$, corresponding to progressively smaller labeled sample sizes, and conduct 500 Monte Carlo replications over random masking patterns. We apply SL, SAEM, and OW-SAEM to the resulting semi-supervised datasets. SL uses only the labeled observations, whereas SAEM and OW-SAEM additionally incorporate the unlabeled observations.

Table~\ref{tab:MIMIC_results_merged} reports the estimation performance of SL, SAEM, and OW-SAEM relative to the full-data OLS estimator, which uses the outcomes from all 15,543 ICU stays. Compared with SL, SAEM consistently reduces the MSE for estimating both $\bbeta_1^*$ and $\btheta^*$. These gains become more pronounced as the missing rate increases while the total sample size remains fixed. As the labeled sample size decreases, both SL and SAEM become less accurate; however, the larger unlabeled sample available at higher missing rates enables SAEM to achieve increasingly substantial improvements over SL. The OW-SAEM estimators of both $\bbeta_1^*$ and $\btheta^*$ perform similarly to their corresponding SAEM estimators, with estimated weights close to one. This suggests that, for the data considered here, directly incorporating the unlabeled information is already effective and additional reweighting provides little further benefit.

\begin{table}[!htbp]
\centering
\caption{MIMIC-III data: percentage of 500 random masks in which SL and SAEM declare each covariate in $\btheta^*$ significant at the $5\%$ level. Covariates are grouped by their significance in the full-data benchmark and ordered by the benchmark $p$-value.}
\label{tab:MIMIC_clinical_comparison}
\small
\setlength{\tabcolsep}{4pt}
\begin{tabular*}{\textwidth}{@{\extracolsep{\fill}}lrrrrrrrr@{}}
\toprule
 & \multicolumn{2}{c}{Missing 10\%} & \multicolumn{2}{c}{Missing 30\%} & \multicolumn{2}{c}{Missing 50\%} & \multicolumn{2}{c}{Missing 70\%}\\
\cmidrule(lr){2-3}\cmidrule(lr){4-5}\cmidrule(lr){6-7}\cmidrule(lr){8-9}
Covariate & SL & SAEM & SL & SAEM & SL & SAEM & SL & SAEM \\
\midrule

\multicolumn{9}{@{}l}{\itshape Significant in the full-data benchmark (20 covariates)}\\
Glucose & 100 & 100 & 100 & 100 & 100 & 100 & 98 & 100 \\
Median vasopressor dose & 100 & 100 & 99 & 100 & 92 & 100 & 77 & 99 \\
Gender & 100 & 100 & 100 & 100 & 100 & 100 & 87 & 100 \\
Hemoglobin & 100 & 100 & 100 & 100 & 98 & 100 & 79 & 99 \\
Total output & 100 & 100 & 100 & 100 & 99 & 100 & 86 & 100 \\
FiO$_2$ & 100 & 100 & 100 & 100 & 98 & 100 & 80 & 97 \\
PTT & 100 & 100 & 100 & 100 & 94 & 100 & 69 & 92 \\
Albumin & 100 & 100 & 98 & 100 & 79 & 96 & 52 & 78 \\
BUN & 98 & 100 & 88 & 100 & 75 & 100 & 52 & 96 \\
Heart rate & 96 & 100 & 69 & 91 & 45 & 73 & 25 & 50 \\
Arterial pH & 93 & 99 & 66 & 81 & 42 & 61 & 29 & 41 \\
Creatinine & 100 & 94 & 83 & 97 & 65 & 100 & 48 & 100 \\
Shock index & 84 & 96 & 54 & 72 & 39 & 55 & 20 & 39 \\
Max.\ vasopressor dose & 92 & 85 & 75 & 71 & 62 & 53 & 55 & 45 \\
Remaining 6 (min.) & 100 & 100 & 100 & 100 & 100 & 100 & 100 & 100 \\

\addlinespace[3pt]
\multicolumn{9}{@{}l}{\itshape Not significant in the full-data benchmark (27 covariates)}\\
Potassium & 19 & 16 & 22 & 23 & 16 & 17 & 8 & 18 \\
Respiratory rate & 16 & 14 & 21 & 22 & 18 & 22 & 13 & 19 \\
PaCO$_2$ & 12 & 6 & 18 & 16 & 18 & 17 & 14 & 17 \\
Lactate & 9 & 6 & 15 & 14 & 13 & 16 & 12 & 13 \\
Diastolic BP & 11 & 3 & 20 & 11 & 16 & 7 & 13 & 6 \\
Mean BP & 3 & 1 & 9 & 7 & 8 & 11 & 8 & 10 \\
Remaining 21 (max.) & 2 & 0 & 4 & 2 & 6 & 5 & 8 & 6 \\
\midrule

TPR & 98.2 & 98.7 & 91.6 & 95.6 & 84.4 & 91.9 & 72.9 & 86.9 \\
FDR & 3.4 & 2.1 & 6.0 & 4.7 & 7.3 & 5.7 & 9.0 & 7.1 \\

\bottomrule
\end{tabular*}
\end{table}

We also compare the covariates identified as significant for the prediction target $\btheta^*$ by each method with those identified by the full-data benchmark, using a $5\%$ significance level across different missing rates. Table~\ref{tab:MIMIC_clinical_comparison} reports the percentage of random masks under which each covariate is declared significant, together with the true positive rate (TPR), defined as the proportion of benchmark-significant covariates that are selected, and the false discovery rate (FDR), defined as the proportion of selected covariates that are not significant under the benchmark. Both metrics are averaged over the random masks. We report results only for SL and SAEM because OW-SAEM yields results very similar to those of SAEM. At the $10\%$ missing rate, both methods select most benchmark-significant covariates while rarely selecting covariates that are nonsignificant under the benchmark. As the missing rate increases, SL loses benchmark-significant covariates and incorrectly selects benchmark-nonsignificant covariates more rapidly than SAEM. Accordingly, SAEM achieves a higher TPR and a lower FDR than SL across all considered missing rates. For example, blood urea nitrogen (BUN) is a marker of renal function, and sepsis-associated acute kidney injury affects approximately $40$--$50\%$ of patients with sepsis and is strongly associated with adverse outcomes \citep{gomez2016sepsisaki,zarbock2023saaki}. BUN is identified as an important predictor by the full-data benchmark. At the $70\%$ missing rate, however, SL selects BUN in only $52\%$ of the random repetitions, whereas SAEM selects it in $96\%$.

We further evaluate the prediction performance of the considered methods. For this analysis, we randomly split the data into training ($80\%$) and test ($20\%$) sets in each of 500 Monte Carlo replications. We then apply the same missing rates by masking outcomes in the training set. SL, SAEM, and OW-SAEM are fitted to the resulting semi-supervised training data, and their prediction MSEs are evaluated against the observed $Y$ values in the held-out test set. Predictions are constructed using only the baseline covariates $\bX$ and the estimated $\btheta^*$. Table~\ref{tab:MIMIC_prediction} reports the results, together with the prediction MSE of the OLS fit that uses the outcomes of all training stays. SAEM and OW-SAEM achieve smaller prediction MSE than SL at every missing rate, particularly when it is high.

\begin{table}[!htbp]
\centering
\caption{MIMIC sepsis data: test-set prediction MSE across missing rates. Full-label OLS is the OLS estimator that uses the outcomes of all training stays.}
\label{tab:MIMIC_prediction}
\small
\setlength{\tabcolsep}{6pt}
\renewcommand{\arraystretch}{1.08}
\begin{tabular}{llrrrr}
\toprule
Quantity & Method & Missing 10\% & Missing 30\% & Missing 50\% & Missing 70\% \\
\midrule
Pred MSE & Full-label OLS & 0.857 & 0.857 & 0.857 & 0.857 \\
 & SL & 0.859 & 0.865 & 0.875 & 0.894 \\
 & SAEM & 0.857 & 0.858 & 0.860 & 0.864 \\
 & OW-SAEM & 0.857 & 0.858 & 0.859 & 0.864 \\
\midrule
Avg. $\hat\lambda_{\btheta}$ & OW-SAEM & 1.003 & 1.000 & 0.997 & 0.998 \\
\bottomrule
\end{tabular}
\end{table}

Additional results for the MIMIC-III analysis, together with an additional real-data analysis using the Mayo Clinic primary biliary cholangitis (PBC) data, are provided in Section~\ref{sec:real_data_supplement} of the Supplementary Material. Unlike the MIMIC-III analysis, the PBC analysis illustrates the application of the proposed methods to an actual semi-supervised dataset without artificially masking the outcomes.

\section{Discussion}
This paper studies semi-supervised linear regression with surrogate variables. We distinguish between surrogates observed before the outcome and those observed after the outcome, and show how this temporal ordering can be exploited for estimation and prediction. We further demonstrate that surrogate misclassification affects the two targets differently: $\bbeta_1^*$ is sensitive to incorrect partitioning or omission of surrogates, whereas $\btheta^*$ is robust to surrogate misclassification but may suffer efficiency loss when relevant surrogates are omitted.

Several directions merit further investigation. First, the current paper focuses on a low-dimensional linear working model, although an optimally weighted strategy is proposed to ensure safe use of the unlabeled samples under model misspecification. It would nevertheless be valuable to extend the proposed framework to more flexible settings, including nonparametric models and high-dimensional regimes. Second, our analysis assumes that the labeled and unlabeled samples are drawn from the same population. In many applications, however, the labeling mechanism may be more complex, and it would be useful to further study how surrogates can be incorporated in the presence of selection bias \citep{zhang2023double,zhang2026decaying,testa2025semiparametric}. Third, the present framework assumes that the outcome $Y$ is missing in the unlabeled sample, while the surrogate variables $\bS_1$ and $\bS_2$ are fully observed. In longitudinal studies, a more challenging setting arises when $Y$, $\bS_1$, and $\bS_2$ are all subject to missingness. Developing estimators for such general incomplete-data structures would require accommodating substantially more complex missing-data mechanisms and represents another important direction for future research.

\section*{Supplementary material}\label{SM}

The Supplementary Material contains additional numerical results and proofs of all main theoretical results. Sections~\ref{sec:numerical} and~\ref{sec:real_data_supplement} present additional simulation and real-data analysis results, respectively. Auxiliary lemmas and proofs of the main results are provided in Sections~\ref{sec:lemmas} and~\ref{sec:proofs}.

\bibliographystyle{plainnat}
\bibliography{paper-ref}

\appendix

\clearpage\newpage

\renewcommand\thesection{S\arabic{section}}
\renewcommand\thesubsection{\thesection.\arabic{subsection}}
\setcounter{equation}{0}
\setcounter{theorem}{0}
\setcounter{lemma}{0}
\setcounter{assumption}{0}
\setcounter{table}{0}
\setcounter{figure}{0}
\renewcommand{\theequation}{S\arabic{equation}}
\renewcommand{\thetheorem}{S\arabic{theorem}}
\renewcommand{\thelemma}{S\arabic{lemma}}
\renewcommand{\theassumption}{S\arabic{assumption}}
\renewcommand{\thetable}{S\arabic{table}}
\renewcommand{\thefigure}{S\arabic{figure}}
\renewcommand{\theHsection}{S\arabic{section}}
\renewcommand{\theHequation}{S\arabic{equation}}
\renewcommand{\theHtheorem}{S\arabic{theorem}}
\renewcommand{\theHlemma}{S\arabic{lemma}}
\renewcommand{\theHassumption}{S\arabic{assumption}}
\renewcommand{\theHtable}{S\arabic{table}}
\renewcommand{\theHfigure}{S\arabic{figure}}

\par\bigskip

\begin{center}
\textbf{\uppercase{Supplementary Material to ``The roles of pre- and post-outcome surrogates in semi-supervised linear regression''}}
\end{center}

The Supplementary Material contains additional numerical results and proofs of all main theoretical results. Sections~\ref{sec:numerical} and~\ref{sec:real_data_supplement} present additional simulation and real-data analysis results, respectively. Auxiliary lemmas and proofs of the main results are provided in Sections~\ref{sec:lemmas} and~\ref{sec:proofs}. All results and notations are numbered and referenced in alignment with the main text, unless explicitly noted otherwise.

\section{Additional Simulation Results}\label{sec:numerical}
This section reports further results for Setting~1 of Section~\ref{sec:simulation} of the main text, each based on 1,000 Monte Carlo replications. In Tables~\ref{tab:sup_setting1_other_ratio} and~\ref{tab:sup_setting1_other_fixed}, MSE is the empirical mean squared error of the estimator of the full parameter vector, CRLB is the corresponding Cram\'er--Rao lower bound, and Volume and Coverage are the average volume and the empirical coverage of the $95\%$ confidence ellipsoid, as in the main text. In Tables~\ref{tab:sim1_comp} and~\ref{tab:sim2_comp}, MSE is the empirical mean squared error of each component, and Length and Coverage (Cov.) are the average length and the empirical coverage, in percent, of the $95\%$ Wald interval for that component. The supervised (SL) estimator uses only the labeled observations, so its results do not depend on $m$; it coincides with SAEM at $m=0$ and is reported once, in the rows or columns labelled $m=0$.

Table~\ref{tab:sup_setting1_other_ratio} fixes $m/n=10$ and varies $n$, and Table~\ref{tab:sup_setting1_other_fixed} fixes $n=500$ and varies $m$; both report the remaining model parameters $\balpha^*$, $\bbeta^*$ and $\bgamma^*$. In both tables, SAEM achieves smaller MSE than SL, with values close to the CRLB, and smaller confidence-ellipsoid volumes.

\begin{table}[!htbp]
\caption{Estimation results for $\balpha^*$, $\bbeta^*$ and $\bgamma^*$ under Setting~1 with $m=10n$.}
\label{tab:sup_setting1_other_ratio}
\tablefont
\setlength{\tabcolsep}{4pt}
\begin{tabular*}{\textwidth}{@{\extracolsep{\fill}}lrrlrrrr@{}}
\toprule
Parameter & $n$ & $m$ & Method & MSE & CRLB & Volume & Coverage (\%) \\
\midrule
$\balpha^*$ & 200 & 2000 & SL & 0.0410 & 0.0400 & 1.48e$-$4 & 93.7 \\
 & 200 & 2000 & SAEM & 0.0037 & 0.0036 & 1.01e$-$8 & 95.0 \\
 & 500 & 5000 & SL & 0.0162 & 0.0160 & 3.71e$-$6 & 93.9 \\
 & 500 & 5000 & SAEM & 0.0015 & 0.0015 & 2.56e$-$10 & 95.2 \\
 & 5000 & 50000 & SL & 0.0016 & 0.0016 & 3.75e$-$10 & 94.5 \\
 & 5000 & 50000 & SAEM & 1.43e$-$4 & 1.45e$-$4 & 2.57e$-$14 & 95.1 \\
\addlinespace[3pt]
$\bbeta^*$ & 200 & 2000 & SL & 0.0713 & 0.0700 & 1.28e$-$3 & 93.5 \\
 & 200 & 2000 & SAEM & 0.0222 & 0.0223 & 3.82e$-$5 & 90.5 \\
 & 500 & 5000 & SL & 0.0290 & 0.0280 & 8.22e$-$5 & 94.2 \\
 & 500 & 5000 & SAEM & 0.0088 & 0.0089 & 2.59e$-$6 & 94.0 \\
 & 5000 & 50000 & SL & 0.0028 & 0.0028 & 8.24e$-$8 & 94.8 \\
 & 5000 & 50000 & SAEM & 8.74e$-$4 & 8.91e$-$4 & 2.65e$-$9 & 94.9 \\
\addlinespace[3pt]
$\bgamma^*$ & 200 & 2000 & SL & 0.3206 & 0.3150 & 8.04e$-$11 & 91.6 \\
 & 200 & 2000 & SAEM & 0.1191 & 0.1122 & 4.86e$-$18 & 92.2 \\
 & 500 & 5000 & SL & 0.1260 & 0.1260 & 4.96e$-$15 & 94.9 \\
 & 500 & 5000 & SAEM & 0.0444 & 0.0449 & 3.33e$-$22 & 94.2 \\
 & 5000 & 50000 & SL & 0.0124 & 0.0126 & 1.60e$-$25 & 95.8 \\
 & 5000 & 50000 & SAEM & 0.0045 & 0.0045 & 1.08e$-$32 & 95.2 \\
\bottomrule
\end{tabular*}
\end{table}

\begin{table}[!htbp]
\caption{Estimation results for $\balpha^*$, $\bbeta^*$ and $\bgamma^*$ under Setting~1 with $n=500$. The row $m=0$ is the supervised estimator.}
\label{tab:sup_setting1_other_fixed}
\tablefont
\setlength{\tabcolsep}{4pt}
\begin{tabular*}{\textwidth}{@{\extracolsep{\fill}}lrlrrrr@{}}
\toprule
Parameter & $m$ & Method & MSE & CRLB & Volume & Coverage (\%) \\
\midrule
$\balpha^*$ & 0 & SL & 0.0162 & 0.0160 & 3.71e$-$6 & 93.9 \\
 & 500 & SAEM & 0.0080 & 0.0080 & 2.33e$-$7 & 94.2 \\
 & 5000 & SAEM & 0.0015 & 0.0015 & 2.56e$-$10 & 95.2 \\
 & 50000 & SAEM & 1.61e$-$4 & 1.58e$-$4 & 3.61e$-$14 & 94.7 \\
\addlinespace[3pt]
$\bbeta^*$ & 0 & SL & 0.0290 & 0.0280 & 8.22e$-$5 & 94.2 \\
 & 500 & SAEM & 0.0179 & 0.0175 & 1.99e$-$5 & 94.9 \\
 & 5000 & SAEM & 0.0088 & 0.0089 & 2.59e$-$6 & 94.0 \\
 & 50000 & SAEM & 0.0071 & 0.0072 & 1.36e$-$6 & 95.1 \\
\addlinespace[3pt]
$\bgamma^*$ & 0 & SL & 0.1260 & 0.1260 & 4.96e$-$15 & 94.9 \\
 & 500 & SAEM & 0.0810 & 0.0814 & 2.92e$-$17 & 94.8 \\
 & 5000 & SAEM & 0.0444 & 0.0449 & 3.33e$-$22 & 94.2 \\
 & 50000 & SAEM & 0.0378 & 0.0376 & 4.06e$-$28 & 94.3 \\
\bottomrule
\end{tabular*}
\end{table}

\clearpage
Tables~\ref{tab:sim1_comp} and~\ref{tab:sim2_comp} report component-wise results for $\balpha^*$, $\bbeta^*$, $\bgamma^*$ and $\btheta^*$ under the two designs. SAEM achieves smaller MSE and shorter intervals than SL for every component in both designs, and the component-wise coverages of both methods lie between $92.9\%$ and $96.7\%$.

{\tablefont
\setlength{\LTleft}{0pt}\setlength{\LTright}{0pt}\setlength{\tabcolsep}{4pt}
\begin{longtable}{@{\extracolsep{\fill}}lrrrrrrr@{}}
\caption{Component-wise results under Setting~1 with $m=10n$.}\label{tab:sim1_comp}\\
\toprule
& & \multicolumn{3}{c}{SL} & \multicolumn{3}{c}{SAEM}\\ \cmidrule(lr){3-5}\cmidrule(lr){6-8}
Component & $n$ & MSE & Length & Coverage (\%) & MSE & Length & Coverage (\%) \\
\midrule
\endfirsthead
\multicolumn{8}{l}{\tablename\ \thetable\ (continued)} \\
\toprule
& & \multicolumn{3}{c}{SL} & \multicolumn{3}{c}{SAEM}\\ \cmidrule(lr){3-5}\cmidrule(lr){6-8}
Component & $n$ & MSE & Length & Coverage (\%) & MSE & Length & Coverage (\%) \\
\midrule
\endhead
\bottomrule
\endfoot
\bottomrule
\endlastfoot
\multicolumn{8}{@{}l}{\itshape Parameter $\balpha^*$}\\*
\addlinespace[1pt]
$\balpha_{1,1}^*$ & 200 & 0.0051 & 0.2773 & 95.3 & 4.73e$-$4 & 0.0836 & 94.1 \\
$\balpha_{2,1}^*$ &  & 0.0051 & 0.2777 & 94.7 & 4.43e$-$4 & 0.0837 & 95.2 \\
$\balpha_{3,1}^*$ &  & 0.0052 & 0.2770 & 94.3 & 4.70e$-$4 & 0.0835 & 94.7 \\
$\balpha_{4,1}^*$ &  & 0.0050 & 0.2778 & 95.6 & 4.79e$-$4 & 0.0837 & 94.7 \\
$\balpha_{1,2}^*$ &  & 0.0050 & 0.2773 & 95.0 & 4.77e$-$4 & 0.0836 & 94.5 \\
$\balpha_{2,2}^*$ &  & 0.0050 & 0.2776 & 95.1 & 4.38e$-$4 & 0.0837 & 96.4 \\
$\balpha_{3,2}^*$ &  & 0.0052 & 0.2770 & 93.8 & 4.51e$-$4 & 0.0836 & 95.0 \\
$\balpha_{4,2}^*$ &  & 0.0053 & 0.2777 & 94.9 & 4.28e$-$4 & 0.0837 & 94.6 \\
\addlinespace[2pt]
$\balpha_{1,1}^*$ & 500 & 0.0018 & 0.1752 & 95.7 & 1.95e$-$4 & 0.0528 & 94.4 \\
$\balpha_{2,1}^*$ &  & 0.0020 & 0.1751 & 94.8 & 1.80e$-$4 & 0.0528 & 94.7 \\
$\balpha_{3,1}^*$ &  & 0.0022 & 0.1750 & 93.8 & 1.75e$-$4 & 0.0528 & 95.9 \\
$\balpha_{4,1}^*$ &  & 0.0020 & 0.1750 & 94.6 & 1.86e$-$4 & 0.0529 & 95.2 \\
$\balpha_{1,2}^*$ &  & 0.0020 & 0.1752 & 94.6 & 1.88e$-$4 & 0.0529 & 93.9 \\
$\balpha_{2,2}^*$ &  & 0.0021 & 0.1751 & 94.2 & 1.84e$-$4 & 0.0528 & 96.0 \\
$\balpha_{3,2}^*$ &  & 0.0020 & 0.1750 & 94.7 & 1.85e$-$4 & 0.0529 & 94.2 \\
$\balpha_{4,2}^*$ &  & 0.0020 & 0.1750 & 94.9 & 1.79e$-$4 & 0.0529 & 95.3 \\
\addlinespace[2pt]
$\balpha_{1,1}^*$ & 5000 & 2.04e$-$4 & 0.0554 & 94.5 & 1.86e$-$5 & 0.0167 & 94.4 \\
$\balpha_{2,1}^*$ &  & 2.05e$-$4 & 0.0554 & 94.1 & 1.70e$-$5 & 0.0167 & 96.5 \\
$\balpha_{3,1}^*$ &  & 2.03e$-$4 & 0.0554 & 94.7 & 1.69e$-$5 & 0.0167 & 95.8 \\
$\balpha_{4,1}^*$ &  & 2.01e$-$4 & 0.0554 & 94.4 & 1.81e$-$5 & 0.0167 & 95.7 \\
$\balpha_{1,2}^*$ &  & 2.05e$-$4 & 0.0554 & 94.7 & 1.77e$-$5 & 0.0167 & 95.7 \\
$\balpha_{2,2}^*$ &  & 2.02e$-$4 & 0.0554 & 95.2 & 1.82e$-$5 & 0.0167 & 94.7 \\
$\balpha_{3,2}^*$ &  & 2.13e$-$4 & 0.0554 & 94.5 & 1.71e$-$5 & 0.0167 & 95.6 \\
$\balpha_{4,2}^*$ &  & 1.93e$-$4 & 0.0554 & 95.7 & 1.93e$-$5 & 0.0167 & 95.7 \\
\addlinespace[2pt]
\midrule
\multicolumn{8}{@{}l}{\itshape Parameter $\bbeta^*=(\bbeta_1^*,\bbeta_2^*)$}\\*
\addlinespace[1pt]
$\bbeta_{1,1}^*$ & 200 & 0.0155 & 0.4798 & 94.1 & 0.0047 & 0.2659 & 94.5 \\
$\bbeta_{1,2}^*$ &  & 0.0150 & 0.4795 & 95.6 & 0.0048 & 0.2660 & 94.9 \\
$\bbeta_{1,3}^*$ &  & 0.0147 & 0.4799 & 95.9 & 0.0049 & 0.2660 & 94.4 \\
$\bbeta_{1,4}^*$ &  & 0.0156 & 0.4799 & 94.8 & 0.0045 & 0.2660 & 95.3 \\
$\bbeta_{2,1}^*$ &  & 0.0049 & 0.2767 & 95.4 & 0.0016 & 0.1536 & 93.5 \\
$\bbeta_{2,2}^*$ &  & 0.0056 & 0.2768 & 94.0 & 0.0017 & 0.1535 & 93.4 \\
\addlinespace[2pt]
$\bbeta_{1,1}^*$ & 500 & 0.0063 & 0.3034 & 93.9 & 0.0018 & 0.1703 & 95.5 \\
$\bbeta_{1,2}^*$ &  & 0.0064 & 0.3036 & 93.8 & 0.0020 & 0.1704 & 94.5 \\
$\bbeta_{1,3}^*$ &  & 0.0058 & 0.3034 & 95.7 & 0.0019 & 0.1703 & 95.3 \\
$\bbeta_{1,4}^*$ &  & 0.0062 & 0.3037 & 94.7 & 0.0018 & 0.1703 & 94.9 \\
$\bbeta_{2,1}^*$ &  & 0.0023 & 0.1753 & 93.1 & 6.98e$-$4 & 0.0984 & 94.0 \\
$\bbeta_{2,2}^*$ &  & 0.0021 & 0.1753 & 93.8 & 6.15e$-$4 & 0.0983 & 94.7 \\
\addlinespace[2pt]
$\bbeta_{1,1}^*$ & 5000 & 5.79e$-$4 & 0.0960 & 95.4 & 1.87e$-$4 & 0.0541 & 95.7 \\
$\bbeta_{1,2}^*$ &  & 6.22e$-$4 & 0.0960 & 94.7 & 1.88e$-$4 & 0.0541 & 94.4 \\
$\bbeta_{1,3}^*$ &  & 5.95e$-$4 & 0.0959 & 95.3 & 1.95e$-$4 & 0.0541 & 94.9 \\
$\bbeta_{1,4}^*$ &  & 5.76e$-$4 & 0.0960 & 94.4 & 1.78e$-$4 & 0.0541 & 94.5 \\
$\bbeta_{2,1}^*$ &  & 2.13e$-$4 & 0.0554 & 93.5 & 6.43e$-$5 & 0.0312 & 94.5 \\
$\bbeta_{2,2}^*$ &  & 2.03e$-$4 & 0.0554 & 95.0 & 6.12e$-$5 & 0.0312 & 95.1 \\
\addlinespace[2pt]
\midrule
\multicolumn{8}{@{}l}{\itshape Parameter $\bgamma^*=(\bgamma_1^*,\bgamma_2^*,\bgamma_3^*)$}\\*
\addlinespace[1pt]
$\bgamma_{1,(1,1)}^*$ & 200 & 0.0211 & 0.5561 & 93.4 & 0.0074 & 0.3235 & 93.6 \\
$\bgamma_{1,(2,1)}^*$ &  & 0.0207 & 0.5552 & 94.8 & 0.0074 & 0.3235 & 94.3 \\
$\bgamma_{1,(3,1)}^*$ &  & 0.0217 & 0.5559 & 94.5 & 0.0075 & 0.3236 & 94.2 \\
$\bgamma_{1,(4,1)}^*$ &  & 0.0197 & 0.5554 & 94.8 & 0.0071 & 0.3233 & 94.6 \\
$\bgamma_{1,(1,2)}^*$ &  & 0.0199 & 0.5555 & 94.7 & 0.0073 & 0.3233 & 93.9 \\
$\bgamma_{1,(2,2)}^*$ &  & 0.0181 & 0.5547 & 96.7 & 0.0069 & 0.3233 & 94.9 \\
$\bgamma_{1,(3,2)}^*$ &  & 0.0189 & 0.5553 & 96.4 & 0.0076 & 0.3234 & 93.3 \\
$\bgamma_{1,(4,2)}^*$ &  & 0.0184 & 0.5548 & 95.0 & 0.0071 & 0.3231 & 94.2 \\
$\bgamma_{1,(1,3)}^*$ &  & 0.0209 & 0.5569 & 94.8 & 0.0073 & 0.3233 & 94.6 \\
$\bgamma_{1,(2,3)}^*$ &  & 0.0206 & 0.5560 & 94.5 & 0.0072 & 0.3233 & 94.2 \\
$\bgamma_{1,(3,3)}^*$ &  & 0.0209 & 0.5566 & 93.7 & 0.0073 & 0.3234 & 93.9 \\
$\bgamma_{1,(4,3)}^*$ &  & 0.0216 & 0.5562 & 94.5 & 0.0073 & 0.3231 & 93.6 \\
$\bgamma_{2,(1,1)}^*$ &  & 0.0100 & 0.3922 & 95.0 & 0.0040 & 0.2381 & 93.0 \\
$\bgamma_{2,(2,1)}^*$ &  & 0.0105 & 0.3921 & 93.8 & 0.0041 & 0.2381 & 93.6 \\
$\bgamma_{2,(1,2)}^*$ &  & 0.0111 & 0.3918 & 93.4 & 0.0042 & 0.2381 & 93.0 \\
$\bgamma_{2,(2,2)}^*$ &  & 0.0108 & 0.3917 & 93.4 & 0.0041 & 0.2380 & 93.6 \\
$\bgamma_{2,(1,3)}^*$ &  & 0.0095 & 0.3928 & 95.2 & 0.0040 & 0.2381 & 93.5 \\
$\bgamma_{2,(2,3)}^*$ &  & 0.0102 & 0.3926 & 94.7 & 0.0041 & 0.2381 & 94.3 \\
$\bgamma_{3,1}^*$ &  & 0.0054 & 0.2777 & 93.3 & 0.0024 & 0.1810 & 93.8 \\
$\bgamma_{3,2}^*$ &  & 0.0054 & 0.2774 & 93.8 & 0.0024 & 0.1812 & 93.8 \\
$\bgamma_{3,3}^*$ &  & 0.0050 & 0.2781 & 95.0 & 0.0024 & 0.1812 & 93.8 \\
\addlinespace[2pt]
$\bgamma_{1,(1,1)}^*$ & 500 & 0.0084 & 0.3496 & 94.1 & 0.0027 & 0.2058 & 96.2 \\
$\bgamma_{1,(2,1)}^*$ &  & 0.0082 & 0.3497 & 94.2 & 0.0028 & 0.2059 & 95.1 \\
$\bgamma_{1,(3,1)}^*$ &  & 0.0083 & 0.3493 & 94.8 & 0.0026 & 0.2058 & 96.0 \\
$\bgamma_{1,(4,1)}^*$ &  & 0.0085 & 0.3496 & 94.4 & 0.0027 & 0.2056 & 95.0 \\
$\bgamma_{1,(1,2)}^*$ &  & 0.0080 & 0.3506 & 95.0 & 0.0027 & 0.2059 & 94.1 \\
$\bgamma_{1,(2,2)}^*$ &  & 0.0080 & 0.3508 & 95.2 & 0.0029 & 0.2060 & 94.5 \\
$\bgamma_{1,(3,2)}^*$ &  & 0.0076 & 0.3503 & 95.2 & 0.0028 & 0.2059 & 94.3 \\
$\bgamma_{1,(4,2)}^*$ &  & 0.0083 & 0.3506 & 94.9 & 0.0027 & 0.2057 & 95.0 \\
$\bgamma_{1,(1,3)}^*$ &  & 0.0075 & 0.3513 & 95.8 & 0.0027 & 0.2057 & 95.5 \\
$\bgamma_{1,(2,3)}^*$ &  & 0.0079 & 0.3515 & 94.2 & 0.0028 & 0.2058 & 94.3 \\
$\bgamma_{1,(3,3)}^*$ &  & 0.0080 & 0.3510 & 95.1 & 0.0027 & 0.2058 & 95.0 \\
$\bgamma_{1,(4,3)}^*$ &  & 0.0079 & 0.3513 & 94.7 & 0.0027 & 0.2056 & 94.9 \\
$\bgamma_{2,(1,1)}^*$ &  & 0.0040 & 0.2471 & 94.7 & 0.0016 & 0.1512 & 94.7 \\
$\bgamma_{2,(2,1)}^*$ &  & 0.0041 & 0.2471 & 95.0 & 0.0015 & 0.1512 & 95.1 \\
$\bgamma_{2,(1,2)}^*$ &  & 0.0041 & 0.2478 & 94.4 & 0.0016 & 0.1513 & 94.7 \\
$\bgamma_{2,(2,2)}^*$ &  & 0.0038 & 0.2478 & 95.8 & 0.0015 & 0.1513 & 94.8 \\
$\bgamma_{2,(1,3)}^*$ &  & 0.0039 & 0.2483 & 95.6 & 0.0014 & 0.1513 & 95.6 \\
$\bgamma_{2,(2,3)}^*$ &  & 0.0038 & 0.2483 & 96.0 & 0.0014 & 0.1513 & 94.9 \\
$\bgamma_{3,1}^*$ &  & 0.0020 & 0.1748 & 94.0 & 8.73e$-$4 & 0.1148 & 94.7 \\
$\bgamma_{3,2}^*$ &  & 0.0019 & 0.1753 & 95.2 & 9.06e$-$4 & 0.1149 & 94.5 \\
$\bgamma_{3,3}^*$ &  & 0.0020 & 0.1756 & 94.5 & 8.02e$-$4 & 0.1150 & 94.9 \\
\addlinespace[2pt]
$\bgamma_{1,(1,1)}^*$ & 5000 & 7.78e$-$4 & 0.1109 & 95.4 & 2.75e$-$4 & 0.0653 & 95.0 \\
$\bgamma_{1,(2,1)}^*$ &  & 8.02e$-$4 & 0.1110 & 96.0 & 2.90e$-$4 & 0.0653 & 94.7 \\
$\bgamma_{1,(3,1)}^*$ &  & 8.20e$-$4 & 0.1109 & 95.2 & 2.79e$-$4 & 0.0653 & 95.3 \\
$\bgamma_{1,(4,1)}^*$ &  & 7.86e$-$4 & 0.1109 & 96.0 & 2.73e$-$4 & 0.0653 & 95.7 \\
$\bgamma_{1,(1,2)}^*$ &  & 7.87e$-$4 & 0.1108 & 95.3 & 2.68e$-$4 & 0.0653 & 94.7 \\
$\bgamma_{1,(2,2)}^*$ &  & 7.86e$-$4 & 0.1108 & 94.3 & 2.81e$-$4 & 0.0653 & 94.9 \\
$\bgamma_{1,(3,2)}^*$ &  & 8.06e$-$4 & 0.1108 & 95.1 & 2.79e$-$4 & 0.0653 & 95.2 \\
$\bgamma_{1,(4,2)}^*$ &  & 7.49e$-$4 & 0.1108 & 95.6 & 2.62e$-$4 & 0.0653 & 95.7 \\
$\bgamma_{1,(1,3)}^*$ &  & 7.37e$-$4 & 0.1109 & 95.1 & 2.64e$-$4 & 0.0653 & 96.1 \\
$\bgamma_{1,(2,3)}^*$ &  & 7.93e$-$4 & 0.1109 & 94.6 & 2.79e$-$4 & 0.0653 & 95.3 \\
$\bgamma_{1,(3,3)}^*$ &  & 7.50e$-$4 & 0.1108 & 95.2 & 2.81e$-$4 & 0.0653 & 95.4 \\
$\bgamma_{1,(4,3)}^*$ &  & 7.32e$-$4 & 0.1108 & 95.8 & 2.58e$-$4 & 0.0653 & 96.0 \\
$\bgamma_{2,(1,1)}^*$ &  & 3.85e$-$4 & 0.0784 & 95.5 & 1.44e$-$4 & 0.0480 & 96.2 \\
$\bgamma_{2,(2,1)}^*$ &  & 4.03e$-$4 & 0.0784 & 94.5 & 1.45e$-$4 & 0.0480 & 94.6 \\
$\bgamma_{2,(1,2)}^*$ &  & 3.94e$-$4 & 0.0784 & 95.2 & 1.56e$-$4 & 0.0480 & 94.5 \\
$\bgamma_{2,(2,2)}^*$ &  & 4.20e$-$4 & 0.0784 & 94.0 & 1.49e$-$4 & 0.0480 & 95.1 \\
$\bgamma_{2,(1,3)}^*$ &  & 4.21e$-$4 & 0.0784 & 94.0 & 1.57e$-$4 & 0.0480 & 94.4 \\
$\bgamma_{2,(2,3)}^*$ &  & 4.06e$-$4 & 0.0784 & 95.0 & 1.52e$-$4 & 0.0480 & 93.9 \\
$\bgamma_{3,1}^*$ &  & 1.94e$-$4 & 0.0555 & 95.7 & 8.68e$-$5 & 0.0364 & 94.8 \\
$\bgamma_{3,2}^*$ &  & 2.06e$-$4 & 0.0554 & 94.1 & 9.07e$-$5 & 0.0364 & 94.1 \\
$\bgamma_{3,3}^*$ &  & 2.04e$-$4 & 0.0554 & 95.0 & 9.01e$-$5 & 0.0364 & 93.6 \\
\addlinespace[2pt]
\midrule
\multicolumn{8}{@{}l}{\itshape Parameter $\btheta^*$}\\*
\addlinespace[1pt]
$\btheta_{1}^*$ & 200 & 0.0145 & 0.4796 & 95.5 & 0.0027 & 0.1939 & 94.0 \\
$\btheta_{2}^*$ &  & 0.0162 & 0.4802 & 94.6 & 0.0028 & 0.1941 & 93.9 \\
$\btheta_{3}^*$ &  & 0.0156 & 0.4792 & 93.8 & 0.0029 & 0.1938 & 93.2 \\
$\btheta_{4}^*$ &  & 0.0154 & 0.4804 & 94.8 & 0.0026 & 0.1940 & 94.3 \\
\addlinespace[2pt]
$\btheta_{1}^*$ & 500 & 0.0058 & 0.3034 & 95.6 & 0.0010 & 0.1235 & 93.9 \\
$\btheta_{2}^*$ &  & 0.0060 & 0.3032 & 93.6 & 0.0010 & 0.1234 & 95.9 \\
$\btheta_{3}^*$ &  & 0.0066 & 0.3030 & 94.5 & 0.0010 & 0.1235 & 94.0 \\
$\btheta_{4}^*$ &  & 0.0060 & 0.3031 & 94.7 & 0.0010 & 0.1235 & 94.8 \\
\addlinespace[2pt]
$\btheta_{1}^*$ & 5000 & 6.31e$-$4 & 0.0960 & 93.9 & 9.89e$-$5 & 0.0392 & 95.8 \\
$\btheta_{2}^*$ &  & 5.80e$-$4 & 0.0960 & 95.5 & 9.81e$-$5 & 0.0392 & 95.5 \\
$\btheta_{3}^*$ &  & 6.05e$-$4 & 0.0960 & 94.9 & 1.01e$-$4 & 0.0392 & 94.9 \\
$\btheta_{4}^*$ &  & 6.07e$-$4 & 0.0960 & 95.4 & 9.75e$-$5 & 0.0392 & 94.6 \\
\addlinespace[2pt]
\end{longtable}
}
{\tablefont
\setlength{\LTleft}{0pt}\setlength{\LTright}{0pt}\setlength{\tabcolsep}{4pt}
\begin{longtable}{@{\extracolsep{\fill}}lrrrrrrrrrrrr@{}}
\caption{Component-wise results under Setting~1 with $n=500$. The columns $m=0$ give the supervised estimator.}\label{tab:sim2_comp}\\
 
\toprule
& \multicolumn{3}{c}{SL ($m=0$)} & \multicolumn{3}{c}{SAEM, $m=500$} & \multicolumn{3}{c}{SAEM, $m=5000$} & \multicolumn{3}{c}{SAEM, $m=50000$}\\ \cmidrule(lr){2-4}\cmidrule(lr){5-7}\cmidrule(lr){8-10}\cmidrule(lr){11-13}
Component & MSE & Length & Cov. & MSE & Length & Cov. & MSE & Length & Cov. & MSE & Length & Cov. \\
 
\midrule
\endfirsthead
\multicolumn{13}{l}{\tablename\ \thetable\ (continued)}\\
\toprule
& \multicolumn{3}{c}{SL ($m=0$)} & \multicolumn{3}{c}{SAEM, $m=500$} & \multicolumn{3}{c}{SAEM, $m=5000$} & \multicolumn{3}{c}{SAEM, $m=50000$}\\ \cmidrule(lr){2-4}\cmidrule(lr){5-7}\cmidrule(lr){8-10}\cmidrule(lr){11-13}
Component & MSE & Length & Cov. & MSE & Length & Cov. & MSE & Length & Cov. & MSE & Length & Cov. \\
 
\midrule
\endhead
\bottomrule
\endfoot
\bottomrule
\endlastfoot
\multicolumn{13}{@{}l}{\itshape Parameter $\balpha^*$}\\*
\addlinespace[1pt]
$\balpha_{1,1}^*$ & 0.0018 & 0.1752 & 95.7 & 0.0010 & 0.1239 & 95.5 & 1.95e$-$4 & 0.0528 & 94.4 & 2.16e$-$5 & 0.0174 & 94.1 \\
$\balpha_{2,1}^*$ & 0.0020 & 0.1751 & 94.8 & 9.27e$-$4 & 0.1238 & 96.3 & 1.80e$-$4 & 0.0528 & 94.7 & 1.91e$-$5 & 0.0174 & 96.0 \\
$\balpha_{3,1}^*$ & 0.0022 & 0.1750 & 93.8 & 0.0010 & 0.1237 & 94.7 & 1.75e$-$4 & 0.0528 & 95.9 & 2.03e$-$5 & 0.0174 & 94.7 \\
$\balpha_{4,1}^*$ & 0.0020 & 0.1750 & 94.6 & 0.0010 & 0.1238 & 95.3 & 1.86e$-$4 & 0.0529 & 95.2 & 1.97e$-$5 & 0.0174 & 94.6 \\
$\balpha_{1,2}^*$ & 0.0020 & 0.1752 & 94.6 & 0.0010 & 0.1240 & 95.3 & 1.88e$-$4 & 0.0529 & 93.9 & 2.03e$-$5 & 0.0174 & 94.1 \\
$\balpha_{2,2}^*$ & 0.0021 & 0.1751 & 94.2 & 0.0011 & 0.1238 & 93.2 & 1.84e$-$4 & 0.0528 & 96.0 & 1.99e$-$5 & 0.0174 & 94.7 \\
$\balpha_{3,2}^*$ & 0.0020 & 0.1750 & 94.7 & 0.0010 & 0.1237 & 94.9 & 1.85e$-$4 & 0.0529 & 94.2 & 2.05e$-$5 & 0.0174 & 94.8 \\
$\balpha_{4,2}^*$ & 0.0020 & 0.1750 & 94.9 & 0.0010 & 0.1238 & 93.9 & 1.79e$-$4 & 0.0529 & 95.3 & 1.97e$-$5 & 0.0174 & 95.1 \\
 
\midrule
\multicolumn{13}{@{}l}{\itshape Parameter $\bbeta^*=(\bbeta_1^*,\bbeta_2^*)$}\\*
\addlinespace[1pt]
$\bbeta_{1,1}^*$ & 0.0063 & 0.3034 & 93.9 & 0.0037 & 0.2396 & 95.4 & 0.0018 & 0.1703 & 95.5 & 0.0015 & 0.1528 & 94.2 \\
$\bbeta_{1,2}^*$ & 0.0064 & 0.3036 & 93.8 & 0.0041 & 0.2398 & 92.9 & 0.0020 & 0.1704 & 94.5 & 0.0016 & 0.1529 & 94.4 \\
$\bbeta_{1,3}^*$ & 0.0058 & 0.3034 & 95.7 & 0.0038 & 0.2396 & 94.4 & 0.0019 & 0.1703 & 95.3 & 0.0015 & 0.1528 & 95.8 \\
$\bbeta_{1,4}^*$ & 0.0062 & 0.3037 & 94.7 & 0.0037 & 0.2397 & 95.1 & 0.0018 & 0.1703 & 94.9 & 0.0015 & 0.1528 & 95.8 \\
$\bbeta_{2,1}^*$ & 0.0023 & 0.1753 & 93.1 & 0.0014 & 0.1384 & 93.2 & 6.98e$-$4 & 0.0984 & 94.0 & 5.45e$-$4 & 0.0882 & 94.0 \\
$\bbeta_{2,2}^*$ & 0.0021 & 0.1753 & 93.8 & 0.0013 & 0.1384 & 95.3 & 6.15e$-$4 & 0.0983 & 94.7 & 4.84e$-$4 & 0.0883 & 95.2 \\
\midrule
\multicolumn{13}{@{}l}{\itshape Parameter $\bgamma^*=(\bgamma_1^*,\bgamma_2^*,\bgamma_3^*)$}\\*
\addlinespace[1pt]
$\bgamma_{1,(1,1)}^*$ & 0.0084 & 0.3496 & 94.1 & 0.0055 & 0.2800 & 94.0 & 0.0027 & 0.2058 & 96.2 & 0.0023 & 0.1876 & 94.8 \\
$\bgamma_{1,(2,1)}^*$ & 0.0082 & 0.3497 & 94.2 & 0.0054 & 0.2802 & 94.0 & 0.0028 & 0.2059 & 95.1 & 0.0024 & 0.1876 & 94.2 \\
$\bgamma_{1,(3,1)}^*$ & 0.0083 & 0.3493 & 94.8 & 0.0055 & 0.2798 & 93.7 & 0.0026 & 0.2058 & 96.0 & 0.0023 & 0.1875 & 95.2 \\
$\bgamma_{1,(4,1)}^*$ & 0.0085 & 0.3496 & 94.4 & 0.0053 & 0.2799 & 93.7 & 0.0027 & 0.2056 & 95.0 & 0.0023 & 0.1874 & 95.0 \\
$\bgamma_{1,(1,2)}^*$ & 0.0080 & 0.3506 & 95.0 & 0.0051 & 0.2804 & 94.9 & 0.0027 & 0.2059 & 94.1 & 0.0023 & 0.1874 & 94.4 \\
$\bgamma_{1,(2,2)}^*$ & 0.0080 & 0.3508 & 95.2 & 0.0051 & 0.2806 & 95.5 & 0.0029 & 0.2060 & 94.5 & 0.0024 & 0.1875 & 94.9 \\
$\bgamma_{1,(3,2)}^*$ & 0.0076 & 0.3503 & 95.2 & 0.0049 & 0.2802 & 95.2 & 0.0028 & 0.2059 & 94.3 & 0.0024 & 0.1874 & 93.9 \\
$\bgamma_{1,(4,2)}^*$ & 0.0083 & 0.3506 & 94.9 & 0.0052 & 0.2803 & 94.9 & 0.0027 & 0.2057 & 95.0 & 0.0023 & 0.1873 & 93.6 \\
$\bgamma_{1,(1,3)}^*$ & 0.0075 & 0.3513 & 95.8 & 0.0047 & 0.2806 & 96.7 & 0.0027 & 0.2057 & 95.5 & 0.0022 & 0.1873 & 95.5 \\
$\bgamma_{1,(2,3)}^*$ & 0.0079 & 0.3515 & 94.2 & 0.0049 & 0.2808 & 95.4 & 0.0028 & 0.2058 & 94.3 & 0.0024 & 0.1874 & 93.7 \\
$\bgamma_{1,(3,3)}^*$ & 0.0080 & 0.3510 & 95.1 & 0.0049 & 0.2803 & 95.7 & 0.0027 & 0.2058 & 95.0 & 0.0023 & 0.1873 & 95.0 \\
$\bgamma_{1,(4,3)}^*$ & 0.0079 & 0.3513 & 94.7 & 0.0048 & 0.2805 & 95.2 & 0.0027 & 0.2056 & 94.9 & 0.0022 & 0.1872 & 95.9 \\
$\bgamma_{2,(1,1)}^*$ & 0.0040 & 0.2471 & 94.7 & 0.0029 & 0.2004 & 94.4 & 0.0016 & 0.1512 & 94.7 & 0.0014 & 0.1395 & 95.2 \\
$\bgamma_{2,(2,1)}^*$ & 0.0041 & 0.2471 & 95.0 & 0.0027 & 0.2004 & 94.3 & 0.0015 & 0.1512 & 95.1 & 0.0013 & 0.1395 & 95.6 \\
$\bgamma_{2,(1,2)}^*$ & 0.0041 & 0.2478 & 94.4 & 0.0028 & 0.2008 & 94.7 & 0.0016 & 0.1513 & 94.7 & 0.0014 & 0.1395 & 93.0 \\
$\bgamma_{2,(2,2)}^*$ & 0.0038 & 0.2478 & 95.8 & 0.0024 & 0.2007 & 96.4 & 0.0015 & 0.1513 & 94.8 & 0.0013 & 0.1395 & 94.6 \\
$\bgamma_{2,(1,3)}^*$ & 0.0039 & 0.2483 & 95.6 & 0.0025 & 0.2009 & 95.9 & 0.0014 & 0.1513 & 95.6 & 0.0013 & 0.1395 & 94.9 \\
$\bgamma_{2,(2,3)}^*$ & 0.0038 & 0.2483 & 96.0 & 0.0025 & 0.2008 & 95.9 & 0.0014 & 0.1513 & 94.9 & 0.0012 & 0.1395 & 95.1 \\
$\bgamma_{3,1}^*$ & 0.0020 & 0.1748 & 94.0 & 0.0015 & 0.1451 & 93.5 & 8.73e$-$4 & 0.1148 & 94.7 & 7.85e$-$4 & 0.1078 & 94.4 \\
$\bgamma_{3,2}^*$ & 0.0019 & 0.1753 & 95.2 & 0.0013 & 0.1453 & 94.4 & 9.06e$-$4 & 0.1149 & 94.5 & 8.16e$-$4 & 0.1078 & 94.1 \\
$\bgamma_{3,3}^*$ & 0.0020 & 0.1756 & 94.5 & 0.0013 & 0.1454 & 95.8 & 8.02e$-$4 & 0.1150 & 94.9 & 7.11e$-$4 & 0.1079 & 95.8 \\
\midrule
\multicolumn{13}{@{}l}{\itshape Parameter $\btheta^*$}\\*
\addlinespace[1pt]
$\btheta_{1}^*$ & 0.0058 & 0.3034 & 95.6 & 0.0031 & 0.2233 & 94.8 & 0.0010 & 0.1235 & 93.9 & 5.75e$-$4 & 0.0916 & 94.1 \\
$\btheta_{2}^*$ & 0.0060 & 0.3032 & 93.6 & 0.0035 & 0.2230 & 94.7 & 0.0010 & 0.1234 & 95.9 & 5.81e$-$4 & 0.0916 & 94.2 \\
$\btheta_{3}^*$ & 0.0066 & 0.3030 & 94.5 & 0.0034 & 0.2229 & 93.9 & 0.0010 & 0.1235 & 94.0 & 5.60e$-$4 & 0.0916 & 95.3 \\
$\btheta_{4}^*$ & 0.0060 & 0.3031 & 94.7 & 0.0030 & 0.2231 & 94.9 & 0.0010 & 0.1235 & 94.8 & 5.14e$-$4 & 0.0916 & 95.8 \\
 
\end{longtable}
}

\clearpage
\section{Additional Real-data Results}\label{sec:real_data_supplement}

\subsection{MIMIC-III sepsis data}\label{sec:mimic_supplement}

Table~\ref{tab:mimic_benchmark} lists the covariates declared significant at the $5\%$ level by the full-data benchmark in Section~\ref{sec:real_data} of the main text for both $\btheta^*$ and $\bbeta_1^*$, together with their estimates and standard errors. Table~\ref{tab:mimic_beta1_selection} reports results analogous to those in Table~\ref{tab:MIMIC_clinical_comparison} of the main text, but focuses on $\bbeta_1^*$ rather than $\btheta^*$. For $\bbeta_1^*$, the benchmark identifies 12 significant covariates, eight of which have $p$-values between $0.001$ and $0.05$. Consequently, both SL and SAEM select these benchmark-significant covariates less frequently than those identified for $\btheta^*$. Nevertheless, SAEM achieves a higher average TPR and a lower average FDR than SL at every missing rate, although the improvements are less pronounced than those observed for $\btheta^*$. This pattern is consistent with the smaller MSE improvements for $\bbeta_1^*$ reported in Table~\ref{tab:MIMIC_results_merged} of the main text.

\begin{table}[!htbp]
\caption{MIMIC-III data: covariates declared significant at the $5\%$ level by the full-data benchmark, with estimates and standard errors.}
\label{tab:mimic_benchmark}
\tablefont
\setlength{\tabcolsep}{4pt}
\begin{tabular*}{\textwidth}{@{\extracolsep{\fill}}lrrlrr@{}}
\toprule
\multicolumn{3}{c}{$\btheta^*$} & \multicolumn{3}{c}{$\bbeta_1^*$}\\ \cmidrule(lr){1-3}\cmidrule(lr){4-6}
Covariate & Estimate & SE & Covariate & Estimate & SE \\
\midrule
Gender & $-$0.041 & 0.008 & Elixhauser score & 0.061 & 0.006 \\
Elixhauser score & 0.164 & 0.008 & GCS & 0.016 & 0.006 \\
GCS & $-$0.121 & 0.008 & Systolic BP & $-$0.046 & 0.020 \\
Heart rate & $-$0.080 & 0.030 & Mean BP & 0.070 & 0.026 \\
FiO$_2$ & 0.053 & 0.011 & Diastolic BP & $-$0.064 & 0.023 \\
Glucose & $-$0.051 & 0.007 & Potassium & $-$0.012 & 0.006 \\
BUN & 0.049 & 0.014 & Glucose & $-$0.025 & 0.006 \\
Creatinine & 0.068 & 0.027 & Platelet count & $-$0.016 & 0.006 \\
Total bilirubin & 0.097 & 0.009 & Arterial pH & $-$0.040 & 0.015 \\
Albumin & $-$0.028 & 0.008 & 4-hourly input & 0.019 & 0.006 \\
Hemoglobin & $-$0.043 & 0.008 & Total output & 0.146 & 0.047 \\
Platelet count & $-$0.102 & 0.008 & Cumulative balance & 0.146 & 0.042 \\
PTT & 0.034 & 0.008 &  &  &  \\
Arterial pH & $-$0.049 & 0.019 &  &  &  \\
Shock index & 0.093 & 0.039 &  &  &  \\
Median vasopressor dose & 0.095 & 0.014 &  &  &  \\
Max.\ vasopressor dose & 0.012 & 0.005 &  &  &  \\
Total output & 0.067 & 0.014 &  &  &  \\
4-hourly output & $-$0.088 & 0.010 &  &  &  \\
Cumulative balance & 0.126 & 0.019 &  &  &  \\
\bottomrule
\end{tabular*}
\end{table}

\begin{table}[!htbp]
\caption{MIMIC-III data: percentage of 500 random repetitions in which SL and SAEM declare each covariate in $\bbeta_1^*$ significant at the $5\%$ level. Covariates are grouped by their significance in the full-data benchmark and ordered by the benchmark $p$-value.}
\label{tab:mimic_beta1_selection}
\tablefont
\setlength{\tabcolsep}{4pt}
\begin{tabular*}{\textwidth}{@{\extracolsep{\fill}}lrrrrrrrr@{}}
\toprule
 & \multicolumn{2}{c}{Missing 10\%} & \multicolumn{2}{c}{Missing 30\%} & \multicolumn{2}{c}{Missing 50\%} & \multicolumn{2}{c}{Missing 70\%}\\
\cmidrule(lr){2-3}\cmidrule(lr){4-5}\cmidrule(lr){6-7}\cmidrule(lr){8-9}
Covariate & SL & SAEM & SL & SAEM & SL & SAEM & SL & SAEM \\
\midrule
\multicolumn{9}{@{}l}{\itshape Significant in the full-data benchmark (12 covariates)}\\
Elixhauser score & 100 & 100 & 100 & 100 & 100 & 100 & 100 & 100 \\
Glucose & 100 & 100 & 100 & 100 & 91 & 96 & 66 & 74 \\
Cumulative balance & 100 & 100 & 96 & 98 & 76 & 83 & 47 & 53 \\
4-hourly input & 100 & 100 & 93 & 97 & 69 & 74 & 44 & 54 \\
Total output & 100 & 100 & 87 & 92 & 65 & 73 & 39 & 43 \\
Diastolic BP & 89 & 87 & 57 & 57 & 32 & 34 & 17 & 16 \\
Mean BP & 98 & 99 & 74 & 79 & 47 & 52 & 29 & 31 \\
Platelet count & 96 & 98 & 68 & 75 & 46 & 52 & 27 & 31 \\
Arterial pH & 97 & 98 & 67 & 74 & 45 & 52 & 29 & 34 \\
GCS & 93 & 96 & 65 & 73 & 43 & 50 & 25 & 32 \\
Systolic BP & 79 & 83 & 51 & 57 & 35 & 37 & 23 & 27 \\
Potassium & 42 & 46 & 38 & 34 & 26 & 29 & 19 & 22 \\
\addlinespace[3pt]
\multicolumn{9}{@{}l}{\itshape Not significant in the full-data benchmark (35 covariates)}\\
PaCO$_2$ & 20 & 21 & 22 & 22 & 19 & 20 & 17 & 18 \\
Total bilirubin & 21 & 22 & 23 & 22 & 17 & 16 & 15 & 13 \\
4-hourly output & 16 & 15 & 17 & 18 & 14 & 15 & 9 & 12 \\
BUN & 11 & 9 & 15 & 18 & 15 & 16 & 11 & 13 \\
PaO$_2$ & 9 & 9 & 16 & 13 & 13 & 14 & 11 & 12 \\
PTT & 6 & 6 & 13 & 12 & 12 & 12 & 8 & 9 \\
WBC count & 8 & 6 & 12 & 13 & 12 & 16 & 15 & 16 \\
Creatinine & 11 & 11 & 31 & 31 & 48 & 49 & 45 & 56 \\
Max.\ vasopressor dose & 8 & 7 & 23 & 23 & 27 & 28 & 34 & 37 \\
Remaining 26 (max.) & 2 & 1 & 6 & 5 & 9 & 8 & 8 & 8 \\
\midrule
TPR & 91.2 & 92.2 & 74.5 & 77.9 & 56.3 & 61.1 & 38.7 & 43.0 \\
FDR & 8.9 & 8.2 & 18.1 & 16.7 & 25.9 & 24.1 & 36.0 & 34.5 \\
\bottomrule
\end{tabular*}
\end{table}

\clearpage
\subsection{Analysis of the primary biliary cholangitis (PBC) data}\label{sec:pbc_supplement}

We analyze the Mayo Clinic PBC data (\texttt{pbcseq}) from the \texttt{survival} R package \citep{therneau2023survival,murtaugh1994primary}. The 312 randomized patients were enrolled between 1974 and 1984, with protocol visits scheduled at 6 months, 1 year, and annually thereafter. The timing of these repeated clinical and laboratory measurements determines the surrogate structure. Table~\ref{tab:pbc_variables} summarizes the variables used in the analysis. Age is standardized, laboratory measurements are log-transformed and standardized, and sex is coded as a binary indicator. After excluding patients with missing covariate or surrogate values, 86 patients have an observed outcome $Y$ and are treated as labeled, whereas 112 patients do not and are treated as unlabeled.

To evaluate prediction performance, we repeat the following procedure 500 times. The 86 labeled patients are randomly divided into a training set ($80\%$) and a test set ($20\%$), while all 112 unlabeled patients are included in the training set. Each method estimates $\btheta^*$ using the training data, and prediction performance is evaluated by the test-set MSE $n_{\rm test}^{-1}\sum_{i\in{\rm test}}(Y_i-\bX_i^\top\widehat{\btheta})^2$. Table~\ref{tab:pbc_prediction_mse} shows that both SAEM and OW-SAEM reduce the average test-set MSE relative to SL, with SAEM achieving a slightly smaller MSE than OW-SAEM.

Table~\ref{tab:pbc_beta1_theta} reports the point estimates and $95\%$ Wald confidence intervals for $\bbeta_1^*$ and $\btheta^*$ using the full semi-supervised dataset, without splitting the labeled observations into training and test sets. Across all considered methods, Bilirubin is the only variable identified as significant for both the direct-effect and prediction coefficients. Moreover, SAEM and OW-SAEM generally yield shorter confidence intervals than SL for most parameters, with the only exceptions being the intercept and Sex coefficients associated with $\bbeta_1^*$.

\begin{table}[!htbp]
\caption{Variable construction in the PBC data analysis.}
\label{tab:pbc_variables}
\tablefont
\setlength{\tabcolsep}{4pt}
\begin{tabular*}{\textwidth}{@{\extracolsep{\fill}}llll@{}}
\toprule
Variable & Time point & Contents & Dimension \\
\midrule
$\bX$ & Baseline (day 0) & Intercept, age, sex, bilirubin, alkaline phosphatase & $p=5$ \\
$\bS_1$ & 6 months ($\pm$90 days) & Bilirubin & $d_1=1$ \\
$Y$ & 1 year ($\pm$10 days) & Bilirubin & 1 \\
$\bS_2$ & 2 years ($\pm$180 days) & Bilirubin, albumin & $d_2=2$ \\
\bottomrule
\end{tabular*}
\end{table}

\begin{table}[!htbp]
\caption{PBC data: test-set prediction MSE across 500 random training--test splits.}
\label{tab:pbc_prediction_mse}
\tablefont
\setlength{\tabcolsep}{4pt}
\begin{tabular*}{\textwidth}{@{\extracolsep{\fill}}lcccc@{}}
\toprule
Method & Mean MSE & MSE reduction relative to SL & Proportion of repetitions with lower MSE than SL \\
\midrule
SL & 0.3349 & -- & -- \\
SAEM & 0.3224 & 3.75\% & 70.8\% \\
OW-SAEM & 0.3250 & 2.95\% & 73.4\% \\
\bottomrule
\end{tabular*}
\end{table}

\begin{table}[!htbp]
\centering
\caption{PBC data: point estimates, $95\%$ confidence intervals (CIs), and CI lengths for $\bbeta_1^*$ and $\btheta^*$. Length denotes the upper endpoint minus the lower endpoint. The estimated weights of OW-SAEM are $\hat\lambda_{\bbeta_1}=0.911$ and $\hat\lambda_{\btheta}=0.694$.}
\label{tab:pbc_beta1_theta}
\tablefont
\setlength{\tabcolsep}{2pt}
\renewcommand{\arraystretch}{1.08}
\begin{tabular*}{\textwidth}{@{\extracolsep{\fill}}llccccc@{}}
\toprule
Method & Statistic & Intercept & Age & Sex & Bilirubin & Alk.\ phos. \\
\midrule
\multicolumn{7}{@{}l}{\itshape Target $\bbeta_1^*$}\\
\addlinespace[2pt]
\multirow{3}{*}{SL}
 & Estimate & 0.219 & $-$0.041 & $-$0.101 & 0.433 & 0.024 \\
 & $95\%$ CI & (0.069, 0.370) & ($-$0.156, 0.075) & ($-$0.299, 0.096) & (0.047, 0.819) & ($-$0.093, 0.142) \\
 & Length & 0.301 & 0.231 & 0.395 & 0.772 & 0.235 \\
\addlinespace[5pt]
\multirow{3}{*}{SAEM}
 & Estimate & 0.239 & $-$0.022 & $-$0.129 & 0.415 & 0.038 \\
 & $95\%$ CI & (0.065, 0.414) & ($-$0.111, 0.067) & ($-$0.330, 0.072) & (0.114, 0.716) & ($-$0.061, 0.137) \\
 & Length & 0.349 & 0.178 & 0.402 & 0.602 & 0.198 \\
 \addlinespace[5pt]
\multirow{3}{*}{OW-SAEM}
 & Estimate & 0.238 & $-$0.024 & $-$0.127 & 0.416 & 0.037 \\
 & $95\%$ CI & (0.068, 0.407) & ($-$0.114, 0.066) & ($-$0.324, 0.071) & (0.112, 0.721) & ($-$0.063, 0.136) \\
 & Length & 0.339 & 0.180 & 0.395 & 0.609 & 0.199 \\
\midrule
\multicolumn{7}{@{}l}{\itshape Target $\btheta^*$}\\
\addlinespace[2pt]
\multirow{3}{*}{SL}
 & Estimate & 0.196 & $-$0.039 & $-$0.094 & 0.905 & $-$0.012 \\
 & $95\%$ CI & (0.002, 0.390) & ($-$0.176, 0.097) & ($-$0.342, 0.155) & (0.762, 1.048) & ($-$0.137, 0.112) \\
 & Length & 0.388 & 0.273 & 0.497 & 0.286 & 0.249 \\
\addlinespace[5pt]
\multirow{3}{*}{SAEM}
 & Estimate & 0.238 & $-$0.016 & $-$0.128 & 0.899 & 0.038 \\
 & $95\%$ CI & (0.047, 0.430) & ($-$0.118, 0.086) & ($-$0.352, 0.097) & (0.785, 1.013) & ($-$0.065, 0.141) \\
 & Length & 0.383 & 0.204 & 0.449 & 0.228 & 0.206 \\
\addlinespace[5pt]
\multirow{3}{*}{OW-SAEM}
 & Estimate & 0.225 & $-$0.023 & $-$0.117 & 0.901 & 0.023 \\
 & $95\%$ CI & (0.049, 0.401) & ($-$0.130, 0.084) & ($-$0.331, 0.097) & (0.785, 1.017) & ($-$0.083, 0.129) \\
 & Length & 0.352 & 0.214 & 0.428 & 0.232 & 0.212 \\
\bottomrule
\end{tabular*}
\end{table}

\clearpage

\section{Auxiliary lemmas}\label{sec:lemmas}
For the Gaussian results in this section and in the proofs of the main Gaussian theorems, we use Assumption~\ref{ass:joint_gaussian} from the main text, under which the coefficients in \eqref{eq:models} are the population least-squares coefficients and the equations in \eqref{eq:models} are the corresponding Gaussian conditional-mean models. In this fixed-dimensional, fixed-distribution setting, the positive-definite covariance matrix and Gaussian fourth moments are finite. In particular, we may choose finite constants $c_1,c_2>0$ such that $\sigma_2^2<c_1$, $\lambda_{\rm min}(\bSigma_3)>1/c_1$, $\|\bgamma_3^*\|_2<c_2$, and $\lambda_{\rm max}(\bSigma_{\bX,2})<c_2$. These constants merely record consequences of the stated Gaussian projection setting; they are not additional asymptotic assumptions. Throughout the Supplementary Material, $\|\cdot\|_F$ denotes the Frobenius norm, and $\|\cdot\|_2$ denotes the Euclidean norm for vectors and the operator norm for matrices.

\begin{lemma}\label{lemma:lemma1}
Under Assumption~\ref{ass:joint_gaussian}, it holds that
$$
E(\bX_i\bvarepsilon_{1i}^\top)=0,\;\; E(\bar{\bX}_{1i}\varepsilon_{2i})=0,\;\; E(\bar{\bX}_{2i}\bvarepsilon_{3i}^\top)=0,\;\; E\{\bar{\bX}_{1i}(\varepsilon_{2i}\bgamma_3^*+\bvarepsilon_{3i})^\top\}=0,
$$
as well as 
\begin{align*}
E(\|\bX_i\bvarepsilon_{1i}^\top\|_F^2)<\infty,\;\; E(\varepsilon_{2i}^2\|\bar{\bX}_{1i}\|_2^2)=O(1),\;\; E(\|\bar{\bX}_{2i}\bvarepsilon_{3i}^\top\|_F^2)=O(1),\;\; E\{\|\bar{\bX}_{1i}(\varepsilon_{2i}\bgamma_3^*+\bvarepsilon_{3i})^\top\|_F^2\}=O(1).
\end{align*}
\end{lemma}

\begin{proof}
The first part of the lemma follows directly from the definition of the population slope \citep{zhang2019semi}. For the second part, it holds that 
\begin{align*}
E(\varepsilon_{2i}^2\|\bar{\bX}_{1i}\|_2^2)=E\{\|\bar{\bX}_{1i}\|_2^2E(\varepsilon_{2i}^2\mid \bar{\bX}_{1i})\}\le c_1 E(\|\bar{\bX}_{1i}\|_2^2)<\infty.
\end{align*}
Under Assumption~\ref{ass:joint_gaussian}, the other terms can be proved in a similar manner.
\end{proof}

\begin{lemma}\label{lemma:lemma2}
Under Assumption~\ref{ass:joint_gaussian}, denote $\bc^*=\sigma_2^2(\bSigma_3+\sigma_2^2\bgamma_3^{*}\bgamma_3^{*\top})^{-1}\bgamma_3^*$ and $\bc^{(0)}=(\sigma_2^{(0)})^2\{\bSigma_3^{(0)}+(\sigma_2^{(0)})^2\bgamma_3^{(0)}\bgamma_3^{(0)\top}\}^{-1}\bgamma_3^{(0)}$, where $\sigma_2^2=E(\varepsilon_{2i}^2)$ and $\bSigma_3=E(\bvarepsilon_{3i}\bvarepsilon_{3i}^\top)$, then 
$$
\|\bc^*\|_2=O(1),\;\; \|\bc^{(0)}-\bc^*\|_2=O_p(n^{-1/2}),\;\; {\rm and }\;\; \|\bc^{(0)}\|_2=O_p(1).
$$
\end{lemma}
\begin{proof}
	\mbox{}\par
	\noindent
Assumption~\ref{ass:joint_gaussian} implies that $\sigma_2^2<c_1$, $\lambda_{\rm min}(\bSigma_3)>1/c_1$. Notice that $(\bSigma_3+\sigma_2^2\bgamma_3^{*}\bgamma_3^{*\top})-\bSigma_3$ is a positive semidefinite matrix, $\lambda_{\rm min}(\bSigma_3+\sigma_2^2\bgamma_3^{*}\bgamma_3^{*\top})\ge \lambda_{\rm min}(\bSigma_3)>1/c_1$, then $\lambda_{\rm max}\{(\bSigma_3+\sigma_2^2\bgamma_3^{*}\bgamma_3^{*\top})^{-1}\}<c_1$. Therefore,
\begin{align*}
\|\bc^*\|_2\le \sigma_2^2\|(\bSigma_3+\sigma_2^2\bgamma_3^{*}\bgamma_3^{*\top})^{-1}\|_2\|\bgamma_3^*\|_2=\sigma_2^2\lambda_{\rm max}\{(\bSigma_3+\sigma_2^2\bgamma_3^{*}\bgamma_3^{*\top})^{-1}\}\|\bgamma_3^*\|_2\le c_1^2c_2=O(1). 
\end{align*}
Under labeled-sample Gaussian OLS regularity, the estimation errors $\bbeta^{(0)}-\bbeta^*$, $(\sigma_2^{(0)})^2-\sigma_2^2$, $\bgamma_{12}^{(0)}-\bgamma_{12}^*$, $\bgamma_3^{(0)}-\bgamma_3^*$, and $\bSigma_3^{(0)}-\bSigma_3$ are all $O_p(n^{-1/2})$ since these estimates are computed using labeled samples, then it holds that 
$$
\|\bSigma_3^{(0)}+(\sigma_2^{(0)})^2\bgamma_3^{(0)}\bgamma_3^{(0)\top}-\bSigma_3-\sigma_2^2\bgamma_3^*\bgamma_3^{*\top}\|_2=O_p(n^{-1/2}).
$$ Moreover, since $\lambda_{\rm min}(\bSigma_3+\sigma_2^2\bgamma_3^*\bgamma_3^{*\top})\ge \lambda_{\rm min}(\bSigma_3)\ge 1/c_1$, $\lambda_{\rm max}\{(\bSigma_3+\sigma_2^2\bgamma_3^*\bgamma_3^{*\top})^{-1}\}\le c_1$, hence $$\|\{\bSigma_3^{(0)}+(\sigma_2^{(0)})^2\bgamma_3^{(0)}\bgamma_3^{(0)\top}\}^{-1}-(\bSigma_3+\sigma_2^2\bgamma_3^*\bgamma_3^{*\top})^{-1}\|_2=O_p(n^{-1/2}),$$ 
$\|\bc^{(0)}-\bc^*\|_2=O_p(n^{-1/2})$, and $\|\bc^{(0)}\|_2=O_p(1)$.
\end{proof}
\begin{lemma}\label{lemma:delta}
Let $\bD=(\bSigma_3+\sigma_2^2\bgamma_3^*\bgamma_3^{*\top})^{-1}$, we obtain
	\begin{align*}
		\bc^{(0)}-\bc^*%
		&=\{\bD\bgamma_3^*-\sigma_2^2(\bgamma_3^{*\top} \bD\bgamma_3^*)\bD\bgamma_3^*\}\dfrac{1}{n}\sum_{i\in\mathcal{I}}(\varepsilon_{2i}^2-\sigma_2^2)-\sigma_2^2 \bD \dfrac{1}{n}\sum_{i\in\mathcal{I}}(\bvarepsilon_{3i} \bvarepsilon_{3i}^\top-\bSigma_3)\bD\bgamma_3^*\\
		&\qquad+\{\bD-\sigma_2^2\bD\bgamma_3^*\bgamma_3^{*\top} \bD-\sigma_2^2(\bgamma_3^{*\top}\bD\bgamma_3^*)\bD\}\dfrac{1}{n}\sum_{i\in\mathcal{I}}\varepsilon_{2i}\bvarepsilon_{3i}+o_p(n^{-1/2}),
		\end{align*}
	and
	\begin{align*}
		(\sigma_i^{\bOmega^{(0)}})^2-(\sigma_i^{\bOmega^{*}})^2%
		&=\{\sigma_2^2(\bgamma_3^{*\top} \bD\bgamma_3^*)-1\}^2\dfrac{1}{n}\sum_{i\in\mathcal{I}}(\varepsilon_{2i}^2-\sigma_2^2)+ \sigma_2^4\bgamma_3^{*\top} \bD \dfrac{1}{n}\sum_{i\in\mathcal{I}}(\bvarepsilon_{3i} \bvarepsilon_{3i}^\top-\bSigma_3)\bD\bgamma_3^*\\
		&\qquad -2\sigma_2^2(1-\sigma_2^2\bgamma_3^{*\top}\bD\bgamma_3^*)\bgamma_3^{*\top}\bD\dfrac{1}{n}\sum_{i\in\mathcal{I}}\varepsilon_{2i}\bvarepsilon_{3i}+o_p(n^{-1/2}).
	\end{align*}
\end{lemma}
\begin{proof}
	\mbox{}\par
	\noindent
Note that
\begin{align*}
(\sigma_2^{(0)})^2-\sigma_2^2&=\dfrac{1}{n}\sum_{i\in\mathcal{I}}(\varepsilon_{2i}^2-\sigma_2^2)+o_p(n^{-1/2})=O_p(n^{-1/2}),\\
\bSigma_3^{(0)}-\bSigma_3&=\dfrac{1}{n}\sum_{i\in\mathcal{I}}(\bvarepsilon_{3i}\bvarepsilon_{3i}^\top-\bSigma_3)+o_p(n^{-1/2})=O_p(n^{-1/2}),\\
\bgamma_3^{(0)}-\bgamma_3^*&=\dfrac{1}{n}\sum_{i\in\mathcal{I}}\varepsilon_{2i}\bvarepsilon_{3i}/\sigma_2^2+o_p(n^{-1/2})=O_p(n^{-1/2}),
\end{align*}
the lemma follows directly from an application of the delta method.
\end{proof}

In the following, we define $\bN_d=(\bI_{d^2}+\bK_{d,d})/2$ as a $d^2\times d^2$ symmetrizer matrix, which acts as the orthogonal projector onto the subspace of vectorized symmetric matrices, here $\bK_{d,d}$ is the commutation matrix satisfying $\bK_{d,d}{\rm vec}(\bA)={\rm vec}(\bA^\top)$ for any $d\times d$ matrix $\bA$.
For a symmetric covariance matrix of size $d$, let $\mathbf U_d\in\mathbb R^{d^2\times d(d+1)/2}$ have orthonormal columns spanning $\mathcal S_d=\{\operatorname{vec}(\bH):\bH=\bH^\top\}$. Then $\mathbf U_d^\top\mathbf U_d=\bI_{d(d+1)/2}$ and $\mathbf U_d\mathbf U_d^\top=\bN_d$.
\begin{lemma}\label{lemma:matrix_inverse}
	Let $\bu = \bSigma_3^{-1}\bgamma_3$ and $t = \bgamma_3^\top\bSigma_3^{-1}\bgamma_3$. Let $a_1, a_2, a_3, b_1, b_2, b_3, c_{31}, c_{32}$ be scalar coefficients. Consider the symmetric block matrix $\bA$ partitioned as
	\[
	\bA = \begin{bmatrix} \bA_1 & \bA_2 \\ \bA_2^\top & \bA_3 \end{bmatrix},
	\]
	where the blocks are defined as:
	\begin{align*}
		\bA_1 &= a_1 \bSigma_3^{-1}+b_1 \bu\bu^\top,\\
		\bA_2 &= a_2 (\bu^\top \otimes \bSigma_3^{-1}+\bSigma_3^{-1}\otimes \bu^\top) + b_2(\bu\bu^\top \otimes \bu^\top+\bu^\top \otimes \bu\bu^\top),\\
		\bA_3 & = \bN_{d_2}\{a_3(\bSigma_3^{-1}\otimes \bSigma_3^{-1})+b_3(\bSigma_3^{-1}\otimes \bu\bu^\top + \bu\bu^\top \otimes \bSigma_3^{-1})\\
		&\qquad\qquad+c_{31}(\bu\bu^\top\otimes \bu\bu^\top)\}\bN_{d_2}+c_{32}(\bu\bu^\top\otimes \bu\bu^\top),
	\end{align*} 
	Let $\bP_A={\rm diag}(\bI_{d_2},\mathbf U_{d_2})$, assume that $\bP_A^\top\bA\bP_A$ and $\mathbf U_{d_2}^\top\bA_3\mathbf U_{d_2}$ are positive definite, and write $\bA_{\cal S}^{[-1]}=\bP_A(\bP_A^\top\bA\bP_A)^{-1}\bP_A^\top$ for the inverse on the admissible symmetric-covariance space. Its $(1,1)$-block is
 	\[
	[\bA_{\cal S}^{[-1]}]_{1,1} = \frac{1}{\delta_1}\bSigma_3 - \frac{\delta_2}{\delta_1(\delta_1 + \delta_2 t)} \bgamma_3\bgamma_3^\top,
 	\]
	where the scalar parameters are defined as: $D_1 = a_3 + b_3 t$, $D_2 = a_3 + 2b_3 t + (c_{31} + c_{32})t^2$
	\begin{align*}
		\delta_1 &= a_1 - \frac{2a_2^2 t}{D_1}, \;\;\delta_2= b_1 - \frac{4(a_2 + b_2 t)^2}{D_2} + \frac{2a_2^2}{D_1}.
	\end{align*}
\end{lemma}

\begin{proof}
	\mbox{}\par
	\noindent
	Let $\bSigma_3=\bL\bL^\top$ be the Cholesky decomposition of $\bSigma_3$, and 	
	\[
	\bT \;=\;
	\begin{bmatrix}
		\bL^\top & 0 \\
		0 & \bL^\top \otimes\bL^\top
	\end{bmatrix}.
	\]
	Then,
	\[
	\tilde \bA = \bT \bA \bT^\top =
	\begin{bmatrix}
		\bL^\top \bA_1 \bL
		&
		\bL^\top \bA_2 (\bL\otimes \bL) \\[0.2em]
		(\bL^\top\otimes \bL^\top) \bA_2^\top \bL
		&
		(\bL^\top\otimes \bL^\top)\, \bA_3\, (\bL\otimes \bL)
	\end{bmatrix}=\begin{bmatrix}
	\tilde{\bA}_1 &\tilde{\bA}_2\\
	\tilde{\bA}_2^\top & \tilde{\bA}_3
\end{bmatrix}
	\]
	where $\tilde{\bgamma}_3=\bL^{-1}\bgamma_3$
	\begin{align*}
		\tilde \bA_1 &= a_1 \bI + b_1\tilde\bgamma_3\tilde\bgamma_3^\top,\\
		\tilde \bA_2 &= a_2(\tilde\bgamma_3^\top\otimes \bI + \bI\otimes\tilde\bgamma_3^\top)
		+ b_2(\tilde\bgamma_3\tilde\bgamma_3^\top\otimes\tilde\bgamma_3^\top
		+ \tilde\bgamma_3^\top\otimes\tilde\bgamma_3\tilde\bgamma_3^\top),\\
		\tilde \bA_3 &= \bN_{d_2}[a_3(\bI\otimes \bI)+ b_3(\bI\otimes\tilde\bgamma_3\tilde\bgamma_3^\top+ \tilde\bgamma_3\tilde\bgamma_3^\top\otimes \bI)+ c_{31}(\tilde\bgamma_3\tilde\bgamma_3^\top
		\otimes\tilde\bgamma_3\tilde\bgamma_3^\top)]\bN_{d_2}+c_{32}(\tilde\bgamma_3\tilde\bgamma_3^\top
		\otimes\tilde\bgamma_3\tilde\bgamma_3^\top),
	\end{align*}
	where the last equality holds, since $\bK_{d_2,d_2}(\bL\otimes \bL)=(\bL\otimes \bL)\bK_{d_2,d_2}$ and thus $\bN_{d_2}(\bL\otimes \bL)=(\bL\otimes \bL)\bN_{d_2}$. Define $\tilde{\bA}_{\cal S}^{[-1]}=\bP_A(\bP_A^\top\tilde{\bA}\bP_A)^{-1}\bP_A^\top$ analogously. Its compressed Schur complement is $\tilde{\bA}_1-\tilde{\bA}_2\mathbf U_{d_2}(\mathbf U_{d_2}^\top\tilde{\bA}_3\mathbf U_{d_2})^{-1}\mathbf U_{d_2}^\top\tilde{\bA}_2^\top$. Thus we solve $\tilde{\bA}_3\bPhi=\tilde{\bA}_2^\top$ with each column of $\bPhi$ in ${\cal S}_{d_2}$. Let %
  	$$
	\tilde{\bA}_3^\prime=a_3(\bI\otimes \bI)+ b_3(\bI\otimes\tilde\bgamma_3\tilde\bgamma_3^\top+ \tilde\bgamma_3\tilde\bgamma_3^\top\otimes \bI)+ (c_{31}+c_{32})(\tilde\bgamma_3\tilde\bgamma_3^\top
	\otimes\tilde\bgamma_3\tilde\bgamma_3^\top).
	$$
	We next show that the solution of $\tilde{\bA}_3\bPhi=\tilde{\bA}_2^\top$ is equal to the solution of $\tilde{\bA}_3^\prime\bPhi=\tilde{\bA}_2^\top$. Note $\tilde{\bA}_2^\top=a_2\bT_1+b_2\bT_2$, where
	\begin{align*}
		\bT_1=\tilde\bgamma_3\otimes \bI + \bI\otimes\tilde\bgamma_3,\;\; \bT_2=\tilde\bgamma_3\tilde\bgamma_3^\top\otimes\tilde\bgamma_3
		+ \tilde\bgamma_3\otimes\tilde\bgamma_3\tilde\bgamma_3^\top
	\end{align*}
	Since $\bN_{d_2}\bT_1=\bT_1$ and $\bN_{d_2}\bT_2=\bT_2$, we have $\tilde{\bA}_2^\top\in {\rm Range}(\bN_{d_2})
	=\{{\rm vec}(\bH):\bH=\bH^\top\}=: \mathcal S$. Consider the linear system $\tilde{\bA}_3\bPhi=\tilde{\bA}_2^\top$ on the symmetric subspace $\mathcal S$. Since $\tilde{\bA}_2^\top\in\mathcal S$ and $\tilde{\bA}_3$ preserves this subspace, invertibility of its restriction gives a unique solution $\bPhi\in\mathcal S$. Write $\tilde{\bA}_3=\bN_{d_2}\bB\bN_{d_2}+c_{32}\bR$, where
	\[
	\bB=a_3(\bI\otimes \bI)+b_3(\bI\otimes\tilde\bgamma_3\tilde\bgamma_3^\top+\tilde\bgamma_3\tilde\bgamma_3^\top\otimes\bI)+c_{31}(\tilde\bgamma_3\tilde\bgamma_3^\top\otimes\tilde\bgamma_3\tilde\bgamma_3^\top),
	\quad \bR=(\tilde\bgamma_3\tilde\bgamma_3^\top)\otimes(\tilde\bgamma_3\tilde\bgamma_3^\top).
	\]
	Next, for any $\bx\in\mathcal S$ we have $\bN_{d_2}\bx=\bx$ and $\bB\bx\in\mathcal S$, $\bR\bx\in\mathcal S$,
	so
	\[
	\tilde{\bA}_3\bx
	= \bN_{d_2}\bB\bN_{d_2}\bx+c_{32}\bR\bx
	= \bB\bx+c_{32}\bR\bx
	= \tilde{\bA}_3'\bx,
	\]
	Hence, on the symmetric subspace
	$\mathcal S$, the operators $\tilde{\bA}_3$ and $\tilde{\bA}_3'$ coincide. Consequently, any
	$\bPhi\in\mathcal S$ satisfies $\tilde{\bA}_3\bPhi=\tilde{\bA}_2^\top$ must be the solution to $\tilde{\bA}_3'\bPhi=\tilde{\bA}_2^\top $. In the following we solve $\tilde{\bA}_3'\bPhi=\tilde{\bA}_2^\top $. Denote $t := \tilde\bgamma_3^\top \tilde\bgamma_3$ and $c_3=c_{31}+c_{32}$, we have
	\begin{align*}
		\tilde \bA_3^\prime \bT_1
		&= (a_3 + b_3 t) \bT_1
		+ (b_3 + c_3 t) \bT_2, \;\;
		\tilde \bA_3^\prime \bT_2
		= (a_3 + 2b_3 t + c_3 t^2)\bT_2.
	\end{align*}
	For $t=0$, $\bgamma_3=\mathbf0$, $\bA_2=0$, and the asserted inverse block is simply $a_1^{-1}\bSigma_3$; hence, in the remaining argument assume $t>0$. We now solve the linear system $\tilde \bA_3^\prime \bPhi = \tilde \bA_2^\top$. $\bPhi$ has the form $x_1\bT_1+x_2\bT_2$, 
  	\begin{align*}
		x_1(a_3 + b_3 t) &= a_2,\;\;
		x_1(b_3 + c_3 t)
		+ x_2(a_3 + 2b_3 t + c_3 t^2) = b_2,
	\end{align*}
	denote $D_1=a_3+b_3t$, $D_2=a_3+2b_3t+c_3t^2$, $E=b_3+c_3t$, which gives
	\begin{align*}
		x_1 &= \frac{a_2}{D_1},\;\;
		x_2 = 
		\frac{b_2(a_3+b_3t)-a_2(b_3+c_3t)}
		{(a_3 + 2b_3 t + c_3 t^2)(a_3+b_3t)}=\dfrac{b_2-x_1E}{D_2}
	\end{align*}
	Then $\bPhi =x_1\bT_1+x_2\bT_2$, 
	\begin{align*}
	\tilde{\bA}_2\bPhi=\lambda_1 \bI+\lambda_2\tilde{\bgamma}_3\tilde{\bgamma}_3^\top,\;\; \lambda_1=\dfrac{2a_2^2t}{D_1},\;\; \lambda_2=\dfrac{4(a_2+b_2t)^2}{D_2}-\dfrac{2a_2^2}{D_1}
 	\end{align*}
	The last expression holds, since $\tilde{\bA}_2$ is the linear combination of $\bT_1^\top$ and $\bT_2^\top$, $\bPhi$ is the linear combination of $\bT_1$ and $\bT_2$, their product must be the linear combination of $\bI$ and $\tilde{\bgamma}_3\tilde{\bgamma}_3^\top$. Then $\lambda_1$ and $\lambda_2$ can be obtained by computing the eigenvalues. When $d_2=1$ and $t>0$, there is no nonzero perpendicular direction; the Schur complement is scalar and its inverse is $(\delta_1+\delta_2t)^{-1}$ in whitened coordinates. The perpendicular-direction calculation below is for $d_2\ge 2$. For any $\bv$ such that $\bv^\top \tilde{\bgamma}_3=0$, we have 
 	$\tilde{\bA}_2\bPhi \bv = \lambda_1 \bv$, and $\tilde{\bA}_2\bPhi \tilde{\bgamma}_3=(\lambda_1+\lambda_2t)\tilde{\bgamma}_3$. Notice that
 	\begin{align*}
		\bT_1\bv = \tilde{\bgamma}_3\otimes \bv+\bv\otimes \tilde{\bgamma}_3,\;\; \bT_2 \bv = 0,
	\end{align*}
	and $\tilde{\bA}_2^\top \bv=a_2\bT_1\bv=a_2(\tilde{\bgamma}_3\otimes \bv+\bv\otimes \tilde{\bgamma}_3)$, $\bPhi \bv=x_1\bT_1\bv=(a_2/D_1)(\tilde{\bgamma}_3\otimes \bv+\bv\otimes \tilde{\bgamma}_3)$, $\tilde{\bA}_2\bPhi \bv=a_2\bT_1^\top\bPhi \bv=(2a_2^2t /D_1)\bv$, so $\lambda_1=2a_2^2t /D_1$. On the other hand, we have $\bT_1\tilde{\bgamma}_3=2(\tilde{\bgamma}_3\otimes \tilde{\bgamma}_3)$, $\bT_2\tilde{\bgamma}_3=2t(\tilde{\bgamma}_3\otimes \tilde{\bgamma}_3)$, $\tilde{\bA}_2^\top \tilde{\bgamma}_3=2(a_2+b_2t)(\tilde{\bgamma}_3\otimes \tilde{\bgamma}_3)$, $\bPhi \tilde{\bgamma}_3=\{2(a_2+b_2t)/D_2\}(\tilde{\bgamma}_3\otimes \tilde{\bgamma}_3)$ (Since we have $\tilde{\bA}_3^\prime(\tilde{\bgamma}_3\otimes \tilde{\bgamma}_3)=D_2(\tilde{\bgamma}_3\otimes \tilde{\bgamma}_3)$, hence $\tilde{\bA}_3^{\prime-1}(\tilde{\bgamma}_3\otimes \tilde{\bgamma}_3)=D_2^{-1}(\tilde{\bgamma}_3\otimes \tilde{\bgamma}_3)$), $\tilde{\bA}_2\bPhi\tilde{\bgamma}_3=4t(a_2+b_2t)^2\tilde{\bgamma}_3/D_2$, so $\lambda_2 =\{4t(a_2+b_2t)^2/D_2-\lambda_1\}/t =4(a_2+b_2t)^2/D_2-2a_2^2/D_1$.
  	Then 
	\begin{align*}
		\tilde{\bA}_1-\tilde{\bA}_2\bPhi&=(a_1-\lambda_1)\bI+(b_1-\lambda_2)\tilde{\bgamma}_3\tilde{\bgamma}_3^\top=\delta_1\bI+\delta_2\tilde{\bgamma}_3\tilde{\bgamma}_3^\top,\\
		[\tilde{\bA}_{\cal S}^{[-1]}]_{1,1}&=(\tilde{\bA}_1-\tilde{\bA}_2\bPhi)^{-1}=\dfrac{1}{\delta_1}\bI-\dfrac{\delta_2}{\delta_1(\delta_1+\delta_2t)}\tilde{\bgamma}_3\tilde{\bgamma}_3^\top.
 		\end{align*}
	On the admissible symmetric-covariance subspace, $\bA_{\cal S}^{[-1]}=\bT^\top\tilde{\bA}_{\cal S}^{[-1]}\bT$,
	\begin{align*}
		[\bA_{\cal S}^{[-1]}]_{1,1}&=\bL	[\tilde{\bA}_{\cal S}^{[-1]}]_{1,1}\bL^\top =\dfrac{1}{\delta_1}\bSigma_3-\dfrac{\delta_2}{\delta_1(\delta_1+\delta_2t)}\bgamma_3\bgamma_3^\top,\;\; {\rm now}\;\; t=\bgamma_3^\top\bSigma_3^{-1}\bgamma_3.
	\end{align*}
This proves the lemma.
\end{proof}
\begin{lemma}[Weighted sums over the unlabeled sample]\label{lemma:lemma5}
Let $f_i$, $i\in \mathcal{J}$, be independent and identically distributed fixed-dimensional random vectors or matrices satisfying $E\|f_i\|^2<\infty$. For $N=n+m$, we have 
\[\frac{1}{N}\sum_{i\in \mathcal{J}}f_i=\frac{m}{N}E(f_i)+O_p\Big(\frac{\sqrt{m}}{N}\Big)\]
When $m=0$, the sum is interpreted as zero.
\end{lemma}
\begin{proof}
Since the observations are independent, 
\[E\Big[\frac{1}{N}\sum_{i\in\mathcal{J}}\{f_i-E(f_i)\}\Big]=0\]
Furthermore,
\[E\Big\|\frac{1}{N}\sum_{i\in\mathcal{J}}\{f_i-E(f_i)\}\Big\|_F^2=\frac{1}{N^2}\sum_{i\in\mathcal{J}}E\|f_i-E(f_i)\|_F^2=\frac{m}{N^2}E\|f_i-E(f_i)\|_F^2\]
Here the cross terms vanish by independence and centering. For matrices, the operator norm is bounded by the Frobenius norm. By Chebyshev's inequality,
\[\frac{1}{N}\sum_{i\in\mathcal{J}}\{f_i-E(f_i)\}=O_p\Big(\frac{\sqrt{m}}{N}\Big)\]
which establishes the results.
\end{proof}

\section{Proofs of the main theorems}\label{sec:proofs}
\begin{proof}[Proof of Theorem \ref{thm:upper_bound}]
We consider the sampling regime specified in Section~\ref{sec:upper_bound} of the main text. The result for $\widehat{\balpha}$ holds for any $m\geq0$, since $N\geq n\to\infty$. For the remaining estimators, we first retain the original asymptotic expansions under $\tau>0$, which implies $m\to\infty$. We then establish the boundary case $\tau=0$, equivalently $m=o(n)$, by comparing the one-step SAEM estimators with their supervised counterparts.
\mbox{}\par
\noindent\textbf{Step 1. Control $\widehat\balpha-\balpha^*$.}
We have
\begin{align*}
\widehat\balpha-\balpha^*&=\Big(\dfrac{1}{N}\sum_{i\in\mathcal{I}\cup\mathcal{J}}\bX_i\bX_i^\top\Big)^{-1}\Big(\dfrac{1}{N}\sum_{i\in\mathcal{I}\cup\mathcal{J}}\bX_i\bvarepsilon_{1i}^\top\Big),\\
\sqrt{N}\{{\rm vec}(\widehat\balpha)-{\rm vec}(\balpha^*)\}&\xrightarrow{\rm d}\mathcal{N}\left(0,(\bI_{d_1}\otimes \bSigma_{\bX}^{-1})E\{(\bvarepsilon_{1i}\bvarepsilon_{1i}^\top)\otimes (\bX_i\bX_i^\top)\}(\bI_{d_1}\otimes \bSigma_{\bX}^{-1})\right).
\end{align*}
Under Assumption~\ref{ass:joint_gaussian}, $\bvarepsilon_{1i}$ is independent of $\bX_i$, so that $E\{(\bvarepsilon_{1i}\bvarepsilon_{1i}^\top)\otimes (\bX_i\bX_i^\top)\}=\bSigma_1\otimes\bSigma_{\bX}$ and the limiting covariance matrix equals $\bSigma_1\otimes\bSigma_{\bX}^{-1}=\bV_{\balpha}$.

\noindent\textbf{Step 2. Control $\widehat\bbeta-\bbeta^*$.}
We denote $\bSigma_{\bX,1}=E(\bar{\bX}_{1i}\bar{\bX}_{1i}^\top)$, $\bv_i=\bgamma_3^*\varepsilon_{2i}+\bvarepsilon_{3i}$, $\bc^{(0)}=(\sigma_2^{(0)})^2\{\bSigma_3^{(0)}+(\sigma_2^{(0)})^2\bgamma_3^{(0)}\bgamma_3^{(0)\top}\}^{-1}\bgamma_3^{(0)}$ and $\bc^*=\sigma_2^2(\bSigma_3+\sigma_2^2\bgamma_3^{*}\bgamma_3^{*\top})^{-1}\bgamma_3^*$. Note that
\begin{align*}
	\bbeta^{(1)} - \bbeta^* &= \bigg( N^{-1} \sum_{i=1}^{N} \bar{\bX}_{1i} \bar{\bX}_{1i}^\top \bigg)^{-1} \bigg\{ \frac{1}{N} \sum_{i \in \mathcal{I}} \bar{\bX}_{1i} \varepsilon_{2i}+ \frac{1}{N} \sum_{i \in \mathcal{J}} \bar{\bX}_{1i} \left( \mu_i^{\bOmega^{(0)}} - \bar{\bX}_{1i}^\top \bbeta^* \right) \bigg\},
\end{align*}
where
\begin{align*}
	\mu_i^{\bOmega^{(0)}}&=\bbeta^{(0)\top}\bar{\bX}_{1i}+\bc^{(0)\top}(\bS_{2i}-\bgamma_{12}^{(0)\top}\bar{\bX}_{1i}-\bbeta^{(0)\top}\bar{\bX}_{1i}\bgamma_3^{(0)}).
 \end{align*}
Notice that
\begin{align}
	&\frac{1}{N} \sum_{i \in \mathcal{J}} \bar{\bX}_{1i} \left( \mu_i^{\bOmega^{(0)}} - \bar{\bX}_{1i}^\top \bbeta^* \right)\nonumber\\
 	=\ &\frac{1}{N} \sum_{i \in \mathcal{J}} \bar{\bX}_{1i} \left\{ \bbeta^{(0)\top}\bar{\bX}_{1i} + \bc^{(0)\top} \left(\bS_{2i}-\bgamma_{12}^{(0)\top}\bar{\bX}_{1i}-\bbeta^{(0)\top}\bar{\bX}_{1i}\bgamma_3^{(0)}\right) \right\}- \frac{1}{N} \sum_{i \in \mathcal{J}} \bar{\bX}_{1i} \bar{\bX}_{1i}^\top \bbeta^*\nonumber\\ 
	=\ &\frac{1}{N} \sum_{i \in \mathcal{J}} \bar{\bX}_{1i} \left[ \bar{\bX}_{1i}^\top (\bbeta^{(0)}-\bbeta^*)+\bc^{(0)\top}\left\{\bv_i+(\bgamma_{12}^*-\bgamma_{12}^{(0)})^\top \bar{\bX}_{1i}+(\bbeta^*\bgamma_3^{*\top}-\bbeta^{(0)}\bgamma_3^{(0)\top})^\top \bar{\bX}_{1i}\right\}\right]\nonumber\\
	=\ &\frac{1}{N} \sum_{i \in \mathcal{J}} \bar{\bX}_{1i} \left[ \bar{\bX}_{1i}^\top (\bbeta^{(0)}-\bbeta^*)+\bc^{*\top}\left\{\bv_i+(\bgamma_{12}^*-\bgamma_{12}^{(0)})^\top \bar{\bX}_{1i}+(\bbeta^*\bgamma_3^{*\top}-\bbeta^{(0)}\bgamma_3^{(0)\top})^\top \bar{\bX}_{1i}\right\}\right]+o_p(n^{-1/2}),\nonumber
\end{align}
where the last equality holds, since $\|\bc^{(0)}-\bc^*\|_2=O_p(n^{-1/2})$ by Lemma \ref{lemma:lemma2}, and by Assumption~\ref{ass:joint_gaussian} and Lemma \ref{lemma:lemma1},
\begin{align*}
\frac{1}{N} \sum_{i \in \mathcal{J}} \bar{\bX}_{1i}\Big\{\bv_i+(\bgamma_{12}^*-\bgamma_{12}^{(0)})^\top \bar{\bX}_{1i}+(\bbeta^*\bgamma_3^{*\top}-\bbeta^{(0)}\bgamma_3^{(0)\top})^\top \bar{\bX}_{1i}\Big\}^\top=O_p(n^{-1/2}).
\end{align*}
Hence, the additional term arising from replacing $\bc^{(0)}$ with $\bc^{*}$ is an $O_p(n^{-1})=o_p(n^{-1/2})$ term. Then the above term becomes 
\begin{align*}
	&\dfrac{1}{N}\sum_{i\in\mathcal{J}} \bar{\bX}_{1i}\bc^{*\top}\bv_i + \dfrac{1}{N} \sum_{i \in \mathcal{J}} \bar{\bX}_{1i}\bar{\bX}_{1i}^\top(\bbeta^{(0)}-\bbeta^*)\\
	&\qquad -\dfrac{1}{N} \sum_{i \in \mathcal{J}} \bar{\bX}_{1i}\bar{\bX}_{1i}^\top\left\{(\bgamma_{12}^{(0)}-\bgamma_{12}^{*})+ \bbeta^*(\bgamma_3^{(0)}-\bgamma_3^*)^\top+(\bbeta^{(0)}-\bbeta^*)\bgamma_3^{*\top}\right\}\bc^*+o_p(n^{-1/2})\\
	=\ &\dfrac{1}{N}\sum_{i\in\mathcal{J}} \bar{\bX}_{1i}\bc^{*\top}\bv_i+\dfrac{1-\bc^{*\top}\bgamma_3^*}{N} \sum_{i \in \mathcal{J}} \bar{\bX}_{1i}\bar{\bX}_{1i}^\top(\bbeta^{(0)}-\bbeta^*)\\
	&\qquad -\dfrac{1}{N}\sum_{i\in\mathcal{J}} \bar{\bX}_{1i}\bar{\bX}_{1i}^\top\Big\{[\bI_{p+d_1},\bbeta^*]\Big(\dfrac{1}{n}\sum_{i\in\mathcal{I}}\bar{\bX}_{2i}\bar{\bX}_{2i}^\top\Big)^{-1}\Big(\dfrac{1}{n}\sum_{i\in\mathcal{I}}\bar{\bX}_{2i}\bvarepsilon_{3i}^\top\Big)\bc^*\Big\}+o_p(n^{-1/2})\\
	=\ &\dfrac{1}{N}\sum_{i\in\mathcal{J}} \bar{\bX}_{1i}\bc^{*\top}\bv_i+\dfrac{1-\bc^{*\top}\bgamma_3^*}{N} \sum_{i \in \mathcal{J}} \bar{\bX}_{1i}\bar{\bX}_{1i}^\top \Big(\bSigma_{\bX,1}^{-1}\dfrac{1}{n}\sum_{i\in\mathcal{I}}\bar{\bX}_{1i}\varepsilon_{2i}\Big)\\
	&\qquad-\dfrac{1}{N}\sum_{i\in\mathcal{J}} \bar{\bX}_{1i}\bar{\bX}_{1i}^\top\Big(\bSigma_{\bX,1}^{-1}\dfrac{1}{n}\sum_{i\in\mathcal{I}}\bar{\bX}_{1i}\bvarepsilon_{3i}^\top \bc^*\Big)+o_p(n^{-1/2})
\end{align*}
where the second equality holds, since $\bgamma^{(0)}-\bgamma^*=(n^{-1}\sum_{i\in\mathcal{I}}\bar{\bX}_{2i}\bar{\bX}_{2i}^\top)^{-1}(n^{-1}\sum_{i\in\mathcal{I}}\bar{\bX}_{2i}\bvarepsilon_{3i}^\top)$, the third equality holds, since $\bbeta^{(0)}-\bbeta^*=(n^{-1}\sum_{i\in\mathcal{I}}\bar{\bX}_{1i}\bar{\bX}_{1i}^\top)^{-1}(n^{-1}\sum_{i\in\mathcal{I}}\bar{\bX}_{1i}\varepsilon_{2i})=\bSigma_{\bX,1}^{-1}n^{-1}\sum_{i\in\mathcal{I}}\bar{\bX}_{1i}\varepsilon_{2i}+o_p(n^{-1/2})$, and
\begin{equation}\label{eq:Ibeta}
	\begin{split}
&[\bI_{p+d_1},\bbeta^*]\Big(\dfrac{1}{n}\sum_{i\in\mathcal{I}}\bar{\bX}_{2i}\bar{\bX}_{2i}^\top\Big)^{-1}\Big(\dfrac{1}{n}\sum_{i\in\mathcal{I}}\bar{\bX}_{2i}\bvarepsilon_{3i}^\top\Big)\\
=\ &\bSigma_{\bX,1}^{-1}[\bSigma_{\bX,1},E(\bar{\bX}_{1i}Y_i)]E(\bar{\bX}_{2i}\bar{\bX}_{2i}^\top)^{-1}\dfrac{1}{n}\sum_{i\in\mathcal{I}}\bar{\bX}_{2i}\bvarepsilon_{3i}^\top+o_p(n^{-1/2})\\
=\ &\bSigma_{\bX,1}^{-1}[\bI_{p+d_1},0]\dfrac{1}{n}\sum_{i\in\mathcal{I}}\bar{\bX}_{2i}\bvarepsilon_{3i}^\top+o_p(n^{-1/2})\\
=\ &\bSigma_{\bX,1}^{-1}\dfrac{1}{n}\sum_{i\in\mathcal{I}}\bar{\bX}_{1i}\bvarepsilon_{3i}^\top+o_p(n^{-1/2}).
	\end{split}
\end{equation}
Assuming $\varepsilon_{2i}\sim \mathcal{N}(0,\sigma_2^2)$ and $\bvarepsilon_{3i}\sim\mathcal{N}(0,\bSigma_3)$, when $m\to \infty$, since $m/N\to \tau$, we obtain
\begin{align*}
\sqrt{n}	(\widehat\bbeta - \bbeta^*) &=
\sqrt{n}  \bSigma_{\bX,1}^{-1}\bigg\{\frac{1}{N} \sum_{i \in \mathcal{I}} \bar{\bX}_{1i} \varepsilon_{2i}+\dfrac{1}{N}\sum_{i\in\mathcal{J}} \bar{\bX}_{1i}\bc^{*\top}\bv_i\\
	&\qquad\qquad\qquad+\dfrac{m(1-\bc^{*\top}\bgamma_3^*)}{Nn}\sum_{i\in\mathcal{I}}\bar{\bX}_{1i} \varepsilon_{2i}-(m/N) \dfrac{1}{n}\sum_{i\in\mathcal{I}}\bar{\bX}_{1i}\bvarepsilon_{3i}^\top \bc^*\bigg\}+o_p(1)\\
	& =\sqrt{n} \bSigma_{\bX,1}^{-1}\bigg\{\dfrac{1}{N}\sum_{i\in\mathcal{J}} \bar{\bX}_{1i}\bc^{*\top}\bv_i+\dfrac{1-(m/N)\bc^{*\top}\bgamma_3^*}{n}\sum_{i\in\mathcal{I}}\bar{\bX}_{1i} \varepsilon_{2i}-\dfrac{m}{nN}\sum_{i\in\mathcal{I}}\bar{\bX}_{1i}\bvarepsilon_{3i}^\top \bc^*\bigg\}+o_p(1)\\
	&\xrightarrow{\rm d}\mathcal{N}(0,\bV_{\bbeta})
\end{align*}
where $\bV_{\bbeta}=\sigma_2^2(1-\tau\kappa)\bSigma_{\bX,1}^{-1}$ as defined in Theorem~\ref{thm:upper_bound}. Since $\bbeta_1^*$ consists of the first $p$ coordinates of $\bbeta^*$, $\sqrt n(\widehat\bbeta_1-\bbeta_1^*)\xrightarrow{\rm d}\mathcal N(\mathbf 0,[\bI_p,\mathbf 0_{p\times d_1}]\bV_{\bbeta}[\bI_p,\mathbf 0_{p\times d_1}]^\top)$. Because $\bSigma_{\bX,1}=\begin{bmatrix}\bI_p&\mathbf 0\\ \balpha^{*\top}&\bI_{d_1}\end{bmatrix}{\rm diag}(\bSigma_{\bX},\bSigma_1)\begin{bmatrix}\bI_p&\balpha^*\\ \mathbf 0&\bI_{d_1}\end{bmatrix}$, we have $[\bI_p,\mathbf 0_{p\times d_1}]\bSigma_{\bX,1}^{-1}[\bI_p,\mathbf 0_{p\times d_1}]^\top=\bSigma_{\bX}^{-1}+\balpha^*\bSigma_1^{-1}\balpha^{*\top}$, and hence $\bV_{\bbeta_1}=\sigma_2^2(1-\tau\kappa)(\bSigma_{\bX}^{-1}+\balpha^*\bSigma_1^{-1}\balpha^{*\top})$.
The remaining case $m=o(n)$, including fixed $m$, is established jointly for all estimators at the end of this proof.

\noindent\textbf{Step 3. Control $\widehat{\btheta}-\btheta^*$.} Note that 
\[\widehat{\btheta}=\widehat{\bbeta}_1+\widehat{\balpha}\widehat{\bbeta}_2,\quad \btheta^*=\bbeta_1^*+\balpha^*\bbeta_2^*\]
Hence
\begin{equation}\label{eq:theta_em_asymp}
	\begin{split}
\widehat{\btheta}-\btheta^*&=(\widehat{\bbeta}_1-\bbeta_1^*)+(\widehat{\balpha}\widehat{\bbeta}_2-\balpha^*\bbeta_2^*)
\\ &=(\widehat{\bbeta}_1-\bbeta_1^*)+\balpha^*(\widehat{\bbeta}_2-\bbeta_2^*)+(\widehat{\balpha}-\balpha^*)\bbeta_2^*+(\widehat{\balpha}-\balpha^*)(\widehat{\bbeta}_2-\bbeta_2^*)
\\ &=[\bI_{p},\balpha^*](\widehat{\bbeta}-\bbeta^*)+(\widehat{\balpha}-\balpha^*)\bbeta_2^*+(\widehat{\balpha}-\balpha^*)(\widehat{\bbeta}_2-\bbeta_2^*)
	\end{split}
\end{equation}
Note that $\widehat{\balpha}-\balpha^*=O_p(N^{-1/2})$, $\widehat{\bbeta}-\bbeta^*=O_p(n^{-1/2})$, thus $(\widehat{\balpha}-\balpha^*)(\widehat{\bbeta}_2-\bbeta_2^*)=o_p(n^{-1/2})$, thus 
\[\widehat{\btheta}-\btheta^*=\underbrace{[\bI_p,\balpha^*](\widehat{\bbeta}-\bbeta^*)}_{\text{first term}}+\underbrace{(\widehat{\balpha}-\balpha^*)\bbeta_2^*}_{\text{second term}}+o_p(n^{-1/2})\]
From Step~2 and the continuous mapping theorem we know that 
\[\sqrt{n}(\widehat{\bbeta}-\bbeta^*)\xrightarrow{\rm d} \mathcal{N}\Big\{0,\bSigma_{\bX,1}^{-1} \sigma_2^2 \Big(1-\tau\times \frac{\sigma_2^2\bgamma_3^{*\top}\bSigma_3^{-1}\bgamma_3^*}{1+\sigma_2^2\bgamma_3^{*\top}\bSigma_3^{-1}\bgamma_3^*}\Big)\Big\}\]
thus 
\[\sqrt{n}[\bI_p,\balpha^*](\widehat{\bbeta}-\bbeta^*)\xrightarrow{\rm d} \mathcal{N}\Big\{0,[\bI_p,\balpha^*] \bV_{\bbeta}[\bI_p,\balpha^*]^{\top}\Big\}\]
Since $\balpha^*=\bSigma_{\bX}^{-1}E(\bX_i \bS_{1i}^{\top})$, we have 
\[[\bI_p, \balpha^*]=[\bI_p,\bSigma_{\bX}^{-1}E(\bX_i \bS_{1i}^{\top})]=\bSigma_{\bX}^{-1}[\bSigma_{\bX},E(\bX_i \bS_{1i}^{\top})]=\bSigma_{\bX}^{-1}E(\bX_i\bar{\bX}_{1i}^{\top})\]
Moreover, since $\bar{\bX}_{1i}=(\bX_i^{\top},\bS_{1i}^{\top})^{\top}$, $\bSigma_{\bX,1}=E(\bar{\bX}_{1i}\bar{\bX}_{1i}^{\top})$ and $E(\bX_i\bar{\bX}_{1i}^{\top})=[\bI_p,\boldsymbol{0}_{p\times d_1}] \bSigma_{\bX,1}$, we have 
\[[\bI_p,\balpha^*] \bSigma_{\bX,1}^{-1}=\bSigma_{\bX}^{-1}[\bI_p,\boldsymbol{0}_{p\times d_1}]\]
Hence 
\[[\bI_p,\balpha^*] \bSigma_{\bX,1}^{-1}[\bI_p,\balpha^*]^{\top}=\bSigma_{\bX}^{-1}[\bI_p,\boldsymbol{0}_{p\times d_1}]\begin{bmatrix}
\bI_{p}\\ \balpha^{* \top}
\end{bmatrix}=\bSigma_{\bX}^{-1}\]
Therefore, the asymptotic variance of first term is 
\[[\bI_p,\balpha^*] \bV_{\bbeta}[\bI_p,\balpha^*]^{\top}=\sigma_2^2 (1-\tau\kappa)\bSigma_{\bX}^{-1}\]
We now consider the second term, since 
\[\widehat{\balpha}-\balpha^*=\Big(\frac{1}{N}\sum_{i=1}^N \bX_i \bX_i^{\top}\Big)^{-1}\Big(\frac{1}{N}\sum_{i=1}^N \bX_i \bvarepsilon_{1i}^{\top}\Big)\]
Therefore,
\[\sqrt{N}(\widehat{\balpha}-\balpha^*)\bbeta_2^*=\Big(\frac{1}{N}\sum_{i=1}^N \bX_i\bX_i^{\top}\Big)^{-1}\Big\{\frac{1}{\sqrt{N}}\sum_{i=1}^N \bX_i (\bvarepsilon_{1i}^{\top}\bbeta_2^*)\Big\}\]
By the law of large numbers, we have $\Big(\frac{1}{N}\sum_{i=1}^N \bX_i \bX_i^{\top}\Big)^{-1}\xrightarrow{\rm p} \bSigma_{\bX}^{-1}$, and under Gaussian design, we further have 
\[\bvarepsilon_{1i}=\bS_{1i}-\balpha^{* \top}\bX_i\]
is Gaussian and independent of $\bX_i$. Hence $E\big\{\bX_i(\bvarepsilon_{1i}^{\top}\bbeta_2^*)\big\}=0$, and 
\[E\Big\{\bX_i \bX_i^{\top}(\bvarepsilon_{1i}^{\top}\bbeta_2^*)^2\Big\}=E(\bX_i\bX_i^{\top})E\big\{(\bvarepsilon_{1i}^{\top}\bbeta_2^*)^2\big\}=\bSigma_{\bX}\bbeta_2^{* \top}\bSigma_{1} \bbeta_2^*\]
Therefore, by the multivariate central limit theorem and Slutsky's theorem,
\[\sqrt{N}(\widehat{\balpha}-\balpha^*)\bbeta_2^*\xrightarrow{\rm d} \mathcal{N}\Big(0,\bSigma_{\bX}^{-1}(\bSigma_{\bX}\bbeta_2^{*\top}\bSigma_{1}\bbeta_2^*)\bSigma_{\bX}^{-1}\Big)=\mathcal{N}\Big(0,(\bbeta_2^{* \top}\bSigma_{1}\bbeta_2^*)\bSigma_{\bX}^{-1}\Big)\]
Since $\sqrt{n}(\widehat{\balpha}-\balpha^*)\bbeta_2^*=\sqrt{n/N} \Big\{\sqrt{N}(\widehat{\balpha}-\balpha^*)\bbeta_2^*\Big\}$, and $n/N\to 1-\tau$, Slutsky's theorem yields
\[\sqrt{n}(\widehat{\balpha}-\balpha^*)\bbeta_2^*\xrightarrow{\rm d}
\mathcal{N}\left(\mathbf{0},(1-\tau)(\bbeta_2^{*\top}\bSigma_1\bbeta_2^*)\bSigma_{\bX}^{-1}
\right)
\]
Finally we consider the covariance between the first term and second term, under Gaussian design assumption, $\bar{\bX}_{2i}=(\bX_{i}^{\top},\bS_{1i}^{\top},Y_i)^{\top}$ is jointly Gaussian, $\varepsilon_{2i}=Y_i-\bar{\bX}_{1i}^{\top}\bbeta^*$ is Gaussian and independent of both $\bar{\bX}_{1i}$ and $\bvarepsilon_{1i}$, therefore, 
\[E\Big[\bar{\bX}_{1i}\bX_{i}^{\top}\varepsilon_{2i}(\bvarepsilon_{1i}^{\top}\bbeta_2^*)\Big]=E\Big[\bar{\bX}_{1i}\bX_{i}^{\top}(\bvarepsilon_{1i}^{\top}\bbeta_2^*)\Big]E(\varepsilon_{2i})=0\]
The same argument applies to the terms involving $\bvarepsilon_{3i}$ and $\bv_i$, since $(\varepsilon_{2i},\bvarepsilon_{3i})$ is independent of $(\bX_i,\bvarepsilon_{1i})$. Hence, the covariance between the first term and the second term equals $\mathbf{0}$. Therefore, we finally derive 
\[\sqrt{n}(\widehat{\btheta}-\btheta^*)\xrightarrow{\rm d} \mathcal{N}(0,\bV_{\btheta})\]
 where 
\[\bV_{\btheta}=\Big[\sigma_2^2(1-\kappa\tau)+(1-\tau)\bbeta_2^{* \top}\bSigma_{1}\bbeta_2^*\Big]\bSigma_{\bX}^{-1}\]

\noindent\textbf{Step 4. Control $\widehat{\bgamma}_3-\bgamma_3^*$.} Note that
\begin{align*}
	\widehat\bgamma-\bgamma^*
	&= \left(\dfrac{1}{N}\sum_{i\in\mathcal{I}}\bar{\bX}_{2i}\bar{\bX}_{2i}^\top +\dfrac{1}{N}\sum_{i\in\mathcal{J}} \bSigma_{\bgamma i}^{\bOmega^{(0)}}\right)^{-1}\left(\dfrac{1}{N}\sum_{i\in\mathcal{I}} \bar{\bX}_{2i}\bS_{2i}^\top + \dfrac{1}{N}\sum_{i\in\mathcal{J}} \bmu_{\bgamma i}^{\bOmega^{(0)}}\bS_{2i}^\top\right)
 	-\bgamma^*\\
	&=\left(\dfrac{1}{N}\sum_{i\in\mathcal{I}}\bar{\bX}_{2i}\bar{\bX}_{2i}^\top +\dfrac{1}{N}\sum_{i\in\mathcal{J}} \bSigma_{\bgamma i}^{\bOmega^{(0)}}\right)^{-1}\left\{\dfrac{1}{N}\sum_{i\in\mathcal{I}}\bar{\bX}_{2i}\bvarepsilon_{3i}^\top +\dfrac{1}{N}\sum_{i\in\mathcal{J}}(\bmu_{\bgamma i}^{\bOmega^{(0)}}\bS_{2i}^\top-\bSigma_{\bgamma i}^{\bOmega^{(0)}}\bgamma^*)\right\}\\
 	&:=\widehat{\bSigma}_{\bX,2}^{-1}(B+C), 
\end{align*}
where $\widehat{\bSigma}_{\bX,2}=N^{-1}\{\sum_{i\in\mathcal I}\bar{\bX}_{2i}\bar{\bX}_{2i}^{\top}+\sum_{i\in\mathcal J}\bSigma_{\bgamma i}^{\bOmega^{(0)}}\}$ and
\begin{align*}
	\bSigma_{\bgamma i}^{\bOmega^{(0)}}&=
	\begin{bmatrix}
		\bar{\bX}_{1i}\bar{\bX}_{1i}^\top & \bar{\bX}_{1i}\mu_i^{\bOmega^{(0)}}\\
		\bar{\bX}_{1i}^\top\mu_i^{\bOmega^{(0)}} & (\mu_i^{\bOmega^{(0)}})^2+(\sigma_i^{\bOmega^{(0)}})^2
 	\end{bmatrix},\;\;
	\bmu_{\bgamma i}^{\bOmega^{(0)}}=
	\begin{bmatrix}
		\bar{\bX}_{1i}\\
		\mu_i^{\bOmega^{(0)}}
	\end{bmatrix},\\
	B&=\dfrac1N\sum_{i\in\mathcal I}\bar{\bX}_{2i}\bvarepsilon_{3i}^{\top},\quad C=\dfrac1N\sum_{i\in\mathcal J}\begin{bmatrix}
		\bar{\bX}_{1i}(\bS_{2i}-\bgamma_{12}^{*\top}\bar{\bX}_{1i}-\mu_i^{\bOmega^{(0)}}\bgamma_3^*)^\top\\
		\mu_i^{\bOmega^{(0)}}(\bS_{2i}-\bgamma_{12}^{*\top}\bar{\bX}_{1i})^\top -\{(\mu_i^{\bOmega^{(0)}})^2+(\sigma_i^{\bOmega^{(0)}})^2\}\bgamma_3^{*\top}
  	\end{bmatrix}.
\end{align*}
We denote $
(\sigma_i^{\bOmega^{*}})^2=\sigma_2^2-\sigma_2^4\bgamma_3^{*\top}(\bSigma_3+\sigma_2^2\bgamma_3^*\bgamma_3^{*\top})^{-1}\bgamma_3^*=\sigma_2^2/\{1+\sigma_2^2\bgamma_3^{*\top}\bSigma_{3}^{-1}\bgamma_3^*\}$. We first consider $\tau>0$, so that $m\to\infty$. In this regime, we show that $\widehat{\bSigma}_{\bX,2}\xrightarrow{\rm p}\bSigma_{\bX,2}=E(\bar{\bX}_{2i}\bar{\bX}_{2i}^\top)$ by establishing the following limits for the unlabeled-sample conditional moments.%
\begin{align*}
	&\mu_{i}^{\bOmega^{(0)}}-Y_i=\bar{\bX}_{1i}^\top\bbeta^{(0)}+\bc^{(0)\top }\big(\bS_{2i}-\bgamma_{12}^{(0)\top}\bar{\bX}_{1i}-\bar{\bX}_{1i}^\top\bbeta^{(0)}\bgamma_{3}^{(0)}\big)-\bbeta^{*\top}\bar{\bX}_{1i}-\varepsilon_{2i}\\
	&=\bar{\bX}_{1i}^\top(\bbeta^{(0)}-\bbeta^*)+\bc^{(0)\top}\big\{\bvarepsilon_{3i}+\varepsilon_{2i}\bgamma_3^*+(\bgamma_{12}^*-\bgamma_{12}^{(0)})^\top \bar{\bX}_{1i}+(\bbeta^*\bgamma_3^{*\top}-\bbeta^{(0)}\bgamma_3^{(0)\top})^\top\bar{\bX}_{1i}\big\}-\varepsilon_{2i}\\
	&=\bar{\bX}_{1i}^\top(\bbeta^{(0)}-\bbeta^*)+\bc^{(0)\top}(\bvarepsilon_{3i}+\varepsilon_{2i}\bgamma_3^*)-\varepsilon_{2i}\\
	&\quad+ \bc^{(0)\top}\big\{ -(\bgamma_{12}^{(0)}-\bgamma_{12}^*)^\top- \bgamma_3^{*}(\bbeta^{(0)}-\bbeta^*)^\top-(\bgamma_{3}^{(0)}-\bgamma_3^*)\bbeta^{*\top} -(\bgamma_{3}^{(0)}-\bgamma_3^*)(\bbeta^{(0)}-\bbeta^*)^\top\big\}\bar{\bX}_{1i}\\
	&:=C_{i1}+\cdots+C_{i7}.
\end{align*}
By Assumption~\ref{ass:joint_gaussian} and the law of large numbers,
\begin{align*}
	m^{-1}\sum_{i\in\mathcal{J}}\bar{\bX}_{1i}\mu_i^{\bOmega^{(0)}}&=m^{-1}\sum_{i\in\mathcal{J}}\bar{\bX}_{1i}Y_i+m^{-1}\sum_{i\in\mathcal{J}}\bar{\bX}_{1i}(\mu_i^{\bOmega^{(0)}}-Y_i)\\
	&=E(\bar{\bX}_{1i}Y_i)+o_p(1)+m^{-1}\sum_{i\in\mathcal{J}}\bar{\bX}_{1i}(C_{i1}+\cdots+C_{i7}),
\end{align*}
We next control $m^{-1}\sum_{i\in\mathcal{J}}\bar{\bX}_{1i}C_{ij}$ for $j=1,2,\cdots,7$. By eigenvalue interlacing theorem and Assumption~\ref{ass:joint_gaussian}, $\lambda_{\rm max}(\bSigma_{\bX,1})\le \lambda_{\rm max}(\bSigma_{\bX,2})<c_2$, together with law of large numbers and Lemma \ref{lemma:lemma1}-\ref{lemma:lemma2},
\begin{equation}\label{eq:Xmu}
\begin{split}
m^{-1}\sum_{i\in\mathcal{J}}\bar{\bX}_{1i}C_{i1}&=\{\bSigma_{\bX,1}+o_p(1)\}(\bbeta^{(0)}-\bbeta^*)+O_p(m^{-1/2})=O_p(n^{-1/2}),\\
m^{-1}\sum_{i\in\mathcal{J}}\bar{\bX}_{1i}(C_{i2}+C_{i3})&=m^{-1}\sum_{i\in\mathcal{J}}\bar{\bX}_{1i}\{\bc^{(0)\top}(\bvarepsilon_{3i}+\varepsilon_{2i}\bgamma_3^*)-\varepsilon_{2i}\}=O_p(m^{-1/2}),\\
m^{-1}\sum_{i\in\mathcal{J}}\bar{\bX}_{1i}C_{i4}&=-\{\bSigma_{\bX,1}+o_p(1)\}(\bgamma_{12}^{(0)}-\bgamma_{12}^*)\bc^{(0)}=O_p(n^{-1/2}),\\
m^{-1}\sum_{i\in\mathcal{J}}\bar{\bX}_{1i}C_{i5}&=-\{\bSigma_{\bX,1}+o_p(1)\}(\bbeta^{(0)}-\bbeta^*)\bgamma_3^{*\top}\bc^{(0)}=O_p(n^{-1/2}),\\
m^{-1}\sum_{i\in\mathcal{J}}\bar{\bX}_{1i}C_{i6}&=-\{\bSigma_{\bX,1}+o_p(1)\}\bbeta^*(\bgamma_3^{(0)}-\bgamma_3^*)^{\top}\bc^{(0)}=O_p(n^{-1/2}),\\
m^{-1}\sum_{i\in\mathcal{J}}\bar{\bX}_{1i}C_{i7}&=-\{\bSigma_{\bX,1}+o_p(1)\}(\bbeta^{(0)}-\bbeta^*)(\bgamma_3^{(0)}-\bgamma_3^*)^{\top}\bc^{(0)}=O_p(n^{-1}),
\end{split}
\end{equation} 
hence $m^{-1}\sum_{i\in\mathcal{J}}\bar{\bX}_{1i}\mu_i^{\bOmega^{(0)}}=E(\bar{\bX}_{1i}Y_i)+o_p(1)$. For the second part, by Lemma \ref{lemma:delta}, we have $(\sigma_i^{\bOmega^{(0)}})^2-(\sigma_i^{\bOmega^{*}})^2=O_p(n^{-1/2})$, 
\begin{align*}
m^{-1}\sum_{i\in\mathcal{J}}(\mu_i^{\bOmega^{(0)}})^2+(\sigma_i^{\bOmega^{(0)}})^2&=m^{-1}\sum_{i\in\mathcal{J}}Y_i^2+m^{-1}\sum_{i\in\mathcal{J}}(\mu_i^{\bOmega^{(0)}}-Y_i)^2\\
&\qquad+2m^{-1}\sum_{i\in\mathcal{J}}Y_i(\mu_i^{\bOmega^{(0)}}-Y_i)+(\sigma_i^{\bOmega^{*}})^2+O_p(n^{-1/2}),
 \end{align*}
the first term $m^{-1}\sum_{i\in\mathcal{J}}Y_i^2\xrightarrow{\rm p}E(Y_i^2)$, we next show that the sum of the second and third terms and the conditional variance term is $o_p(1)$. For the second term, we have
\begin{align}
	m^{-1}\sum_{i\in\mathcal{J}}(\mu_i^{\bOmega^{(0)}}-Y_i)^2&=m^{-1}\sum_{i\in\mathcal{J}}(C_{i1}+\cdots+C_{i7})^2\nonumber\\
	&=m^{-1}\sum_{i\in\mathcal{J}}\{\bc^{(0)\top}(\bvarepsilon_{3i}+\varepsilon_{2i}\bgamma_3^*)-\varepsilon_{2i}\}^2+o_p(1/\sqrt{n}).\label{eq:Ys_1}
\end{align}
To establish the second equality, we next show that all remaining terms are $o_p(n^{-1/2})$ terms, we denote $\bv_i=\varepsilon_{2i}\bgamma_3^*+\bvarepsilon_{3i}$, together with Lemma \ref{lemma:lemma1}-\ref{lemma:lemma2} and Assumption~\ref{ass:joint_gaussian},
\begin{align*}
m^{-1}\sum_{i\in\mathcal{J}}C_{i1}^2&=(\bbeta^{(0)}-\bbeta^*)^\top \{\bSigma_{\bX,1}+O_p(m^{-1/2})\}(\bbeta^{(0)}-\bbeta^*)=O_p(n^{-1}),\\
2m^{-1}\sum_{i\in\mathcal{J}}C_{i1}(C_{i2}+C_{i3})&=2(\bbeta^{(0)}-\bbeta^*)^\top m^{-1}\sum_{i\in\mathcal{J}}\bar{\bX}_{1i}(\bv_i^\top\bc^{(0)}-\varepsilon_{2i})=O_p(1/\sqrt{mn}),\\
2m^{-1}\sum_{i\in\mathcal{J}}C_{i1}C_{i4}&=2(\bbeta^{(0)}-\bbeta^*)^\top m^{-1}\sum_{i\in\mathcal{J}}\bar{\bX}_{1i}\bar{\bX}_{1i}^\top(\bgamma_{12}^*-\bgamma_{12}^{(0)})\bc^{(0)}=O_p(n^{-1}),\\
2m^{-1}\sum_{i\in\mathcal{J}}C_{i1}C_{i5}&=2(\bbeta^{(0)}-\bbeta^*)^\top m^{-1}\sum_{i\in\mathcal{J}}\bar{\bX}_{1i}\bar{\bX}_{1i}^\top(\bbeta^*-\bbeta^{(0)})\bgamma_3^{*\top}\bc^{(0)}=O_p(n^{-1}),\\
2m^{-1}\sum_{i\in\mathcal{J}}C_{i1}C_{i6}&=2(\bbeta^{(0)}-\bbeta^*)^\top m^{-1}\sum_{i\in\mathcal{J}}\bar{\bX}_{1i}\bar{\bX}_{1i}^\top\bbeta^*(\bgamma_3^{*}-\bgamma_3^{(0)})^\top\bc^{(0)}=O_p(n^{-1}),\\
2m^{-1}\sum_{i\in\mathcal{J}}C_{i1}C_{i7}&=2(\bbeta^{(0)}-\bbeta^*)^\top m^{-1}\sum_{i\in\mathcal{J}}\bar{\bX}_{1i}\bar{\bX}_{1i}^\top(\bbeta^{(0)}-\bbeta^*)(\bgamma_3^{*}-\bgamma_3^{(0)})^\top\bc^{(0)}=O_p(n^{-3/2}),
\end{align*}
and
\begin{align*}
2m^{-1}\sum_{i\in\mathcal{J}}(C_{i2}+C_{i3})C_{i4}&=2m^{-1}\sum_{i\in\mathcal{J}}(\bc^{(0)\top}\bv_i-\varepsilon_{2i})\bar{\bX}_{1i}^\top(\bgamma_{12}^*-\bgamma_{12}^{(0)})\bc^{(0)}=O_p(1/\sqrt{mn}),\\
2m^{-1}\sum_{i\in\mathcal{J}}(C_{i2}+C_{i3})C_{i5}&=2m^{-1}\sum_{i\in\mathcal{J}}(\bc^{(0)\top}\bv_i-\varepsilon_{2i})\bar{\bX}_{1i}^\top(\bbeta^*-\bbeta^{(0)})\bgamma_3^{*\top}\bc^{(0)}=O_p(1/\sqrt{mn}),\\
2m^{-1}\sum_{i\in\mathcal{J}}(C_{i2}+C_{i3})C_{i6}&=2m^{-1}\sum_{i\in\mathcal{J}}(\bc^{(0)\top}\bv_i-\varepsilon_{2i})\bar{\bX}_{1i}^\top\bbeta^*(\bgamma_3^{*}-\bgamma_3^{(0)})^\top\bc^{(0)}=O_p(1/\sqrt{mn}),\\
2m^{-1}\sum_{i\in\mathcal{J}}(C_{i2}+C_{i3})C_{i7}&=2m^{-1}\sum_{i\in\mathcal{J}}(\bc^{(0)\top}\bv_i-\varepsilon_{2i})\bar{\bX}_{1i}^\top(\bbeta^{(0)}-\bbeta^*)(\bgamma_3^{*}-\bgamma_3^{(0)})^\top\bc^{(0)}=O_p(1/\sqrt{mn^2}),
\end{align*}
and
\begin{align*}
m^{-1}\sum_{i\in\mathcal{J}}C_{i4}^2&=\bc^{(0)\top}(\bgamma_{12}^*-\bgamma_{12}^{(0)})^\top m^{-1}\sum_{i\in\mathcal{J}}\bar{\bX}_{1i}\bar{\bX}_{1i}^\top (\bgamma_{12}^*-\bgamma_{12}^{(0)})\bc^{(0)}=O_p(n^{-1}),\\
2m^{-1}\sum_{i\in\mathcal{J}}C_{i4}C_{i5}&=2\bc^{(0)\top}(\bgamma_{12}^*-\bgamma_{12}^{(0)})^\top m^{-1}\sum_{i\in\mathcal{J}}\bar{\bX}_{1i}\bar{\bX}_{1i}^\top (\bbeta^*-\bbeta^{(0)})\bgamma_3^{*\top}\bc^{(0)}=O_p(n^{-1}),\\
2m^{-1}\sum_{i\in\mathcal{J}}C_{i4}C_{i6}&=O_p(n^{-1}),\qquad 2m^{-1}\sum_{i\in\mathcal{J}}C_{i4}C_{i7}=O_p(n^{-3/2}),
\end{align*}
and
\begin{align*}
&m^{-1}\sum_{i\in\mathcal{J}}C_{i5}^2=\bc^{(0)\top}\bgamma_3^*(\bbeta^*-\bbeta^{(0)})^\top m^{-1}\sum_{i\in\mathcal{J}}\bar{\bX}_{1i}\bar{\bX}_{1i}^\top (\bbeta^*-\bbeta^{(0)})\bgamma_3^{*\top} \bc^{(0)}=O_p(n^{-1}),\\
&2m^{-1}\sum_{i\in\mathcal{J}}C_{i5}C_{i6}=O_p(n^{-1}),\qquad 2m^{-1}\sum_{i\in\mathcal{J}}C_{i5}C_{i7}=O_p(n^{-3/2}),
\end{align*}
and
\begin{align*}
m^{-1}\sum_{i\in\mathcal{J}}C_{i6}^2&=O_p(n^{-1}),\;\;2m^{-1}\sum_{i\in\mathcal{J}}C_{i6}C_{i7}=O_p(n^{-3/2}),\;\; m^{-1}\sum_{i\in\mathcal{J}}C_{i7}^2=O_p(n^{-2}).
\end{align*}
For the third part,
\begin{align}
2m^{-1}\sum_{i\in\mathcal{J}}Y_i(\mu_i^{\bOmega^{(0)}}-Y_i)&=2m^{-1}\sum_{i\in\mathcal{J}}(\bbeta^{*\top}\bar{\bX}_{1i}+\varepsilon_{2i})(\mu_i^{\bOmega^{(0)}}-Y_i)\nonumber\\
&=2m^{-1}\sum_{i\in\mathcal{J}}\varepsilon_{2i}(\mu_i^{\bOmega^{(0)}}-Y_i)+o_p(1)\nonumber\\
&=2m^{-1}\sum_{i\in\mathcal{J}}\varepsilon_{2i}\{\bc^{(0)\top}(\bvarepsilon_{3i}+\varepsilon_{2i}\bgamma_3^*)-\varepsilon_{2i}\}+o_p(1).\label{eq:Ys_2}
\end{align}
where the second equality is by \eqref{eq:Xmu} and Assumption~\ref{ass:joint_gaussian}. The third equality holds, since 
\begin{align*}
&2m^{-1}\sum_{i\in\mathcal{J}}\varepsilon_{2i}(\mu_i^{\bOmega^{(0)}}-Y_i)\\
=\ &2m^{-1}\sum_{i\in\mathcal{J}}\varepsilon_{2i}(C_{i2}+C_{i3})+2m^{-1}\sum_{i\in\mathcal{J}}\varepsilon_{2i}(C_{i1}+C_{i4}+C_{i5}+C_{i6}+C_{i7})\\
=\ &2m^{-1}\sum_{i\in\mathcal{J}}\varepsilon_{2i}(C_{i2}+C_{i3})+2m^{-1}\sum_{i\in\mathcal{J}}\varepsilon_{2i}\bar{\bX}_{1i}^\top\bDelta_{\bOmega^{(0)}}\\
=\ &2m^{-1}\sum_{i\in\mathcal{J}}\varepsilon_{2i}(C_{i2}+C_{i3})+O_p(1/\sqrt{mn})
\end{align*}
where
\begin{align}
\bDelta_{\bOmega^{(0)}} &=(\bbeta^{(0)}-\bbeta^*)+\Big\{(\bgamma_{12}^*-\bgamma_{12}^{(0)})+(\bbeta^*-\bbeta^{(0)})\bgamma_3^{*\top}\nonumber\\
&\qquad\qquad\qquad\qquad+\bbeta^*(\bgamma_3^*-\bgamma_3^{(0)})^{\top}+(\bbeta^{(0)}-\bbeta^*)(\bgamma_3^*-\bgamma_3^{(0)})^{\top}\Big\}\bc^{(0)}.\label{def:delta_theta}
\end{align}
Combining \eqref{eq:Ys_1} and \eqref{eq:Ys_2}, by Lemma \ref{lemma:lemma2}, 
\begin{align*}
&m^{-1}\sum_{i\in\mathcal{J}}(\mu_i^{\bOmega^{(0)}}-Y_i)^2+2m^{-1}\sum_{i\in\mathcal{J}}Y_i(\mu_i^{\bOmega^{(0)}}-Y_i)+(\sigma_i^{\bOmega^{*}})^2\\
=\  &m^{-1}\sum_{i\in\mathcal{J}}[\{\bc^{(0)\top}(\bvarepsilon_{3i}+\varepsilon_{2i}\bgamma_3^*)-\varepsilon_{2i}\}^2+2\varepsilon_{2i}\{\bc^{(0)\top}(\bvarepsilon_{3i}+\varepsilon_{2i}\bgamma_3^*)-\varepsilon_{2i}\}]+(\sigma_i^{\bOmega^{*}})^2+o_p(1)\\
=\ &m^{-1}\sum_{i\in\mathcal{J}}[\{(\bc^*+\bc^{(0)}-\bc^*)^\top(\bvarepsilon_{3i}+\varepsilon_{2i}\bgamma_3^*)-\varepsilon_{2i}\}^2+2\varepsilon_{2i}\{(\bc^*+\bc^{(0)}-\bc^*)^\top(\bvarepsilon_{3i}+\varepsilon_{2i}\bgamma_3^*)-\varepsilon_{2i}\}]\\
&\qquad+(\sigma_i^{\bOmega^{*}})^2+o_p(1)\\
=\ &E[\{\bc^{*\top}(\bvarepsilon_{3i}+\varepsilon_{2i}\bgamma_3^*)-\varepsilon_{2i}\}^2+2\varepsilon_{2i}\{\bc^{*\top}(\bvarepsilon_{3i}+\varepsilon_{2i}\bgamma_3^*)-\varepsilon_{2i}\}]\\
&\qquad +(\sigma_i^{\bOmega^{*}})^2+O_p(n^{-1/2})+o_p(1)=o_p(1),
\end{align*}
where the third equality is by $\bc^{(0)}-\bc^*=O_p(n^{-1/2})$ and the law of large numbers, the last equality holds, since $E[\{\bc^{*\top}(\bvarepsilon_{3i}+\varepsilon_{2i}\bgamma_3^*)-\varepsilon_{2i}\}^2+2\varepsilon_{2i}\{\bc^{*\top}(\bvarepsilon_{3i}+\varepsilon_{2i}\bgamma_3^*)-\varepsilon_{2i}\}]=-\sigma_2^2/\{1+\sigma_2^2\bgamma_3^{*\top}\bSigma_{3}^{-1}\bgamma_3^*\}=-(\sigma_i^{\bOmega^{*}})^2$. Then it is shown that $m^{-1}\sum_{i\in\mathcal{J}}(\mu_i^{\bOmega^{(0)}})^2+(\sigma_i^{\bOmega^{(0)}})^2\xrightarrow{\rm p}E(Y_i^2)$ and thus $\widehat{\bSigma}_{\bX,2}\xrightarrow{\rm p}\bSigma_{\bX,2}$. In the following, we focus on controlling $\bgamma_3^{(1)}-\bgamma_3^*$. Note that the last row of $\bSigma_{\bX,2}^{-1}$ has the form of $(-\bbeta^{*\top},1)/\sigma_2^2$, then 
\begin{align}
	&\widehat{\bgamma}_3-\bgamma_3^*=\left[(-\bbeta^{*\top},1)(B+C)/\sigma_2^2\right]^\top+o_p(n^{-1/2})\nonumber\\
    &=\dfrac{1}{N}\sum_{i\in\mathcal{I}}\dfrac{\varepsilon_{2i}\bvarepsilon_{3i}}{\sigma_2^2}+\dfrac{1}{N\sigma_2^2}\sum_{i\in\mathcal{J}}\left\{(\mu_i^{\bOmega^{(0)}}-\bar{\bX}_{1i}^\top\bbeta^*)(\bS_{2i}-\bgamma_{12}^{*\top}\bar{\bX}_{1i}-\mu_i^{\bOmega^{(0)}}\bgamma_3^*) - (\sigma_i^{\bOmega^{(0)}})^2\bgamma_3^{*}\right\}+o_p(n^{-1/2})\nonumber\\
	&=\dfrac{1}{N}\sum_{i\in\mathcal{I}}\dfrac{\varepsilon_{2i}\bvarepsilon_{3i}}{\sigma_2^2}+\dfrac{1}{N\sigma_2^2}\sum_{i\in\mathcal{J}}\left\{(\varepsilon_{2i}+\mu_i^{\bOmega^{(0)}}-Y_i)(\bvarepsilon_{3i}+Y_i\bgamma_3^*-\mu_i^{\bOmega^{(0)}}\bgamma_3^*)- (\sigma_i^{\bOmega^{(0)}})^2\bgamma_3^{*}\right\}+o_p(n^{-1/2})\nonumber\\
	&=\dfrac{1}{N}\sum_{i\in\mathcal{I}\cup\mathcal{J}}\dfrac{\varepsilon_{2i}\bvarepsilon_{3i}}{\sigma_2^2}+\dfrac{1}{N}\sum_{i\in\mathcal{J}}\dfrac{(\mu_i^{\bOmega^{(0)}}-Y_i)(\bvarepsilon_{3i}-\varepsilon_{2i}\bgamma_3^*)}{\sigma_2^2}-\dfrac{1}{N}\sum_{i\in\mathcal{J}}\dfrac{(\mu_i^{\bOmega^{(0)}}-Y_i)^2\bgamma_3^*+(\sigma_i^{\bOmega^{(0)}})^2\bgamma_3^{*}}{\sigma_2^2}+o_p(n^{-1/2}).\label{gamma_3_diff}
 \end{align}
We have
\begin{align*}
	&\dfrac{1}{N}\sum_{i\in\mathcal{J}}\dfrac{(\mu_{i}^{\bOmega^{(0)}}-Y_i)(\bvarepsilon_{3i}-\varepsilon_{2i}\bgamma_3^*)}{\sigma_2^2}\\
	=\ &\dfrac{1}{N}\sum_{i\in\mathcal{J}}\dfrac{(C_{i1}+\cdots+C_{i7})(\bvarepsilon_{3i}-\varepsilon_{2i}\bgamma_3^*)}{\sigma_2^2}\\
 	=\ &\dfrac{1}{N}\sum_{i\in\mathcal{J}}\dfrac{\bc^{(0)\top}(\bvarepsilon_{3i}+\varepsilon_{2i}\bgamma_3^*)(\bvarepsilon_{3i}-\varepsilon_{2i}\bgamma_3^*)-\varepsilon_{2i}(\bvarepsilon_{3i}-\varepsilon_{2i}\bgamma_3^*)}{\sigma_2^2}+o_p(n^{-1/2}),
\end{align*}
the last equality holds, since by Lemma \ref{lemma:lemma1} and Assumption~\ref{ass:joint_gaussian}, 
\begin{align*}
\dfrac{1}{N}\sum_{i\in\mathcal{J}}\dfrac{(C_{i1}+C_{i4}+\cdots+C_{i7})(\bvarepsilon_{3i}-\varepsilon_{2i}\bgamma_3^*)}{\sigma_2^2}&=\dfrac{1}{N\sigma_2^2}\sum_{i\in\mathcal{J}}(\bvarepsilon_{3i}-\varepsilon_{2i}\bgamma_3^*)\bar{\bX}_{1i}^\top \bDelta_{\bOmega^{(0)}}\\
&=O_p(1/\sqrt{mn})=o_p(n^{-1/2})
\end{align*}
where $\bDelta_{\bOmega^{(0)}}$ is defined in \eqref{def:delta_theta}. By \eqref{eq:Ys_1},
\begin{align*}
	&\dfrac{1}{N}\sum_{i\in\mathcal{J}}\dfrac{(\mu_{i}^{\bOmega^{(0)}}-Y_i)^2\bgamma_3^*}{\sigma_2^2}=\dfrac{1}{N}\sum_{i\in\mathcal{J}}\dfrac{\{\bc^{(0)\top}(\bvarepsilon_{3i}+\varepsilon_{2i}\bgamma_3^*)-\varepsilon_{2i}\}^2\bgamma_3^*}{\sigma_2^2}+o_p(n^{-1/2}).
\end{align*}
 	Combining the upper bounds above, we have
\begin{align*}
	\widehat{\bgamma}_3-\bgamma_3^*&=\dfrac{1}{N}\sum_{i\in\mathcal{I}\cup\mathcal{J}}\dfrac{\varepsilon_{2i}\bvarepsilon_{3i}}{\sigma_2^2}+\dfrac{1}{N}\sum_{i\in\mathcal{J}}\dfrac{\bc^{(0)\top}(\bvarepsilon_{3i}+\varepsilon_{2i}\bgamma_3^*)(\bvarepsilon_{3i}-\varepsilon_{2i}\bgamma_3^*)-\varepsilon_{2i}(\bvarepsilon_{3i}-\varepsilon_{2i}\bgamma_3^*)}{\sigma_2^2}\\
	&\quad -\dfrac{1}{N}\sum_{i\in\mathcal{J}} \dfrac{\{\bc^{(0)\top}(\bvarepsilon_{3i}+\varepsilon_{2i}\bgamma_3^*)-\varepsilon_{2i}\}^2\bgamma_3^*+(\sigma_i^{\bOmega^{(0)}})^2\bgamma_3^*}{\sigma_2^2}+o_p(n^{-1/2})\\
 	&=\dfrac{1}{N}\sum_{i\in\mathcal{I}}\dfrac{\varepsilon_{2i}\bvarepsilon_{3i}}{\sigma_2^2}+\dfrac{1}{N}\sum_{i\in\mathcal{J}}\dfrac{(\bI_{d_2}-\bgamma_3^*\bc^{(0)\top})\bv_i\bv_i^\top \bc^{(0)}-(\sigma_i^{\bOmega^{(0)}})^2\bgamma_3^{*}}{\sigma_2^2}+o_p(n^{-1/2})\\
	&=\dfrac{1}{N}\sum_{i\in\mathcal{I}}\dfrac{\varepsilon_{2i}\bvarepsilon_{3i}}{\sigma_2^2}+\dfrac{1}{N}\sum_{i\in\mathcal{J}}\dfrac{(\bI_{d_2}-\bgamma_3^*\bc^{*\top})\bv_i\bv_i^\top\bc^*-(\sigma_i^{\bOmega^{*}})^2\bgamma_3^{*}}{\sigma_2^2}\\
	&\quad+\dfrac{1}{N}\sum_{i\in\mathcal{J}}\dfrac{\{\bv_i-2\bgamma_3^*(\bv_i^\top\bc^*)\}\bv_i^\top(\bc^{(0)}-\bc^*)+\bgamma_3^*\{(\sigma_i^{\bOmega^{*}})^2-(\sigma_i^{\bOmega^{(0)}})^2\}}{\sigma_2^2}+o_p(n^{-1/2}),
\end{align*}
where the second equality holds, since
\begin{align*}
	&(\bc^{(0)\top}\bv_i-\varepsilon_{2i})(\bvarepsilon_{3i}-\varepsilon_{2i}\bgamma_3^*)-\bgamma_3^*(\bc^{(0)\top}\bv_i-\varepsilon_{2i})^2+\varepsilon_{2i}\bvarepsilon_{3i}\\
	=\ &\bc^{(0)\top}\bv_i\bvarepsilon_{3i}-\varepsilon_{2i}\bvarepsilon_{3i}-\bc^{(0)\top}\bv_i\varepsilon_{2i}\bgamma_3^*+\varepsilon_{2i}^2\bgamma_3^*-\bgamma_3^*(\bc^{(0)\top}\bv_i)^2+2\bgamma_3^*\bc^{(0)\top}\bv_i\varepsilon_{2i}-\bgamma_3^*\varepsilon_{2i}^2+\varepsilon_{2i}\bvarepsilon_{3i}\\
	=\ &\bc^{(0)\top}\bv_i\bvarepsilon_{3i}+\bc^{(0)\top}\bv_i\varepsilon_{2i}\bgamma_3^*-\bgamma_3^*(\bc^{(0)\top}\bv_i)^2\\
	=\ &\bc^{(0)\top}\bv_i\bv_{i}-\bgamma_3^*(\bc^{(0)\top}\bv_i)^2\\
 	=\ & (\bI_{d_2}-\bgamma_3^*\bc^{(0)\top})\bv_i\bv_i^\top \bc^{(0)}.
\end{align*}
 	By Lemma \ref{lemma:delta}, let $\bD=(\bSigma_3+\sigma_2^2\bgamma_3^*\bgamma_3^{*\top})^{-1}$, we obtain
\begin{equation}\label{eq:gamma}
	\begin{split}
	\widehat{\bgamma}_3-\bgamma_3^*
	&=\dfrac{1}{N}\sum_{i\in\mathcal{I}}\dfrac{\varepsilon_{2i}\bvarepsilon_{3i}}{\sigma_2^2}+\dfrac{1}{N}\sum_{i\in\mathcal{J}}\dfrac{(\bI_{d_2}-\bgamma_3^*\bc^{*\top})\bv_i\bv_i^\top\bc^*-(\sigma_i^{\bOmega^{*}})^2\bgamma_3^{*}}{\sigma_2^2}\\
	&\qquad+ \dfrac{m E[\{\bv_i-2\bgamma_3^*(\bv_i^\top\bc^*)\}\bv_i^\top]}{N\sigma_2^2}(\bc^{(0)}-\bc^*)+\dfrac{m\bgamma_3^*}{N\sigma_2^2}\{(\sigma_i^{\bOmega^{*}})^2-(\sigma_i^{\bOmega^{(0)}})^2\}+o_p(n^{-1/2})\\
	&=\dfrac{1}{N}\sum_{i\in\mathcal{I}}\dfrac{\varepsilon_{2i}\bvarepsilon_{3i}}{\sigma_2^2}+\dfrac{1}{N}\sum_{i\in\mathcal{J}}\dfrac{(\bI_{d_2}-\bgamma_3^*\bc^{*\top})\bv_i\bv_i^\top\bc^*-(\sigma_i^{\bOmega^{*}})^2\bgamma_3^{*}}{\sigma_2^2}\\
 	&\qquad + (m/N) (\bSigma_3/\sigma_2^2-\bgamma_3^*\bgamma_3^{*\top})\{\bD\bgamma_3^*-\sigma_2^2(\bgamma_3^{*\top} \bD\bgamma_3^*)\bD\bgamma_3^*\}\dfrac{1}{n}\sum_{i\in\mathcal{I}}(\varepsilon_{2i}^2-\sigma_2^2)\\
	&\qquad-(m/N) (\bSigma_3/\sigma_2^2-\bgamma_3^*\bgamma_3^{*\top})\sigma_2^2 \bD \dfrac{1}{n}\sum_{i\in\mathcal{I}}(\bvarepsilon_{3i} \bvarepsilon_{3i}^\top-\bSigma_3)\bD\bgamma_3^*\\
	&\qquad+(m/N) (\bSigma_3/\sigma_2^2-\bgamma_3^*\bgamma_3^{*\top})\{\bD-\sigma_2^2\bD\bgamma_3^*\bgamma_3^{*\top} \bD-\sigma_2^2(\bgamma_3^{*\top}\bD\bgamma_3^*)\bD\}\dfrac{1}{n}\sum_{i\in\mathcal{I}}\varepsilon_{2i}\bvarepsilon_{3i}\\
	&\qquad -(m/N)( \bgamma_3^*/\sigma_2^2)\{\sigma_2^2(\bgamma_3^{*\top} \bD\bgamma_3^*)-1\}^2\dfrac{1}{n}\sum_{i\in\mathcal{I}}(\varepsilon_{2i}^2-\sigma_2^2)\\
	&\qquad-(m/N)( \bgamma_3^*/\sigma_2^2) \sigma_2^4\bgamma_3^{*\top} \bD \dfrac{1}{n}\sum_{i\in\mathcal{I}}(\bvarepsilon_{3i} \bvarepsilon_{3i}^\top-\bSigma_3)\bD\bgamma_3^*\\
 	&\qquad +(m/N)( \bgamma_3^*/\sigma_2^2)2\sigma_2^2(1-\sigma_2^2\bgamma_3^{*\top}\bD\bgamma_3^*)\bgamma_3^{*\top}\bD\dfrac{1}{n}\sum_{i\in\mathcal{I}}\varepsilon_{2i}\bvarepsilon_{3i}+o_p(n^{-1/2})\\
	&=E_1+E_2+E_3+E_4+o_p(n^{-1/2}),
		\end{split}
\end{equation}
where
\begin{align*}
E_1&=-\dfrac{(m/N)\bgamma_3^{*\top} \bSigma_{3}^{-1}\bgamma_3^*}{(1+\sigma_2^2\bgamma_3^{*\top} \bSigma_{3}^{-1}\bgamma_3^*)^2}\bgamma_3^*\dfrac{1}{n}\sum_{i\in\mathcal{I}}(\varepsilon_{2i}^2-\sigma_2^2),\;\;
E_2=- (m/N)\bSigma_3\bD\dfrac{1}{n}\sum_{i\in\mathcal{I}}(\bvarepsilon_{3i} \bvarepsilon_{3i}^\top-\bSigma_3)\bD\bgamma_3^*,\\
E_3&=\dfrac{1}{N}\sum_{i\in\mathcal{I}}\dfrac{\varepsilon_{2i}\bvarepsilon_{3i}}{\sigma_2^2}+\dfrac{m}{N}\left\{\dfrac{1}{\sigma_2^2}\dfrac{1}{1+\sigma_2^2\bgamma_3^{*\top} \bSigma_{3}^{-1}\bgamma_3^*}\bI_{d_2}- \dfrac{1-\sigma_2^2\bgamma_3^{*\top} \bSigma_{3}^{-1}\bgamma_3^*}{(1+\sigma_2^2\bgamma_3^{*\top} \bSigma_{3}^{-1}\bgamma_3^*)^2}\bgamma_3^*\bgamma_3^{*\top}\bSigma_{3}^{-1}\right\}\dfrac{1}{n}\sum_{i\in\mathcal{I}}\varepsilon_{2i}\bvarepsilon_{3i},\\
E_4&=\dfrac{1}{N}\sum_{i\in\mathcal{J}}\dfrac{(\bI_{d_2}-\bgamma_3^*\bc^{*\top})\bv_i\bv_i^\top\bc^*-(\sigma_i^{\bOmega^{*}})^2\bgamma_3^{*}}{\sigma_2^2}.
\end{align*}
Notice that 
\begin{align*}
E\left\{(\bI_{d_2}-\bgamma_3^*\bc^{*\top})\bv_i\bv_i^\top\bc^*\right\}&=(\bI_{d_2}-\bgamma_3^*\bc^{*\top})(\bSigma_3+\sigma_2^2\bgamma_3^*\bgamma_3^{*\top})\bc^*\\
&=\sigma_2^2\bgamma_3^*-\sigma_2^2(\bc^{*\top}\bgamma_3^*)\bgamma_3^*=(\sigma_i^{\bOmega^{*}})^2\bgamma_3^{*},
\end{align*}
hence all of $E_1,\cdots,E_4$ are mean-zero. Assuming $\varepsilon_{2i}\sim \mathcal{N}(0,\sigma_2^2)$ and $\bvarepsilon_{3i}\sim\mathcal{N}(0,\bSigma_3)$, $E_1$, $E_2$, $E_3$, and $E_4$ are pairwise uncorrelated. Denote $t=\bgamma_3^{*\top}\bSigma_3^{-1}\bgamma_3^*$, the variance 
\begin{align*}
{\rm var}(E_1)&=(m/N)^2\dfrac{2\sigma_2^4}{n}\dfrac{t^2}{(1+\sigma_2^2t)^4}\bgamma_3^*\bgamma_3^{*\top}\\
{\rm var}(E_2)&=(m/N)^2\dfrac{t}{n(1+\sigma_2^2t)^2}\bSigma_3+(m/N)^2\dfrac{2-(1+\sigma_2^2t)^2}{n(1+\sigma_2^2t)^4}\bgamma_3^*\bgamma_3^{*\top},\\
{\rm var}(E_3)&=  \dfrac{1}{n\sigma_2^2}\Big\{1-\dfrac{(m/N) \sigma_2^2t}{1+\sigma_2^2t}\Big\}^2\bSigma_3-\dfrac{(m/N)(1-\sigma_2^2t)\{2(1+\sigma_2^2t)^2-(m/N) \sigma_2^2t(3+\sigma_2^2t)\}}{n(1+\sigma_2^2t)^4}\bgamma_3^*\bgamma_3^{*\top}\\
{\rm var}(E_4)&=\dfrac{m}{N^2}\Big\{\dfrac{t}{1+\sigma_2^2t}\bSigma_3+\dfrac{1-\sigma_2^2t}{(1+\sigma_2^2t)^2}\bgamma_3^*\bgamma_3^{*\top}\Big\}
\end{align*}
the second equality holds, since by Isserlis's theorem, for normal vector $\bvarepsilon_{3i}\sim N(0,\bSigma_3)$, it holds that ${\rm var}\{\bSigma_3\bD(\bvarepsilon_{3i}\bvarepsilon_{3i}^\top-\bSigma_3)\bD\bgamma_3^*\}=\bSigma_3\bD\{(\bgamma_{3}^{*\top}\bD\bSigma_3 \bD\bgamma_3^*)\bSigma_3+\bSigma_3  \bD\bgamma_3^*\bgamma_3^{*\top}\bD \bSigma_3\}\bD\bSigma_3$. By Sherman–Morrison formula, $$\bD=\bSigma_3^{-1}-\dfrac{\sigma_2^2}{1+\sigma_2^2t}\bSigma_3^{-1}\bgamma_3^*\bgamma_3^{*\top}\bSigma_3^{-1},$$
Combining the covariance contributions of $E_1,\ldots,E_4$ and taking $m/N\to\tau$, we obtain
\[
\sqrt{n}(\widehat{\bgamma}_3-\bgamma_3^*)\xrightarrow{\rm d}\mathcal{N}(\mathbf{0},\bV_{\bgamma,1}),
\]
where
\begin{align}\label{def:V1}
\bV_{\bgamma,1}
=\frac{1-\tau\kappa}{\sigma_2^2}\bSigma_3 -\tau \frac{1-\sigma_2^2t}
{(1+\sigma_2^2t)^2}
\bgamma_3^*\bgamma_3^{*\top}.
\end{align}

\noindent\textbf{Step 5. Control $\widehat{\bgamma}_{12}-\bgamma_{12}^*$.} Note that the first $(p+d_1)$ rows of $\bSigma_{\bX,2}^{-1}$ are 
\begin{align*}
	[\bSigma_{\bX,2}^{-1}]_{1:(p+d_1),:}=
	\begin{bmatrix}
		\bSigma_{\bX,1}^{-1}+\dfrac{\bbeta^*\bbeta^{*\top}}{\sigma_2^2} &-\dfrac{\bbeta^*}{\sigma_2^2}
	\end{bmatrix}.
\end{align*}
 	Therefore, 
\begin{align*}
	\widehat{\bgamma}_{12}-\bgamma_{12}^*&=(\bSigma_{\bX,1}^{-1}+\bbeta^*\bbeta^{*\top}/\sigma_2^2,-\bbeta^*/\sigma_2^2)(B+C)+o_p(n^{-1/2})\\
 	&=\dfrac{1}{N}\sum_{i\in\mathcal{I}}(\bSigma_{\bX,1}^{-1}\bar{\bX}_{1i}\bvarepsilon_{3i}^\top-\bbeta^*\varepsilon_{2i}\bvarepsilon_{3i}^\top/\sigma_2^2)\\
    &\quad +\dfrac{1}{N}\sum_{i\in\mathcal{J}}\Big[\bSigma_{\bX,1}^{-1}\bar{\bX}_{1i}\bvarepsilon_{3i}^\top+(Y_i-\mu_i^{\bOmega^{(0)}})\bSigma_{\bX,1}^{-1}\bar{\bX}_{1i}\bgamma_3^{*\top}-\varepsilon_{2i}\bbeta^*\bvarepsilon_{3i}^\top/\sigma_2^2\\
	&\qquad\qquad\qquad+(Y_i-\mu_i^{\bOmega^{(0)}})\bbeta^*(\bvarepsilon_{3i}-\varepsilon_{2i}\bgamma_3^{*})^\top/\sigma_2^2+(Y_i-\mu_i^{\bOmega^{(0)}})^2\bbeta^*\bgamma_3^{*\top}/\sigma_2^2\\
	&\qquad\qquad\qquad+(\sigma_i^{\bOmega^{(0)}})^2\bbeta^*\bgamma_3^{*\top}/\sigma_2^2\Big]+o_p(n^{-1/2}).
 \end{align*}
For the term $ N^{-1}\sum_{i\in\mathcal{J}}(Y_i-\mu_i^{\bOmega^{(0)}})\bSigma_{\bX,1}^{-1}\bar{\bX}_{1i}\bgamma_3^{*\top}$, we have 
\begin{align*}
	&\dfrac{1}{N}\sum_{i\in\mathcal{J}}(Y_i-\mu_i^{\bOmega^{(0)}})\bSigma_{\bX,1}^{-1}\bar{\bX}_{1i}\bgamma_3^{*\top}=-\dfrac{1}{N}\sum_{i\in\mathcal{J}}(C_{i1}+\cdots+C_{i7})\bSigma_{\bX,1}^{-1}\bar{\bX}_{1i}\bgamma_3^{*\top}\\
	&=-\dfrac{1}{N}\sum_{i\in\mathcal{J}}\Big[\bar{\bX}_{1i}^\top(\bbeta^{(0)}-\bbeta^*)+\bc^{*\top}(\bvarepsilon_{3i}+\varepsilon_{2i}\bgamma_3^*)-\varepsilon_{2i}\\
 	&\quad+ \bc^{*\top}\big\{ -(\bgamma_{12}^{(0)}-\bgamma_{12}^*)^\top- \bgamma_3^{*}(\bbeta^{(0)}-\bbeta^*)^\top-(\bgamma_{3}^{(0)}-\bgamma_3^*)\bbeta^{*\top} \big\}\bar{\bX}_{1i}\Big]\bSigma_{\bX,1}^{-1}\bar{\bX}_{1i}\bgamma_3^{*\top}+o_p(n^{-1/2})\\
	&=\dfrac{1}{N}\sum_{i\in\mathcal{J}}\bSigma_{\bX,1}^{-1}\bar{\bX}_{1i}\bgamma_3^{*\top}\varepsilon_{2i}-(m/N) (1-\bgamma_3^{*\top}\bc^*)(\bbeta^{(0)}-\bbeta^*)\bgamma_3^{*\top}\\
	&\quad-\dfrac{1}{N}\sum_{i\in\mathcal{J}}\bc^{*\top}(\bvarepsilon_{3i}+\varepsilon_{2i}\bgamma_3^*)\bSigma_{\bX,1}^{-1}\bar{\bX}_{1i}\bgamma_3^{*\top} +(m/N)\{(\bgamma_{12}^{(0)}-\bgamma_{12}^*)+\bbeta^{*}(\bgamma_{3}^{(0)}-\bgamma_3^*)^\top\}\bc^*\bgamma_3^{*\top}+o_p(n^{-1/2})\\
 	&=\dfrac{1}{N}\sum_{i\in\mathcal{J}}\bSigma_{\bX,1}^{-1}\bar{\bX}_{1i}\bgamma_3^{*\top}\varepsilon_{2i}-(m/N) (1-\bgamma_3^{*\top}\bc^*)\bSigma_{\bX,1}^{-1}\dfrac{1}{n}\sum_{i\in\mathcal{I}}\bar{\bX}_{1i}\varepsilon_{2i}\bgamma_3^{*\top}\\
	&\quad-\dfrac{1}{N}\sum_{i\in\mathcal{J}}\bc^{*\top}(\bvarepsilon_{3i}+\varepsilon_{2i}\bgamma_3^*)\bSigma_{\bX,1}^{-1}\bar{\bX}_{1i}\bgamma_3^{*\top} +(m/N)\bSigma_{\bX,1}^{-1}\dfrac{1}{n}\sum_{i\in\mathcal{I}}\bar{\bX}_{1i}\bvarepsilon_{3i}^\top\bc^*\bgamma_3^{*\top}+o_p(n^{-1/2}),
\end{align*}
where the second equality holds, since by Lemma \ref{lemma:lemma2},
\begin{align*}
&\dfrac{1}{N}\sum_{i\in\mathcal{J}}\bSigma_{\bX,1}^{-1} \bar{\bX}_{1i} \bar{\bX}_{1i}^\top  (\bbeta^{(0)}-\bbeta^*)(\bgamma_3^{(0)}-\bgamma_3^*)^\top \bc^{(0)}\bgamma_3^{*\top}=O_p(n^{-1}),\\
&\dfrac{1}{N}\sum_{i\in\mathcal{J}}\bSigma_{\bX,1}^{-1} \bar{\bX}_{1i}\bar{\bX}_{1i}^\top \{(\bgamma_{12}^{(0)}-\bgamma_{12}^*)+\bbeta^*(\bgamma_3^{(0)}-\bgamma_3^*)^\top+(\bbeta^{(0)}-\bbeta^*)\bgamma_3^{*\top}\}(\bc^{(0)}-\bc^*)\bgamma_3^{*\top}=O_p(n^{-1}).
\end{align*}
The last equality holds, since by \eqref{eq:Ibeta},
\begin{align*}
&(m/N)\{(\bgamma_{12}^{(0)}-\bgamma_{12}^*)+\bbeta^{*}(\bgamma_{3}^{(0)}-\bgamma_3^*)^\top\}\bc^*\bgamma_3^{*\top}\\
=\ &(m/N)[\bI_{p+d_1},\bbeta^*](\dfrac{1}{n}\sum_{i\in\mathcal{I}}\bar{\bX}_{2i}\bar{\bX}_{2i}^\top)^{-1}\dfrac{1}{n}\sum_{i\in\mathcal{I}}\bar{\bX}_{2i}\bvarepsilon_{3i}^\top\bc^*\bgamma_3^{*\top}\\
=\ &(m/N)\bSigma_{\bX,1}^{-1}\dfrac{1}{n}\sum_{i\in\mathcal{I}}\bar{\bX}_{1i}\bvarepsilon_{3i}^\top\bc^*\bgamma_3^{*\top}+o_p(n^{-1/2}).
\end{align*}

Then we obtain 
\begin{align*}
	&\widehat{\bgamma}_{12}-\bgamma_{12}^* \\	
	=\ &\dfrac{1}{N}\sum_{i\in\mathcal{I}}(\bSigma_{\bX,1}^{-1}\bar{\bX}_{1i}\bvarepsilon_{3i}^\top-\bbeta^*\varepsilon_{2i}\bvarepsilon_{3i}^\top/\sigma_2^2)+\bSigma_{\bX,1}^{-1}\dfrac{m}{nN}\sum_{i\in\mathcal{I}}\bar{\bX}_{1i}\bvarepsilon_{3i}^\top\bc^*\bgamma_3^{*\top}- (1-\bgamma_3^{*\top}\bc^*)\bSigma_{\bX,1}^{-1}\dfrac{m}{nN}\sum_{i\in\mathcal{I}}\bar{\bX}_{1i}\varepsilon_{2i}\bgamma_3^{*\top}\\
 	&\qquad +\dfrac{1}{N}\sum_{i\in\mathcal{J}}\Big[\bSigma_{\bX,1}^{-1}\bar{\bX}_{1i}\bvarepsilon_{3i}^\top-\varepsilon_{2i}\bbeta^*\bvarepsilon_{3i}^\top/\sigma_2^2+(Y_i-\mu_i^{\bOmega^{(0)}})\bbeta^*(\bvarepsilon_{3i}-\varepsilon_{2i}\bgamma_3^{*})^\top/\sigma_2^2\\
	&\qquad\qquad\qquad\quad+(Y_i-\mu_i^{\bOmega^{(0)}})^2\bbeta^*\bgamma_3^{*\top}/\sigma_2^2+(\sigma_i^{\bOmega^{(0)}})^2\bbeta^*\bgamma_3^{*\top}/\sigma_2^2\Big]\\
 	&\qquad +\dfrac{1}{N}\sum_{i\in\mathcal{J}}\bSigma_{\bX,1}^{-1}\bar{\bX}_{1i}\bgamma_3^{*\top}\varepsilon_{2i}-\dfrac{1}{N}\sum_{i\in\mathcal{J}}\bc^{*\top}(\bvarepsilon_{3i}+\varepsilon_{2i}\bgamma_3^*)\bSigma_{\bX,1}^{-1}\bar{\bX}_{1i}\bgamma_3^{*\top}+o_p(n^{-1/2})\\
	=\ &\dfrac{1}{N}\sum_{i\in\mathcal{I}}(\bSigma_{\bX,1}^{-1}\bar{\bX}_{1i}\bvarepsilon_{3i}^\top-\bbeta^*\varepsilon_{2i}\bvarepsilon_{3i}^\top/\sigma_2^2)+\bSigma_{\bX,1}^{-1}\dfrac{m}{nN}\sum_{i\in\mathcal{I}}\bar{\bX}_{1i}\bvarepsilon_{3i}^\top\bc^*\bgamma_3^{*\top}- (1-\bgamma_3^{*\top}\bc^*)\bSigma_{\bX,1}^{-1}\dfrac{m}{nN}\sum_{i\in\mathcal{I}}\bar{\bX}_{1i}\varepsilon_{2i}\bgamma_3^{*\top}\\
 	&\qquad +\dfrac{1}{N}\sum_{i\in\mathcal{J}}\bSigma_{\bX,1}^{-1}\bar{\bX}_{1i}\bvarepsilon_{3i}^\top+\dfrac{1}{N}\sum_{i\in\mathcal{J}}\dfrac{-\bc^{(0)\top}\bv_i\bbeta^*\bv_i^\top+(\bc^{(0)\top}\bv_i)^2\bbeta^*\bgamma_3^{*\top}+(\sigma_i^{\bOmega^{(0)}})^2\bbeta^*\bgamma_3^{*\top}}{\sigma_2^2}\\
 	&\qquad +\dfrac{1}{N}\sum_{i\in\mathcal{J}}\bSigma_{\bX,1}^{-1}\bar{\bX}_{1i}\bgamma_3^{*\top}\varepsilon_{2i}-\dfrac{1}{N}\sum_{i\in\mathcal{J}}\bc^{*\top}(\bvarepsilon_{3i}+\varepsilon_{2i}\bgamma_3^*)\bSigma_{\bX,1}^{-1}\bar{\bX}_{1i}\bgamma_3^{*\top}+o_p(n^{-1/2})\\
	=\
	&\dfrac{1}{N}\sum_{i\in\mathcal{I}}(\bSigma_{\bX,1}^{-1}\bar{\bX}_{1i}\bvarepsilon_{3i}^\top-\bbeta^*\varepsilon_{2i}\bvarepsilon_{3i}^\top/\sigma_2^2)+\bSigma_{\bX,1}^{-1}\dfrac{m}{nN}\sum_{i\in\mathcal{I}}\bar{\bX}_{1i}\bvarepsilon_{3i}^\top\bc^*\bgamma_3^{*\top}- (1-\bgamma_3^{*\top}\bc^*)\bSigma_{\bX,1}^{-1}\dfrac{m}{nN}\sum_{i\in\mathcal{I}}\bar{\bX}_{1i}\varepsilon_{2i}\bgamma_3^{*\top}\\
 	&\qquad +\dfrac{1}{N}\sum_{i\in\mathcal{J}}\bSigma_{\bX,1}^{-1}\bar{\bX}_{1i}\bvarepsilon_{3i}^\top+\dfrac{1}{N}\sum_{i\in\mathcal{J}}\dfrac{-\bc^{*\top}\bv_i\bbeta^*\bv_i^\top+(\bc^{*\top}\bv_i)^2\bbeta^*\bgamma_3^{*\top}+(\sigma_i^{\bOmega^{*}})^2\bbeta^*\bgamma_3^{*\top}}{\sigma_2^2}\\
 	&\qquad +\dfrac{1}{N}\sum_{i\in\mathcal{J}}\bSigma_{\bX,1}^{-1}\bar{\bX}_{1i}\bgamma_3^{*\top}\varepsilon_{2i}-\dfrac{1}{N}\sum_{i\in\mathcal{J}}\bc^{*\top}(\bvarepsilon_{3i}+\varepsilon_{2i}\bgamma_3^*)\bSigma_{\bX,1}^{-1}\bar{\bX}_{1i}\bgamma_3^{*\top}\\
	&\qquad +\dfrac{1}{N}\sum_{i\in\mathcal{J}}\{(\sigma_i^{\bOmega^{(0)}})^2- (\sigma_i^{\bOmega^{*}})^2\}\bbeta^*\bgamma_3^{*\top}/\sigma_2^2+\dfrac{1}{N}\sum_{i\in\mathcal{J}} \bbeta^*  (\bc^{(0)}-\bc^*)^\top \bv_i\bv_i^\top (2\bc^*\bgamma_3^{*\top}-\bI)/\sigma_2^2+o_p(n^{-1/2})
\end{align*}
where the second equality holds, since by \eqref{gamma_3_diff}, 
\begin{align*}
	&\dfrac{1}{N}\sum_{i\in\mathcal{J}}\{-\varepsilon_{2i}\bbeta^*\bvarepsilon_{3i}^\top/\sigma_2^2+(Y_i-\mu_i^{\bOmega^{(0)}})\bbeta^*(\bvarepsilon_{3i}-\varepsilon_{2i}\bgamma_3^{*})^\top/\sigma_2^2+(Y_i-\mu_i^{\bOmega^{(0)}})^2\bbeta^*\bgamma_3^{*\top}/\sigma_2^2\}\\
 	=\ &\dfrac{1}{N}\sum_{i\in\mathcal{J}}-\varepsilon_{2i}\bbeta^*\bvarepsilon_{3i}^\top/\sigma_2^2+\dfrac{1}{N}\sum_{i\in\mathcal{J}}\{\bc^{(0)\top}(\bvarepsilon_{3i}+\varepsilon_{2i}\bgamma_3^*)-\varepsilon_{2i}\}\bbeta^*(\varepsilon_{2i}\bgamma_3^{*}-\bvarepsilon_{3i})^\top/\sigma_2^2\\
 	&\quad +\dfrac{1}{N}\sum_{i\in\mathcal{J}}\dfrac{\{\bc^{(0)\top}(\bvarepsilon_{3i}+\varepsilon_{2i}\bgamma_3^*)\}^2\bbeta^*\bgamma_3^{*\top}-2\bc^{(0)\top}(\bvarepsilon_{3i}+\varepsilon_{2i}\bgamma_3^*)\varepsilon_{2i}\bbeta^*\bgamma_3^{*\top}+\varepsilon_{2i}^2\bbeta^*\bgamma_3^{*\top}}{\sigma_2^2}+o_p(n^{-1/2})\\
	=\ &\dfrac{1}{N}\sum_{i\in\mathcal{J}}\dfrac{-\bc^{(0)\top}\bv_i\bbeta^*\bv_i^\top+(\bc^{(0)\top}\bv_i)^2\bbeta^*\bgamma_3^{*\top}}{\sigma_2^2}+o_p(n^{-1/2}).
\end{align*}
Therefore, it follows that $\widehat{\bgamma}_{12}-\bgamma_{12}^*=F_1+F_2+o_p(n^{-1/2})$, where
\begin{align*}
F_1&=\dfrac{1}{n}\sum_{i\in\mathcal{I}}\bSigma_{\bX,1}^{-1}\bar{\bX}_{1i}\bvarepsilon_{3i}^\top[(n/N)\bI_{d_2}+(m/N)\bc^*\bgamma_3^{*\top}]-(m/N) (1-\bgamma_3^{*\top}\bc^*)\bSigma_{\bX,1}^{-1}\dfrac{1}{n}\sum_{i\in\mathcal{I}}\bar{\bX}_{1i}\varepsilon_{2i}\bgamma_3^{*\top}\\
	&\qquad +\dfrac{1}{N}\sum_{i\in\mathcal{J}}\bSigma_{\bX,1}^{-1}\bar{\bX}_{1i}\bvarepsilon_{3i}^\top +\dfrac{1}{N}\sum_{i\in\mathcal{J}}\bSigma_{\bX,1}^{-1}\bar{\bX}_{1i}\bgamma_3^{*\top}\varepsilon_{2i}-\dfrac{1}{N}\sum_{i\in\mathcal{J}}\bc^{*\top}(\bvarepsilon_{3i}+\varepsilon_{2i}\bgamma_3^*)\bSigma_{\bX,1}^{-1}\bar{\bX}_{1i}\bgamma_3^{*\top}
\end{align*}
and
\begin{align*}
F_2&=\dfrac{1}{N}\sum_{i\in\mathcal{I}}-\bbeta^*\varepsilon_{2i}\bvarepsilon_{3i}^\top/\sigma_2^2 +\dfrac{1}{N}\sum_{i\in\mathcal{J}}\dfrac{-\bc^{*\top}\bv_i\bbeta^*\bv_i^\top+(\bc^{*\top}\bv_i)^2\bbeta^*\bgamma_3^{*\top}+(\sigma_i^{\bOmega^{*}})^2\bbeta^*\bgamma_3^{*\top}}{\sigma_2^2}\\
	&\qquad +\dfrac{1}{N}\sum_{i\in\mathcal{J}}\{(\sigma_i^{\bOmega^{(0)}})^2- (\sigma_i^{\bOmega^{*}})^2\}\bbeta^*\bgamma_3^{*\top}/\sigma_2^2+\dfrac{1}{N}\sum_{i\in\mathcal{J}} \bbeta^*  (\bc^{(0)}-\bc^*)^\top \bv_i\bv_i^\top (2\bc^*\bgamma_3^{*\top}-\bI_{d_2})/\sigma_2^2\\
	&=\bbeta^*\Big[ \dfrac{1}{N}\sum_{i\in\mathcal{I}}-\varepsilon_{2i}\bvarepsilon_{3i}/\sigma_2^2-\dfrac{1}{N}\sum_{i\in\mathcal{J}}\dfrac{(\bI_{d_2}-\bgamma_3^*\bc^{*\top})\bv_i\bv_i^\top \bc^*-(\sigma_i^{\bOmega^{*}})^2\bgamma_3^*}{\sigma_2^2}\\
	&\quad \qquad -(m/N) \{ (\sigma_i^{\bOmega^{*}})^2-(\sigma_i^{\bOmega^{(0)}})^2\}\bgamma_3^{*}/\sigma_2^2 - (m/N) (\bI_{d_2}-2\bgamma_3^*\bc^{*\top})E\{\bv_i \bv_i^\top\} (\bc^{(0)}-\bc^*) /\sigma_2^2 \Big]^\top+o_p(n^{-1/2}),
\end{align*}
Following \eqref{eq:gamma}, we have \[
F_2=-\bbeta^*
(\widehat{\bgamma}_3-\bgamma_3^*)^\top
+o_p(n^{-1/2})
\]
Therefore $\sqrt{n}\,{\rm vec}(F_2)\xrightarrow{\rm d}\mathcal{N}(0,\bV_{\bgamma,1}\otimes \bbeta^*\bbeta^{*\top})$. %
For the term $F_1$, rearranging the finite-sample covariance expression and using $m/N\to\tau$ and $n/N\to1-\tau$, we obtain
\[\sqrt n\operatorname{vec}(F_1)\xrightarrow{\rm d}
\mathcal N(\mathbf 0,\bV_{\bgamma,2}),\]
where
\[
\bV_{\bgamma,2}=\left[(1-\tau)\bSigma_3+\tau\frac{\sigma_2^2\bgamma_3^*\bgamma_3^{*\top}}{1+\sigma_2^2t}\right]\otimes\bSigma_{\bX,1}^{-1}.\]
Note that under Gaussian assumption, $F_1$ and $F_2$ have zero limiting cross-covariance and a joint Gaussian limit, therefore,%
\begin{align*}
	&\sqrt{n}\{{\rm vec}(\widehat{\bgamma}_{12})-{\rm vec}(\bgamma_{12}^*)\}\xrightarrow{\rm d}\mathcal{N}\Big(0,\bV_{\bgamma,2}+\{\bV_{\bgamma,1}\otimes (\bbeta^*\bbeta^{*\top})\}\Big)
\end{align*}
where $\bV_{\bgamma,1}$ is defined in \eqref{def:V1}. The upper-right block of the joint limiting covariance matrix of $\sqrt n\,\{\operatorname{vec}(\widehat{\bgamma}_{12}-\bgamma_{12}^*)^\top,(\widehat{\bgamma}_3-\bgamma_3^*)^\top\}^\top$ is $-\bV_{\bgamma,1}\otimes\bbeta^*$.
In conclusion, 
\begin{align*}
\sqrt{n}\Big( ({\rm vec}(\widehat{\bgamma}_{12})^\top,\widehat{\bgamma}_3^\top)^\top  -({\rm vec}(\bgamma_{12}^*)^\top,\bgamma_3^{*\top})^\top        \Big)\xrightarrow{\rm d}\mathcal{N}\left(0,
\begin{bmatrix}
\bV_{\bgamma,2}+\bV_{\bgamma,1}\otimes (\bbeta^*\bbeta^{*\top}) & -\bV_{\bgamma,1}\otimes \bbeta^*\\
-\bV_{\bgamma,1}\otimes \bbeta^{*\top} &\bV_{\bgamma,1}
\end{bmatrix}
\right),
\end{align*}
The supervised variance is 
\begin{align*}
\begin{bmatrix}
\bV_{\bgamma,2}^{\rm sup}+\bV_{\bgamma,1}^{\rm sup}\otimes (\bbeta^*\bbeta^{*\top}) & -\bV_{\bgamma,1}^{\rm sup}\otimes \bbeta^*\\
-\bV_{\bgamma,1}^{\rm sup}\otimes \bbeta^{*\top} &\bV_{\bgamma,1}^{\rm sup}
\end{bmatrix}
\end{align*}
where
\begin{align*}
\bV_{\bgamma,2}^{\rm sup}&=\bSigma_3\otimes \bSigma_{\bX,1}^{-1},\;\; \bV_{\bgamma,1}^{\rm sup}=\bSigma_3/\sigma_2^2
\end{align*}
Moreover, $\bDelta_{\bV,2}= \bV_{\bgamma,2}^{\rm sup}-\bV_{\bgamma,2}\succeq 0$, $\bDelta_{\bV,1}=\bV_{\bgamma,1}^{\rm sup}-\bV_{\bgamma,1}\succeq 0$
\begin{align*}
\begin{bmatrix}
	\bDelta_{\bV,2}+\bDelta_{\bV,1}\otimes (\bbeta^*\bbeta^{*\top}) & -\bDelta_{\bV,1}\otimes \bbeta^*\\
	-\bDelta_{\bV,1}\otimes \bbeta^{*\top} & \bDelta_{\bV,1}
\end{bmatrix}&=\begin{bmatrix}
	\bDelta_{\bV,2} & 0\\
 0 & 0
\end{bmatrix}+
\begin{bmatrix}
\bI_{d_2}\otimes\bbeta^*\\
-\bI_{d_2}
\end{bmatrix}
\bDelta_{\bV,1}
\begin{bmatrix}
\bI_{d_2}\otimes\bbeta^{*\top} & -\bI_{d_2}
\end{bmatrix}\succeq \mathbf{0},
\end{align*}
The remaining regime when $m=o(n)$. We now establish the asymptotic results when $m=o(n)$, including fixed $m$. We compare the one-step SAEM estimators with their supervised counterparts and show that their differences are $o_p(n^{-1/2})$. The labeled-sample OLS initial estimators and the associated Gaussian parameter estimates are root-$n$ consistent under Assumption~\ref{ass:joint_gaussian} and the nonsingularity conditions used above.
\[\sqrt{n}\Big(\frac{\sqrt{m}}{N}+\frac{m}{N\sqrt{n}}\Big)=\frac{\sqrt{mn}}{N}+\frac{m}{N}\leq \sqrt{\frac{m}{n}}+\frac{m}{n}\to 0\]
where the inequality follows from $N=n+m\geq n$. Consequently,
\[\frac{\sqrt{m}}{N}+\frac{m}{N\sqrt{n}}=o(n^{-1/2})\]
\textbf{First we show} $\widehat{\balpha}$ is first-order equivalent to supervised estimator. Define
\[\widehat{\balpha}_{L}=\Big(\sum_{i\in\mathcal{I}}{\bX}_{i}{\bX}_{i}^{\top}\Big)^{-1}\sum_{i\in\mathcal{I}}\bX_i \bS_{1i}^{\top}\]
By subtracting two normal equations, we have 
\[\widehat{\balpha}-\widehat{\balpha}_L=\Big(\sum_{i\in\mathcal{I}\cup\mathcal{J}}\bX_i\bX_i^{\top}\Big)^{-1}\sum_{i\in\mathcal{J}}\bX_i (\bS_{1i}-\widehat{\balpha}_L^{\top}\bX_i)^{\top}\]
Substituting $\bS_{1i}-\widehat{\balpha}_L^{\top}\bX_{i}=\bvarepsilon_{1i}+(\balpha^*-\widehat{\balpha}_L)^{\top}\bX_i$, and since $E(\bX_{i}\bvarepsilon_{1i}^{\top})=\mathbf{0}$, and $\widehat{\balpha}_L-\balpha^*=O_p(n^{-1/2})$, we get 
\[\widehat{\balpha}-\widehat{\balpha}_L=O_p\Big(\frac{\sqrt{m}}{N}+\frac{m}{N\sqrt{n}}\Big)=o_p(n^{-1/2})\]
Since $m=o(n)$, we have $N/n\to 1$, thus $\sqrt{N}(\widehat{\balpha}-\widehat{\balpha}_L)\xrightarrow{\rm p} \mathbf{0}$.
 
\textbf{Second we show} $\widehat{\bbeta}-\bbeta^{(0)}=o_p(n^{-1/2})$. For one-step SAEM estimator,
\[\Big(\sum_{i\in\mathcal{I}\cup\mathcal{J}}\bar{\bX}_{1i}\bar{\bX}_{1i}^{\top}\Big)\widehat{\bbeta}=\sum_{i\in\mathcal{I}}\bar{\bX}_{1i}Y_{i}+\sum_{i\in\mathcal{J}}\bar{\bX}_{1i}\mu_{i}^{\bOmega^{(0)}}\]
For supervised OLS initialization,
\[\Big(\sum_{i\in\mathcal{I}}\bar{\bX}_{1i}\bar{\bX}_{1i}^{\top}\Big)\bbeta^{(0)}=\sum_{i\in\mathcal{I}}\bar{\bX}_{1i}Y_{i}\]
Taking subtraction, we get
\[\widehat{\bbeta}-\bbeta^{(0)}=\Big(\sum_{i\in\mathcal{I}\cup\mathcal{J}}\bar{\bX}_{1i}\bar{\bX}_{1i}^{\top}\Big)^{-1}\times\sum_{i\in\mathcal{J}}\bar{\bX}_{1i}\big(\mu_i^{\bOmega^{(0)}}-\bar{\bX}_{1i}^{\top}\bbeta^{(0)}\big)\]
 Under true parameter $\bOmega^*$, we already defined $\bc^*=\sigma_2^2(\bSigma_3+\sigma_2^2\bgamma_3^*\bgamma_3^{*\top})^{-1}\bgamma_3^*$ and $\bv_i=\varepsilon_{2i}\bgamma_3^*+\bvarepsilon_{3i}$, therefore
\[\mu_i^{\bOmega^*}-\bar{\bX}_{1i}^{\top}\bbeta^*=\bc^{*\top}\bv_i\]
Since $E\big(\bar{\bX}_{1i}\bv_i^{\top}\big)=0$, by Lemma~\ref{lemma:lemma5}, we have
\[\frac{1}{N}\sum_{i\in\mathcal{J}}\bar{\bX}_{1i}\bc^{*\top}\bv_i=O_p\Big(\frac{\sqrt{m}}{N}\Big)\]
We next control the error induced by replacing the true parameters with the labeled-sample initial estimates. By the Gaussian conditional mean formula,
\[\big\{\mu_i^{\bOmega^{(0)}}-\bar{\bX}_{1i}^{\top}\bbeta^{(0)}\big\}-\big\{\mu_i^{\bOmega^*}-\bar{\bX}_{1i}^{\top}\bbeta^*\big\}=(\bc^{(0)}-\bc^*)^\top \bv_i-\bc^{(0)\top}
\Big[
\big\{
\bgamma_{12}^{(0)}-\bgamma_{12}^*
+\bbeta^{(0)}\bgamma_3^{(0)\top}
-\bbeta^*\bgamma_3^{*\top}
\big\}^{\top}\bar{\bX}_{1i}
\Big]\]
Consequently, the corresponding weighted sum equals
\begin{align*}
&\frac{1}{N}\sum_{i\in \mathcal{J}}\bar{\bX}_{1i}
\Big[
\big\{\mu_i^{\bOmega^{(0)}}-
 \bar{\bX}_{1i}^{\top}\bbeta^{(0)}\big\}
-\big\{\mu_i^{\bOmega^*}-
 \bar{\bX}_{1i}^{\top}\bbeta^*\big\}
\Big]\\
&=
\left(\frac{1}{N}\sum_{i\in \mathcal{J}}
\bar{\bX}_{1i}\bv_i^\top\right)(\bc^{(0)}-\bc^*)-
\left(\frac{1}{N}\sum_{i\in \mathcal{J}}
\bar{\bX}_{1i}\bar{\bX}_{1i}^\top\right)
\big\{
\bgamma_{12}^{(0)}-\bgamma_{12}^*
+\bbeta^{(0)}\bgamma_3^{(0)\top}
-\bbeta^*\bgamma_3^{*\top}
\big\}\bc^{(0)}.
\end{align*}
By Lemma~\ref{lemma:lemma5} and the population orthogonality $E(\bar {\bX}_{1i}\bv_i^\top)=\mathbf{0}$,
\[
\frac{1}{N}\sum_{i\in\mathcal{J}}\bar{\bX}_{1i}\bv_i^{\top}=O_p\left(\frac{\sqrt m}{N}\right)
\]
Moreover, the root-$n$ consistency of the labeled-sample initial estimators and Lemma~\ref{lemma:lemma2} imply
\[
\bc^{(0)}-\bc^*=O_p(n^{-1/2}),\quad \bc^{(0)}=O_p(1),
\]
and
\[\bgamma_{12}^{(0)}-\bgamma_{12}^*
+\bbeta^{(0)}\bgamma_3^{(0)\top}
-\bbeta^*\bgamma_3^{*\top}
=O_p(n^{-1/2})\]
In addition, since $E\|\bar{\bX}_{1i}\|^2<\infty$, Markov's inequality yields
\[\frac{1}{N}\sum_{i\in \mathcal{J}}\bar{\bX}_{1i}\bar{\bX}_{1i}^\top
=O_p\left(\frac{m}{N}\right)\]
without requiring $m\to\infty$. It follows that the first term in the weighted-sum expansion is $O_p\{\sqrt m/(N\sqrt n)\}$, whereas the second term is $O_p\{m/(N\sqrt n)\}$. Since $\sqrt m\le m$ for $m\ge1$, we obtain
\[
\frac{1}{N}\sum_{i\in \mathcal{J}}\bar{\bX}_{1i}
\Big[\big\{\mu_i^{\bOmega^{(0)}}-
\bar{\bX}_{1i}^{\top}\bbeta^{(0)}\big\}
-\big\{\mu_i^{\bOmega^*}-
\bar{\bX}_{1i}^{\top}\bbeta^*\big\}
\Big]=O_p\Big(\frac{m}{N\sqrt n}\Big)\]
This bound also holds when $m=0$, in which case the weighted sum is identically zero. Therefore,
\[\widehat{\bbeta}-\bbeta^{(0)}=O_p\Big(\frac{\sqrt{m}}{N}+\frac{m}{N\sqrt{n}}\Big)=o_p(n^{-1/2})\]
Consequently,
\[
\sqrt n(\widehat{\bbeta}-\bbeta^*)=\sqrt n(\bbeta^{(0)}-\bbeta^*)+o_p(1)
\]
By the central limit theorem for labeled-sample OLS and Slutsky's theorem,
\[
\sqrt n(\widehat{\bbeta}-\bbeta^*)\xrightarrow{\rm d}\mathcal N(\mathbf 0,\sigma_2^2\bSigma_{\bX,1}^{-1}),
\]
which coincides with $\bV_{\bbeta}$ in Theorem~\ref{thm:upper_bound} at $\tau=0$.
 
\textbf{Third we show that} $\widehat{\bgamma}-\bgamma^{(0)}=o_p(n^{-1/2})$. Recall that for each unlabeled observation, the E-step quantities $\bmu_{\bgamma i}^{\bOmega^{(0)}}$ and $\bSigma_{\bgamma i}^{\bOmega^{(0)}}$ are the conditional first and second moments of $\bar{\bX}_{2i}$ given $(\bX_i,\bS_{1i},\bS_{2i})$, evaluated at the initial parameter estimates.
The same conditional-moment formulas can also be evaluated at the true parameter $\bOmega^*$.
\[
\Big[\frac{1}{N}\sum_{i\in \mathcal{I}}\bar{\bX}_{2i}\bar{\bX}_{2i}^{\top}
+\frac{1}{N}\sum_{i\in\mathcal{J}}\bSigma_{\bgamma i}^{\bOmega^{(0)}}\Big]\widehat{\bgamma}
=\frac{1}{N}\sum_{i\in \mathcal{I}}\bar{\bX}_{2i}\bS_{2i}^{\top}+\frac{1}{N}\sum_{i\in\mathcal{J}}\boldsymbol{\mu}_{\bgamma i}^{\bOmega^{(0)}}
\bS_{2i}^{\top}\]
Meanwhile, the labeled-sample OLS estimator $\bgamma^{(0)}$ satisfies
\[
\sum_{i\in \mathcal{I}}\bar{\bX}_{2i}\Big(\bS_{2i}-\bgamma^{(0)\top}\bar{\bX}_{2i}\Big)^{\top}
=\mathbf{0}
\]
Subtracting the labeled normal equations from the one-step normal equations gives the exact identity
\[\widehat{\bgamma}-\bgamma^{(0)}=\left[\frac{1}{N}\sum_{i\in \mathcal{I}}\bar{\bX}_{2i}\bar{\bX}_{2i}^{\top}
+\frac{1}{N}\sum_{i\in \mathcal{J}}\bSigma_{\bgamma i}^{\bOmega^{(0)}}\right]^{-1}\times
\frac1N\sum_{i\in\mathcal{J}}\left\{\bmu_{\bgamma i}^{\bOmega^{(0)}}
\bS_{2i}^{\top}-\bSigma_{\bgamma i}^{\bOmega^{(0)}}
\bgamma^{(0)}\right\}\]
Since $m=o(n)$, the law of large numbers and the consistency of $\bOmega^{(0)}$ yield
\[
\frac{1}{N}\sum_{i\in \mathcal{I}}\bar{\bX}_{2i}\bar{\bX}_{2i}^{\top}+\frac{1}{N}\sum_{i\in\mathcal{J}}
\bSigma_{\bgamma i}^{\bOmega^{(0)}}=\bSigma_{\bX,2}+o_p(1)
\]
where the unlabeled contribution is $O_p(m/N)=o_p(1)$. At the true parameters, the conditional score satisfies
\[
E\left(\bmu_{\bgamma i}^{\bOmega^*}\bS_{2i}^{\top}-\bSigma_{\bgamma i}^{\bOmega^*}
\bgamma^*\right)=E(\bar{\bX}_{2i}\bvarepsilon_{3i}^{\top})=\mathbf{0}
\]
Thus, by Lemma~\ref{lemma:lemma5},
\[
\frac1N\sum_{i\in \mathcal{J}}
\left(\bmu_{\bgamma i}^{\bOmega^*}\bS_{2i}^{\top}-\bSigma_{\bgamma i}^{\bOmega^*}
\bgamma^*\right)=O_p\left(\frac{\sqrt m}{N}\right)
\]
Moreover, by \eqref{eq:mu_sig}--\eqref{eq:gamma_mu_sig}, the conditional moments and the corresponding conditional score are continuously differentiable in a neighborhood of $\bOmega^*$, where the relevant covariance matrices remain uniformly positive definite. Their derivatives are bounded by
\[
C(1+\|\bar{\bX}_{1i}\|^2+\|\bS_{2i}\|^2).
\]
Since $\bOmega^{(0)}-\bOmega^*=O_p(n^{-1/2})$ and the Gaussian moments are finite, the mean-value theorem gives
\[
\frac{1}{N}\sum_{i\in \mathcal{J}}\left(
\bmu_{\bgamma i}^{\bOmega^{(0)}}\bS_{2i}^{\top}
-\bSigma_{\bgamma i}^{\bOmega^{(0)}}
\bgamma^{(0)}\right)-\frac1N\sum_{i\in \mathcal{J}}\left(\bmu_{\bgamma i}^{\bOmega^*}\bS_{2i}^{\top}
-\bSigma_{\bgamma i}^{\bOmega^*}
\bgamma^*\right)=O_p\left(\frac{m}{N\sqrt n}\right)
\]
Combining these results,
\[
\widehat{\bgamma}-\bgamma^{(0)}=O_p\left(\frac{\sqrt m}{N}+\frac{m}{N\sqrt n}
\right)=o_p(n^{-1/2})
\]
Therefore,
\[
\sqrt n\operatorname{vec}(\widehat{\boldsymbol{\gamma}}-\boldsymbol{\gamma}^*)=\sqrt n\operatorname{vec}
(\boldsymbol{\gamma}^{(0)}-\boldsymbol{\gamma}^*)
+o_p(1)
\]
By the Gaussian OLS central limit theorem,
\[
\sqrt n\operatorname{vec}
(\boldsymbol{\gamma}^{(0)}-\boldsymbol{\gamma}^*)
\xrightarrow{\rm d}
\mathcal{N}\left(\mathbf 0,
\bSigma_3\otimes\bSigma_{\bX,2}^{-1}\right),
\]
whose block representation coincides with $\bV_{\bgamma}$ in Theorem~\ref{thm:upper_bound} at $\tau=0$. This establishes the result.
 
\textbf{Finally we show that} $\widehat{\btheta}-\widehat{\btheta}_{\rm sup}=o_p(n^{-1/2})$. Define 
\[\widehat{\btheta}_{\rm sup}=\bbeta_1^{(0)}+\widehat{\balpha}_L\bbeta_2^{(0)}\]
By normal equations,
\[\sum_{i\in\mathcal{I}}\bX_i(Y_i-\bX_i^{\top}\bbeta_1^{(0)}-\bS_{1i}^{\top}\bbeta_2^{(0)})=\mathbf{0}\]
and
\[\sum_{i\in\mathcal{I}}\bX_i (\bS_{1i}-\widehat{\balpha}^{\top}_L\bX_i)^{\top}=\mathbf{0}\]
We get 
\[\widehat{\btheta}_{\rm sup}=\Big(\sum_{i\in\mathcal{I}}\bX_i\bX_i^{\top}\Big)^{-1}\sum_{i\in\mathcal{I}}\bX_i Y_i\]
Subtracting SAEM and labeled-only estimator,
\[\widehat{\btheta}-\widehat{\btheta}_{\rm sup}=(\widehat{\bbeta}_1-\bbeta^{(0)}_1)+(\widehat{\balpha}-\widehat{\balpha}_L)\bbeta_2^{(0)}+\widehat{\balpha}(\widehat{\bbeta}_2-\bbeta_2^{(0)})\]
Since we already showed that $\widehat{\bbeta}-{\bbeta}^{(0)}=o_p(n^{-1/2})$ and $\widehat{\balpha}-\widehat{\balpha}_L=o_p(n^{-1/2})$, hence
\[\sqrt{n}(\widehat{\btheta}-\btheta^*)=\sqrt{n}(\widehat{\btheta}_{\rm sup}-\btheta^{*})+o_p(1)\]
By the central limit theorem for the labeled marginal OLS estimator and Slutsky's theorem,
\[\sqrt{n}(\widehat{\btheta}-\btheta^*)\xrightarrow{\rm d}\mathcal{N}(\mathbf{0},(\sigma_2^2+\bbeta_2^{*\top}\bSigma_1\bbeta_2^*)\bSigma_{\bX}^{-1})\]
which coincides with $\bV_{\btheta}$ at $\tau=0$.

When $m=0$, all unlabeled sums vanish and the one-step EM update returns the supervised regression estimators exactly. When $m$ is a fixed positive integer, the preceding bounds further imply
\[
\widehat{\bbeta}-\bbeta^{(0)}=O_p(n^{-1}),\quad \widehat{\bgamma}-\bgamma^{(0)}=O_p(n^{-1})
\]
together with the corresponding $O_p(n^{-1})$ differences for $\widehat\balpha$ and $\widehat\btheta$ relative to their labeled-only counterparts. Thus, the SAEM and supervised estimators have the same first-order limiting distributions whenever $m=o(n)$, including fixed $m$. These distributions agree with the covariance matrices in Theorem~\ref{thm:upper_bound} evaluated at $\tau=0$. This completes the proof.
\end{proof}

\begin{proof}[Proof of Theorem \ref{thm:lower_bound}]
		\mbox{}\par
	\noindent
Denote $\mathcal{I}=\{1,2,\cdots,n\}$, $\mathcal{J}=\{n+1,n+2,\cdots,n+m\}$, and $N=n+m$. The following Fisher information calculations use the actual sample sizes $n$ and $m$, with the
observed-data information obtained by combining the labeled and unlabeled contributions. These calculations do not require $m\to\infty$. After appropriate normalization, the resulting
Cram\'er--Rao bounds converge to the limiting covariance matrices in Theorem~\ref{thm:upper_bound} as $m/N\to\tau$. For the labeled data, we first have
\begin{align*}
	\bS_{1i}&\mid \bX_i\sim \mathcal{N}(\balpha^{\top}\bX_i,\bSigma_1),\\ 
	Y_i&\mid \bX_i,\bS_{1i}\sim \mathcal{N}(\bbeta_1^{\top}\bX_i+\bbeta_2^{\top}\bS_{1i},\sigma_2^2),\\ 
	\bS_{2i}&\mid \bX_i,\bS_{1i},Y_i\sim\mathcal{N}(\bgamma_1^{\top}\bX_i+\bgamma_2^{\top}\bS_{1i}+\bgamma_3 Y_i,\bSigma_3).
\end{align*}
For the unlabeled data, it follows that
\begin{align*}
	\bS_{1i}&\mid \bX_i\sim \mathcal{N}(\balpha^{\top}\bX_i,\bSigma_1),\;\; 
	\bS_{2i}\mid \bX_i,\bS_{1i}\sim\mathcal{N}((\bgamma_1+\bbeta_1\bgamma_3^{\top})^\top \bX_i+(\bgamma_2+\bbeta_2\bgamma_3^{\top})^\top \bS_{1i},\bSigma_3+\sigma_2^2\bgamma_3\bgamma_3^{\top}).
\end{align*}
Let $p(\cdot)$ denote the density functions, the likelihood function of the observed data is 
\begin{align*}
	L(\theta)&=\prod_{i=1}^{n}p(\bX_i)p(\bS_{1i}\mid \bX_i)p(Y_i\mid \bX_i,\bS_{1i})p(\bS_{2i}\mid \bX_i,\bS_{1i},Y_i)\\
	&\qquad\times \prod_{i=n+1}^{n+m}p(\bX_i)p(\bS_{1i}\mid \bX_i)p(\bS_{2i}\mid \bX_i,\bS_{1i}).
\end{align*}
Denote $\bar{\bX}_{1i}=(\bX_i^\top,\bS_{1i}^\top)^\top\in\mathbb{R}^{p+d_1}$, $\bbeta=(\bbeta_1^\top,\bbeta_2^\top)^\top\in\mathbb{R}^{p+d_1}$, $\bgamma_{12}=(\bgamma_1^\top,\bgamma_2^\top)^\top\in\mathbb{R}^{(p+d_1)\times d_2}$ , the log-likelihood function is 
\begin{align*}
	l(\theta)&=C+\sum_{i=1}^{n+m}\log p(\bX_i)-\sum_{i=1}^{n+m}\bigg\{\dfrac{\log(|\bSigma_1|)}{2}+\dfrac{(\bS_{1i}-\balpha^{\top}\bX_i)^\top\bSigma_1^{-1}(\bS_{1i}-\balpha^{\top}\bX_i)}{2}\bigg\}\\
	&\qquad-\sum_{i=1}^n\bigg\{\dfrac{\log(\sigma_2^2)}{2}+\dfrac{(Y_{i}-\bbeta^{\top}\bar{\bX}_{1i})^2}{2\sigma_2^2}\bigg\}\\	&\qquad-\sum_{i=1}^n\bigg\{\dfrac{\log(|\bSigma_3|)}{2}+\dfrac{(\bS_{2i}-\bgamma_{12}^{\top}\bar{\bX}_{1i}-\bgamma_3 Y_i)^\top \bSigma_3^{-1}(\bS_{2i}-\bgamma_{12}^{\top}\bar{\bX}_{1i}-\bgamma_3 Y_i)}{2}\bigg\}\\
	&\qquad-\dfrac{m\log(|\bSigma_3+\sigma_2^2\bgamma_3\bgamma_3^{\top}|)}{2}\\
	&\qquad-\sum_{i=n+1}^{n+m}\dfrac{\{\bS_{2i}-(\bgamma_{12}+\bbeta\bgamma_3^{\top})^\top \bar{\bX}_{1i}\}^\top(\bSigma_3+\sigma_2^2\bgamma_3\bgamma_3^{\top})^{-1}\{\bS_{2i}-(\bgamma_{12}+\bbeta\bgamma_3^{\top})^\top \bar{\bX}_{1i}\} }{2}.
\end{align*}
Here, $C$ represents constant terms in the log-likelihood (e.g., $\log(2\pi)$), which are grouped as $C$ since they do not affect differentiation. We calculate the first derivatives with respect to ${\rm vec}(\balpha)$ and ${\rm vec}(\bSigma_1)$ since they are matrix parameters
\begin{align*}
\dfrac{\partial l}{\partial{\rm vec}(\balpha)}&=\sum_{i=1}^{N}\{\bSigma_{1}^{-1}(\bS_{1i}-\balpha^\top \bX_i)\}\otimes \bX_i,\\
\dfrac{\partial l}{\partial{\rm vec}(\bSigma_1)}&=-\dfrac{N}{2}{\rm vec}(\bSigma_{1}^{-1})+\dfrac{1}{2}\sum_{i=1}^N{\rm vec}\{\bSigma_{1}^{-1}(\bS_{1i}-\balpha^\top \bX_i)(\bS_{1i}-\balpha^\top \bX_i)^\top \bSigma_{1}^{-1}\}.
\end{align*}
The first derivatives with respect to $\bbeta$ and $\sigma_2^2$ are
\begin{align*}
	\dfrac{\partial l}{\partial \bbeta}&=\dfrac{\sum_{i=1}^n(Y_i-\bbeta^\top\bar{\bX}_{1i})\bar{\bX}_{1i}}{\sigma_2^2}+\sum_{i=n+1}^{n+m}\bar{\bX}_{1i}\bgamma_3^\top(\bSigma_3+\sigma_2^2\bgamma_3\bgamma_3^\top)^{-1}\{\bS_{2i}-(\bgamma_{12}+\bbeta\bgamma_3^{\top})^\top \bar{\bX}_{1i}\},\\
	\dfrac{\partial l}{\partial \sigma_2^2}&=-\dfrac{n}{2\sigma_2^2}+\dfrac{\sum_{i=1}^n(Y_i-\bbeta^\top\bar{\bX}_{1i})^2}{2\sigma_2^4}-\dfrac{m\bgamma_3^\top(\bSigma_3+\sigma_2^2\bgamma_3\bgamma_3^\top)^{-1}\bgamma_3}{2}\\
	&\qquad+\dfrac{1}{2}\sum_{i=n+1}^{n+m}\left[\{\bS_{2i}-(\bgamma_{12}+\bbeta\bgamma_3^{\top})^\top \bar{\bX}_{1i}\}^\top(\bSigma_3+\sigma_2^2\bgamma_3\bgamma_3^\top)^{-1}\bgamma_3\right]^2.
\end{align*}
The first derivatives with respect to ${\rm vec}(\bgamma_{12})$ and $\bgamma_3$, and ${\rm vec}(\bSigma_3)$ are given by
\begin{align*}
	&\dfrac{\partial l}{\partial {\rm vec}(\bgamma_{12})}=\sum_{i=1}^n \{\bSigma_3^{-1}(\bS_{2i}-\bgamma_{12}^\top\bar{\bX}_{1i}-\bgamma_3Y_i)\}\otimes\bar{\bX}_{1i}\\
	&\qquad\qquad\qquad+\sum_{i=n+1}^{n+m} [ (\bSigma_3+\sigma_2^2\bgamma_3\bgamma_3^\top)^{-1}\{\bS_{2i}-(\bgamma_{12}+\bbeta\bgamma_3^{\top})^\top \bar{\bX}_{1i}\}]\otimes \bar{\bX}_{1i},\\
	&\dfrac{\partial l}{\partial \bgamma_{3}}=\sum_{i=1}^nY_i\bSigma_3^{-1}(\bS_{2i}-\bgamma_{12}^\top\bar{\bX}_{1i}-\bgamma_3Y_i)-m\sigma_2^2(\bSigma_3+\sigma_2^2\bgamma_3\bgamma_3^\top)^{-1}\bgamma_3\\
	&\qquad\qquad\qquad+\sum_{i=n+1}^{n+m}(\bbeta^\top\bar{\bX}_{1i})(\bSigma_3+\sigma_2^2\bgamma_3\bgamma_3^\top)^{-1}\{\bS_{2i}-(\bgamma_{12}+\bbeta\bgamma_3^{\top})^\top \bar{\bX}_{1i}\}\\
	&\qquad\qquad\qquad +\sum_{i=n+1}^{n+m}\sigma_2^2[\bgamma_3^\top(\bSigma_3+\sigma_2^2\bgamma_3\bgamma_3^\top)^{-1}\{\bS_{2i}-(\bgamma_{12}+\bbeta\bgamma_3^{\top})^\top \bar{\bX}_{1i}\}]\\
	&\qquad\qquad\qquad\qquad\qquad\times (\bSigma_3+\sigma_2^2\bgamma_3\bgamma_3^\top)^{-1}\{\bS_{2i}-(\bgamma_{12}+\bbeta\bgamma_3^{\top})^\top \bar{\bX}_{1i}\},\\
	&\dfrac{\partial l}{\partial {\rm vec}(\bSigma_3)}=-\dfrac{n}{2}{\rm vec}(\bSigma_3^{-1})+\dfrac{1}{2}\sum_{i=1}^n{\rm vec}\{\bSigma_3^{-1}(\bS_{2i}-\bgamma_{12}^\top\bar{\bX}_{1i}-\bgamma_3Y_i)(\bS_{2i}-\bgamma_{12}^\top\bar{\bX}_{1i}-\bgamma_3Y_i)^\top\bSigma_3^{-1}\}\\
	&\qquad\qquad\qquad-\dfrac{m}{2}{\rm vec}\{(\bSigma_3+\sigma_2^2\bgamma_3\bgamma_3^\top)^{-1}\}\\
	&\qquad\qquad\qquad+\dfrac{1}{2}\sum_{i=n+1}^{n+m}{\rm vec}\Big[(\bSigma_3+\sigma_2^2\bgamma_3\bgamma_3^\top)^{-1}\{\bS_{2i}-(\bgamma_{12}+\bbeta\bgamma_3^{\top})^\top \bar{\bX}_{1i}\}\\
	&\qquad\qquad\qquad\qquad\qquad\qquad\qquad\{\bS_{2i}-(\bgamma_{12}+\bbeta\bgamma_3^{\top})^\top \bar{\bX}_{1i}\}^\top(\bSigma_3+\sigma_2^2\bgamma_3\bgamma_3^\top)^{-1}\Big].
\end{align*}
In the following, denote $\br_{1i}=\bS_{1i}-\balpha^\top \bX_i$, $\br_{2i}=\bS_{2i}-\bgamma_{12}^\top\bar{\bX}_{1i}-\bgamma_3Y_i$,  $\tilde{\br}_{2i}=\bS_{2i}-(\bgamma_{12}+\bbeta\bgamma_3^{\top})^\top \bar{\bX}_{1i}$, $\bD=(\bSigma_{3}+\sigma_2^2\bgamma_3\bgamma_3^\top)^{-1}$. In the following, we define $\bN_d=(\bI_{d^2}+\bK_{d,d})/2$ as the $d^2\times d^2$ symmetrizer matrix, which acts as the orthogonal projector onto the subspace of vectorized symmetric matrices, here $\bK_{d,d}$ is the commutation matrix satisfying $\bK_{d,d}{\rm vec}(\bA)={\rm vec}(\bA^\top)$ for any $d\times d$ matrix $\bA$. 
For a symmetric covariance matrix of size $d$, let $\mathbf U_d\in\mathbb R^{d^2\times d(d+1)/2}$ have orthonormal columns spanning ${\cal S}_d=\{{\rm vec}(\bH):\bH=\bH^\top\}$. Thus $\mathbf U_d\mathbf U_d^\top=\bN_d$. Whenever an information block contains ${\rm vec}(\bSigma)$, its inverse is taken in the identifiable coordinates $\mathbf U_d^\top{\rm vec}(\bSigma)$, not in the redundant $d^2$-dimensional space. For a matrix $\bH$ acting on a space with such a covariance block and an orthonormal embedding $\bP$, write $\bH_{\cal S}^{[-1]}=\bP(\bP^\top\bH\bP)^{-1}\bP^\top$ for the lifted inverse on the admissible tangent space. This notation is not an ordinary inverse of the full matrix.
We compute the mixed second-order partial derivatives with respect to $({\rm vec}(\balpha),{\rm vec}(\balpha))$, $({\rm vec}(\balpha),{\rm vec}(\bSigma_1))$, and $({\rm vec}(\bSigma_1),{\rm vec}(\bSigma_1))$
\begin{align*}
\dfrac{\partial^2l}{\partial{\rm vec}(\balpha)\partial{\rm vec}(\balpha)^\top}&=-\bSigma_{1}^{-1}\otimes \sum_{i=1}^{N}\bX_i\bX_i^\top,\\
\dfrac{\partial^2l}{\partial{\rm vec}(\balpha)\partial{\rm vec}(\bSigma_1)^\top}&=-\dfrac{1}{2}\sum_{i=1}^N[\{(\br_{1i}^\top \bSigma_{1}^{-1})\otimes \bSigma_{1}^{-1}\}\otimes \bX_i+\{\bSigma_{1}^{-1}\otimes (\br_{1i}^\top \bSigma_{1}^{-1})\}\otimes \bX_i],\\
\dfrac{\partial^2l}{\partial{\rm vec}(\bSigma_1)\partial{\rm vec}(\bSigma_1)^\top}&=\bN_{d_1}\Big\{\dfrac{N}{2}\bSigma_{1}^{-1}\otimes \bSigma_{1}^{-1}-\dfrac{1}{2}\sum_{i=1}^N(\bSigma_{1}^{-1}\br_{1i}\br_{1i}^\top \bSigma_{1}^{-1})\otimes \bSigma_{1}^{-1} \\
&\qquad\qquad-\dfrac{1}{2}\sum_{i=1}^N\bSigma_{1}^{-1}\otimes (\bSigma_{1}^{-1}\br_{1i}\br_{1i}^\top \bSigma_{1}^{-1})\Big\}\bN_{d_1}.
\end{align*}
The mixed second-order partial derivatives with respect to $(\bbeta,\bbeta)$, $(\bbeta,\sigma_2^2)$, $(\bbeta,{\rm vec}(\bgamma_{12}))$, $(\bbeta,\bgamma_{3})$, and $(\bbeta,{\rm vec}(\bSigma_3))$ are given by
\begin{align*}
	&\dfrac{\partial^2 l}{\partial \bbeta\partial\bbeta^\top}=-\dfrac{\sum_{i=1}^n\bar{\bX}_{1i}\bar{\bX}_{1i}^\top}{\sigma_2^2}-\sum_{i=n+1}^{n+m}(\bgamma_3^\top\bD \bgamma_3)\bar{\bX}_{1i}\bar{\bX}_{1i}^\top,\\
	&\dfrac{\partial^2 l}{\partial \bbeta\partial\sigma_2^2}=-\dfrac{\sum_{i=1}^n(Y_i-\bbeta^\top\bar{\bX}_{1i})\bar{\bX}_{1i}}{\sigma_2^4}-\sum_{i=n+1}^{n+m}(\bgamma_3^\top\bD\bgamma_3)(\bgamma_3^\top\bD\tilde{\br}_{2i})\bar{\bX}_{1i},\\
	&\dfrac{\partial^2 l}{\partial \bbeta\partial{\rm vec}(\bgamma_{12})^\top}=-\sum_{i=n+1}^{n+m} (\bgamma_3^\top\bD)\otimes (\bar{\bX}_{1i}\bar{\bX}_{1i}^\top),\\
	&\dfrac{\partial^2 l}{\partial \bbeta\partial\bgamma_{3}^\top}=\sum_{i=n+1}^{n+m}\bigg[\bar{\bX}_{1i}\tilde{\br}_{2i}^\top \bD-(\bbeta^{\top} \bar{\bX}_{1i})\bar{\bX}_{1i} \bgamma_3^{\top} \bD -\sigma_2^2(\tilde{\br}_{2i}^\top \bD\bgamma_{3})\bar{\bX}_{1i}\bgamma_3^\top  \bD - \sigma_2^2(\bgamma_3^\top\bD\bgamma_3)\bar{\bX}_{1i}\tilde{\br}_{2i}^\top \bD\bigg],\\
	&\dfrac{\partial^2 l}{\partial \bbeta\partial{\rm vec}(\bSigma_3)^\top}=-\dfrac{1}{2}\sum_{i=n+1}^{n+m}\bar{\bX}_{1i}\left[(\bD\tilde{\br}_{2i})\otimes (\bD\bgamma_3)+(\bD\bgamma_3)\otimes (\bD\tilde{\br}_{2i})\right]^\top.
\end{align*}
The mixed second-order partial derivatives with respect to $(\sigma_2^2,\sigma_2^2)$, $(\sigma_2^2,{\rm vec}(\bgamma_{12}))$, $(\sigma_2^2,\bgamma_{3})$, and $(\sigma_2^2,{\rm vec}(\bSigma_3))$ are given by
\begin{align*}
	&\dfrac{\partial^2 l}{\partial (\sigma_2^2)^2}=\dfrac{n}{2\sigma_2^4}-\dfrac{\sum_{i=1}^n(Y_i-\bbeta^\top\bar{\bX}_{1i})^2}{\sigma_2^6}+\dfrac{m}{2}(\bgamma_3^\top\bD\bgamma_3)^2-\sum_{i=n+1}^{n+m}(\tilde{\br}_{2i}^\top \bD\bgamma_{3})^2(\bgamma_3^\top\bD\bgamma_3),\\
	&\dfrac{\partial^2 l}{\partial \sigma_2^2\partial {\rm vec}(\bgamma_{12})^\top}=-\sum_{i=n+1}^{n+m}\tilde{\br}_{2i}^\top\bD\bgamma_3{\rm vec}(\bar{\bX}_{1i}\bgamma_3^\top\bD)^\top,\\
	&\dfrac{\partial^2 l}{\partial \sigma_2^2\partial \bgamma_{3}^\top}=\Big\{-m\bD\bgamma_3+m\sigma_2^2(\bgamma_3^\top\bD\bgamma_3)\bD\bgamma_3+\sum_{i=n+1}^{n+m}(\tilde{\br}_{2i}^\top\bD\bgamma_3)\bD\tilde{\br}_{2i}-\sigma_2^2\sum_{i=n+1}^{n+m}(\tilde{\br}_{2i}^\top \bD\bgamma_3)(\bgamma_3^\top\bD\bgamma_3)\bD \tilde{\br}_{2i}\\
	&\qquad\qquad\qquad-\sum_{i=n+1}^{n+m}(\tilde{\br}_{2i}^\top \bD\bgamma_3)(\bar{\bX}_{1i}^\top\bbeta)\bD\bgamma_3-\sigma_2^2\sum_{i=n+1}^{n+m}(\tilde{\br}_{2i}^\top\bD\bgamma_3)^2\bD\bgamma_3 \Big\}^\top,\\
	&\dfrac{\partial^2 l}{\partial \sigma_2^2\partial {\rm vec}(\bSigma_{3})^\top}=\dfrac{m}{2}{\rm vec}(\bD\bgamma_3\bgamma_3^\top\bD)^\top-\sum_{i=n+1}^{n+m}\tilde{\br}_{2i}^\top\bD\bgamma_3\times \dfrac{1}{2} \Big[{\rm vec}(\bD\tilde{\br}_{2i}\bgamma_3^\top\bD) + {\rm vec}(\bD\bgamma_3\tilde{\br}_{2i}^\top\bD)\Big]^\top
\end{align*}
The mixed second-order partial derivatives with respect to $\{{\rm vec}(\bgamma_{12}),{\rm vec}(\bgamma_{12})\}$, $\{{\rm vec}(\bgamma_{12}),\bgamma_{3}\}$, and $\{{\rm vec}(\bgamma_{12}),{\rm vec}(\bSigma_3)\}$ are given by
\begin{align*}
	&\dfrac{\partial^2 l}{\partial {\rm vec}(\bgamma_{12})\partial {\rm vec}(\bgamma_{12})^\top}=-\sum_{i=1}^n\bSigma_3^{-1}\otimes (\bar{\bX}_{1i}\bar{\bX}_{1i}^\top)-\sum_{i=n+1}^{n+m}\bD\otimes (\bar{\bX}_{1i}\bar{\bX}_{1i}^\top),\\
	&\dfrac{\partial^2 l}{\partial {\rm vec}(\bgamma_{12})\partial \bgamma_3^\top}=-\sum_{i=1}^n(Y_i\bSigma_3^{-1})\otimes \bar{\bX}_{1i}-\sigma_2^2\sum_{i=n+1}^{n+m}\{(\bgamma_3^\top\bD\tilde{\br}_{2i})\bD\}\otimes \bar{\bX}_{1i}\\
	&\qquad\qquad\qquad\qquad -\sigma_2^2\sum_{i=n+1}^{n+m}(\bD\bgamma_{3}\tilde{\br}_{2i}^\top\bD)\otimes \bar{\bX}_{1i} -\sum_{i=n+1}^{n+m}\{(\bbeta^\top\bar{\bX}_{1i})\bD\}\otimes \bar{\bX}_{1i},\\
	&\dfrac{\partial^2 l}{\partial {\rm vec}(\bgamma_{12})\partial {\rm vec}(\bSigma_3)^\top}=-\sum_{i=1}^n(\bI_{d_2}\otimes\bar{\bX}_{1i})[\bSigma_3^{-1}\otimes ( \br_{2i}^\top\bSigma_3^{-1})+(\br_{2i}^\top\bSigma_3^{-1})\otimes \bSigma_{3}^{-1}]/2\\
	&\qquad\qquad\qquad\qquad\qquad\qquad -\sum_{i=n+1}^{n+m}(\bI_{d_2}\otimes\bar{\bX}_{1i})[\bD\otimes (\tilde{\br}_{2i}^\top \bD)+(\tilde{\br}_{2i}^\top \bD)\otimes \bD]/2.
\end{align*}
The mixed second-order partial derivatives with respect to $(\bgamma_{3},\bgamma_3)$, and $\{\bgamma_{3},{\rm vec}(\bSigma_{3})\}$ are given by
 \begin{align*}
	&\dfrac{\partial^2 l}{\partial \bgamma_3\partial \bgamma_3^\top}=-\sum_{i=1}^nY_i^2\bSigma_3^{-1}-m\sigma_2^2\bD+m\sigma_2^4(\bgamma_3^\top \bD \bgamma_3)\bD+m\sigma_2^4\bD\bgamma_3\bgamma_3^\top \bD\\
	&\qquad\qquad -\sum_{i=n+1}^{n+m} (\bbeta^\top\bar{\bX}_{1i}+\sigma_2^2\bgamma_3^\top \bD\tilde{\br}_{2i})^2\bD +\sum_{i=n+1}^{n+m}(\sigma_2^2-\sigma_2^4\bgamma_3^\top \bD\bgamma_3)\bD\tilde{\br}_{2i}\tilde{\br}_{2i}^\top \bD\\
	&\qquad\qquad -\sigma_2^2\sum_{i=n+1}^{n+m}(\bbeta^\top\bar{\bX}_{1i}+\sigma_2^2\bgamma_3^\top \bD\tilde{\br}_{2i})(\bD\bgamma_3\tilde{\br}_{2i}^\top \bD+\bD\tilde{\br}_{2i}\bgamma_3^\top \bD),\\
	&\dfrac{\partial^2 l}{\partial\bgamma_3\partial {\rm vec}(\bSigma_3)^\top}={\rm Sym}\Big[-\sum_{i=1}^n Y_i(\bSigma_3^{-1}\br_{2i})^\top\otimes \bSigma_3^{-1}+m\sigma_2^2(\bD\bgamma_3)^\top \otimes\bD\\
	&\qquad\qquad\qquad\qquad\qquad\qquad-\sum_{i=n+1}^{n+m}(\bbeta^\top\bar{\bX}_{1i})(\bD\tilde{\br}_{2i})^\top \otimes \bD -\sigma_2^2\sum_{i=n+1}^{n+m}(\bD\tilde{\br}_{2i})^\top \otimes (\bD\tilde{\br}_{2i}\bgamma_3^\top \bD)\\
	&\qquad\qquad\qquad\qquad\qquad\qquad-\sigma_2^2\sum_{i=n+1}^{n+m}(\bD\tilde{\br}_{2i})^\top \otimes \{(\bgamma_3^\top \bD \tilde{\br}_{2i})\bD\}\Big].
\end{align*}
The operation Sym() proceeds as follows for each row of a matrix: first, the row is transposed into a column vector; this column vector is then reshaped into a square matrix following column‑major ordering; the resulting square matrix $\bA$ is then symmetrized by computing $(\bA+\bA^\top)/2$; next, the symmetrized matrix is flattened column‑wise into a column vector; finally, this column vector is transposed to form the corresponding row of the new matrix. Finally, we compute
\begin{align*}
	&\dfrac{\partial^2 l}{\partial {\rm vec}(\bSigma_3)\partial {\rm vec}(\bSigma_3)^\top}=\bN_{d_2}\Big\{\dfrac{n}{2}(\bSigma_3^{-1}\otimes \bSigma_3^{-1})+\dfrac{m}{2}(\bSigma_3+\sigma_2^2\bgamma_3\bgamma_3^{\top})^{-1}\otimes (\bSigma_3+\sigma_2^2\bgamma_3\bgamma_3^{\top})^{-1}\\
	&\qquad\qquad\qquad\qquad\qquad\qquad\quad-\dfrac{1}{2}\sum_{i=1}^n(\bSigma_3^{-1}\br_{2i}\br_{2i}^\top\bSigma_3^{-1})\otimes \bSigma_3^{-1} -\dfrac{1}{2}\sum_{i=1}^n\bSigma_3^{-1}\otimes (\bSigma_3^{-1}\br_{2i}\br_{2i}^\top\bSigma_3^{-1})\\
	&\qquad\qquad\qquad\qquad\qquad\qquad\quad-\dfrac{1}{2}\sum_{i=n+1}^{n+m}(\bD\tilde{\br}_{2i}\tilde{\br}_{2i}^\top\bD)\otimes \bD-\dfrac{1}{2}\sum_{i=n+1}^{n+m}\bD\otimes(\bD\tilde{\br}_{2i}\tilde{\br}_{2i}^\top\bD)\Big\}\bN_{d_2}.
 \end{align*}

Then we can compute the expectation of all second-order partial derivatives
\begin{align*}
	E\left\{\dfrac{\partial^2 l}{\partial {\rm vec}(\balpha)\partial {\rm vec}(\balpha)^\top}\right\}&=-N\bSigma_{1}^{-1}\otimes \bSigma_{\bX},\;\; E\left\{\dfrac{\partial^2 l}{\partial {\rm vec}(\balpha)\partial {\rm vec}(\bSigma_1)^\top}\right\}=0\\
		E\left\{\dfrac{\partial^2 l}{\partial {\rm vec}(\bSigma_1)\partial {\rm vec}(\bSigma_1)^\top}\right\}&=-\bN_{d_1}\Big(\dfrac{N}{2}\bSigma_{1}^{-1}\otimes \bSigma_{1}^{-1}\Big)\bN_{d_1},
\end{align*}
and
\begin{align*}
	E\left(\dfrac{\partial^2 l}{\partial \bbeta\partial \bbeta^\top}\right)&=-\bSigma_{\bX,1}\left\{\dfrac{n}{\sigma_2^2}+m\bgamma_3^\top\bD\bgamma_3\right\},\\
	E\left(\dfrac{\partial^2 l}{\partial \bbeta\partial\sigma_2^2}\right)&=0,\;\; 
	E\left\{\dfrac{\partial^2 l}{\partial \bbeta\partial{\rm vec}(\bgamma_{12})^\top}\right\}=-m (\bgamma_3^\top\bD)\otimes \bSigma_{\bX,1},\\
	E\left(\dfrac{\partial^2 l}{\partial \bbeta\partial\bgamma_{3}^\top}\right)&=-m\bSigma_{\bX,1}\bbeta\bgamma_3^\top\bD,\;\; E\left\{\dfrac{\partial^2 l}{\partial \bbeta\partial{\rm vec}(\bSigma_3)^\top}\right\}=0.
\end{align*}
and
\begin{align*}
	E\left\{\dfrac{\partial^2 l}{\partial (\sigma_2^2)^2}\right\}&=-\dfrac{n}{2\sigma_2^4}-\dfrac{m}{2}(\bgamma_3^\top\bD\bgamma_3)^2, \;\; E\left\{\dfrac{\partial^2 l}{\partial \sigma_2^2\partial {\rm vec}(\bgamma_{12})^\top}\right\}=0,\\
	E\left(\dfrac{\partial^2 l}{\partial\sigma_2^2\partial\bgamma_3^\top}\right)&=\left\{-m\sigma_2^2(\bgamma_3^{\top}\bD\bgamma_3) \bD\bgamma_3\right\}^\top,\;\;
	E\left\{\dfrac{\partial^2 l}{\partial\sigma_2^2\partial{\rm vec}(\bSigma_3)^\top}\right\}
	=-\dfrac{m}{2}(\bgamma_3^\top\bD)\otimes (\bgamma_3^\top\bD),
\end{align*}
and
\begin{align*}
	&E\left\{\dfrac{\partial^2 l}{\partial {\rm vec}(\bgamma_{12})\partial {\rm vec}(\bgamma_{12})^\top}\right\}=-n\bSigma_3^{-1}\otimes \bSigma_{\bX,1}-m\bD\otimes \bSigma_{\bX,1},\\
	&E\left\{\dfrac{\partial^2 l}{\partial {\rm vec}(\bgamma_{12})\partial\bgamma_3^\top}\right\}=-n\bSigma_3^{-1}\otimes (\bSigma_{\bX,1}\bbeta)-m\bD\otimes \bSigma_{\bX,1}\bbeta,\;\;
	E\left\{\dfrac{\partial^2 l}{\partial {\rm vec}(\bgamma_{12})\partial {\rm vec}(\bSigma_3)^\top}\right\}=0,
\end{align*}
and
\begin{align*}
	&E\left(\dfrac{\partial^2 l}{\partial \bgamma_3\partial \bgamma_3^\top}\right)=-nE(Y_i^2)\bSigma_3^{-1}-m(\bbeta^\top\bSigma_{\bX,1}\bbeta)\bD-m\sigma_2^4\bD\bgamma_3\bgamma_3^\top\bD-m\sigma_2^4\bgamma_3^\top\bD\bgamma_3\bD,\\
	&E\left\{\dfrac{\partial^2 l}{\partial \bgamma_3\partial{\rm vec}(\bSigma_3)^\top}\right\}=-m\sigma_2^2\{\bD\otimes (\bgamma_3^\top \bD)+(\bgamma_3^\top \bD)\otimes \bD \} /2,\\
	&E\left\{\dfrac{\partial^2 l}{\partial {\rm vec}(\bSigma_3)\partial {\rm vec}(\bSigma_3)^\top}\right\}=\bN_{d_2}\Big(-\dfrac{n}{2}\bSigma_3^{-1}\otimes \bSigma_{3}^{-1}-\dfrac{m}{2}\bD\otimes \bD\Big)\bN_{d_2}.
\end{align*}
Since the expected mixed second derivatives of the log-likelihood with respect to ${\rm vec}(\balpha)$ and the other parameters are all $\mathbf{0}$, $\balpha$ is orthogonal to the other parameters, and its CR lower bound is 
$$
(N\bSigma_{1}^{-1}\otimes \bSigma_{\bX})^{-1}=\dfrac{1}{N}\bSigma_{1}\otimes \bSigma_{\bX}^{-1}.
$$
For the remaining parameters, the Fisher information matrix is 
\begin{align*}
	&\bM=\\
	&\!\!\!\!\!\!\!\!\!\!\!\!\!\!\!\!\!\!\!\!\!\!\!\!\!\!\!
	\begin{bmatrix}
		\bSigma_{\bX,1}\left(\dfrac{n}{\sigma_2^2}+m\bgamma_3^\top \bD\bgamma_3\right) & \mathbf{0}_{(p+d_1)\times 1} & m (\bgamma_3^\top \bD)\otimes \bSigma_{\bX,1} & m\bSigma_{\bX,1}\bbeta\bgamma_3^\top \bD & \mathbf{0}_{(p+d_1)\times d_2^2}\\
		\mathbf{0}_{1\times(p+d_1)}  & \dfrac{n}{2\sigma_2^4}+\dfrac{m(\bgamma_3^\top \bD\bgamma_3)^2}{2} & \mathbf{0}_{1\times (p+d_1)d_2}  & m\sigma_2^2\bgamma_3^\top \bD\bgamma_3 \bgamma_3^\top \bD & \bM_{2,5}\\
		m(\bD\bgamma_3)\otimes \bSigma_{\bX,1} & \mathbf{0}_{(p+d_1)d_2\times 1}  & (n\bSigma_3^{-1}+m\bD)\otimes \bSigma_{\bX,1} & (n\bSigma_3^{-1}+m\bD)\otimes (\bSigma_{\bX,1}\bbeta) & \mathbf{0}_{(p+d_1)d_2\times d_2^2}\\
		m\bD\bgamma_3\bbeta^\top \bSigma_{\bX,1} & m\sigma_2^2\bgamma_3^\top \bD\bgamma_3\bD\bgamma_3 & (n\bSigma_3^{-1}+m\bD)\otimes (\bbeta^\top \bSigma_{\bX,1}) & \bM_{4,4} & \bM_{4,5}\\
		\mathbf{0}_{d_2^2\times (p+d_1)}  & \bM_{2,5}^\top & \mathbf{0}_{d_2^2\times (p+d_1)d_2}  & \bM_{4,5}^\top & \bM_{5,5}
	\end{bmatrix}
\end{align*}
where
	\begin{align*}
	\bM_{2,5}&=\dfrac{m(\bgamma_3^\top \bD)\otimes (\bgamma_3^\top \bD)}{2} \\
	\bM_{4,5}&=m\sigma_2^2 \dfrac{(\bD\bgamma_3)^\top \otimes \bD+\bD\otimes (\bD\bgamma_3)^\top}{2}\\
	\bM_{5,5}&=\bN_{d_2}\Big(\dfrac{n\bSigma_3^{-1}\otimes\bSigma_3^{-1}}{2}+\dfrac{m\bD\otimes \bD}{2}\Big)\bN_{d_2} \\
	\bM_{4,4}&=nE(Y_i^2)\bSigma_3^{-1}+m(\bbeta^\top\bSigma_{\bX,1}\bbeta) \bD+m\sigma_2^4\bD\bgamma_3\bgamma_3^\top \bD+m\sigma_2^4(\bgamma_3^\top \bD\bgamma_3)\bD.
\end{align*}
Let $q=p+d_1$ and $\mathbf P_M={\rm diag}(\bI_q,1,\bI_{qd_2},\bI_{d_2},\mathbf U_{d_2})$. We use $\bM_{\cal S}^{[-1]}=\mathbf P_M(\mathbf P_M^\top\bM\mathbf P_M)^{-1}\mathbf P_M^\top$ throughout the following block calculation, and similarly for each congruent information matrix $\bM_k$ (with the same parameter ordering). We compute $[\bM_{\cal S}^{[-1]}]_{1,1}$, $[\bM_{\cal S}^{[-1]}]_{3,3}$ and $[\bM_{\cal S}^{[-1]}]_{4,4}$. The compression changes only the coordinates of the symmetric covariance nuisance parameter; the first, third and fourth inverse blocks give the CR lower bounds for $\bbeta$, ${\rm vec}(\bgamma_{12})$, and $\bgamma_3$. Since $n>0$ and the labeled Gaussian model has nonsingular covariance and design matrices, the compressed information is positive definite. Let 
\begin{align*}
	\bL_1=
	\begin{bmatrix}
		\bI & 0 & 0 & 0 & 0\\
		0 & \bI & 0 & 0 & 0\\
		-\dfrac{m}{n/\sigma_2^2+m\bgamma_3^\top \bD\bgamma_3}(\bD\bgamma_3)\otimes \bI_{p+d_1} & 0 & \bI & 0 & 0\\
		-\dfrac{m\bD\bgamma_3\bbeta^\top}{n/\sigma_2^2+m\bgamma_3^\top \bD\bgamma_3} & 0 & 0 & \bI & 0\\
		0 & 0 & 0 & 0 & \bI
	\end{bmatrix}.
\end{align*}
We first have 
\begin{align*}
	\bM_1=\bL_1\bM\bL_1^\top=
	&\begin{bmatrix}
		\bSigma_{\bX,1}\left(\dfrac{n}{\sigma_2^2}+m\bgamma_3^\top \bD\bgamma_3\right) & 0 & 0 & 0 & 0\\
		0 & \dfrac{n}{2\sigma_2^4}+\dfrac{m(\bgamma_3^\top \bD\bgamma_3)^2}{2} & 0 & m\sigma_2^2\bgamma_3^\top \bD\bgamma_3 \bgamma_3^\top \bD & \bM_{2,5}\\
 		0 & 0 & \bM_{1,3,3} & \bM_{1,3,4} & 0\\
		0 & m\sigma_2^2\bgamma_3^\top \bD\bgamma_3\bD\bgamma_3 & \bM_{1,4,3} & \bM_{1,4,4} & \bM_{4,5}\\
		0 & \bM_{2,5}^\top & 0 & \bM_{4,5}^\top & \bM_{5,5}
	\end{bmatrix}
\end{align*}
where
\begin{align*}
	\bM_{1,3,3}&=(n\bSigma_3^{-1}+m\bD-\dfrac{m^2}{n/\sigma_2^2+m\bgamma_3^\top \bD\bgamma_3}\bD\bgamma_3\bgamma_3^\top \bD)\otimes \bSigma_{\bX,1},\\
	\bM_{1,3,4}&=(n\bSigma_3^{-1}+m\bD)\otimes \bSigma_{\bX,1}\bbeta-\dfrac{m^2}{n/\sigma_2^2+m\bgamma_3^\top \bD\bgamma_3}(\bD\bgamma_3)\otimes \bSigma_{\bX,1}\bbeta\bgamma_3^\top \bD,\\
	&=(n\bSigma_3^{-1}+m\bD)\otimes \bSigma_{\bX,1}\bbeta-\dfrac{m^2}{n/\sigma_2^2+m\bgamma_3^\top \bD\bgamma_3}(\bD\bgamma_3\bgamma_3^\top \bD)\otimes \bSigma_{\bX,1}\bbeta,\\
	\bM_{1,4,3}&=(n\bSigma_3^{-1}+m\bD)\otimes \bbeta^\top \bSigma_{\bX,1}-\dfrac{m^2}{n/\sigma_2^2+m\bgamma_3^\top \bD\bgamma_3}(\bgamma_3^\top \bD)\otimes \bD\bgamma_3\bbeta^\top\bSigma_{\bX,1},\\
	&=(n\bSigma_3^{-1}+m\bD)\otimes \bbeta^\top \bSigma_{\bX,1}-\dfrac{m^2}{n/\sigma_2^2+m\bgamma_3^\top \bD\bgamma_3}(\bD\bgamma_3\bgamma_3^\top \bD)\otimes \bbeta^\top\bSigma_{\bX,1},\\
	\bM_{1,4,4}&=\bM_{4,4}-\dfrac{m^2}{n/\sigma_2^2+m\bgamma_3^\top \bD\bgamma_3}(\bbeta^\top\bSigma_{\bX,1}\bbeta)\bD\bgamma_3\bgamma_3^\top \bD.
\end{align*}
Notice that for any vectors $\bu$ and $\bv$, it holds that
$$
(\bu\bu^\top)\otimes \bv^\top = \bu^\top \otimes (\bu\bv^\top)\;\; {\rm and }\;\; \bu\otimes \bv\bu^\top = (\bu\bu^\top)\otimes \bv,
$$
which gives the third and fifth equalities.

Let $\tilde{\bM}_{1,1}=\bSigma_{\bX,1}(n/\sigma_2^2+m\bgamma_3^\top \bD\bgamma_3)$, and
\begin{align*}
	\bL_2=\begin{bmatrix}
		\bI & 0 & 0 & 0 & 0\\
		0 & \bI & 0 & 0 & 0\\
		0 & 0 & \bI & 0 & 0\\
		0 & -\dfrac{m\sigma_2^2\bgamma_3^\top \bD\bgamma_3}{n/(2\sigma_2^4)+m(\bgamma_3^\top \bD\bgamma_3)^2/2}\bD\bgamma_3 & 0 & \bI & 0\\
		0 & -\dfrac{(m/2)\{(\bD\bgamma_3)\otimes (\bD\bgamma_3)\}}{n/(2\sigma_2^4)+m(\bgamma_3^\top \bD\bgamma_3)^2/2} & 0 & 0 & \bI 
	\end{bmatrix},
\end{align*} 
then it holds that
\begin{align*}
	&\bM_2=\bL_2\bM_1\bL_2^\top=
	\begin{bmatrix}
		\tilde{\bM}_{1,1} & 0 & 0 & 0 & 0\\
		0 & \dfrac{n}{2\sigma_2^4}+\dfrac{m(\bgamma_3^\top \bD\bgamma_3)^2}{2} & 0 & 0 & 0\\
		0 & 0 & \bM_{1,3,3} & \bM_{1,3,4} & 0\\
		0 & 0 & \bM_{1,4,3} & \bM_{2,4,4} & \bM_{2,4,5}\\
		0 & 0 & 0 & \bM_{2,5,4} & \bM_{2,5,5}
	\end{bmatrix}
\end{align*}
where
\begin{align*}
	\bM_{2,4,4}&=\bM_{1,4,4}-\dfrac{(m\sigma_2^2\bgamma_3^\top \bD\bgamma_3)^2}{n/(2\sigma_2^4)+m(\bgamma_3^\top \bD\bgamma_3)^2/2}\bD\bgamma_3\bgamma_3^\top \bD\\
	&=\bM_{4,4}-\dfrac{m^2}{n/\sigma_2^2+m\bgamma_3^\top \bD\bgamma_3}(\bbeta^\top\bSigma_{\bX,1}\bbeta)\bD\bgamma_3\bgamma_3^\top \bD-\dfrac{(m\sigma_2^2\bgamma_3^\top \bD\bgamma_3)^2}{n/(2\sigma_2^4)+m(\bgamma_3^\top \bD\bgamma_3)^2/2}\bD\bgamma_3\bgamma_3^\top \bD,\\
	\bM_{2,4,5}&=m\sigma_2^2 \dfrac{(\bD\bgamma_3)^\top\otimes \bD+\bD\otimes (\bD\bgamma_3)^\top}{2} -\dfrac{m^2\sigma_2^2\bgamma_3^\top \bD\bgamma_3/2}{n/(2\sigma_2^4)+m(\bgamma_3^\top \bD\bgamma_3)^2/2}(\bD\bgamma_3\bgamma_3^\top \bD)\otimes (\bgamma_3^\top \bD)\\
	\bM_{2,5,4}&=m\sigma_2^2\Big\{\dfrac{(\bD\bgamma_3)^\top\otimes \bD+\bD\otimes (\bD\bgamma_3)^\top}{2}\Big\}^\top-\dfrac{m^2\sigma_2^2\bgamma_3^\top \bD\bgamma_3/2}{n/(2\sigma_2^4)+m(\bgamma_3^\top \bD\bgamma_3)^2/2}  (\bD\bgamma_3\bgamma_3^\top \bD)\otimes (\bD\bgamma_3)\\
	\bM_{2,5,5}&=\bN_{d_2}\Big(\dfrac{n\bSigma_3^{-1}\otimes\bSigma_3^{-1}}{2}+\dfrac{m\bD\otimes \bD}{2}\Big)\bN_{d_2}-\dfrac{(m/2)^2}{n/(2\sigma_2^4)+m(\bgamma_3^\top \bD\bgamma_3)^2/2}\{(\bD\bgamma_3\bgamma_3^\top \bD)\otimes (\bD\bgamma_3\bgamma_3^\top \bD)\}.
\end{align*}

Let $\tilde{\bM}_{2,2}=n/(2\sigma_2^4)+m(\bgamma_3^\top \bD\bgamma_3)^2/2$, it follows that 
\begin{align*}
	&\bL_3=
	\begin{bmatrix}
		\bI & 0 & 0 & 0 & 0 \\
		0 & \bI & 0 & 0 & 0 \\
		0 & 0 & \bI & 0 & 0 \\
		0 & 0 & -\bI_{d_2}\otimes \bbeta^\top & \bI & 0\\
		0 & 0 & 0 & 0 & \bI   
	\end{bmatrix}, \;\;	\bM_3=\bL_3\bM_2\bL_3^\top=
\begin{bmatrix}
\tilde{\bM}_{1,1} & 0 & 0 & 0 & 0\\
0 & \tilde{\bM}_{2,2} & 0 & 0 & 0\\
0 & 0 & \bM_{1,3,3} & 0 & 0\\
0 & 0 & 0 & \bM_{3,4,4} & \bM_{2,4,5}\\
0 & 0 & 0 & \bM_{2,5,4} & \bM_{2,5,5}
\end{bmatrix}.
\end{align*}
where
\begin{align*}
	\bM_{3,4,4}&=\bM_{2,4,4}-(n\bSigma_3^{-1}+m\bD) \bbeta^\top\bSigma_{\bX,1}\bbeta+\dfrac{m^2}{n/\sigma_2^2+m\bgamma_3^\top \bD\bgamma_3}(\bD\bgamma_3\bgamma_3^\top \bD)\bbeta^\top\bSigma_{\bX,1}\bbeta\\
	&=nE(Y_i^2)\bSigma_3^{-1}+m\bbeta^\top\bSigma_{\bX,1}\bbeta \bD+m\sigma_2^4\bD\bgamma_3\bgamma_3^\top \bD+m\sigma_2^4\bgamma_3^\top \bD\bgamma_3\bD\\
	&\qquad -\dfrac{m^2}{n/\sigma_2^2+m\bgamma_3^\top \bD\bgamma_3}\bD\bgamma_3\bbeta^\top\bSigma_{\bX,1}\bbeta\bgamma_3^\top \bD-\dfrac{(m\sigma_2^2\bgamma_3^\top \bD\bgamma_3)^2}{n/(2\sigma_2^4)+m(\bgamma_3^\top \bD\bgamma_3)^2/2}\bD\bgamma_3\bgamma_3^\top \bD\\
	&\qquad-(n\bSigma_3^{-1}+m\bD) (\bbeta^\top\bSigma_{\bX,1}\bbeta)+\dfrac{m^2}{n/\sigma_2^2+m\bgamma_3^\top \bD\bgamma_3}(\bD\bgamma_3\bgamma_3^\top \bD)(\bbeta^\top\bSigma_{\bX,1}\bbeta)\\
	&=n\sigma_2^2\bSigma_{3}^{-1}+m\sigma_2^4\bD\bgamma_3\bgamma_3^\top \bD+m\sigma_2^4\bgamma_3^\top \bD\bgamma_3\bD-\dfrac{(m\sigma_2^2\bgamma_3^\top \bD\bgamma_3)^2}{n/(2\sigma_2^4)+m(\bgamma_3^\top \bD\bgamma_3)^2/2}\bD\bgamma_3\bgamma_3^\top \bD,
\end{align*}
We next compute the inverse of the following $2\times 2$ block matrix $[\bM_{3,4,4},\bM_{2,4,5};\bM_{2,5,4},\bM_{2,5,5}]$. We first simplify the expressions, denote $t=\bgamma_3^\top\bSigma_3^{-1}\bgamma_3$, by Sherman-Morrison Formula,
\begin{align*}
	\bD&=\bSigma_3^{-1}-\dfrac{\sigma_2^2}{1+\sigma_2^2t}\bSigma_3^{-1}\bgamma_3\bgamma_3^\top\bSigma_3^{-1},\;\;
	\bD\bgamma_3=\dfrac{1}{1+\sigma_2^2t}\bSigma_3^{-1}\bgamma_3,\;\;\;\; \bgamma_3^\top \bD\bgamma_3=\dfrac{t}{1+\sigma_2^2t},
\end{align*}
We can rewrite the above expressions as 
\begin{align*}
	\bM_{3,4,4}&=\Big(n\sigma_2^2+\dfrac{m\sigma_2^4t}{1+\sigma_2^2t}\Big)\bSigma_3^{-1}+\dfrac{m\sigma_2^4}{(1+\sigma_2^2t)}\dfrac{n-(n+m)\sigma_2^4t^2}{n(1+\sigma_2^2t)^2+mt^2\sigma_2^4}\bSigma_3^{-1}\bgamma_3\bgamma_3^\top\bSigma_3^{-1},\\
	\bM_{2,4,5}&=\dfrac{m\sigma_2^2}{2}\{(\bD\bgamma_3)^\top\otimes \bD+\bD\otimes (\bD\bgamma_3)^\top\}\\ &\qquad-\dfrac{m^2\sigma_2^2\bgamma_3^\top \bD\bgamma_3/2}{n/(\sigma_2^4)+m(\bgamma_3^\top \bD\bgamma_3)^2}\{(\bD\bgamma_3\bgamma_3^\top \bD)\otimes (\bgamma_3^\top \bD)+(\bgamma_3^\top \bD)\otimes (\bD\bgamma_3\bgamma_3^\top \bD)\}\\
	&=\dfrac{m\sigma_2^2}{2(1+\sigma_2^2t)}\{(\bSigma_3^{-1}\bgamma_3)^\top\otimes \bSigma_3^{-1}+\bSigma_3^{-1}\otimes (\bSigma_3^{-1}\bgamma_3)^\top\}\\
	&\qquad-\dfrac{m\sigma_2^4}{2(1+\sigma_2^2t)^2}\Big\{1+\dfrac{m\sigma_2^2t}{n(1+\sigma_2^2t)^2+m\sigma_2^4t^2}\Big\}\\
	&\qquad\qquad\times \{(\bSigma_3^{-1}\bgamma_3\bgamma_3^\top \bSigma_3^{-1})\otimes (\bgamma_3^\top \bSigma_3^{-1})+(\bgamma_3^\top \bSigma_3^{-1})\otimes (\bSigma_3^{-1}\bgamma_3\bgamma_3^\top \bSigma_3^{-1})\},
	\end{align*}
	$\bM_{2,5,4}=\bM_{2,4,5}^\top$, and
	\begin{align*}
	\bM_{2,5,5}&=\bN_{d_2}\Big[\dfrac{n+m}{2}\bSigma_3^{-1}\otimes \bSigma_3^{-1}-\dfrac{m\sigma_2^2}{2(1+\sigma_2^2t)}\{\bSigma_3^{-1}\otimes (\bSigma_3^{-1}\bgamma_3\bgamma_3^\top \bSigma_3^{-1})+(\bSigma_3^{-1}\bgamma_3\bgamma_3^\top \bSigma_3^{-1})\otimes \bSigma_3^{-1}\}\\
	&\qquad\qquad+\dfrac{m}{2}\Big(\dfrac{\sigma_2^2}{1+\sigma_2^2t}\Big)^2\{ (\bSigma_3^{-1}\bgamma_3\bgamma_3^\top \bSigma_3^{-1})\otimes (\bSigma_3^{-1}\bgamma_3\bgamma_3^\top \bSigma_3^{-1})\}\Big] \bN_{d_2}\\
	&\qquad-\dfrac{m^2\sigma_2^4}{2(1+\sigma_2^2t)^2\{n(1+\sigma_2^2t)^2+m\sigma_2^4t^2\}}\{ (\bSigma_3^{-1}\bgamma_3\bgamma_3^\top \bSigma_3^{-1})\otimes (\bSigma_3^{-1}\bgamma_3\bgamma_3^\top \bSigma_3^{-1})\},
\end{align*}
Denote $\tilde{\bM}_{3,3}=\bM_{1,3,3}$. Then, by Lemma \ref{lemma:matrix_inverse} and substituting the specific coefficients, that is,  
\begin{align*}
a_1&=n\sigma_2^2+\dfrac{m\sigma_2^4t}{1+\sigma_2^2t},\;\; b_1=\dfrac{m\sigma_2^4}{(1+\sigma_2^2t)}\dfrac{n-(n+m)\sigma_2^4t^2}{n(1+\sigma_2^2t)^2+mt^2\sigma_2^4},\;\; a_2=\dfrac{m\sigma_2^2}{2(1+\sigma_2^2t)},\\
b_2&=-\dfrac{m\sigma_2^4}{2(1+\sigma_2^2t)^2}\Big\{1+\dfrac{m\sigma_2^2t}{n(1+\sigma_2^2t)^2+m\sigma_2^4t^2}\Big\},\;\; a_3=\dfrac{n+m}{2},\;\; b_3=-\dfrac{m\sigma_2^2}{2(1+\sigma_2^2t)},\\
c_{31}&=\dfrac{m}{2}\Big(\dfrac{\sigma_2^2}{1+\sigma_2^2t}\Big)^2,\;\;{\rm and}\;\; c_{32}=-\dfrac{m^2\sigma_2^4}{2(1+\sigma_2^2t)^2\{n(1+\sigma_2^2t)^2+m\sigma_2^4t^2\}}.
\end{align*}
We can compute 
	\begin{align*}
	\begin{bmatrix}
		\bM_{3,4,4} & \bM_{2,4,5}\\
		\bM_{2,5,4} & \bM_{2,5,5}
	\end{bmatrix}_{\mathcal S}^{[-1]}
=\begin{bmatrix}
\bV_{\bgamma,1}^{\rm low} &*\\
 * & *
\end{bmatrix},
\end{align*}
where $*$ represents a collection of unspecified terms whose explicit forms may vary across different positions, but none of them affects the final result. For brevity, they are all denoted by the same symbol, and 
\begin{align*}
\bV_{\bgamma,1}^{\rm low}=\dfrac{1}{n\sigma_2^2}\left(1-\dfrac{m}{N}\dfrac{\sigma_2^2t}{1+\sigma_2^2t}\right)\bSigma_3-\dfrac{m}{nN}\dfrac{1-\sigma_2^2t}{(1+\sigma_2^2t)^2}\bgamma_3\bgamma_3^\top.
\end{align*}
Finally, we obtain $\bM_3=\bL\bM\bL^\top$, where
\begin{align*}
	&\bL=\bL_3\bL_2\bL_1
	=
	\begin{bmatrix}
		\bI & 0 & 0 & 0 & 0\\
		0 & \bI & 0 & 0 & 0\\
		-\dfrac{m(\bD\bgamma_3)\otimes \bI_{p+d_1}}{n/\sigma_2^2+m\bgamma_3^\top \bD\bgamma_3} & 0 & \bI & 0 & 0\\
		0 & -\dfrac{m\sigma_2^2(\bgamma_3^\top \bD\bgamma_3)\bD\bgamma_3}{n/(2\sigma_2^4)+m(\bgamma_3^\top \bD\bgamma_3)^2/2} & -\bI_{d_2}\otimes \bbeta^\top & \bI & 0\\
		0 & -\dfrac{(m/2)(\bD\bgamma_3)\otimes (\bD\bgamma_3)}{n/(2\sigma_2^4)+m(\bgamma_3^\top \bD\bgamma_3)^2/2} & 0 & 0 & \bI 
	\end{bmatrix}.
\end{align*}
Therefore, it follows that $\bM_{\cal S}^{[-1]}=\bL^\top \bM_{3,\cal S}^{[-1]}\bL$, we first compute
 
\begin{align*}
	\bL^\top \bM_{3,\cal S}^{[-1]}&=\bL^\top
	\times \begin{bmatrix}
		\tilde{\bM}_{1,1}^{-1} & 0 & 0 & 0 & 0\\
		0 & \tilde{\bM}_{2,2}^{-1} & 0 & 0 & 0\\
		0 & 0 & \tilde{\bM}_{3,3}^{-1} & 0 & 0\\
		0 & 0 & 0 & \bV_{\bgamma,1}^{\rm low} & *\\
		0 & 0 & 0 & * & *
	\end{bmatrix}\\
&=\begin{bmatrix}
		\tilde{\bM}_{1,1}^{-1} & 0 & -\dfrac{m(\bgamma_3^\top \bD)\otimes \bI_{p+d_1}}{n/\sigma_2^2+m\bgamma_3^\top \bD\bgamma_3}\tilde{\bM}_{3,3}^{-1} & 0 & 0\\
 		* & * & * & * & *\\
		0 & 0 & \tilde{\bM}_{3,3}^{-1} & (-\bI_{d_2}\otimes \bbeta)\bV_{\bgamma,1}^{\rm low} & *\\
		0 & 0 & 0 & \bV_{\bgamma,1}^{\rm low} & *\\
		* & * & * & * & *
	\end{bmatrix},
\end{align*}
and
\begin{align*}
	\bM_{\cal S}^{[-1]}=&\bL^\top \bM_{3,\cal S}^{[-1]}\bL
	=\begin{bmatrix}
	\bV_{\bbeta}^{\rm low} & * & * & * & *\\
		* & * & * & * & *\\
		* & * & \tilde{\bM}_{3,3}^{-1}+(\bI_{d_2}\otimes \bbeta)\bV_{\bgamma,1}^{\rm low}(\bI_{d_2}\otimes \bbeta^\top) & -\bV_{\bgamma,1}^{\rm low}\otimes \bbeta & *\\
 		* & * & -\bV_{\bgamma,1}^{\rm low}\otimes \bbeta^\top & \bV_{\bgamma,1}^{\rm low} & *\\
		* & * & * & * & *
	\end{bmatrix}
\end{align*}
where the CR lower bound for $\bbeta$ is 
\begin{align*}
\bV_{\bbeta}^{\rm low}&=\tilde{\bM}_{1,1}^{-1}+\dfrac{m(\bgamma_3^\top \bD)\otimes \bI_{p+d_1}}{n/\sigma_2^2+m\bgamma_3^\top \bD\bgamma_3}\tilde{\bM}_{3,3}^{-1}\dfrac{m(\bD\bgamma_3)\otimes \bI_{p+d_1}}{n/\sigma_2^2+m\bgamma_3^\top \bD\bgamma_3}\\ &=\bSigma_{\bX,1}^{-1}(n/\sigma_2^2+m\bgamma_3^\top \bD\bgamma_3)^{-1}\\
&\qquad+\dfrac{m(\bgamma_3^\top \bD)\otimes \bI_{p+d_1}}{n/\sigma_2^2+m\bgamma_3^\top \bD\bgamma_3}(n\bSigma_3^{-1}+m\bD-\dfrac{m^2\bD\bgamma_3\bgamma_3^\top \bD}{n/\sigma_2^2+m\bgamma_3^\top \bD\bgamma_3})^{-1}\otimes\bSigma_{\bX,1}^{-1}\dfrac{m(\bD\bgamma_3)\otimes \bI_{p+d_1}}{n/\sigma_2^2+m\bgamma_3^\top \bD\bgamma_3}\\ &= \bSigma_{\bX,1}^{-1}(n/\sigma_2^2+m\bgamma_3^\top \bD\bgamma_3)^{-1}\\
&\qquad+\dfrac{m^2}{(n/\sigma_2^2+m\bgamma_3^\top \bD\bgamma_3)^2}\times \bgamma_3^\top \bD(n\bSigma_3^{-1}+m\bD-\dfrac{m^2\bD\bgamma_3\bgamma_3^\top \bD}{n/\sigma_2^2+m\bgamma_3^\top \bD\bgamma_3})^{-1} \bD\bgamma_3 \bSigma_{\bX,1}^{-1}\\
  &= \bSigma_{\bX,1}^{-1}\dfrac{\sigma_2^2(n+m-m\sigma_2^2\bgamma_3^\top \bD\bgamma_3)}{n(n+m)}\\
&= \bSigma_{\bX,1}^{-1}\dfrac{\sigma_2^2}{n}\left\{1-\dfrac{m\sigma_2^2\bgamma_3^\top(\bSigma_3+\sigma_2^2\bgamma_3\bgamma_3^\top)^{-1} \bgamma_3}{n+m}\right\}.
\end{align*}
Then we can simplify the following expression
\begin{align*}
&\tilde{\bM}_{3,3}^{-1}+(\bI_{d_2}\otimes \bbeta)\bV_{\bgamma,1}^{\rm low}(\bI_{d_2}\otimes \bbeta^\top)\\
=\ &(n\bSigma_3^{-1}+m\bD-\dfrac{m^2}{n/\sigma_2^2+m\bgamma_3^\top \bD\bgamma_3}\bD\bgamma_3\bgamma_3^\top \bD)^{-1}\otimes \bSigma_{\bX,1}^{-1}+\bV_{\bgamma,1}^{\rm low}\otimes(\bbeta\bbeta^\top)\\
=\ &\dfrac{1}{m+n}\left[\bSigma_3+\dfrac{m}{n}\times \dfrac{\sigma_2^2}{1+\sigma_2^2 \bgamma_3^{\top}\bSigma_{3}^{-1}\bgamma_3}\bgamma_3\bgamma_3^{\top}\right]\otimes \bSigma_{\bX,1}^{-1}+\bV_{\bgamma,1}^{\rm low}\otimes(\bbeta\bbeta^\top):= \bV_{\bgamma,2}^{\rm low}+\bV_{\bgamma,1}^{\rm low}\otimes(\bbeta\bbeta^\top),
 \end{align*}
then the CR lower bound for $({\rm vec}(\bgamma_{12})^\top,\bgamma_3^\top)^\top$ is
\begin{align*}
\begin{bmatrix}
\bV_{\bgamma,2}^{\rm low}+\bV_{\bgamma,1}^{\rm low}\otimes(\bbeta\bbeta^\top) & -\bV_{\bgamma,1}^{\rm low}\otimes \bbeta\\
-\bV_{\bgamma,1}^{\rm low}\otimes \bbeta^\top & \bV_{\bgamma,1}^{\rm low}
 \end{bmatrix}.
\end{align*}
We now consider the lower bound for estimating $\btheta^*$. Let $\boldsymbol{\eta}=(\bbeta^{\top},\text{vec}(\balpha)^{\top})^\top$, $g(\boldsymbol{\eta})=\btheta=\bbeta_1+\balpha \bbeta_2$, its Jacobian with regard to $\bbeta$ and $\balpha$ is 
\[G_{\bbeta}=[\bI_p, \balpha^*],\quad G_{\balpha}=\bbeta_2^{*\top}\otimes \bI_p\]
respectively, which is because
\[d\btheta=[\bI_p,\balpha^{*}]d\bbeta+(d\balpha)\bbeta_2^{*}=[\bI_p,\balpha^{*}]d\bbeta+(\bbeta_2^{*\top}\otimes \bI_p)d\text{vec}(\balpha).\]
The score for the marginal distribution of $(\bX_i,\bS_{1i})$ is orthogonal to the scores of the subsequent conditional likelihood factors, since the latter have conditional mean zero. Thus the inverse-information cross-block between ${\rm vec}(\balpha)$ and $\bbeta$ vanishes, and the delta method gives 
 \[\text{CRLB}_{\btheta}=G_{\bbeta}\text{CRLB}_{\bbeta}G_{\bbeta}^{\top}+G_{\balpha}\text{CRLB}_{\text{vec}(\balpha)}G_{\balpha}^{\top}.\]
According to former proofs, we know 
\[\text{CRLB}_{\text{vec}({\balpha})}=\frac{1}{N}(\bSigma_{1}\otimes \bSigma_{\bX}^{-1}),\quad nG_{\bbeta}\text{CRLB}_{\bbeta}G_{\bbeta}^{\top}=\sigma_2^2\Big(1-\frac{m}{N} \frac{\sigma_2^2\bgamma_3^{*\top}\bSigma_3^{-1}\bgamma_3^{*}}{1+\sigma_2^2\bgamma_3^{*\top}\bSigma_{3}^{-1}\bgamma_3^{*}}\Big)\bSigma_{\bX}^{-1},\]
and 
\[nG_{\balpha}\text{CRLB}_{\text{vec}(\balpha)}G_{\balpha}^{\top}=\frac{n}{N}(\bbeta_2^{*\top}\otimes \bI_p)(\bSigma_1\otimes \bSigma_{\bX}^{-1})(\bbeta_2^*\otimes \bI_p)=\frac{n}{N}(\bbeta_2^{*\top}\bSigma_1 \bbeta_2^*)\bSigma_{\bX}^{-1}.\]
Therefore, we finally derive 
\[n\text{CRLB}_{\btheta}=\bigg\{\sigma_2^2 \bigg(1-\frac{m}{N}\frac{\sigma_2^2\bgamma_3^{*\top}\bSigma_3^{-1}\bgamma_3^*}{1+\sigma_2^2\bgamma_3^{*\top}\bSigma_{3}^{-1}\bgamma_3^{*}}\bigg)+\frac{n}{N}\bbeta_2^{*\top}\bSigma_{1}\bbeta_2^*\bigg\}\bSigma_{\bX}^{-1},\]
which gives the proof.
\end{proof}

\begin{lemma}[Gaussian closure for a working surrogate partition]
\label{lem:working_gaussian}
Suppose Assumption~\ref{ass:joint_gaussian} holds. For any fixed surrogate-selection scheme, the no-intercept sequential population projections in \eqref{eq:models_mis} coincide with the corresponding Gaussian conditional means, and the resulting residuals $\bvarepsilon_{1i}^\prime$, $\varepsilon_{2i}^\prime$, and $\bvarepsilon_{3i}^\prime$ are mutually independent Gaussian innovations, with positive (scalar or matrix) variances for nonempty blocks.
\end{lemma}
\begin{proof}
We prove the result for nonempty pre- and post-outcome surrogate blocks. The cases $d_1'=0$ or $d_2'=0$ follow by deleting the corresponding blocks and statements.

For a fixed selection scheme ${\rm p}$, write
\[
    \bA_i=\bS_{1i}^{\prime},
    \qquad
    \bB_i=\bS_{2i}^{\prime},
\]
and define
\[
    \bU_i
    =
    (\bA_i^{\top},Y_i,\bB_i^{\top})^{\top}.
\]
Because $\bA_i$ and $\bB_i$ are obtained by selecting and reordering coordinates of
\[
    \bS_i=(\bS_{1i}^{\top},\bS_{2i}^{\top})^{\top},
\]
there exists a row-selection matrix $\bR_{\rm p}$ such that
\[
    \bU_i=\bR_{\rm p}\bW_i,
	\qquad
    \bW_i=(\bS_{1i}^{\top},Y_i,\bS_{2i}^{\top})^{\top}.
 \]
Let $\bSigma_W={\rm Var}(\bW_i\mid\bX_i)$. By \eqref{eq:models}, $E(\bW_i\mid\bX_i=\bx)=\bA_W^{\top}\bx$ with $\bA_W=\bSigma_{\bX}^{-1}E(\bX_i\bW_i^{\top})$. Under Assumption~\ref{ass:joint_gaussian}, Gaussian conditioning therefore implies
\[
    \bU_i\mid\bX_i=\bx
    \sim
    \mathcal{N}
    \left(
        \bR_{\rm p}\bA_W^{\top}\bx,\,
        \bSigma_{\rm p}
    \right),
    \qquad
    \bSigma_{\rm p}
    =
    \bR_{\rm p}\bSigma_W\bR_{\rm p}^{\top}.
\]
Since the rows selected by $\bR_{\rm p}$ are distinct and
$\bSigma_W$ is positive definite, $\bSigma_{\rm p}$ is also positive
definite.

Write the conditional mean as
\[
    E(\bU_i\mid\bX_i=\bx)
    =
    \begin{pmatrix}
        \bmu_A(\bx)\\
        \mu_Y(\bx)\\
        \bmu_B(\bx)
    \end{pmatrix},
\]
where each component is linear in $\bx$, and partition the
conditional covariance matrix as
\[
    \bSigma_{\rm p}
    =
    \begin{pmatrix}
        \bSigma_{AA}
        &
        \bSigma_{AY}
        &
        \bSigma_{AB}
        \\
        \bSigma_{YA}
        &
        \sigma_{YY}
        &
        \bSigma_{YB}
        \\
        \bSigma_{BA}
        &
        \bSigma_{BY}
        &
        \bSigma_{BB}
    \end{pmatrix}.
\]

First, since $\bmu_A(\bX_i)=E(\bA_i\mid\bX_i)$ is linear in
$\bX_i$, it coincides with the population least-squares projection
of $\bA_i$ onto $\bX_i$. Hence the first residual in
Model~\eqref{eq:models_mis} satisfies
\[
    \bvarepsilon_{1i}^{\prime}
    =
    \bA_i-E(\bA_i\mid\bX_i).
\]
It follows that
\[
    \bvarepsilon_{1i}^{\prime}\mid\bX_i
    \sim
    \mathcal{N}
    \left(
        \bzero,\bSigma_{AA}
    \right).
\]
Because this conditional distribution does not depend on
$\bX_i$,
\[
    \bvarepsilon_{1i}^{\prime}\ind\bX_i.
\]
Thus, we may take
\[
    \bSigma_1^{\prime}=\bSigma_{AA}.
\]

Second, the conditional-distribution formula for a multivariate
Gaussian vector gives
\[
\begin{split}
    E(Y_i\mid\bX_i,\bA_i)
    =
    \mu_Y(\bX_i)
    +
    \bSigma_{YA}\bSigma_{AA}^{-1}
    \left\{
        \bA_i-\bmu_A(\bX_i)
    \right\}.
\end{split}
\]
The right-hand side is linear in $(\bX_i,\bA_i)$ and therefore
coincides with the population least-squares projection of $Y_i$
onto $(\bX_i,\bA_i)$. Consequently,
\[
    \varepsilon_{2i}^{\prime}
    =
    Y_i-E(Y_i\mid\bX_i,\bA_i).
\]
Moreover,
\[
    \varepsilon_{2i}^{\prime}
    \mid
    (\bX_i,\bA_i)
    \sim
    \mathcal{N}
    \left(
        0,\sigma_2^{\prime 2}
    \right),
\]
where
\[
    \sigma_2^{\prime 2}
    =
    \sigma_{YY}
    -
    \bSigma_{YA}
    \bSigma_{AA}^{-1}
    \bSigma_{AY}.
\]
Since $\bSigma_{\rm p}$ is positive definite,
$\sigma_2^{\prime 2}>0$. Because this conditional distribution
does not depend on $(\bX_i,\bA_i)$,
\[
    \varepsilon_{2i}^{\prime}
    \ind
    (\bX_i,\bA_i).
\]
In particular,
$\bvarepsilon_{1i}^{\prime}$ is measurable with respect to
$(\bX_i,\bA_i)$, and hence
\[
    \varepsilon_{2i}^{\prime}
    \ind
    \bvarepsilon_{1i}^{\prime}.
\]

Finally, let
\[
    \bC_i=(\bA_i^{\top},Y_i)^{\top},
\]
and define
\[
    \bSigma_{CC}
    =
    \begin{pmatrix}
        \bSigma_{AA} & \bSigma_{AY}\\
        \bSigma_{YA} & \sigma_{YY}
    \end{pmatrix},
    \qquad
    \bSigma_{BC}
    =
    \begin{pmatrix}
        \bSigma_{BA} & \bSigma_{BY}
    \end{pmatrix}.
\]
Again by the Gaussian conditional-distribution formula,
\[
\begin{split}
    E(\bB_i\mid\bX_i,\bA_i,Y_i)
    &=
    \bmu_B(\bX_i)
    +
    \bSigma_{BC}\bSigma_{CC}^{-1}
    \left\{
        \bC_i-E(\bC_i\mid\bX_i)
    \right\}.
\end{split}
\]
This is linear in $(\bX_i,\bA_i,Y_i)$ and therefore coincides
with the population least-squares projection of $\bB_i$ onto
$(\bX_i,\bA_i,Y_i)$. Thus,
\[
    \bvarepsilon_{3i}^{\prime}
    =
    \bB_i
    -
    E(\bB_i\mid\bX_i,\bA_i,Y_i).
\]
Furthermore,
\[
    \bvarepsilon_{3i}^{\prime}
    \mid
    (\bX_i,\bA_i,Y_i)
    \sim
    \mathcal{N}
    \left(
        \bzero,\bSigma_3^{\prime}
    \right),
\]
where
\[
    \bSigma_3^{\prime}
    =
    \bSigma_{BB}
    -
    \bSigma_{BC}
    \bSigma_{CC}^{-1}
    \bSigma_{BC}^{\top}.
\]
Because $\bSigma_{\rm p}$ is positive definite, the Schur complement
$\bSigma_3^{\prime}$ is positive definite. The conditional
distribution above does not depend on
$(\bX_i,\bA_i,Y_i)$, and therefore
\[
    \bvarepsilon_{3i}^{\prime}
    \ind
    (\bX_i,\bA_i,Y_i).
\]

Both $\bvarepsilon_{1i}^{\prime}$ and
$\varepsilon_{2i}^{\prime}$ are measurable functions of
$(\bX_i,\bA_i,Y_i)$. Hence
\[
    \bvarepsilon_{3i}^{\prime}
    \ind
    \left(
        \bvarepsilon_{1i}^{\prime},
        \varepsilon_{2i}^{\prime}
    \right).
\]
Together with
\[
    \varepsilon_{2i}^{\prime}
    \ind
    \bvarepsilon_{1i}^{\prime},
\]
this proves that the three induced residuals are mutually
independent Gaussian innovations.
\end{proof}

\begin{proof}[Proof of Theorem \ref{thm:thm_beta1}]
	
By Lemma~\ref{lem:working_gaussian}, for any fixed surrogate-selection scheme ${\rm p}$, the induced Gaussian working model satisfies the conditions required for Theorem~\ref{thm:upper_bound}.
Applying Theorem~\ref{thm:upper_bound} under the sampling regime specified in Section~\ref{sec:upper_bound} of the main text, including $\tau=0$, yields
	\begin{equation}\label{eq:beta1_working_an}
		\sqrt{n}
		\left(
		\widehat{\bbeta}_1^{\rm p}
		-\bar{\bbeta}_1^{\rm p}
		\right)
		\xrightarrow{\rm d}
		\mathcal{N}
		\left(
		\mathbf{0},
		\bV_{\bbeta_1}^{\rm p}
		\right).
	\end{equation}
	It remains to characterize the bias term
	$\bar{\bbeta}_1^{\rm p}-\bbeta_1^*$. Notice that
	\begin{align*}
		\bar{\bbeta}_1^{\rm p}-\bbeta_1^*
		={}&
		(\bI_p,\mathbf{0}_{p\times d_1^\prime})
		\left\{
		E(\bar{\bX}_{1i}^\prime
		\bar{\bX}_{1i}^{\prime\top})
		\right\}^{-1}
		E(\bar{\bX}_{1i}^\prime Y_i)
		-\bbeta_1^*\\
		={}&
		(\bI_p,\mathbf{0}_{p\times d_1^\prime})
		\left\{
		E(\bar{\bX}_{1i}^\prime
		\bar{\bX}_{1i}^{\prime\top})
		\right\}^{-1}
		E\left[
		\bar{\bX}_{1i}^\prime
		\left(
		\bbeta^{*\top}\bar{\bX}_{1i}
		+\varepsilon_{2i}
		\right)
		\right]
		-\bbeta_1^*\\
		={}&
		(\bI_p,\mathbf{0}_{p\times d_1^\prime})
		\begin{pmatrix}
			E(\bX_i\bX_i^\top)
			&
			E(\bX_i\bS_{1i}^{\prime\top})\\
			E(\bS_{1i}^\prime\bX_i^\top)
			&
			E(\bS_{1i}^\prime\bS_{1i}^{\prime\top})
		\end{pmatrix}^{-1}\\
		&\quad\times
		\begin{pmatrix}
			E(\bX_i\bX_i^\top\bbeta_1^*)
			+
			E(\bX_i\bS_{1i}^\top\bbeta_2^*)
			+
			E(\bX_i\varepsilon_{2i})\\
			E(\bS_{1i}^\prime\bX_i^\top\bbeta_1^*)
			+
			E(\bS_{1i}^\prime\bS_{1i}^\top\bbeta_2^*)
			+
			E(\bS_{1i}^\prime\varepsilon_{2i})
		\end{pmatrix}
		-\bbeta_1^*.
	\end{align*}
	Write $\bP_1=(\bP_{11},\bP_{12})$ with $\bP_{11}\in\mathbb R^{d_1^\prime\times d_1}$ and $\bP_{12}\in\mathbb R^{d_1^\prime\times d_2}$, so that $\bS_{1i}^\prime=\bP_{11}\bS_{1i}+\bP_{12}\bS_{2i}$. By Lemma~\ref{lemma:lemma1}, $E(\bX_i\varepsilon_{2i})=\mathbf{0}$, and
	\begin{align}
		E(\bS_{1i}^\prime\varepsilon_{2i})
		&=
		\bP_{11}E(\bS_{1i}\varepsilon_{2i})
		+
		\bP_{12}E(\bS_{2i}\varepsilon_{2i})
		=
		\bP_{12}\bgamma_3^*\sigma_2^2.
		\label{eq:S1prime_epsilon2}
	\end{align}
	Moreover, write
	\begin{equation}\label{eq:block_inverse_beta1}
		(\bI_p,\mathbf{0}_{p\times d_1^\prime})
		\begin{pmatrix}
			E(\bX_i\bX_i^\top)
			&
			E(\bX_i\bS_{1i}^{\prime\top})\\
			E(\bS_{1i}^\prime\bX_i^\top)
			&
			E(\bS_{1i}^\prime\bS_{1i}^{\prime\top})
		\end{pmatrix}^{-1}
		=
		(\bG_1,\bG_2),
	\end{equation}
	where $\bG_1$ equals $\bG$ in Theorem~\ref{thm:thm_beta1} and
	\begin{align*}
		\bG_1
		&=
		\left\{
		E(\bX_i\bX_i^\top)
		-
		E(\bX_i\bS_{1i}^{\prime\top})
		E(\bS_{1i}^\prime\bS_{1i}^{\prime\top})^{-1}
		E(\bS_{1i}^\prime\bX_i^\top)
		\right\}^{-1},\\
		\bG_2
		&=
		-\bG_1
		E(\bX_i\bS_{1i}^{\prime\top})
		E(\bS_{1i}^\prime\bS_{1i}^{\prime\top})^{-1}.
	\end{align*}
	Combining the above expressions, we obtain
	\begin{align*}
		\bar{\bbeta}_1^{\rm p}-\bbeta_1^*
		={}&
		\bG_1
		\left\{
		E(\bX_i\bX_i^\top\bbeta_1^*)
		+
		E(\bX_i\bS_{1i}^\top\bbeta_2^*)
		\right\}\\
		&+
		\bG_2
		\left\{
		E(\bS_{1i}^\prime\bX_i^\top\bbeta_1^*)
		+
		E(\bS_{1i}^\prime\bS_{1i}^\top\bbeta_2^*)
		+
		E(\bS_{1i}^\prime\varepsilon_{2i})
		\right\}
		-\bbeta_1^*\\
		={}&
		\left\{
		\bG_1E(\bX_i\bX_i^\top)
		+
		\bG_2E(\bS_{1i}^\prime\bX_i^\top)
		\right\}\bbeta_1^*\\
		&+
		\left\{
		\bG_1E(\bX_i\bS_{1i}^\top)
		+
		\bG_2E(\bS_{1i}^\prime\bS_{1i}^\top)
		\right\}\bbeta_2^*\\
		&+
		\bG_2E(\bS_{1i}^\prime\varepsilon_{2i})
		-\bbeta_1^*\\
		={}&
		\left\{
		\bG_1E(\bX_i\bS_{1i}^\top)
		+
		\bG_2E(\bS_{1i}^\prime\bS_{1i}^\top)
		\right\}\bbeta_2^*
		+
		\bG_2E(\bS_{1i}^\prime\varepsilon_{2i}),
	\end{align*}
	where the last equality follows from
	\begin{align*}
		\bG_1E(\bX_i\bX_i^\top)
		+
		\bG_2E(\bS_{1i}^\prime\bX_i^\top)
		&=
		(\bI_p,\mathbf{0}_{p\times d_1^\prime})
		\left\{
		E(\bar{\bX}_{1i}^\prime
		\bar{\bX}_{1i}^{\prime\top})
		\right\}^{-1}
		E(\bar{\bX}_{1i}^\prime\bX_i^\top)=
		\bI_p.
	\end{align*}
	
	We continue to deal with the two terms above. Define the population
	regression slope of $\bX_i$ on $\bS_{1i}^\prime$ as
	\[
	\bdelta
	=
	E(\bS_{1i}^\prime\bS_{1i}^{\prime\top})^{-1}
	E(\bS_{1i}^\prime\bX_i^\top)
	\in\mathbb{R}^{d_1^\prime\times p}.
	\]
	Then
	\begin{align*}
		&\left\{
		\bG_1E(\bX_i\bS_{1i}^\top)
		+
		\bG_2E(\bS_{1i}^\prime\bS_{1i}^\top)
		\right\}\bbeta_2^*\\
		&\quad=
		\left\{
		\bG_1E(\bX_i\bS_{1i}^\top)
		-
		\bG_1E(\bX_i\bS_{1i}^{\prime\top})
		E(\bS_{1i}^\prime\bS_{1i}^{\prime\top})^{-1}
		E(\bS_{1i}^\prime\bS_{1i}^\top)
		\right\}\bbeta_2^*\\
		&\quad=
		\bG_1
		\left\{
		E(\bX_i\bS_{1i}^\top)
		-
		E(\bdelta^\top\bS_{1i}^\prime\bS_{1i}^\top)
		\right\}\bbeta_2^*\\
		&\quad=
		\bG_1
		E\left[
		\left(
		\bX_i-\bdelta^\top\bS_{1i}^\prime
		\right)\bS_{1i}^\top
		\right]\bbeta_2^*.
	\end{align*}
	Note that $\bdelta^\top\bS_{1i}^\prime$ is the best linear predictor
	of $\bX_i$ given $\bS_{1i}^\prime$. For the second term, by
	\eqref{eq:block_inverse_beta1} and \eqref{eq:S1prime_epsilon2},
	\begin{align*}
		\bG_2E(\bS_{1i}^\prime\varepsilon_{2i})
		&=
		-\bG_1\bdelta^\top
		E(\bS_{1i}^\prime\varepsilon_{2i})=
		-\bG_1\bdelta^\top
		\bP_{12}\bgamma_3^*\sigma_2^2.
	\end{align*}
	Therefore,
	\begin{align}
		\bar{\bbeta}_1^{\rm p}-\bbeta_1^*
		={}&
		\bG_1
		E\left[
		\left(
		\bX_i-\bdelta^\top\bS_{1i}^\prime
		\right)\bS_{1i}^\top
		\right]\bbeta_2^*
		-
		\bG_1\bdelta^\top
		\bP_{12}\bgamma_3^*\sigma_2^2=
		\bB_{\bbeta_1}^{\rm p}.
		\label{eq:beta1_population_bias}
	\end{align}
	Combining \eqref{eq:beta1_working_an} and
	\eqref{eq:beta1_population_bias}, we obtain
	\[
	\sqrt{n}
	\left(
	\widehat{\bbeta}_1^{\rm p}
	-\bbeta_1^*
	-\bB_{\bbeta_1}^{\rm p}
	\right)
	\xrightarrow{\rm d}
	\mathcal{N}
	\left(
	\mathbf{0},
	\bV_{\bbeta_1}^{\rm p}
	\right),
	\]
	which gives the proof.
\end{proof}

\begin{proof}[Proof of Theorem \ref{thm:thm_theta}]
We first verify the population target. Set $\bar{\btheta}^{\rm p}=\bar{\bbeta}_1^{\rm p}+\bar{\balpha}^{\rm p}\bar{\bbeta}_2^{\rm p}$.
	By the normal equations associated with the population projections,
	\begin{align*}
		E(\bX_i\bS_{1i}^{\prime\top})
		&=
		\bSigma_{\bX}\bar{\balpha}^{\rm p},\\
		E(\bX_iY_i)
		&=
		\bSigma_{\bX}\bar{\bbeta}_1^{\rm p}
		+
		E(\bX_i\bS_{1i}^{\prime\top})
		\bar{\bbeta}_2^{\rm p}\\
		&=
		\bSigma_{\bX}
		\left(
		\bar{\bbeta}_1^{\rm p}
		+
		\bar{\balpha}^{\rm p}\bar{\bbeta}_2^{\rm p}
		\right)
		=
		\bSigma_{\bX}\bar{\btheta}^{\rm p}.
	\end{align*}
	Hence,
	\[
	\bar{\btheta}^{\rm p}
	=
	\bSigma_{\bX}^{-1}E(\bX_iY_i)
	=
	\btheta^*.
	\]
	By Lemma~\ref{lem:working_gaussian}, the induced Gaussian working model satisfies the conditions of Theorem~\ref{thm:upper_bound}. Applying that theorem under selection scheme ${\rm p}$ gives
\begin{equation}\label{eq:theta_selected_an}
	\sqrt{n}
	\left(
	\widehat{\btheta}^{\rm p}-\btheta^*
	\right)
		\xrightarrow{\rm d}
		\mathcal{N}
		\left(
		\mathbf{0},
		\widetilde{\bV}_{\btheta}^{\rm p}
		\right),
	\end{equation}
	where
	\[
	\widetilde{\bV}_{\btheta}^{\rm p}
	=
	\left\{
	\sigma_2^{\prime 2}
	\left(
	1-\tau\kappa^{\rm p}
	\right)
	+
	(1-\tau)
	\bar{\bbeta}_2^{{\rm p}\top}
	\bSigma_1^\prime
	\bar{\bbeta}_2^{\rm p}
	\right\}
	\bSigma_{\bX}^{-1}
	\]
	and
	\[
	\kappa^{\rm p}
	=
	\frac{
		\sigma_2^{\prime 2}
		\bar{\bgamma}_3^{{\rm p}\top}
		\bSigma_3^{\prime-1}
		\bar{\bgamma}_3^{\rm p}
	}{
		1+
		\sigma_2^{\prime 2}
		\bar{\bgamma}_3^{{\rm p}\top}
		\bSigma_3^{\prime-1}
		\bar{\bgamma}_3^{\rm p}
	}.
	\]
	By Lemma~\ref{lem:working_gaussian},
	\begin{align*}
		\nu^2
		&=
		\bar{\bbeta}_2^{{\rm p}\top}
		\bSigma_1^\prime
		\bar{\bbeta}_2^{\rm p}
		+
		\sigma_2^{\prime 2},\\
		\nu_{\rm p}^2
		&=
		\sigma_2^{\prime 2}
		(1-\kappa^{\rm p}).
	\end{align*}
	The second identity follows from the Gaussian conditional-variance
	formula,
	\begin{align*}
		&{\rm Var}
		\left(
		\varepsilon_{2i}^\prime
		\mid
		\bar{\bgamma}_3^{\rm p}\varepsilon_{2i}^\prime
		+
		\bvarepsilon_{3i}^\prime
		\right)\\
		&\quad=
		\sigma_2^{\prime 2}
		-
		\sigma_2^{\prime 4}
		\bar{\bgamma}_3^{{\rm p}\top}
		\left(
		\bSigma_3^\prime
		+
		\sigma_2^{\prime 2}
		\bar{\bgamma}_3^{\rm p}
		\bar{\bgamma}_3^{{\rm p}\top}
		\right)^{-1}
		\bar{\bgamma}_3^{\rm p}\\
		&\quad=
		\sigma_2^{\prime 2}
		(1-\kappa^{\rm p}).
	\end{align*}
	Consequently, using the preceding conditional variance identities and $m/N\to\tau$, we obtain
	\begin{align*}
		\widetilde{\bV}_{\btheta}^{\rm p}
		&=
		\left\{\sigma_2'^2(1-\tau\kappa^{\rm p})+(1-\tau)\bar{\bbeta}_2^{{\rm p}\top}\bSigma_1'\bar{\bbeta}_2^{\rm p}\right\}\bSigma_{\bX}^{-1}\\ &=\left\{(1-\tau)\nu^2+\tau\nu_{\rm p}^2\right\}\bSigma_{\bX}^{-1}
	\end{align*}
	
	We next prove the finite-sample invariance. For scheme $\rm p$, let
	\[
	\bar{\bX}_{1i}^{\rm p}
	=
	(\bX_i^\top,\bS_{1i}^{\prime\top})^\top,
	\]
	and let $\bA^{\rm p}$ and $\bB^{\rm p}$ be the labeled-sample matrices
	whose $i$th rows are
	$\bar{\bX}_{1i}^{{\rm p}\top}$ and
	$\bS_{2i}^{\prime\top}$, respectively. Let
	\[
	\bY=(Y_i)_{i\in\mathcal I},
	\]
	and define
	\[
	\bM_{\bA}^{\rm p}
	=
	\bI_n
	-
	\bA^{\rm p}
	(\bA^{{\rm p}\top}\bA^{\rm p})^{-1}
	\bA^{{\rm p}\top},
	\]
	\[
	\widetilde{\bY}^{\rm p}
	=
	\bM_{\bA}^{\rm p}\bY,
	\qquad
	\widetilde{\bB}^{\rm p}
	=
	\bM_{\bA}^{\rm p}\bB^{\rm p}.
	\]
	The OLS initial estimators used by Algorithm~\ref{alg:EM} satisfy
	\begin{align*}
		\bbeta^{{\rm p},(0)}
		&=
		(\bA^{{\rm p}\top}\bA^{\rm p})^{-1}
		\bA^{{\rm p}\top}\bY,\\
		\bgamma_3^{{\rm p},(0)}
		&=
		(\widetilde{\bY}^{{\rm p}\top}
		\widetilde{\bY}^{\rm p})^{-1}
		\widetilde{\bB}^{{\rm p}\top}
		\widetilde{\bY}^{\rm p},\\
		(\sigma_2^{{\rm p},(0)})^2
		&=
		n^{-1}
		\widetilde{\bY}^{{\rm p}\top}
		\widetilde{\bY}^{\rm p}.
	\end{align*}
	Moreover,
	\[
	\bSigma_3^{{\rm p},(0)}
	=
	\frac{1}{n}
	\left(
	\widetilde{\bB}^{\rm p}
	-
	\widetilde{\bY}^{\rm p}
	\bgamma_3^{{\rm p},(0)\top}
	\right)^\top
	\left(
	\widetilde{\bB}^{\rm p}
	-
	\widetilde{\bY}^{\rm p}
	\bgamma_3^{{\rm p},(0)\top}
	\right).
	\]
	It follows that
	\[
	\bSigma_3^{{\rm p},(0)}
	+
	(\sigma_2^{{\rm p},(0)})^2
	\bgamma_3^{{\rm p},(0)}
	\bgamma_3^{{\rm p},(0)\top}
	=
	\frac{1}{n}
	\widetilde{\bB}^{{\rm p}\top}
	\widetilde{\bB}^{\rm p}.
	\]
	Therefore, the E-step coefficient satisfies
	\begin{align*}
		\bc^{{\rm p},(0)}
		&=
		(\sigma_2^{{\rm p},(0)})^2
		\left\{
		\bSigma_3^{{\rm p},(0)}
		+
		(\sigma_2^{{\rm p},(0)})^2
		\bgamma_3^{{\rm p},(0)}
		\bgamma_3^{{\rm p},(0)\top}
		\right\}^{-1}
		\bgamma_3^{{\rm p},(0)}\\
		&=
		(\widetilde{\bB}^{{\rm p}\top}
		\widetilde{\bB}^{\rm p})^{-1}
		\widetilde{\bB}^{{\rm p}\top}
		\widetilde{\bY}^{\rm p}.
	\end{align*}
	Thus, by the Frisch--Waugh--Lovell theorem,
	$\bc^{{\rm p},(0)}$ is exactly the coefficient on $\bB^{\rm p}$ in
	the OLS regression of $\bY$ on
	$(\bA^{\rm p},\bB^{\rm p})$.
	
	The normal equations for the initial regression of $\bB^{\rm p}$ on
	$(\bA^{\rm p},\bY)$ also give
	\[
	\bgamma_{12}^{{\rm p},(0)}
	+
	\bbeta^{{\rm p},(0)}
	\bgamma_3^{{\rm p},(0)\top}
	=
	(\bA^{{\rm p}\top}\bA^{\rm p})^{-1}
	\bA^{{\rm p}\top}\bB^{\rm p}.
	\]
	Therefore, the coefficient on $\bA^{\rm p}$ in the OLS regression of
	$\bY$ on $(\bA^{\rm p},\bB^{\rm p})$ is
	\[
	\bbeta^{{\rm p},(0)}
	-
	\left(
	\bgamma_{12}^{{\rm p},(0)}
	+
	\bbeta^{{\rm p},(0)}
	\bgamma_3^{{\rm p},(0)\top}
	\right)
	\bc^{{\rm p},(0)}.
	\]
	For every $j\in\mathcal J$, the E-step imputation is
	\begin{align*}
		\widehat{\mu}_j^{\rm p}
		={}&
		\bar{\bX}_{1j}^{{\rm p}\top}
		\left\{
		\bbeta^{{\rm p},(0)}
		-
		\left(
		\bgamma_{12}^{{\rm p},(0)}
		+
		\bbeta^{{\rm p},(0)}
		\bgamma_3^{{\rm p},(0)\top}
		\right)
		\bc^{{\rm p},(0)}
		\right\}\\
		&+
		\bS_{2j}^{\prime\top}
		\bc^{{\rm p},(0)}.
	\end{align*}
	Consequently,
	\begin{equation}\label{eq:theta_ols_imputation_selected}
		\widehat{\mu}_j^{\rm p}
		=
		\bZ_j^{{\rm p}\top}
		\left(
		\sum_{i\in\mathcal I}
		\bZ_i^{\rm p}\bZ_i^{{\rm p}\top}
		\right)^{-1}
		\left(
		\sum_{i\in\mathcal I}
		\bZ_i^{\rm p}Y_i
		\right),
	\end{equation}
	where
	\[
	\bZ_i^{\rm p}
	=
	(\bX_i^\top,
	\bS_{1i}^{\prime\top},
	\bS_{2i}^{\prime\top})^\top.
	\]
	
	If schemes $\rm p$ and $\rm q$ retain the same surrogate variables,
	there exists a permutation matrix $\mathbf R$ such that
	\[
	\bZ_i^{\rm q}
	=
	\mathbf R\bZ_i^{\rm p}.
	\]
	Using \eqref{eq:theta_ols_imputation_selected} and
	$\mathbf R^{-1}=\mathbf R^\top$, we obtain
	\begin{align*}
		\widehat{\mu}_j^{\rm q}
		&=
		\bZ_j^{{\rm p}\top}
		\mathbf R^\top
		\left\{
		\mathbf R
		\left(
		\sum_{i\in\mathcal I}
		\bZ_i^{\rm p}\bZ_i^{{\rm p}\top}
		\right)
		\mathbf R^\top
		\right\}^{-1}
		\mathbf R
		\left(
		\sum_{i\in\mathcal I}
		\bZ_i^{\rm p}Y_i
		\right)\\
		&=
		\widehat{\mu}_j^{\rm p}.
	\end{align*}
	
	Let $\bX_{\rm f}$ and $\bS_{1,{\rm f}}^{\rm p}$ denote the full-sample
	matrices of $\bX_i$ and $\bS_{1i}^\prime$, respectively, and define
	\[
	\widehat{\bY}_{\bmu}^{\rm p}
	=
	(Y_1,\ldots,Y_n,
	\widehat{\mu}_{n+1}^{\rm p},\ldots,
	\widehat{\mu}_{N}^{\rm p})^\top.
	\]
	The M-step normal equations give
	\begin{align*}
		\bX_{\rm f}^\top\bX_{\rm f}\widehat{\balpha}^{\rm p}
		&=
		\bX_{\rm f}^\top\bS_{1,{\rm f}}^{\rm p},\\
		\bX_{\rm f}^\top\bX_{\rm f}\widehat{\bbeta}_1^{\rm p}
		+
		\bX_{\rm f}^\top\bS_{1,{\rm f}}^{\rm p}
		\widehat{\bbeta}_2^{\rm p}
		&=
		\bX_{\rm f}^\top
		\widehat{\bY}_{\bmu}^{\rm p}.
	\end{align*}
	Combining the two equations yields
	\begin{equation}\label{eq:theta_mstep_selected}
		\widehat{\btheta}^{\rm p}
		=
		\widehat{\bbeta}_1^{\rm p}
		+
		\widehat{\balpha}^{\rm p}
		\widehat{\bbeta}_2^{\rm p}
		=
		(\bX_{\rm f}^\top\bX_{\rm f})^{-1}
		\bX_{\rm f}^\top
		\widehat{\bY}_{\bmu}^{\rm p}.
	\end{equation}
	Since
	\[
	\widehat{\bY}_{\bmu}^{\rm p}
	=
	\widehat{\bY}_{\bmu}^{\rm q},
	\]
	equation \eqref{eq:theta_mstep_selected} implies
	\[
	\widehat{\btheta}^{\rm p}
	=
	\widehat{\btheta}^{\rm q}.
	\]
	
	Finally, suppose that the selected set under scheme $\rm p$ is contained in that under scheme $\rm q$. Let $\bS_{{\rm s},i}^{\rm p}=(\bS_{1i}^{\prime\top},\bS_{2i}^{\prime\top})^\top$ denote the vector of surrogates retained under scheme $\rm p$, and define $\bS_{{\rm s},i}^{\rm q}$ analogously. By the conditional variance decomposition,
	\begin{align*}
		\nu_{\rm p}^2
		&=
		E
		\left\{
		{\rm Var}
		\left(
		Y_i
		\mid
		\bX_i,\bS_{{\rm s},i}^{\rm q}
		\right)
		\mid
		\bX_i,\bS_{{\rm s},i}^{\rm p}
		\right\}\\
		&\quad+
		{\rm Var}
		\left\{
		E
		\left(
		Y_i
		\mid
		\bX_i,\bS_{{\rm s},i}^{\rm q}
		\right)
		\mid
		\bX_i,\bS_{{\rm s},i}^{\rm p}
		\right\}\\
		&=
		\nu_{\rm q}^2
		+
		{\rm Var}
		\left\{
		E
		\left(
		Y_i
		\mid
		\bX_i,\bS_{{\rm s},i}^{\rm q}
		\right)
		\mid
		\bX_i,\bS_{{\rm s},i}^{\rm p}
		\right\}\\
		&\geq
		\nu_{\rm q}^2.
	\end{align*}
	Therefore,
	\[
	\bV_{\btheta}^{\rm p}
	-
	\bV_{\btheta}^{\rm q}
	=
	\tau
	\left(
	\nu_{\rm p}^2-\nu_{\rm q}^2
	\right)
	\bSigma_{\bX}^{-1}
	\succeq
	\mathbf{0}.
	\]
	The reduction is strictly positive definite if and only if $\tau>0$ and $\nu_{\rm p}^2>\nu_{\rm q}^2$. When $\tau=0$, the limiting covariance difference vanishes for any fixed pair of selection schemes. This completes the proof.
 \end{proof}

The following supplementary theorem records the full-vector $\bbeta^*$ result as an auxiliary calculation and gives the detailed optimal-weight results for the two main targets, $\bbeta_1^*$ and $\btheta^*$. The latter results establish the assertions of Theorem~\ref{thm:weighted_targets} in the main text. We denote
\begin{align*}
	\bV_{\varepsilon_2}=E(\bar{\bX}_{1i}\bar{\bX}_{1i}^\top\varepsilon_{2i}^2),\;\; \bV_{\bv}=E\{\bar{\bX}_{1i}\bar{\bX}_{1i}^\top(\bc^{*\top}\bv_i)^2\},\;\; \bV_{\bv\varepsilon_2}=E\{\bar{\bX}_{1i}\bar{\bX}_{1i}^\top(\bc^{*\top}\bv_i)\varepsilon_{2i}\},
\end{align*} 
and $\widehat{\bSigma}_{\bX}^{-1}=(N^{-1}\sum_{i\in\mathcal{I}\cup\mathcal{J}}\bX_i\bX_i^\top)^{-1}$, $\widehat{\bSigma}_{\bX,L}^{-1}=(n^{-1}\sum_{i\in\mathcal{I}}\bX_i\bX_i^\top)^{-1}$, $\widehat\bSigma_{\bX,1}^{-1}=(n^{-1}\sum_{i\in\mathcal{I}}\bar{\bX}_{1i}\bar{\bX}_{1i}^\top)^{-1}$, $\widehat\bV_{\bv\varepsilon_{2}}=n^{-1}\sum_{i\in\mathcal{I}}\bar{\bX}_{1i}\bar{\bX}_{1i}^\top\bc^{(0)\top}\hat{\bv}_i\hat{\varepsilon}_{2i}$, and $\widehat\bV_{\bv}=n^{-1}\sum_{i\in\mathcal{I}}\bar{\bX}_{1i}\bar{\bX}_{1i}^\top(\bc^{(0)\top}\hat{\bv}_i)^2$ where $\hat{\varepsilon}_{2i}=Y_i-\bar{\bX}_{1i}^\top \bbeta^{(0)}$, $\hat{\bvarepsilon}_{3i}=\bS_{2i}-\bgamma^{(0)\top}\bar{\bX}_{2i}$, and $\hat{\bv}_i=\hat{\varepsilon}_{2i}\bgamma_3^{(0)}+\hat{\bvarepsilon}_{3i}$.

\begin{theorem}\label{thm:optimal_lambda}
	Suppose Assumption~\ref{ass:non_gaussian} holds. Assume further that $E(\|\bar{\bX}_{2i}\|_2^4)<\infty$, $E(\varepsilon_{2i}^4\mid\bar{\bX}_{1i})\le C$, and $E(\|\bvarepsilon_{3i}\|_2^4\mid\bar{\bX}_{2i})\le C$ almost surely for a finite constant $C$. Let $a_n\to0$ and $na_n\to\infty$, and suppose the sampling conditions in Section~\ref{sec:upper_bound} of the main text hold, with $m/N\to\tau\in[0,1]$. Let
	$\widehat\bbeta_\lambda=\bbeta^{(0)}+\lambda(\widehat\bbeta-\bbeta^{(0)})$
	and
	$\widehat\btheta_\lambda=\widehat\btheta_{\rm sup}+\lambda(\widehat\btheta-\widehat\btheta_{\rm sup})$
	be the estimators defined in \eqref{eq:weighted_est}. Then the following results hold.
 	
	\medskip
	\noindent
	\textnormal{(1) Estimation of $\bbeta^*$.}
	When $\|\bgamma_3^*\|_2>0$ and $\tau>0$, the unique minimizer of ${\rm tr}[
	{\rm Avar}
	\{
	\sqrt n(\widehat\bbeta_\lambda-\bbeta^*)
	\}]
	$ over $\lambda\in\mathbb R$ is
	\begin{align}
		\lambda^*
		={\rm tr}(\bSigma_{\bX,1}^{-1}\bV_{\bv\varepsilon_2}\bSigma_{\bX,1}^{-1})/
			{\rm tr}
			(
			\bSigma_{\bX,1}^{-1}
			\bV_{\bv}
			\bSigma_{\bX,1}^{-1}
			).
		\label{eq:lambda1_star}
	\end{align}
	When $\tau=0$, the limiting covariance is independent of $\lambda$, and we retain the same oracle ratio as a convention. If $\bgamma_3^*=\mathbf{0}$, we set $\lambda^*=0$.
	Write $\widehat L_2={\rm tr}(\widehat\bSigma_{\bX,1}^{-1}\widehat\bV_{\bv}\widehat\bSigma_{\bX,1}^{-1})$. Define
	\begin{align}
		\hat\lambda
		=
		\begin{cases}
				{\rm tr}
				(
				\widehat\bSigma_{\bX,1}^{-1}
				\widehat\bV_{\bv\varepsilon_2}
				\widehat\bSigma_{\bX,1}^{-1}
				)/
				{\rm tr}(
				\widehat\bSigma_{\bX,1}^{-1}
				\widehat\bV_{\bv}
				\widehat\bSigma_{\bX,1}^{-1}),
			&
			\widehat L_2>a_n,\\[4mm]
			0,
			&
			\widehat L_2\le a_n.
		\end{cases}
		\label{eq:lambda1_hat}
	\end{align}
	Then
	\begin{align*}
		\sqrt n
		(\widehat\bbeta_{\hat\lambda}-\bbeta^*)
		\xrightarrow{\rm d}
	    N
		\left(
		0,
		{\rm Avar}
		\left\{
		\sqrt n
		(\widehat\bbeta_{\lambda^*}-\bbeta^*)
		\right\}
		\right).
	\end{align*}
	Moreover, when $\|\bgamma_3^*\|_2>0$,
	\begin{align*}
		&{\rm tr}
		\left[
		{\rm Avar}
		\left\{
		\sqrt n
		(\widehat\bbeta_{\lambda^*}-\bbeta^*)
		\right\}
		\right]=
		{\rm tr}
		\left[
		{\rm Avar}
		\left\{
		\sqrt n
		(\widehat\bbeta_{0}-\bbeta^*)
		\right\}
		\right]-
		\tau
		\frac{
			\left\{
			{\rm tr}
			\left(
			\bSigma_{\bX,1}^{-1}
			\bV_{\bv\varepsilon_2}
			\bSigma_{\bX,1}^{-1}
			\right)
			\right\}^2
		}{
			{\rm tr}
			\left(
			\bSigma_{\bX,1}^{-1}
			\bV_{\bv}
			\bSigma_{\bX,1}^{-1}
			\right)
		}.
	\end{align*}
	The optimal weighting does not increase the limiting covariance trace. The reduction is strict when $\tau>0$ and the numerator of~\eqref{eq:lambda1_star} is nonzero. When $\tau=0$, the reduction vanishes.

	\medskip
	\noindent
	\textnormal{(2) Estimation of $\bbeta_1^*$.}
	When $\|\bgamma_3^*\|_2>0$ and $\tau>0$, the unique minimizer of ${\rm tr}[{\rm Avar}\{
	\sqrt n
	(\widehat\bbeta_{\lambda,1}-\bbeta_1^*)\}]$
	over $\lambda\in\mathbb R$ is
	\begin{align}
		\lambda_2^*
		=
		\frac{
			{\rm tr}
			\left\{
			[\bI_p,\mathbf 0_{p\times d_1}]
			\bSigma_{\bX,1}^{-1}
			\bV_{\bv\varepsilon_2}
			\bSigma_{\bX,1}^{-1}
			[\bI_p,\mathbf 0_{p\times d_1}]^\top
			\right\}
		}{
			{\rm tr}
			\left\{
			[\bI_p,\mathbf 0_{p\times d_1}]
			\bSigma_{\bX,1}^{-1}
			\bV_{\bv}
			\bSigma_{\bX,1}^{-1}
			[\bI_p,\mathbf 0_{p\times d_1}]^\top
			\right\}
		}.
		\label{eq:lambda2_star}
	\end{align}
	When $\tau=0$, every fixed $\lambda$ has the same limiting covariance, and we retain the displayed ratio as a convention whenever its denominator is positive. When $\|\bgamma_3^*\|_2=0$, define $\lambda_2^*=0$. Write $\widehat D_{\bbeta_1}={\rm tr}\{[\bI_p,\mathbf 0_{p\times d_1}]\widehat\bSigma_{\bX,1}^{-1}\widehat\bV_{\bv}\widehat\bSigma_{\bX,1}^{-1}[\bI_p,\mathbf 0_{p\times d_1}]^\top\}$. Define
	\begin{align}
		\hat\lambda_2
		=
		\begin{cases}
			\displaystyle
			\frac{
				{\rm tr}
				\left\{
				[\bI_p,\mathbf 0_{p\times d_1}]
				\widehat\bSigma_{\bX,1}^{-1}
				\widehat\bV_{\bv\varepsilon_2}
				\widehat\bSigma_{\bX,1}^{-1}
				[\bI_p,\mathbf 0_{p\times d_1}]^\top
				\right\}
			}{
				{\rm tr}
				\left\{
				[\bI_p,\mathbf 0_{p\times d_1}]
				\widehat\bSigma_{\bX,1}^{-1}
				\widehat\bV_{\bv}
				\widehat\bSigma_{\bX,1}^{-1}
				[\bI_p,\mathbf 0_{p\times d_1}]^\top
				\right\}
			},
			&
			\widehat D_{\bbeta_1}>a_n,\\[5mm]
			0,
			&
			\widehat D_{\bbeta_1}\le a_n.
		\end{cases}
		\label{eq:lambda2_hat}
	\end{align}
	Then
	\begin{align*}
		\sqrt n
		(\widehat\bbeta_{\hat\lambda_2,1}-\bbeta_1^*)
		\xrightarrow{\rm d}
		N
		\left(
		0,
		{\rm Avar}
		\left\{
		\sqrt n
		(\widehat\bbeta_{\lambda_2^*,1}-\bbeta_1^*)
		\right\}
		\right).
	\end{align*}
	Moreover, when $\|\bgamma_3^*\|_2>0$,
	\begin{align*}
		{\rm tr}
		\left[
		{\rm Avar}
		\left\{
		\sqrt n
		(\widehat\bbeta_{\lambda_2^*,1}-\bbeta_1^*)
		\right\}
		\right]&=
		{\rm tr}
		\left[
		{\rm Avar}
		\left\{
		\sqrt n
		(\widehat\bbeta_{0,1}-\bbeta_1^*)
		\right\}
		\right]\\
		&\qquad-\tau
		\frac{
			\left[
			{\rm tr}
			\left\{
			[\bI_p,\mathbf 0_{p\times d_1}]
			\bSigma_{\bX,1}^{-1}
			\bV_{\bv\varepsilon_2}
			\bSigma_{\bX,1}^{-1}
			[\bI_p,\mathbf 0_{p\times d_1}]^\top
			\right\}
			\right]^2
		}{
			{\rm tr}
			\left\{
			[\bI_p,\mathbf 0_{p\times d_1}]
			\bSigma_{\bX,1}^{-1}
			\bV_{\bv}
			\bSigma_{\bX,1}^{-1}
			[\bI_p,\mathbf 0_{p\times d_1}]^\top
			\right\}
		}.
	\end{align*}
	The reduction is strict when $\tau>0$ and the numerator of~\eqref{eq:lambda2_star} is nonzero. When \(\tau=0\), no first-order trace improvement is obtained.

	\medskip
	\noindent
	\textnormal{(3) Estimation of $\btheta^*$.}
	Let $u_i=\bvarepsilon_{1i}^\top\bbeta_2^*$, $e_i=\varepsilon_{2i}+u_i$, $q_i=u_i+\bc^{*\top}\bv_i$, $\bV_q=E(\bX_i\bX_i^\top q_i^2)$, and $\bV_{eq}=E(\bX_i\bX_i^\top e_iq_i)$. Write $D_{\btheta}=\operatorname{tr}(\bSigma_{\bX}^{-1}\bV_q\bSigma_{\bX}^{-1})$ and $C_{\btheta}=\operatorname{tr}(\bSigma_{\bX}^{-1}\bV_{eq}\bSigma_{\bX}^{-1})$. When $D_{\btheta}>0$ and $\tau>0$, the unique minimizer of $\operatorname{tr}[\operatorname{Avar}\{\sqrt n(\widehat\btheta_\lambda-\btheta^*)\}]$ over $\lambda\in\mathbb R$ is
	\begin{align}
	\lambda_3^*=\frac{C_{\btheta}}{D_{\btheta}}.
	\label{eq:lambda3_star}
	\end{align}
	When $\tau=0$, the limiting covariance does not depend on $\lambda$, and the ratio is retained as a convention if $D_{\btheta}>0$. If $D_{\btheta}=0$, define $\lambda_3^*=0$. For $i\in\mathcal I$, set $\widehat\balpha_L=\widehat\bSigma_{\bX,L}^{-1}n^{-1}\sum_{j\in\mathcal I}\bX_j\bS_{1j}^\top$, $\hat u_i=(\bS_{1i}-\widehat\balpha_L^\top\bX_i)^\top\bbeta_2^{(0)}$, $\hat e_i=Y_i-\bX_i^\top(\bbeta_1^{(0)}+\widehat\balpha_L\bbeta_2^{(0)})$, and $\hat q_i=\hat u_i+\bc^{(0)\top}\hat\bv_i$. Define $\widehat\bV_q=n^{-1}\sum_{i\in\mathcal I}\bX_i\bX_i^\top\hat q_i^2$, $\widehat\bV_{eq}=n^{-1}\sum_{i\in\mathcal I}\bX_i\bX_i^\top\hat e_i\hat q_i$, and $\widehat D_{\btheta}=\operatorname{tr}(\widehat\bSigma_{\bX,L}^{-1}\widehat\bV_q\widehat\bSigma_{\bX,L}^{-1})$. Define
 	\begin{align}
		\hat\lambda_3
		=\begin{cases}
		\displaystyle\frac{\operatorname{tr}(\widehat\bSigma_{\bX,L}^{-1}\widehat\bV_{eq}\widehat\bSigma_{\bX,L}^{-1})}{\widehat D_{\btheta}},&\widehat D_{\btheta}>a_n,\\[3mm]
		0,&\widehat D_{\btheta}\le a_n.
		\end{cases}
		\label{eq:lambda3_hat}
	\end{align}
	Then
	\begin{align*}
		\sqrt n
		(\widehat\btheta_{\hat\lambda_3}-\btheta^*)
		\xrightarrow{\rm d}
		 \mathcal{N}
		\left(
		 0,
		{\rm Avar}
		\left\{
		\sqrt n
		(\widehat\btheta_{\lambda_3^*}-\btheta^*)
		\right\}
		\right).
	\end{align*}
Moreover, when $D_{\btheta}>0$,
\begin{align*}
\operatorname{tr}\{\operatorname{Avar}[\sqrt n(\widehat\btheta_{\lambda_3^*}-\btheta^*)]\}
&=\operatorname{tr}(\bV_{\btheta,0})-\tau\frac{C_{\btheta}^2}{D_{\btheta}}
\le\operatorname{tr}(\bV_{\btheta,0}).
\end{align*}
When $D_{\btheta}=0$, the same trace comparison holds with equality. The improvement is strict when $\tau>0$, $D_{\btheta}>0$, and $C_{\btheta}\neq0$.
\end{theorem}

\begin{proof}[Proof of Theorem~\ref{thm:weighted_targets} and Supplementary Theorem~\ref{thm:optimal_lambda}]
	
\medskip We first consider the regime $\tau>0$. In the following covariance calculations,
the actual sample fractions are retained until their limits are taken. This yields the auxiliary full-vector covariance $\bV_{\lambda}$ derived below.
The case $\tau=0$, equivalently $m=o(n)$, is treated separately below. The consistency arguments for the feasible
weight rely only on labeled observations and do not require $m\to\infty$.

\textbf{Auxiliary result for $\bbeta^*$.} By \eqref{eq:weighted_est}, $\widehat{\bbeta}_{0}=\bbeta^{(0)}$ and $\widehat{\bbeta}_{1}=\widehat{\bbeta}$. For $\lambda\in\{0,1\}$, the labeled OLS normal equations and the update \eqref{eq:bg_update} give
	\begin{align*}
		&\dfrac{1-\lambda}{n}\sum_{i\in\mathcal{I}}
		(Y_i-\bar{\bX}_{1i}^\top\widehat{\bbeta}_{\lambda})
		\bar{\bX}_{1i}
		+\dfrac{\lambda}{N}\sum_{i\in\mathcal{I}}
		(Y_i-\bar{\bX}_{1i}^\top\widehat{\bbeta}_{\lambda})
		\bar{\bX}_{1i}
		+\dfrac{\lambda}{N}\sum_{i\in\mathcal{J}}
		(\mu_i^{\bOmega^{(0)}}-\bar{\bX}_{1i}^\top
		\widehat{\bbeta}_{\lambda})\bar{\bX}_{1i}=0.
	\end{align*}
	Therefore,
	\begin{align*}
		\widehat{\bbeta}_{\lambda}
		&=
		\Big\{
		\dfrac{1-\lambda}{n}\sum_{i\in\mathcal{I}}
		\bar{\bX}_{1i}\bar{\bX}_{1i}^\top
		+\dfrac{\lambda}{N}\sum_{i\in\mathcal{I}}
		\bar{\bX}_{1i}\bar{\bX}_{1i}^\top
		+\dfrac{\lambda}{N}\sum_{i\in\mathcal{J}}
		\bar{\bX}_{1i}\bar{\bX}_{1i}^\top
		\Big\}^{-1}\\
		&\quad\times
		\Big\{
		\dfrac{1-\lambda}{n}\sum_{i\in\mathcal{I}}
		\bar{\bX}_{1i}Y_i
		+\dfrac{\lambda}{N}\sum_{i\in\mathcal{I}}
		\bar{\bX}_{1i}Y_i
		+\dfrac{\lambda}{N}\sum_{i\in\mathcal{J}}
		\bar{\bX}_{1i}\mu_i^{\bOmega^{(0)}}
		\Big\}.
	\end{align*}
	Repeating the argument in the proof of
	Theorem~\ref{thm:upper_bound}, we obtain
	\begin{align*}
	\widehat{\bbeta}_{\lambda}-\bbeta^*&=\Big\{\dfrac{1-\lambda}{n}\sum_{i\in\mathcal{I}}\bar{\bX}_{1i}\bar{\bX}_{1i}^\top+\dfrac{\lambda}{N}\sum_{i\in\mathcal{I}}\bar{\bX}_{1i}\bar{\bX}_{1i}^\top+\dfrac{\lambda}{N}\sum_{i\in\mathcal{J}}\bar{\bX}_{1i}\bar{\bX}_{1i}^\top\Big\}^{-1}\\
		&\qquad\times \Big\{\dfrac{1-\lambda}{n}\sum_{i\in\mathcal{I}}\bar{\bX}_{1i}\varepsilon_{2i}+\dfrac{\lambda}{N}\sum_{i\in\mathcal{I}}\bar{\bX}_{1i}\varepsilon_{2i}+\dfrac{\lambda}{N}\sum_{i\in\mathcal{J}}\bar{\bX}_{1i}(\mu_i^{\bOmega^{(0)}}-\bar{\bX}_{1i}^\top\bbeta^*)\Big\}\\
	&=\bSigma_{\bX,1}^{-1}\Big\{\dfrac{1-\lambda}{n}\sum_{i\in\mathcal{I}}\bar{\bX}_{1i}\varepsilon_{2i}+\dfrac{\lambda}{N}\sum_{i\in\mathcal{I}}\bar{\bX}_{1i}\varepsilon_{2i}+\dfrac{\lambda}{N}\sum_{i\in\mathcal{J}}\bar{\bX}_{1i}(\mu_i^{\bOmega^{(0)}}-\bar{\bX}_{1i}^\top\bbeta^*)\Big\}+o_p(n^{-1/2})\\
	&=\bSigma_{\bX,1}^{-1}\Big[\Big(\dfrac{1-\lambda}{n}+\dfrac{\lambda}{N}\Big)\sum_{i\in\mathcal{I}}\bar{\bX}_{1i}\varepsilon_{2i}+\dfrac{\lambda}{N}\sum_{i\in\mathcal{J}} \bar{\bX}_{1i}\bc^{*\top}\bv_i\\
		&\qquad\qquad+\dfrac{m\lambda }{nN}\sum_{i\in\mathcal{I}}\bar{\bX}_{1i} \{(1-\bc^{*\top}\bgamma_3^*)\varepsilon_{2i}-\bc^{*\top}\bvarepsilon_{3i}\}\Big]+o_p(n^{-1/2})\\
		&=\bSigma_{\bX,1}^{-1}\Big[\dfrac{1}{n}\sum_{i\in\mathcal{I}}\bar{\bX}_{1i}\{\varepsilon_{2i}-\lambda(m/N)\bc^{*\top}\bv_i\}+\dfrac{\lambda}{N}\sum_{i\in\mathcal{J}} \bar{\bX}_{1i}\bc^{*\top}\bv_i\Big]+o_p(n^{-1/2}).
	\end{align*}
	Where the third equality holds, as repeating the proof of Theorem \ref{thm:upper_bound} implies that
	\begin{align*}
		&\dfrac{\lambda}{N}\sum_{i\in\mathcal{J}}\bar{\bX}_{1i}(\mu_i^{\bOmega^{(0)}}-\bar{\bX}_{1i}^\top\bbeta^*)\\
		=\ &\dfrac{\lambda}{N}\sum_{i\in\mathcal{J}} \bar{\bX}_{1i}\bc^{*\top}\bv_i+\dfrac{m\lambda (1-\bc^{*\top}\bgamma_3^*)}{nN}\sum_{i\in\mathcal{I}}\bar{\bX}_{1i} \varepsilon_{2i}- \dfrac{m\lambda}{nN}\sum_{i\in\mathcal{I}}\bar{\bX}_{1i}\bvarepsilon_{3i}^\top \bc^*+o_p(n^{-1/2}).
	\end{align*} 
	Let $r_{n,0}$ and $r_{n,1}$ denote the $o_p(n^{-1/2})$ remainders in the above expansion of $\widehat{\bbeta}_{\lambda}-\bbeta^*$ for $\lambda=0$ and $\lambda=1$. For general $\lambda\in\mathbb R$, \eqref{eq:weighted_est} gives $\widehat{\bbeta}_{\lambda}-\bbeta^*=(1-\lambda)(\widehat{\bbeta}_{0}-\bbeta^*)+\lambda(\widehat{\bbeta}_{1}-\bbeta^*)$, and the leading term of the expansion is affine in $\lambda$. Hence, for every $\lambda\in\mathbb R$,
	\begin{align}
		\widehat{\bbeta}_{\lambda}-\bbeta^*
		&=\bSigma_{\bX,1}^{-1}\Big[\dfrac{1}{n}\sum_{i\in\mathcal{I}}\bar{\bX}_{1i}\{\varepsilon_{2i}-\lambda(m/N)\bc^{*\top}\bv_i\}+\dfrac{\lambda}{N}\sum_{i\in\mathcal{J}} \bar{\bX}_{1i}\bc^{*\top}\bv_i\Big]\nonumber\\
		&\qquad+(1-\lambda)r_{n,0}+\lambda r_{n,1}.
		\label{eq:beta_affine_expansion}
	\end{align}
	Since $r_{n,0}$ and $r_{n,1}$ do not depend on $\lambda$, \eqref{eq:beta_affine_expansion} also holds, with an $o_p(n^{-1/2})$ remainder, when $\lambda$ is replaced by a random weight that is $O_p(1)$. It follows from the central limit theorem that
	\begin{align}
		\sqrt n
		(\widehat{\bbeta}_{\lambda}-\bbeta^*)
		\xrightarrow{\rm d}
		\mathcal N(0,\bV_\lambda),
		\label{eq:weighted_beta_fixed_lambda}
	\end{align}
	where the asymptotic variance is 
	\[\bV_\lambda=\bSigma_{\bX,1}^{-1}\left\{\bV_{\varepsilon_2}
+\tau\lambda^2 \bV_{\bv}-2\tau\lambda \bV_{\bv\varepsilon_2}\right\}\bSigma_{\bX,1}^{-1}\]
	Consequently,
	\begin{align*}
		{\rm tr}(\bV_\lambda)
		&=
		{\rm tr}
		(\bSigma_{\bX,1}^{-1}
		\bV_{\varepsilon_2}
		\bSigma_{\bX,1}^{-1})
		+\tau\lambda^2
		{\rm tr}
		(\bSigma_{\bX,1}^{-1}
		\bV_{\bv}
		\bSigma_{\bX,1}^{-1})
		-2\tau\lambda
		{\rm tr}
		(\bSigma_{\bX,1}^{-1}
		\bV_{\bv\varepsilon_2}
		\bSigma_{\bX,1}^{-1}).
	\end{align*}
	
	We now consider the two possible cases for $\bgamma_3^*$.
	
	\medskip
	\noindent
	\textbf{Case (a) $\|\bgamma_3^*\|_2>0$.} Since $\|\bgamma_3^*\|_2>0$ and the underlying distribution is fixed,
	we have $\|\bc^*\|_2>0$. Hence,
	\begin{align*}
		E\{(\bc^{*\top}\bv_i)^2\mid\bar{\bX}_{1i}\}
		&=
		\bc^{*\top}
		E(\bv_i\bv_i^\top\mid\bar{\bX}_{1i})
		\bc^*\ge
		\lambda_{\rm min}
		\{E(\bv_i\bv_i^\top\mid\bar{\bX}_{1i})\}
		\|\bc^*\|_2^2
		>0
	\end{align*}
	almost surely.
	Therefore,
	\begin{align*}
		L_2
		&=
		{\rm tr}
		(\bSigma_{\bX,1}^{-1}\bV_{\bv}\bSigma_{\bX,1}^{-1})=
		E\left\{
		(\bc^{*\top}\bv_i)^2
		\bar{\bX}_{1i}^{\top}
		\bSigma_{\bX,1}^{-2}
		\bar{\bX}_{1i}
		\right\}
		>0.
	\end{align*}
	When $\tau>0$, the coefficient of $\lambda^2$ in $\operatorname{tr}(\bV_\lambda)$ is strictly positive. Consequently, the unique minimizer is the auxiliary oracle weight $\lambda^*$ defined in \eqref{eq:lambda1_star}. When $\tau=0$, the limiting covariance is independent of $\lambda$, and the same ratio is retained as a convention.
	\begin{align*}
		\lambda^*
		=
		\dfrac{
			{\rm tr}
			(\bSigma_{\bX,1}^{-1}
			\bV_{\bv\varepsilon_2}
			\bSigma_{\bX,1}^{-1})
		}{
			{\rm tr}
			(\bSigma_{\bX,1}^{-1}
			\bV_{\bv}
			\bSigma_{\bX,1}^{-1})
		}.
	\end{align*}
	We denote $L_1=
		{\rm tr}
		(\bSigma_{\bX,1}^{-1}
		\bV_{\bv\varepsilon_2}
		\bSigma_{\bX,1}^{-1})$, $\widehat L_1=
		{\rm tr}
		(\widehat\bSigma_{\bX,1}^{-1}
		\widehat\bV_{\bv\varepsilon_2}
		\widehat\bSigma_{\bX,1}^{-1})$, and $\widehat L_2=
		{\rm tr}
		(\widehat\bSigma_{\bX,1}^{-1}
		\widehat\bV_{\bv}
		\widehat\bSigma_{\bX,1}^{-1})$. To show the consistency of $\hat{L}_1$ and $\hat{L}_2$, we next prove that 
		$\|\widehat\bV_{\bv}-\bV_{\bv}\|_2=O_p(n^{-1/2})$, $
		\|\widehat\bV_{\bv\varepsilon_2}
		-\bV_{\bv\varepsilon_2}\|_2=O_p(n^{-1/2})$, and
		$\|\widehat\bSigma_{\bX,1}
		-\bSigma_{\bX,1}\|_2=O_p(n^{-1/2})$. Under Assumption~\ref{ass:non_gaussian}
		and the moment conditions assumed in Supplementary Theorem~\ref{thm:optimal_lambda}, standard
		fixed-dimensional OLS theory gives
		\begin{align}
			&\|\bbeta^{(0)}-\bbeta^*\|_2
			+\|\bgamma^{(0)}-\bgamma^*\|_2
			+\|\bgamma_3^{(0)}-\bgamma_3^*\|_2\nonumber\\
			&\qquad+|(\sigma_2^{(0)})^2-\sigma_2^2|
			+\|\bSigma_3^{(0)}-\bSigma_3\|_2
			=O_p(n^{-1/2}).
			\label{eq:initial_root_n}
		\end{align}
		Since $\bc^*=
		\sigma_2^2
		(\bSigma_3+\sigma_2^2\bgamma_3^*\bgamma_3^{*\top})^{-1}
		\bgamma_3^*$
		is a smooth function of
		$(\sigma_2^2,\bSigma_3,\bgamma_3^*)$, and
		$\bSigma_3$ is positive definite under
		Assumption~\ref{ass:non_gaussian}, it follows from
		\eqref{eq:initial_root_n} that
		\begin{align}
			\|\bc^{(0)}-\bc^*\|_2=O_p(n^{-1/2}),
			\qquad
			\|\bc^{(0)}\|_2=O_p(1),
			\qquad
			\|\bc^*\|_2=O(1).
			\label{eq:c_root_n}
		\end{align}
		Moreover, since
		$\bv_i=\varepsilon_{2i}\bgamma_3^*+\bvarepsilon_{3i}$,
		the conditional fourth-moment assumptions imply
		\begin{align}
			E(\|\bv_i\|_2^4\mid\bar{\bX}_{1i})
			&\le
			8\|\bgamma_3^*\|_2^4
			E(\varepsilon_{2i}^4\mid\bar{\bX}_{1i})
			+
			8E(\|\bvarepsilon_{3i}\|_2^4\mid\bar{\bX}_{1i})
			=O(1),
			\label{eq:v_fourth}
		\end{align}
		where
		\[
		E(\|\bvarepsilon_{3i}\|_2^4\mid\bar{\bX}_{1i})
		=
		E\{
		E(\|\bvarepsilon_{3i}\|_2^4\mid\bar{\bX}_{2i})
		\mid\bar{\bX}_{1i}\}
		=O(1).
		\]
		Together with $E(\|\bar{\bX}_{2i}\|_2^4)=O(1)$, this yields
		\begin{align}
			&\dfrac{1}{n}\sum_{i\in\mathcal I}
			\|\bar{\bX}_{1i}\|_2^2\|\bv_i\|_2^2=O_p(1),
			\qquad
			\dfrac{1}{n}\sum_{i\in\mathcal I}
			\|\bar{\bX}_{1i}\|_2^4=O_p(1),
			\nonumber\\
			&\dfrac{1}{n}\sum_{i\in\mathcal I}
			\|\bar{\bX}_{1i}\|_2^2\|\bar{\bX}_{2i}\|_2^2=O_p(1),
			\qquad
			\dfrac{1}{n}\sum_{i\in\mathcal I}
			\|\bar{\bX}_{1i}\|_2^2\varepsilon_{2i}^2=O_p(1).
			\label{eq:empirical_moments}
		\end{align}
		We first consider $\widehat{\bV}_{\bv}$. By the definitions of $\hat{\varepsilon}_{2i}$,
		$\hat{\bvarepsilon}_{3i}$, and $\hat{\bv}_i$,
		\begin{align*}
			\hat{\bv}_i-\bv_i
			&=
			\varepsilon_{2i}(\bgamma_3^{(0)}-\bgamma_3^*)
			-\bgamma_3^{(0)}
			\bar{\bX}_{1i}^{\top}(\bbeta^{(0)}-\bbeta^*)
			-(\bgamma^{(0)}-\bgamma^*)^{\top}\bar{\bX}_{2i}.
		\end{align*}
		Hence, by \eqref{eq:initial_root_n} and
		\eqref{eq:empirical_moments},
		\begin{align}
			\dfrac{1}{n}\sum_{i\in\mathcal I}
			\|\bar{\bX}_{1i}\|_2^2
			\|\hat{\bv}_i-\bv_i\|_2^2&\le
			C\|\bgamma_3^{(0)}-\bgamma_3^*\|_2^2
			\dfrac{1}{n}\sum_{i\in\mathcal I}
			\|\bar{\bX}_{1i}\|_2^2\varepsilon_{2i}^2+
			C\|\bgamma_3^{(0)}\|_2^2\|\bbeta^{(0)}-\bbeta^*\|_2^2
			\dfrac{1}{n}\sum_{i\in\mathcal I}
			\|\bar{\bX}_{1i}\|_2^4
			\nonumber\\
			&\qquad+
			C\|\bgamma^{(0)}-\bgamma^*\|_2^2
			\dfrac{1}{n}\sum_{i\in\mathcal I}
			\|\bar{\bX}_{1i}\|_2^2
			\|\bar{\bX}_{2i}\|_2^2
			=O_p(n^{-1}).
			\label{eq:vhat_v_rate}
		\end{align}
		Since $\bc^{(0)\top}\hat{\bv}_i
		-\bc^{*\top}\bv_i=
		(\bc^{(0)}-\bc^*)^\top\bv_i
		+
		\bc^{(0)\top}(\hat{\bv}_i-\bv_i)$, 
		and therefore, by \eqref{eq:c_root_n},
		\eqref{eq:vhat_v_rate}, and \eqref{eq:empirical_moments},
		\begin{align}
			\dfrac{1}{n}\sum_{i\in\mathcal I}
			\|\bar{\bX}_{1i}\|_2^2(\bc^{(0)\top}\hat{\bv}_i
			-\bc^{*\top}\bv_i)^2
			&\le
			2\|\bc^{(0)}-\bc^*\|_2^2
			\dfrac{1}{n}\sum_{i\in\mathcal I}
			\|\bar{\bX}_{1i}\|_2^2\|\bv_i\|_2^2
			\nonumber\\
			&\quad+
			2\|\bc^{(0)}\|_2^2
			\dfrac{1}{n}\sum_{i\in\mathcal I}
			\|\bar{\bX}_{1i}\|_2^2
			\|\hat{\bv}_i-\bv_i\|_2^2
			=O_p(n^{-1}).
			\label{eq:D_rate}
		\end{align}
		We decompose
		\begin{align*}
			\widehat{\bV}_{\bv}-\bV_{\bv}
			&=
			\bigg\{
			\dfrac{1}{n}\sum_{i\in\mathcal I}
			\bar{\bX}_{1i}\bar{\bX}_{1i}^{\top}
			(\bc^{*\top}\bv_i)^2-\bV_{\bv}
			\bigg\}\\
			&\quad+
			\dfrac{1}{n}\sum_{i\in\mathcal I}
			\bar{\bX}_{1i}\bar{\bX}_{1i}^{\top}(\bc^{(0)\top}\hat{\bv}_i
			-\bc^{*\top}\bv_i)^2
			+
			\dfrac{2}{n}\sum_{i\in\mathcal I}
			\bar{\bX}_{1i}\bar{\bX}_{1i}^{\top}
			\bc^{*\top}\bv_i(\bc^{(0)\top}\hat{\bv}_i
			-\bc^{*\top}\bv_i)\\
			&=:H_2+H_3+H_4.
		\end{align*}
		By \eqref{eq:v_fourth} and
		$E(\|\bar{\bX}_{1i}\|_2^4)=O(1)$,
		\begin{align*}
			E\{
			\left\|
			\bar{\bX}_{1i}\bar{\bX}_{1i}^{\top}
			(\bc^{*\top}\bv_i)^2
			\right\|_2^2
			\}
			&\le
			\|\bc^*\|_2^4
			E\{
			\|\bar{\bX}_{1i}\|_2^4
			E(\|\bv_i\|_2^4\mid\bar{\bX}_{1i})
			\}
			=O(1).
		\end{align*}
		Since the dimension is fixed, Chebyshev's inequality applied
		entrywise gives $\|H_2\|_2=O_p(n^{-1/2})$.
		By \eqref{eq:D_rate},
		\[
		\|H_3\|_2
		\le
		\dfrac{1}{n}\sum_{i\in\mathcal I}
		\|\bar{\bX}_{1i}\|_2^2(\bc^{(0)\top}\hat{\bv}_i
		-\bc^{*\top}\bv_i)^2
		=O_p(n^{-1}).
		\]
		Furthermore, by Cauchy--Schwarz,
		\begin{align*}
			\|H_4\|_2
			&\le
			2\left\{
			\dfrac{1}{n}\sum_{i\in\mathcal I}
			\|\bar{\bX}_{1i}\|_2^2
			(\bc^{*\top}\bv_i)^2
			\right\}^{1/2}
			\left\{
			\dfrac{1}{n}\sum_{i\in\mathcal I}
			\|\bar{\bX}_{1i}\|_2^2(\bc^{(0)\top}\hat{\bv}_i
			-\bc^{*\top}\bv_i)^2
			\right\}^{1/2}=O_p(n^{-1/2}).
		\end{align*}
		Consequently,
		\begin{align}
			\|\widehat{\bV}_{\bv}-\bV_{\bv}\|_2
			=O_p(n^{-1/2}).
			\label{eq:Vv_rate}
		\end{align}
		We next consider
		$\widehat{\bV}_{\bv\varepsilon_2}$,
		\begin{align}
			\dfrac{1}{n}\sum_{i\in\mathcal I}
			\|\bar{\bX}_{1i}\|_2^2(\hat{\varepsilon}_{2i}-\varepsilon_{2i})^2
			&\le
			\|\bbeta^{(0)}-\bbeta^*\|_2^2
			\dfrac{1}{n}\sum_{i\in\mathcal I}
			\|\bar{\bX}_{1i}\|_2^4
			=O_p(n^{-1}).
			\label{eq:R_rate}
		\end{align}
		We have
		\begin{align*}
			\widehat{\bV}_{\bv\varepsilon_2}
			-\bV_{\bv\varepsilon_2}
			&=
			\bigg(
			\dfrac{1}{n}\sum_{i\in\mathcal I}
			\bar{\bX}_{1i}\bar{\bX}_{1i}^{\top}
			\bc^{*\top}\bv_i\varepsilon_{2i}
			-\bV_{\bv\varepsilon_2}
			\bigg)\\
			&\quad+
			\dfrac{1}{n}\sum_{i\in\mathcal I}
			\bar{\bX}_{1i}\bar{\bX}_{1i}^{\top}
			(\bc^{(0)\top}\hat{\bv}_i
			-\bc^{*\top}\bv_i)\varepsilon_{2i}\\
			&\quad+
			\dfrac{1}{n}\sum_{i\in\mathcal I}
			\bar{\bX}_{1i}\bar{\bX}_{1i}^{\top}
			\bc^{*\top}\bv_i(\hat{\varepsilon}_{2i}-\varepsilon_{2i})\\
			&\quad+
			\dfrac{1}{n}\sum_{i\in\mathcal I}
			\bar{\bX}_{1i}\bar{\bX}_{1i}^{\top}
			(\bc^{(0)\top}\hat{\bv}_i
			-\bc^{*\top}\bv_i)(\hat{\varepsilon}_{2i}-\varepsilon_{2i})\\
			&=:H_5+H_6+H_7+H_8.
		\end{align*}
		By conditional Cauchy--Schwarz and the fourth-moment
		conditions,
		\begin{align*}
			E\left(
			\|\bv_i\|_2^2\varepsilon_{2i}^2
			\mid\bar{\bX}_{1i}
			\right)
			&\le
			\left\{
			E(\|\bv_i\|_2^4\mid\bar{\bX}_{1i})
			E(\varepsilon_{2i}^4\mid\bar{\bX}_{1i})
			\right\}^{1/2}
			=O(1).
		\end{align*}
		Hence
		\begin{align*}
			E\left(
			\left\|
			\bar{\bX}_{1i}\bar{\bX}_{1i}^{\top}
			\bc^{*\top}\bv_i\varepsilon_{2i}
			\right\|_2^2
			\right)
			&\le
			\|\bc^*\|_2^2
			E\left\{
			\|\bar{\bX}_{1i}\|_2^4
			E(\|\bv_i\|_2^2\varepsilon_{2i}^2
			\mid\bar{\bX}_{1i})
			\right\}
			=O(1),
		\end{align*}
		and therefore, by Chebyshev's inequality, $\|H_5\|_2=O_p(n^{-1/2})$. For $H_6$, Cauchy--Schwarz, \eqref{eq:D_rate}, and
		\eqref{eq:empirical_moments} yield
		\begin{align*}
			\|H_6\|_2
			&\le
			\left\{
			\dfrac{1}{n}\sum_{i\in\mathcal I}
			\|\bar{\bX}_{1i}\|_2^2(\bc^{(0)\top}\hat{\bv}_i
			-\bc^{*\top}\bv_i)^2
			\right\}^{1/2}
			\left(
			\dfrac{1}{n}\sum_{i\in\mathcal I}
			\|\bar{\bX}_{1i}\|_2^2\varepsilon_{2i}^2
			\right)^{1/2}=O_p(n^{-1/2}).
		\end{align*}
		Similarly,
		\begin{align*}
			\|H_7\|_2
			&\le
			\left\{
			\dfrac{1}{n}\sum_{i\in\mathcal I}
			\|\bar{\bX}_{1i}\|_2^2
			(\bc^{*\top}\bv_i)^2
			\right\}^{1/2}
			\left\{
			\dfrac{1}{n}\sum_{i\in\mathcal I}
			\|\bar{\bX}_{1i}\|_2^2(\hat{\varepsilon}_{2i}-\varepsilon_{2i})^2
			\right\}^{1/2}=O_p(n^{-1/2}),
		\end{align*}
		whereas
		\begin{align*}
			\|H_8\|_2
			&\le
			\left\{
			\dfrac{1}{n}\sum_{i\in\mathcal I}
			\|\bar{\bX}_{1i}\|_2^2(\bc^{(0)\top}\hat{\bv}_i
			-\bc^{*\top}\bv_i)^2
			\right\}^{1/2}
			\left\{
			\dfrac{1}{n}\sum_{i\in\mathcal I}
			\|\bar{\bX}_{1i}\|_2^2(\hat{\varepsilon}_{2i}-\varepsilon_{2i})^2
			\right\}^{1/2}=O_p(n^{-1}).
		\end{align*}
		Combining the bounds for $H_5,\ldots,H_8$, we obtain
		\begin{align}
			\|\widehat{\bV}_{\bv\varepsilon_2}
			-\bV_{\bv\varepsilon_2}\|_2
			=O_p(n^{-1/2}).
			\label{eq:Vveps_rate}
		\end{align}
		Finally, since
		$
		E\left(
		\|\bar{\bX}_{1i}\bar{\bX}_{1i}^{\top}\|_2^2
		\right)
		=
		E(\|\bar{\bX}_{1i}\|_2^4)
		=O(1)$, Chebyshev's inequality and the fixed dimensionality imply
		\begin{align}
			\|\widehat{\bSigma}_{\bX,1}
			-\bSigma_{\bX,1}\|_2
			=O_p(n^{-1/2}).
			\label{eq:SigmaX1_rate}
		\end{align}
	Therefore, by standard matrix norm inequalities,
	\[
	\hat L_1-L_1=O_p(n^{-1/2}),
	\qquad
	\hat L_2-L_2=O_p(n^{-1/2}).
	\]
	Since $L_2>0$ and the underlying distribution is fixed, $L_2$ is bounded away from zero. Hence,
	\[
	\frac{\hat L_1}{\hat L_2}-\frac{L_1}{L_2}=\dfrac{\widehat L_1}{\widehat L_2}
	-\lambda^*
	=O_p(n^{-1/2}).
	\]
	Since $\widehat L_2\xrightarrow{\rm p}L_2>0$ and $a_n\to0$,
	\begin{align*}
	P(\widehat L_2>a_n)\to1.
	\end{align*}
	It follows from the definition of $\hat\lambda$ that
	\begin{align*}
		P(
		\hat\lambda=\widehat L_1/\widehat L_2
		)\to1,
	\end{align*}
	and therefore
	\begin{align}
		\hat\lambda-\lambda^*
		=
		O_p(n^{-1/2}).
		\label{eq:lambda_rate_nonzero}
	\end{align}
	
	Since $\hat\lambda=O_p(1)$ by \eqref{eq:lambda_rate_nonzero}, the expansion \eqref{eq:beta_affine_expansion} applies with $\lambda=\hat\lambda$ and gives
	\begin{align*}
		\widehat{\bbeta}_{\hat\lambda}-\bbeta^*
		&=
		\bSigma_{\bX,1}^{-1}
		\Big[
		\dfrac{1}{n}\sum_{i\in\mathcal{I}}
		\bar{\bX}_{1i}
		\{\varepsilon_{2i}
		-\lambda^*(m/N)\bc^{*\top}\bv_i\}
		+\dfrac{\lambda^*}{N}
		\sum_{i\in\mathcal{J}}
		\bar{\bX}_{1i}\bc^{*\top}\bv_i
		\Big]
		+H_1+o_p(n^{-1/2}),
	\end{align*}
	where
	\begin{align*}
		H_1
		&=
		\dfrac{(\lambda^*-\hat\lambda)m}{nN}
		\bSigma_{\bX,1}^{-1}
		\sum_{i\in\mathcal{I}}
		\bar{\bX}_{1i}\bc^{*\top}\bv_i
		+\dfrac{\hat\lambda-\lambda^*}{N}
		\bSigma_{\bX,1}^{-1}
		\sum_{i\in\mathcal{J}}
		\bar{\bX}_{1i}\bc^{*\top}\bv_i.
	\end{align*}
	We have
	$
	E(\bar{\bX}_{1i}\bc^{*\top}\bv_i)=0$. Moreover, by \eqref{eq:c_root_n} and \eqref{eq:v_fourth},
	\begin{align*}
		E\left\{
		\|\bar{\bX}_{1i}\bc^{*\top}\bv_i\|_2^2
		\right\}
		&\le
		\|\bc^*\|_2^2
		E\left[
		\|\bar{\bX}_{1i}\|_2^2
		E(\|\bv_i\|_2^2\mid\bar{\bX}_{1i})
		\right]\le
		C E(\|\bar{\bX}_{1i}\|_2^2)
		=O(1).
	\end{align*}
	Therefore, by Chebyshev's inequality and the fixed dimensionality,
	\[
	\dfrac{1}{n}\sum_{i\in\mathcal I}
	\bar{\bX}_{1i}\bc^{*\top}\bv_i
	=O_p(n^{-1/2}),
	\qquad
	\dfrac{1}{m}\sum_{i\in\mathcal J}
	\bar{\bX}_{1i}\bc^{*\top}\bv_i
	=O_p(m^{-1/2}).
	\]
	It follows that
	\begin{align*}
		\dfrac{m}{nN}
		\sum_{i\in\mathcal I}
		\bar{\bX}_{1i}\bc^{*\top}\bv_i
		&=
		\dfrac{m}{N}
		O_p(n^{-1/2})
		=
		O_p(n^{-1/2}),\\
		\dfrac{1}{N}
		\sum_{i\in\mathcal J}
		\bar{\bX}_{1i}\bc^{*\top}\bv_i
		&=
		\dfrac{m}{N}
		O_p(m^{-1/2})
		=
		O_p(\sqrt m/N).
	\end{align*}
	Together with \eqref{eq:lambda_rate_nonzero}, this gives $H_1=o_p(n^{-1/2})$. Therefore,
	\begin{align*}
		\sqrt n
		(\widehat{\bbeta}_{\hat\lambda}-\bbeta^*)
		\xrightarrow{\rm d}
		\mathcal N(0,\bV_{\lambda^*}),
	\end{align*}
	where
	\begin{align*}
		\bV_{\lambda^*}
		=
		\bSigma_{\bX,1}^{-1}
		\left(
		\bV_{\varepsilon_2}
		+\tau\lambda^{*2}\bV_{\bv}
		-2\tau\lambda^*\bV_{\bv\varepsilon_2}
		\right)
		\bSigma_{\bX,1}^{-1}.
	\end{align*}
	In particular,
	\begin{align*}
		{\rm tr}(\bV_{\lambda^*})
		&=
		{\rm tr}
		(\bSigma_{\bX,1}^{-1}
		\bV_{\varepsilon_2}
		\bSigma_{\bX,1}^{-1})
		-\tau
		\dfrac{
			\{
			{\rm tr}
			(\bSigma_{\bX,1}^{-1}
			\bV_{\bv\varepsilon_2}
			\bSigma_{\bX,1}^{-1})
			\}^2
		}{
			{\rm tr}
			(\bSigma_{\bX,1}^{-1}
			\bV_{\bv}
			\bSigma_{\bX,1}^{-1})
		}.
	\end{align*}
	
	\medskip
	\noindent
	\textbf{Case (b) $\|\bgamma_3^*\|_2=0$.}
	In this case, $\bgamma_3^*=0$ and $\bc^*=0$. Since
	$\bc^{(0)}=O_p(n^{-1/2})$ by \eqref{eq:c_root_n}, and $n^{-1}\sum_{i\in\mathcal I}\|\bar{\bX}_{1i}\|_2^2\|\hat\bv_i\|_2^2=O_p(1)$, we have $\widehat L_2=O_p(n^{-1})$. Hence $na_n\to\infty$ implies
	\begin{align}
		P(\widehat L_2\le a_n)\to1.
		\label{eq:lambda_zero_event}
	\end{align}
	which implies $P(\hat\lambda=0)\to1$. By \eqref{eq:weighted_est}, $\widehat{\bbeta}_{0}$ is exactly the supervised estimator. Hence, with probability tending to one,
	\[
	\widehat{\bbeta}_{\hat\lambda}=	\widehat{\bbeta}_{0}
	=
	\bbeta^{(0)}.
	\]
	The standard least-squares expansion gives
	\begin{align*}
		\bbeta^{(0)}-\bbeta^*
		=
		\bSigma_{\bX,1}^{-1}
		\dfrac{1}{n}
		\sum_{i\in\mathcal{I}}
		\bar{\bX}_{1i}\varepsilon_{2i}
		+o_p(n^{-1/2}),
	\end{align*}
	and therefore
	\begin{align*}
		\sqrt n
		(\widehat{\bbeta}_{\hat\lambda}-\bbeta^*)
		\xrightarrow{\rm d}
		\mathcal N(0,\bV_{\rm sup}),
	\end{align*}
	where $\bV_{\rm sup}
		=
		\bSigma_{\bX,1}^{-1}
		\bV_{\varepsilon_2}
		\bSigma_{\bX,1}^{-1}$. Combining the two cases proves that
	\begin{align*}
		\sqrt n
		(\widehat{\bbeta}_{\hat\lambda}-\bbeta^*)
		\xrightarrow{\rm d}
		\mathcal N(0,\bV_{\rm opt}),
	\end{align*}
	where
	\begin{align*}
		\bV_{\rm opt}
		=
		\begin{cases}
			\bV_{\lambda^*},
			& \|\bgamma_3^*\|_2>0,\\[1mm]
			\bV_{\rm sup},
			& \|\bgamma_3^*\|_2=0.
		\end{cases}
	\end{align*}

We now consider the remaining regime $m=o(n)$. By \eqref{eq:weighted_est}, the update \eqref{eq:bg_update}, and the labeled OLS normal equations,
for any $\lambda\in\mathbb R$,
\[
\widehat{\bbeta}_\lambda-\bbeta^{(0)}
=\lambda(\widehat{\bbeta}-\bbeta^{(0)})
=\lambda\left[
\frac{1}{N}
\sum_{i\in\mathcal I\cup\mathcal J}
\bar{\bX}_{1i}\bar{\bX}_{1i}^{\top}
\right]^{-1}\times
\frac{1}{N}
\sum_{i\in\mathcal J}
\bar{\bX}_{1i}
\left(
\mu_i^{\bOmega^{(0)}}-
\bar{\bX}_{1i}^{\top}\bbeta^{(0)}
\right)
\]
At the population parameters,
\[
\mu_i^{\bOmega^*}
-\bar{\bX}_{1i}^{\top}\bbeta^*
=\bc^{*\top}\bv_i,
\qquad
E(\bar{\bX}_{1i}\bc^{*\top}\bv_i)=0.
\]
By Lemma~\ref{lemma:lemma5} and the moment conditions
of Supplementary Theorem~\ref{thm:optimal_lambda},
\[
\frac1N\sum_{i\in\mathcal J}
\bar{\bX}_{1i}\bc^{*\top}\bv_i
=
O_p\left(\frac{\sqrt m}{N}\right).
\]
Moreover, the root-$n$ consistency of the labeled initial estimates and the explicit imputation formula imply
\[
\frac1N\sum_{i\in\mathcal J}
\bar{\bX}_{1i}
\Big[
\{\mu_i^{\bOmega^{(0)}}-
\bar{\bX}_{1i}^{\top}\bbeta^{(0)}\}-
\{\mu_i^{\bOmega^*}-
\bar{\bX}_{1i}^{\top}\bbeta^*\}
\Big]
=
O_p\left(\frac{m}{N\sqrt n}\right)
\]
The matrix in the inverse above does not depend on $\lambda$ and converges to $\bSigma_{\bX,1}$ in probability, so its inverse is $O_p(1)$. Consequently,
\[
\sup_{|\lambda|\leq M}
\|\hat{\bbeta}_\lambda-\bbeta^{(0)}\|
=
O_p\left(
\frac{\sqrt m}{N}
+\frac{m}{N\sqrt n}
\right)
\]
for every fixed $M<\infty$. The consistency and threshold arguments for the feasible weight established above
imply $\hat\lambda=O_p(1)$.
Restricting $\hat\lambda$ to a sufficiently
large compact interval with arbitrarily
high probability therefore gives
\[
\begin{aligned}
\hat{\bbeta}_{\hat\lambda}
-\bbeta^{(0)}
&=
O_p\left(
\frac{\sqrt m}{N}
+\frac{m}{N\sqrt n}
\right)\\
&=o_p(n^{-1/2}).
\end{aligned}
\]
It follows from the labeled OLS
central limit theorem and Slutsky's theorem
that
\[
\sqrt n
(\hat{\bbeta}_{\hat\lambda}-\bbeta^*)
\xrightarrow{\rm d}
\mathcal N(\mathbf 0,\bV_{\mathrm{sup}}).
\]
For $m=0$, the SAEM estimator $\widehat{\bbeta}$ coincides with $\bbeta^{(0)}$, so that
$\hat{\bbeta}_{\hat\lambda}=\bbeta^{(0)}$
exactly. This completes the auxiliary full-vector argument for $\tau=0$.

\medskip
\noindent
\textbf{Estimation of $\bbeta_1^*$.} For any fixed $\lambda\in\mathbb R$, the fixed-$\lambda$ expansion
established above implies
\begin{align*}
	\sqrt n
	(\widehat\bbeta_{\lambda,1}-\bbeta_1^*)
	\xrightarrow{\rm d}
	 N\Big(
	 \mathbf0,
[\bI_p,\mathbf0_{p\times d_1}]
\bV_\lambda
[\bI_p,\mathbf0_{p\times d_1}]^\top
	\Big).
\end{align*}
When $\tau>0$ and $\bgamma_3^*\neq\mathbf0$, the unique trace minimizer is
$\lambda_2^*$ in \eqref{eq:lambda2_star}. When $\tau=0$, the limiting covariance is independent of $\lambda$. Substitution gives the corresponding optimal trace stated in the theorem.

We next consider the feasible weight. Under Assumption~\ref{ass:non_gaussian}, $E(\bv_i\bv_i^\top\mid\bar\bX_{1i})$ is positive definite almost surely. Since $[\bI_p,\mathbf0_{p\times d_1}]\bSigma_{\bX,1}^{-1}$ has full row rank and $E(\bar\bX_{1i}\bar\bX_{1i}^\top)$ is positive definite, the population denominator in~\eqref{eq:lambda2_star} is positive when $\bgamma_3^*\neq\mathbf0$. The sample denominator satisfies $\widehat D_{\bbeta_1}\xrightarrow{\rm p}D_{\bbeta_1}>0$, where $D_{\bbeta_1}$ denotes this population denominator. Therefore, $P(\widehat D_{\bbeta_1}>a_n)\to1$, and the moment rates established for $\widehat L_1,\widehat L_2$ yield
\begin{align}
	\hat\lambda_2-\lambda_2^*=O_p(n^{-1/2}).
	\label{eq:lambda2_rate}
\end{align}
Using the same stochastic expansion as in the proof for $\bbeta^*$,
\begin{align*}
	\widehat\bbeta_{\hat\lambda_2}
	-\widehat\bbeta_{\lambda_2^*}
	&=
	(\hat\lambda_2-\lambda_2^*)
	\bSigma_{\bX,1}^{-1}
	\left\{
	-\frac{m}{nN}
	\sum_{i\in\mathcal I}
	\bar{\bX}_{1i}\bc^{*\top}\bv_i
	+
	\frac{1}{N}
	\sum_{i\in\mathcal J}
	\bar{\bX}_{1i}\bc^{*\top}\bv_i
	\right\}
	+o_p(n^{-1/2}).
\end{align*}
The term in braces is $O_p(n^{-1/2})$, as established in the proof
for $\bbeta^*$. Together with \eqref{eq:lambda2_rate}, this gives
$
\widehat\bbeta_{\hat\lambda_2}
-\widehat\bbeta_{\lambda_2^*}
=
o_p(n^{-1/2})$,
and therefore $
\widehat\bbeta_{\hat\lambda_2,1}
-\widehat\bbeta_{\lambda_2^*,1}
=
o_p(n^{-1/2})$.
The asserted asymptotic normality follows from the fixed-$\lambda$
central limit theorem and Slutsky's theorem.

Finally, suppose $\|\bgamma_3^*\|_2=0$. As shown in the proof for $\bbeta^*$, $\widehat L_2=O_p(n^{-1})$. Since $0\le\widehat D_{\bbeta_1}\le\widehat L_2$, we also have $\widehat D_{\bbeta_1}=O_p(n^{-1})$. Thus $na_n\to\infty$ gives $P(\widehat D_{\bbeta_1}\le a_n)\to1$ and $P(\hat\lambda_2=0)\to1$. Since $\lambda_2^*=0$ by definition, $\sqrt n
(\widehat\bbeta_{\hat\lambda_2,1}
-\widehat\bbeta_{\lambda_2^*,1})
\xrightarrow{\rm p} 0$. This proves Part~(2).

\medskip
\noindent
\textbf{Estimation of $\btheta^*$.} Recall that
$\btheta^*=\bbeta_1^*+\balpha^*\bbeta_2^*$ and
$\widehat\btheta_\lambda
=\widehat\btheta_{\rm sup}+\lambda(\widehat\btheta-\widehat\btheta_{\rm sup})$, so that $\widehat\btheta_0=\widehat\btheta_{\rm sup}$ and $\widehat\btheta_1=\widehat\btheta$.
Let $\widehat\balpha_0=\widehat\balpha_L$ and $\widehat\balpha_1=\widehat\balpha$ denote the labeled-only and full-sample OLS estimators of $\balpha^*$. For $\lambda\in\{0,1\}$, with the observation weights $w_{i,\lambda}=(1-\lambda)/n+\lambda/N$ for $i\in\mathcal I$ and $w_{i,\lambda}=\lambda/N$ for $i\in\mathcal J$, the normal equations for $\widehat\bbeta_\lambda$ and $\widehat\balpha_\lambda$ imply the exact identity
\[
\widehat\btheta_\lambda
=\widehat\bbeta_{\lambda,1}+\widehat\balpha_\lambda\widehat\bbeta_{\lambda,2}
=\left(\sum_{i\in\mathcal I\cup\mathcal J}w_{i,\lambda}\bX_i\bX_i^\top\right)^{-1}
\left(\sum_{i\in\mathcal I}w_{i,\lambda}\bX_iY_i+
\sum_{i\in\mathcal J}w_{i,\lambda}\bX_i\mu_i^{\bOmega^{(0)}}\right).
\]
A first-order expansion gives, for $\lambda\in\{0,1\}$,
\begin{align}
\widehat\btheta_\lambda-\btheta^*
&=[\bI_p,\balpha^*](\widehat\bbeta_\lambda-\bbeta^*)
+(\widehat\balpha_\lambda-\balpha^*)\bbeta_2^*+o_p(n^{-1/2}).
\label{eq:theta_linearization}
\end{align}
Since $\bS_{1i}=\balpha^{*\top}\bX_i+\bvarepsilon_{1i}$, the normal equations for $\widehat\balpha_\lambda$ give, for $\lambda\in\{0,1\}$,
\[
\widehat\balpha_\lambda-\balpha^*
=\bSigma_{\bX}^{-1}\left\{\frac1n\sum_{i\in\mathcal I}\bX_i\bvarepsilon_{1i}^\top
+\lambda\left(\frac1N\sum_{i\in\mathcal I\cup\mathcal J}\bX_i\bvarepsilon_{1i}^\top
-\frac1n\sum_{i\in\mathcal I}\bX_i\bvarepsilon_{1i}^\top\right)\right\}
+o_p(n^{-1/2}).
\]
The block identity
$[\bI_p,\balpha^*]\bSigma_{\bX,1}^{-1}=[\bSigma_{\bX}^{-1},\mathbf0_{p\times d_1}]$,
together with \eqref{eq:beta_affine_expansion}, yields, for $\lambda\in\{0,1\}$,
\begin{align}
\widehat\btheta_\lambda-\btheta^*
&=\bSigma_{\bX}^{-1}\left[\frac1n\sum_{i\in\mathcal I}\bX_i e_i
+\lambda\left\{\frac1N\sum_{i\in\mathcal I\cup\mathcal J}\bX_iq_i
-\frac1n\sum_{i\in\mathcal I}\bX_iq_i\right\}\right]
+o_p(n^{-1/2}),
\label{eq:theta_fixed_lambda}
\end{align}
where $u_i=\bvarepsilon_{1i}^\top\bbeta_2^*$, $e_i=\varepsilon_{2i}+u_i$, and $q_i=u_i+\bc^{*\top}\bv_i$. Since $\widehat\btheta_\lambda-\btheta^*=(1-\lambda)(\widehat\btheta_0-\btheta^*)+\lambda(\widehat\btheta_1-\btheta^*)$ by \eqref{eq:weighted_est} and the leading term of \eqref{eq:theta_fixed_lambda} is affine in $\lambda$, \eqref{eq:theta_fixed_lambda} holds for every $\lambda\in\mathbb R$, and also when $\lambda$ is replaced by a random weight that is $O_p(1)$. The supervised influence function is $\bSigma_{\bX}^{-1}\bX_ie_i$; therefore,
\begin{align}
\bV_{\btheta,0}=\bSigma_{\bX}^{-1}E(\bX_i\bX_i^\top e_i^2)\bSigma_{\bX}^{-1}.
\label{eq:theta_supervised_comparison}
\end{align}
Similarly, $\bV_{\bbeta_1,0}$ is the limiting covariance matrix of the supervised estimator of $\bbeta_1^*$. Writing $\bV_q=E(\bX_i\bX_i^\top q_i^2)$ and $\bV_{eq}=E(\bX_i\bX_i^\top e_iq_i)$, the independence of subjects and $m/N\to\tau$ give
\[\bV_{\btheta,\lambda}
=\bV_{\btheta,0}
+\tau\lambda^2\bSigma_{\bX}^{-1}\bV_q\bSigma_{\bX}^{-1}
-2\tau\lambda\bSigma_{\bX}^{-1}\bV_{eq}\bSigma_{\bX}^{-1}.
\]
Set $C_{\btheta}=\operatorname{tr}(\bSigma_{\bX}^{-1}\bV_{eq}\bSigma_{\bX}^{-1})$ and $D_{\btheta}=\operatorname{tr}(\bSigma_{\bX}^{-1}\bV_q\bSigma_{\bX}^{-1})$. Taking traces yields
\begin{align}
\operatorname{tr}(\bV_{\btheta,\lambda})
&=\operatorname{tr}(\bV_{\btheta,0})
+\tau\lambda^2D_{\btheta}-2\tau\lambda C_{\btheta}.
\label{eq:theta_trace_lambda}
\end{align}
For $\tau>0$ and $D_{\btheta}>0$, the unique minimizer is $\lambda_3^*=C_{\btheta}/D_{\btheta}$, and
\begin{align}
\operatorname{tr}(\bV_{\btheta,0})
-\operatorname{tr}(\bV_{\btheta,\lambda_3^*})
&=\tau\frac{C_{\btheta}^2}{D_{\btheta}}\ge0.
\label{eq:theta_supervised_trace_gain}
\end{align}
If $D_{\btheta}=0$, then $E\|\bSigma_{\bX}^{-1}\bX_iq_i\|_2^2=0$, so the covariance is independent of $\lambda$ and $\lambda_3^*=0$ is a valid convention. When $\tau=0$, the limiting covariance is also independent of $\lambda$.
For the feasible weight, the labeled OLS residuals and sample moments defined in Part~(3) of Supplementary Theorem~\ref{thm:optimal_lambda} satisfy $\widehat D_{\btheta}\xrightarrow{\rm p}D_{\btheta}$ and, when $D_{\btheta}>0$,
\begin{align}
\hat\lambda_3-\lambda_3^*=o_p(1).
\label{eq:lambda3_rate}
\end{align}
Indeed, the fourth-moment assumptions and consistency of the labeled OLS estimates imply convergence of $\widehat\bV_q$, $\widehat\bV_{eq}$, and $\widehat\bSigma_{\bX,L}$ to their population counterparts. By \eqref{eq:weighted_est}, $\widehat\btheta_{\hat\lambda_3}-\widehat\btheta_{\lambda_3^*}=(\hat\lambda_3-\lambda_3^*)(\widehat\btheta-\widehat\btheta_{\rm sup})$, and $\widehat\btheta-\widehat\btheta_{\rm sup}=O_p(n^{-1/2})$ by \eqref{eq:theta_fixed_lambda}. Consequently, $\widehat\btheta_{\hat\lambda_3}-\widehat\btheta_{\lambda_3^*}=o_p(n^{-1/2})$. Hence the feasible and oracle estimators have the same limiting distribution.

If $D_{\btheta}=0$, then $\bX_iq_i=0$ almost surely. The root-$n$ rates of the labeled OLS estimates, together with \eqref{eq:c_root_n}, the definition of $\hat q_i$, and the empirical moment bounds in \eqref{eq:empirical_moments}, yield
\[\frac1n\sum_{i\in\mathcal I}\|\bX_i\|_2^2(\hat q_i-q_i)^2=O_p(n^{-1}).\]
Indeed, writing $\widehat\balpha_L-\balpha^*$ for the labeled OLS error,
\[
\hat q_i-q_i={}\bvarepsilon_{1i}^\top(\bbeta_2^{(0)}-\bbeta_2^*)
-\bX_i^\top(\widehat\balpha_L-\balpha^*)\bbeta_2^{(0)}+(\bc^{(0)}-\bc^*)^\top\bv_i
+\bc^{(0)\top}(\hat\bv_i-\bv_i).
\]
Together with the expression for $\hat\bv_i-\bv_i$ given above, the root-$n$ rates and the weighted empirical moment bounds in \eqref{eq:empirical_moments} establish the claimed $O_p(n^{-1})$ rate. Since $\|\widehat\bSigma_{\bX,L}^{-1}\|_2=O_p(1)$,
\[
\widehat D_{\btheta}
=\frac1n\sum_{i\in\mathcal I}
\hat q_i^2\|\widehat\bSigma_{\bX,L}^{-1}\bX_i\|_2^2
\le
\|\widehat\bSigma_{\bX,L}^{-1}\|_2^2
\frac1n\sum_{i\in\mathcal I}\|\bX_i\|_2^2(\hat q_i-q_i)^2
=O_p(n^{-1}).
\]
Consequently, $na_n\to\infty$ implies $P(\widehat D_{\btheta}\le a_n)\to1$ and hence $P(\hat\lambda_3=0)\to1$; the feasible and oracle estimators coincide with probability tending to one. Note that $\bgamma_3^*=0$ does not necessarily imply $D_{\btheta}=0$: in that case $q_i=u_i$, and $\lambda_3^*$ may be nonzero.

Finally, consider $\tau=0$, equivalently $m=o(n)$. By \eqref{eq:weighted_est}, the identity for $\widehat\btheta_1$ above, and the labeled OLS normal equations, for any $\lambda\in\mathbb R$,
\[
\widehat\btheta_\lambda-\widehat\btheta_{\rm sup}
=\lambda(\widehat\btheta-\widehat\btheta_{\rm sup})
=\lambda\widehat\bSigma_{\bX}^{-1}\frac{1}{N}
\sum_{i\in\mathcal J}\bX_i
\big(\mu_i^{\bOmega^{(0)}}-\bX_i^\top\widehat\btheta_{\rm sup}\big),
\]
where $\widehat\bSigma_{\bX}=N^{-1}\sum_{i\in\mathcal I\cup\mathcal J}\bX_i\bX_i^\top$. At the population parameters,
$\mu_i^{\bOmega^*}-\bX_i^\top\btheta^*=q_i$ and $E(\bX_iq_i)=0$.
The finite second moments and independence of subjects therefore give
\[
\frac1N\sum_{i\in\mathcal J}\bX_iq_i
=O_p\left(\frac{\sqrt m}{N}\right).
\]
The root-$n$ consistency of $\bOmega^{(0)}$ and $\widehat\btheta_{\rm sup}$, together with the smooth imputation formula and the stated moment conditions, imply
\[
\frac1N\sum_{i\in\mathcal J}\bX_i
\Big\{(\mu_i^{\bOmega^{(0)}}-\mu_i^{\bOmega^*})
-\bX_i^\top(\widehat\btheta_{\rm sup}-\btheta^*)\Big\}
=O_p\left(\frac{m}{N\sqrt n}\right).
\]
Since $\widehat\bSigma_{\bX}\xrightarrow{\rm p}\bSigma_{\bX}$, its inverse is $O_p(1)$. Consequently, for every fixed $M<\infty$,
\[
\sup_{|\lambda|\le M}
\|\widehat\btheta_\lambda-\widehat\btheta_{\rm sup}\|_2
=O_p\left(\frac{\sqrt m}{N}+\frac{m}{N\sqrt n}\right)
=o_p(n^{-1/2}).
\]
The positive- and zero-denominator arguments above show that $\hat\lambda_3=O_p(1)$. Restricting it to a sufficiently large compact interval with arbitrarily high probability yields
$\widehat\btheta_{\hat\lambda_3}-\widehat\btheta_{\rm sup}=o_p(n^{-1/2})$.
When $m=0$, the estimators coincide exactly. This proves Part~(3) and completes the proof.
 
\end{proof}

\end{document}